\documentclass[fleqn,usenatbib]{mnras}
\usepackage{newtxtext,newtxmath}

\usepackage[T1]{fontenc}
\usepackage{ae,aecompl}
\usepackage{graphicx}	
\usepackage{amsmath}	
\usepackage{mathtools}
\usepackage{cases}
\usepackage{cuted}
\usepackage{bigstrut}
\usepackage{subfig}
\usepackage{paralist}
\usepackage{scrextend}
\usepackage{lscape}
\usepackage{wrapfig}
\usepackage[T1]{fontenc}
\usepackage{aecompl}
\usepackage{adjustbox} 
\usepackage{underscore}
\usepackage{float}
\usepackage{amsfonts}
\usepackage{color}
\usepackage{booktabs}
\usepackage{longtable}
\usepackage[maxfloats=256]{morefloats}
\usepackage[flushleft]{threeparttable}
\usepackage[english]{babel}
\usepackage{lastpage}
\usepackage{stmaryrd}
\usepackage{subfig}
\usepackage{xcolor}
\newcommand{\CI}{\mbox{C\,{\sc i}}}
\newcommand{\CII}{\mbox{C\,{\sc ii}}}
\newcommand{\CIV}{\mbox{C\,{\sc iv}}}
\newcommand{\FeII}{\mbox{Fe\,{\sc ii}}}
\newcommand{\MgII}{\mbox{Mg\,{\sc ii}}}
\newcommand{\NI}{\mbox{N\,{\sc i}}}
\newcommand{\NV}{\mbox{N\,{\sc v}}}
\newcommand{\OI}{\mbox{O\,{\sc i}}}
\newcommand{\OVI}{\mbox{O\,{\sc vi}}}
\newcommand{\PII}{\mbox{P\,{\sc ii}}}
\newcommand{\SI}{\mbox{S\,{\sc i}}}
\newcommand{\SII}{\mbox{S\,{\sc ii}}}
\newcommand{\SiI}{\mbox{Si\,{\sc i}}}
\newcommand{\SiII}{\mbox{Si\,{\sc ii}}}
\newcommand{\SiIII}{\mbox{Si\,{\sc iii}}}
\newcommand{\SiIV}{\mbox{Si\,{\sc iv}}}
\newcommand{\AlII}{\mbox{Al\,{\sc ii}}}

\newcommand{\cavii}{\mbox{Ca\,{\sc vii}}}
\newcommand{\mgx}{\mbox{Mg\,{\sc x}}}
\newcommand{\mnii}{\mbox{Mn\,{\sc ii}}}
\newcommand{\hii}{\mbox{H\,{\sc ii}}}
\newcommand{\neviii}{\mbox{Ne\,{\sc viii}}}
\newcommand{\tiii}{\mbox{Ti\,{\sc ii}}}
\newcommand{\civ}{\mbox{C\,{\sc iv}}}
\newcommand{\ciii}{\mbox{C\,{\sc iii}}}
\newcommand{\cii}{\mbox{C\,{\sc ii}}}
\newcommand{\ci}{\mbox{C\,{\sc i}}}

\newcommand{\Sii}{\mbox{S\,{\sc ii}}}
\newcommand{\nitri}{\mbox{N\,{\sc i}}}
\newcommand{\nii}{\mbox{N\,{\sc ii}}}

\newcommand{\feii}{\mbox{Fe\,{\sc ii}}}

\newcommand{\caii}{\mbox{Ca\,{\sc ii}}}
\newcommand{\siiv}{\mbox{Si\,{\sc iv}}}
\newcommand{\siiii}{\mbox{Si\,{\sc iii}}}
\newcommand{\siii}{\mbox{Si\,{\sc ii}}}

\newcommand{\oi}{\mbox{O\,{\sc i}}}

\newcommand{\oiii}{\mbox{O\,{\sc iii}}}
\newcommand{\oiv}{\mbox{O\,{\sc iv}]}}
\newcommand{\oxyv}{\mbox{O\,{\sc v}]}}
\newcommand{\ovi}{\mbox{O\,{\sc vi}}}
\newcommand{\nv}{\mbox{N\,{\sc v}}}

\newcommand{\CLOUDY}{\textsc{cloudy}}
\newcommand{\mgii}{\ifmmode {\rm Mg}{\textsc{ii}} \else Mg\,{\sc ii}\fi}
\newcommand{\heii}{\ifmmode {\rm He}{\textsc{ii}} \else He\,{\sc ii}\fi}

\newcommand{\angstrom}{\text{\normalfont\AA}}
  
\newcommand{\hi}{\mbox{H\,{\sc i}}}
\newcommand{\mgi}{\ifmmode {\rm Mg}{\textsc{i}} \else Mg\,{\sc i}\fi}

\newcommand{\hst}{\it HST}
\newcommand{\lya}{Ly$\alpha$}

\newcommand{\metallicity}{$\log Z/Z_{\sun}$}
\newcommand{\hden}{$\log n_{\H}$}
\newcommand{\colden}{$\log N(\hi)$}

\newcommand{\temp}{$\log T$}
\newcommand{\rvir}{$R_{vir}$}

\newcommand{\degree}{\ensuremath{^\circ}}
\def\kms{\hbox{km~s$^{-1}$}}
\def\cmsq{\hbox{cm$^{-2}$}}
\def\cc{\hbox{cm$^{-3}$}}
\def\H{\hbox{{\rm H}}}
\def\HI{\hbox{{\rm H~}\kern 0.1em{\sc i}}}
\definecolor{darkgreen}{rgb}{0.0, 0.2, 0.13}

\setcitestyle{notesep={; },round,aysep={},yysep={,}}

\title[Properties of the Leo Ring]{Probing the physicochemical properties of the Leo Ring and the \hbox{Leo I} group}

\author[Sameer et al.]
{Sameer,$^{1}$\thanks{E-mail: sxx15@psu.edu} 
Jane C. Charlton,$^{1}$
Glenn G. Kacprzak,$^{2,3}$
Anand Narayanan,$^{4}$
Sriram Sankar,$^{5}$
\newauthor
Philipp Richter,$^{6}$
Bart P. Wakker,$^{7}$
Nikole M. Nielsen,$^{2,3}$
and
Christopher W. Churchill$^{8}$
\\
$^{1}$Department of Astronomy \& Astrophysics, 525 Davey Lab,
The Pennsylvania State University, University Park, PA 16802, USA\\
$^{2}$Centre for Astrophysics and Supercomputing, Swinburne University of Technology, Hawthorn, Victoria 3122, Australia\\
$^{3}$ ARC Centre of Excellence for All Sky Astrophysics in 3 Dimensions (ASTRO 3D)\\
$^{4}$Department of Earth and Space Sciences, Indian Institute of Space Science \& Technology, Thiruvananthapuram 695547, Kerala, INDIA\\
$^{5}$ South African Astronomical Observatory, Cape Town (SAAO)\\
$^{6}$Institut für Physik und Astronomie, Universität Potsdam, Haus 28, Karl-Liebknecht-Str. 24/25, D-14476, Potsdam, Germany\\
$^{7}$Department of Astronomy, University of Wisconsin-Madison, 475 N. Charter Street, Madison, WI 53706, USA\\
$^{8}$Department of Astronomy, New Mexico State University, Las Cruces, NM 88003, USA
}
\date{Accepted 2022 January 5. Received 2022 January 5; in original form 2021 October 13}

\pubyear{2021}

\begin{document}
\label{firstpage}
\maketitle

\begin{abstract}

\noindent We present an absorption line study of the physical and chemical properties of the Leo {\hi} Ring and the Leo I Group as traced by 11 quasar sightlines spread over a $\approx$600$\times$800 kpc$^{2}$ region. Using {\hst}/COS G130/G160 archival observations as constraints, we couple cloud-by-cloud, multiphase, Bayesian ionization modeling with galaxy property information to determine the plausible origin of the absorbing gas along these sightlines. We search for absorption in the range \hbox{600 {\kms} $< cz <$  1400 {\kms}} consistent with the kinematics of the Leo Ring/Group. We find absorption plausibly associated with the Leo Ring towards five sightlines. Along three other sightlines, we find absorption likely to be associated with individual galaxies, intragroup gas, and/or large scale filamentary structure.  The absorption along these five sightlines is stronger in metal lines than expected from individual galaxies, indicative of multiple contributions, and of the complex kinematics of the region. We also identify three sightlines within a 7\degree $\times$ 6\degree field around the Leo Ring, along which we do not find any absorption. We find that the metallicities associated with the Leo Ring are generally high, with values between solar and several times solar. The inferred high metallicities are consistent with an origin of the ring as tidal debris from a major galaxy merger.

\end{abstract}



\renewcommand*{\thefootnote}{\textsuperscript{\arabic{footnote}}}

\section{INTRODUCTION}
\label{sec:Intro}

The circumgalactic medium (CGM) is an important and dynamic interface between a galaxy and its surroundings. This medium is host to a significant fraction of the baryons in galaxies~\citep{wakker2009,werk2014}, and therefore harbours most of the cosmic-fuel that forms stars in the galaxies. It is also the site where the energy output from stellar feedback and AGN activity is deposited. The CGM, therefore, is intertwined with the evolution of galaxies~\citep{tumlinson2017circumgalactic}. 

\smallskip

The CGM is influenced by a wealth of processes such as inflows, outflows, recycled accretion, tidal interactions, and by the effects of cosmic rays and magnetic fields. These processes determine the properties of the CGM such as density, temperature, and metallicity, which are complex and non-uniform even in a single galaxy, and which globally depend on galaxy properties such as mass, star formation rate, and environment. Spectra of background quasars are sensitive probes of the gas along their sightlines, allowing us to resolve the different structures and processes shaping the CGM of intervening galaxies.

\smallskip

The low surface density of bright quasars often results in only one quasar sight-line passing through a given galaxy. Though statistically useful to describe an ensemble, a single data-point per galaxy can limit our understanding of the physical processes within the CGM. \citet{keeney2013} presented a study of three QSOs probing the galaxy ESO 157-49, an edge-on spiral, at impact parameters $\rho <$  200 h$_{70}^{-1}$ kpc, with absorption along the two major axis sightlines being consistent with a ``galactic
fountain'' of recycled gas; while the absorption along the other minor axis sightline was found to trace outflows. \citet{bowen2016} identified four sightlines passing through the halo of the spiral galaxy, NGC 1097, within 0.2-0.6 virial radii disk of gas rotating roughly planar to, but more slowly than, the inner disk observed from 21 cm data, and which also includes gas infalling from the IGM, recreates the data well.
. \citet{chen2014}, using a quadruply-lensed quasar, demonstrated that multiple-QSO probes offer crucial insights into physical properties of the CGM around galaxies. \citet{peroux2018} studied a region of $\approx30$ kpc$^{2}$ associated with the CGM of a z $\approx$1 galaxy using an intervening foreground absorber, aided by the identification of bright, extended background galaxies. Their study suggests efficient metal mixing on kpc-scales in the galaxy's CGM.

\smallskip

\citet{lehner2015}, using a sample of 18 sightlines probing the M31 halo within a projected distance of $ \approx2 R_{vir}$ ({\rvir} = 300 kpc for M31), associate the absorption probed by these sightlines with the CGM of M31 since the absorption centroids are within \hbox{40–90 {\kms}} of the systemic velocity of M31. For 7/18 sightlines which are within {\rvir}, they detect absorption in low ions
(\cii, \siii) through intermediate (\siiii, \siiv) and high ions (\civ, \ovi), suggesting that the CGM of M31 is extended out to at least its virial radius. In a followup study, \citet{lehner2020} identified several additional sightlines yielding a total of 43 QSOs probing the CGM of M31 from 25 to 569 kpc. They do not find an azimuthal dependence of CGM properties, however there is a dependence on the projected distance. 

\smallskip

FIRE simulations of $z$=0 Milky Way analogs~\citep{alcazar2017} suggest that galaxy-galaxy interactions and intergalactic transfer of mass from other galaxies via galactic winds dominates the gas accretion for a considerable period of a galaxy's evolution. It is therefore expected that gas transfer could be detected in CGM observations of galaxy groups. {\hi} 21-cm maps have revealed telltale signatures of such interactions, which are seen, for example, in the form of hydrogen gas extending from the disk of the M31 galaxy to M33~\citep{wolfe2013}. \citet{deblok2018} presented a study of an extended {\hi} distribution centered on the M81 triplet - M81, M82, and NGC 3077. They detect small {\hi} clouds and complexes which are spatially proximal to the main {\hi} tidal features of the triplet, suggesting they are all debris of the interaction that formed the triplet. This debris is similar to the tidal features arising in simulations, indicative of gas transfer from the satellites onto the more massive galaxy. Moreover, it is plausible that these interactions have resulted in the strong star-formation driven molecular outflows observed in M82~\citep{leroy2015}. 

\smallskip

Association of absorption components with particular galaxies in galaxy groups is challenging. There is often an ambiguity about which galaxy might be responsible for the absorption, and in some cases the gas may not be bound to a particular galaxy. Studies hint that the CGM of groups is more extended, and their kinematics broadened as compared to those of isolated galaxies due to a superposition of individual galaxy halos in the group~\citep{Bordoloi2011}, or by intergalactic gas transfer between the group members~\citep{Nielsen2018}.

\smallskip

\citet{kacprzak2010} in their work on the CGM of galaxy groups found two galaxies associated with a DLA at $z$=0.313, with three more galaxies in close vicinity. The authors posit that metal-enriched winds are an unlikely origin for the DLA because of its low metallicity, $\log Z/Z_{\sun} \approx -1.0$. They favor an origin of the absorption from tidal debris and/or a diffuse intragroup medium based on perturbed morphologies for three galaxies and extended optical tidal tails. In their study of the same DLA system at $z$=0.313, \citet{guber2018} demonstrate by help of optical spectroscopy with UVES that the strongly
varying dust depletion in the different absorber components (as traced by the \mnii, \caii, and {\tiii}
ratios) represents the best parameter to discriminate between dust-rich disk components and
(dust-poor) tidal features in group environments. To gather a better understanding of the origin of this DLA, \citet{chen2019} mapped the CGM using nebular emission-lines with VLT/MUSE. However, due to the redshift and surface brightness limits, only a partial view of the CGM could be obtained. Nevertheless, by combining the quasar absorption line data, MUSE gas kinematics, ionization conditions, and group galaxy motions, they concluded that the DLA originates in streams of gas stripped from sub-L$^{\star}$group members. Their study demonstrates that interactions in low-mass galaxy groups can pollute the intragroup medium with metal-enriched gas from star-forming regions. \citet{richter2018} studied the chemical and physical conditions in one of the tidal gas streams (Spur 2) around the Whale galaxy, NGC 4631. They combined $HST$/COS observations of a background quasar with the 21-cm data from the HALOGAS project, and conclude that the observed properties of region probed by the sightline favor a tidal disruption scenario in which a now closely separated ($\approx$6 kpc) satellite galaxy, NGC 4627, stripped metal- and dust-poor gas out of the outer-disk of NGC 4631. 

\smallskip

Despite these recent efforts to better understand the nature of the CGM in galaxy groups, there remains many questions. Recent results disagree as to whether the CGM metallicity is elevated in group environments~\citep{lehner2017,pointon2020}. Compounding the disagreement, all of these studies were done on a limited number of galaxy groups lacking measured velocity dispersions or quantified group properties. Therefore, a general characterization of the gas distribution in galaxy groups remains an open question. 

\smallskip

The Leo Ring is a vast region of {\hi} gas detected in 21-cm emission, nearly 200 kpc in diameter. It is in the environment of the Leo-I group, and because of its proximity to the Milky Way, the combined region is an ideal place to study the distribution of gas in a galaxy group.  The Leo-I group is so nearby ($D \leq 15$ Mpc) that even faint dwarf galaxies can be detected. \citet{muller2018} identified 36 new potential dwarf galaxy candidates in their 500\degree $\times$ 500\degree survey of $gr$ SDSS images within the extended region of the Leo-I group. Very Large Array (VLA) 21-cm maps show the presence of gas clumps in the Leo Ring~\citep{schneider1985}. The origin of the Leo Ring has been disputed - either inflowing material or tidal debris from an interaction. \citet{rosenberg2014}, using three QSO sightlines behind the Leo Ring, measured a metallicity $\approx$ 0.1 Z$_{\sun}$ which is higher than expected for pristine gas but lower than expected from a major galaxy interaction. Using numerical simulations, \citet{dansac2010} showed that the ring can plausibly be produced by a head on collision between M96 and NGC 3384.

\smallskip

Evidence is now mounting in favor of a tidal origin of the Leo Ring. ~\citet{corbelli2021} have performed Multi Unit Spectroscopic Explorer ($MUSE$) optical integral field spectroscopy of three {\hi} clumps in the Leo Ring, detected ionized hydrogen in the ring, and identified four nebular regions fueled by massive stars. Using the metal emission lines of [\oiii], [\nii], and [\Sii] they inferred metallicities $\gtrsim$ solar, inconsistent with an origin from primordial inflowing gas. The origin of the Leo Ring via such a galaxy-galaxy encounter is in agreement with hydrodynamical simulations~\citep{bekki2005MNRAS,dansac2010}.

\smallskip

In this work, we identify three lines of sight passing directly through the Leo {\hi} Ring and associated regions of interaction and five lines of sight passing through the same Leo Group at larger distances from the {\hi} ring detected in 21cm. We also identify three sightlines in the same field which do not show any absorption within the heliocentric velocity range of 600-1400{\kms} related to the group. For each sightline with detected absorption, we conduct cloud-by-cloud, multiphase, Bayesian ionization modeling analysis~\citep{sameer21}[S21] and consider whether absorption is due to tidal debris, and/or an individual dwarf galaxy, and/or the CGM of one or more galaxies, and/or the intragroup medium at large. The paper is organized as follows: in Section~\ref{sec:observations} we describe the spectral observations that are analysed in this work; in Section~\ref{sec:methodology} we describe the methodology used to determine the physical conditions of an absorber; in Section~\ref{sec:Results} we present our results; in Section~\ref{sec:discussion} we summarize and discuss conclusions about the physical conditions in the Leo Group region based on the absorption seen along the sightlines. We conclude in Section~\ref{sec:conclusion}. Throughout this work, we assume a cosmology with $H_{0}\,=\,70$~km~s$^{-1}$~Mpc$^{-1}$, $\Omega_{\rm M}\,=\,0.3$, and $\Omega_{\Lambda}\,=\,0.7.$  Abundances of heavy elements are given in the notation $\rm [X/Y] = \log (X/Y) - \log (X/Y)_{\odot}$ with solar relative abundances taken from \citet{grevesse2011chemical}. All the distances given are in physical units. All the logarithmic values are presented in base-10. 

\section{OBSERVATIONS AND DATA}
\label{sec:observations}

\begin{table*}
\renewcommand\thetable{1}

\caption{\bf List of the twelve sightlines with COS observations in the 7$\degree$ $\times$ 6$\degree$ field containing the Leo Ring}
\begin{tabular}{c@{\hspace{1.0\tabcolsep}}c@{\hspace{1.0\tabcolsep}}c@{\hspace{1.0\tabcolsep}}c@{\hspace{1.0\tabcolsep}}c@{\hspace{1.0\tabcolsep}}c@{\hspace{1.0\tabcolsep}}c@{\hspace{1.0\tabcolsep}}c@{\hspace{1.0\tabcolsep}}c@{\hspace{1.0\tabcolsep}}c@{\hspace{1.0\tabcolsep}}c@{\hspace{1.0\tabcolsep}}c@{\hspace{1.0\tabcolsep}}} \hline \hline
 Sightline ID & QSO.$^{a}$ & RA  &  Dec & $z_{qso}^{b}$ & $z_{abs}^{c}$ & $cz_{abs}^{d}$ & PID$^{e}$ & Observed  & $S/N^{f}$ & Observation \\
& & (J2000) & (J2000) & &  & ({\kms}) & & Wavelength (\angstrom) &  & Date \\

\midrule

 SA & SDSS J104029.17+105318.1 & 160.12 & 10.89  & 0.1363 & - & & 12533 & 1135-1435 & 0.1 & 2011-11-11\\

SB & SDSS J104244.24+164656.1 & 160.68 & 16.78 & 0.9780 & 0.0024 & 720 & 14777 & 1132-1795 & 6.0 & 2017-11-13\\

SC & SDSS J104335.87+115129.0 & 160.90 & 11.86 & 0.7940 & 0.0024 & 720 & 14071 & 1132-1434 & 11.7 & 2015-11-23\\

SD & SDSS J104709.83+130454.6 & 161.79 & 13.08 & 0.4006 & 0.0028 & 840 & 12198 \& 14071 & 1132-1443 & 6.7 & 2011-05-30 \& 2015-12-01\\

SE & SDSS J104741.74+151332.2 & 161.92 & 15.23 & 0.3858 & 0.0023 & 690 & 13342 \& 13833 & 1132-1777 & 23.2 & 2014-12-21 \& 2015-02-05\\

SF & SDSS J104816.24+120734.8 & 162.07	& 12.13 & 0.2911 & 0.0030 & 900 &  12198 & 1142-1443 & 3.7 & 2011-06-03\\

SG & SDSS J104843.49+130605.8 & 162.18	& 13.10 & 0.2185 & 0.0027& 810 &  12198 & 1142-1766 & 4.2 & 2011-07-02\\

SH & SDSS J105125.73+124746.2 & 162.86	& 12.80 & 1.2828 & 0.0027& 810  &  12603 \& 14777 & 1147-1795 & 7.3 & 2013-05-17 \& 2018-01-04\\

SI & SDSS J105220.63+101751.7 & 163.09	& 10.30 & 0.2462 & -  & & 14071 & 1142-1444 & 4.9 & 2002-12-19\\

SJ & SDSS J105945.23+144142.9 & 164.94	& 14.70 & 0.6317 & 0.0024 & 720 &  12248 & 1135-1795 & 9.1 & 2011-06-01\\

SK & SDSS J105956.14+121151.1 & 164.98	& 12.20 & 0.9927 & -  & & 12603 & 1142-1468 & 7.5 & 2012-11-11\\

SL & 4C10.30 & 165.20	& 10.77 & 0.4230 & -  & & 12603 & 1133-1465 & 4.8 & 2012-11-15\\\hline

\end{tabular} \\
\label{tab:sample}
\begin{flushleft} \small
Notes-- $^{a}$ QSO SDSS name; $^{b}$ QSO redshift from NED; $^{c}$ Absorber redshift determined from the centroid of the strongest metal-line absorption or the {\hi} {\lya} absorption when no metal-lines are observed. The redshift corresponding to {\hi} {\lya} centroid is presented when the metal-line absorption is absent; $^{d}$ Heliocentric velocity corresponding to the absorber redshift;  $^{e}$ proposal ID. $^{f}$ Median signal to noise ratio of the spectrum. We do not analyse sightline SA because of the poor S/N ratio of $\approx$0.1.
\end{flushleft}     

\end{table*}

\begin{table*}
\centering
\caption{Candidate galaxies responsible for the absorption along the sightlines}
\begin{tabular}{cccccccc}
\hline\hline
Galaxy & RA & DEC & z$^{a}$ & Morphological$^{b}$ & M$_{r}^{c}$    & Reference$^{d}$               \\
 &  (J2000)  &  (J2000)  &  & Type & (mag) &  \\
\hline
M95                     & 10 43 57.69 & +11 42 13.63 & 0.00256 & SB(r)b                  & $-19.34$ & H$\alpha$ rotation map from \citet{walter2008} \\
M96                     & 10 46 45.74 & +11 49 11.78 & 0.00301& SAB(rs)ab                                                      & $-20.47$ &  H$\alpha$ rotation map from \citet{Silchenko2003} \\
M105                    & 10 47 49.60 & +12 34 53.87 & 0.00292 & E1                                                        & $-20.91$ & Stellar line-of-sight velocity map from \citet{Silchenko2003}       \\
NGC3384                 & 10 48 16.90                                     & +12 37 45.48 & 0.00244 & SB0                                                     & $-19.46$ & Stellar line-of-sight velocity map from \citet{Silchenko2003}   \\  
NGC3412                 & 10 50 53.31                                     &  +13 24 43.71 & 0.00288 & SB0                                                     & $-19.53$ & Major axis values from \citet{aguerri2003}   \\
NGC3489 &   11 00 18.57 & +13 54 04.40 & 0.00232 & SAB0 & $-19.12$& Systemic velocity adopted from NED database&\\

NGC 3377 &  10 47 42.33 & +13 59 09.30 & 0.00232 & E5-6 & $-18.30$ & Systemic velocity adopted from NED database&\\\hline

\end{tabular}
\label{tab:majorgalaxies}
\begin{flushleft} \small
Notes-- $^{a}$ Redshift of the galaxy; $^{b}$ Morphological classification of Galaxy from NED database; $^{c}$ Absolute Magnitude determined from $r$-band SDSS AB magnitude; $^{d}$ Reference paper from which the galaxy velocities in the direction of sightline are determined.
\end{flushleft}     

\end{table*}

\begin{figure*}
    \centering
    \subfloat{{\includegraphics[scale=0.55]{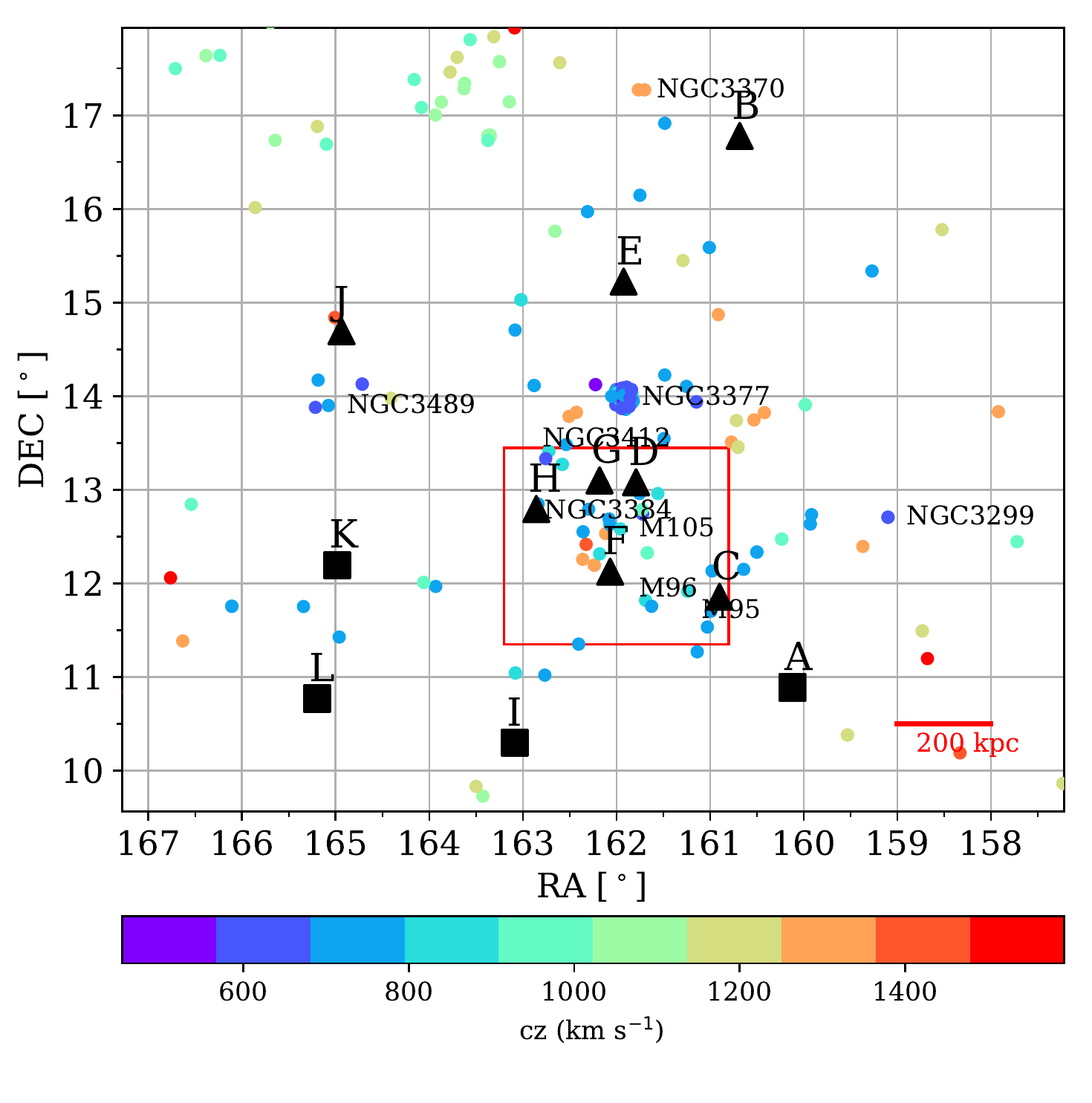} }}%
    \qquad
    \subfloat{{\includegraphics[scale=0.55]{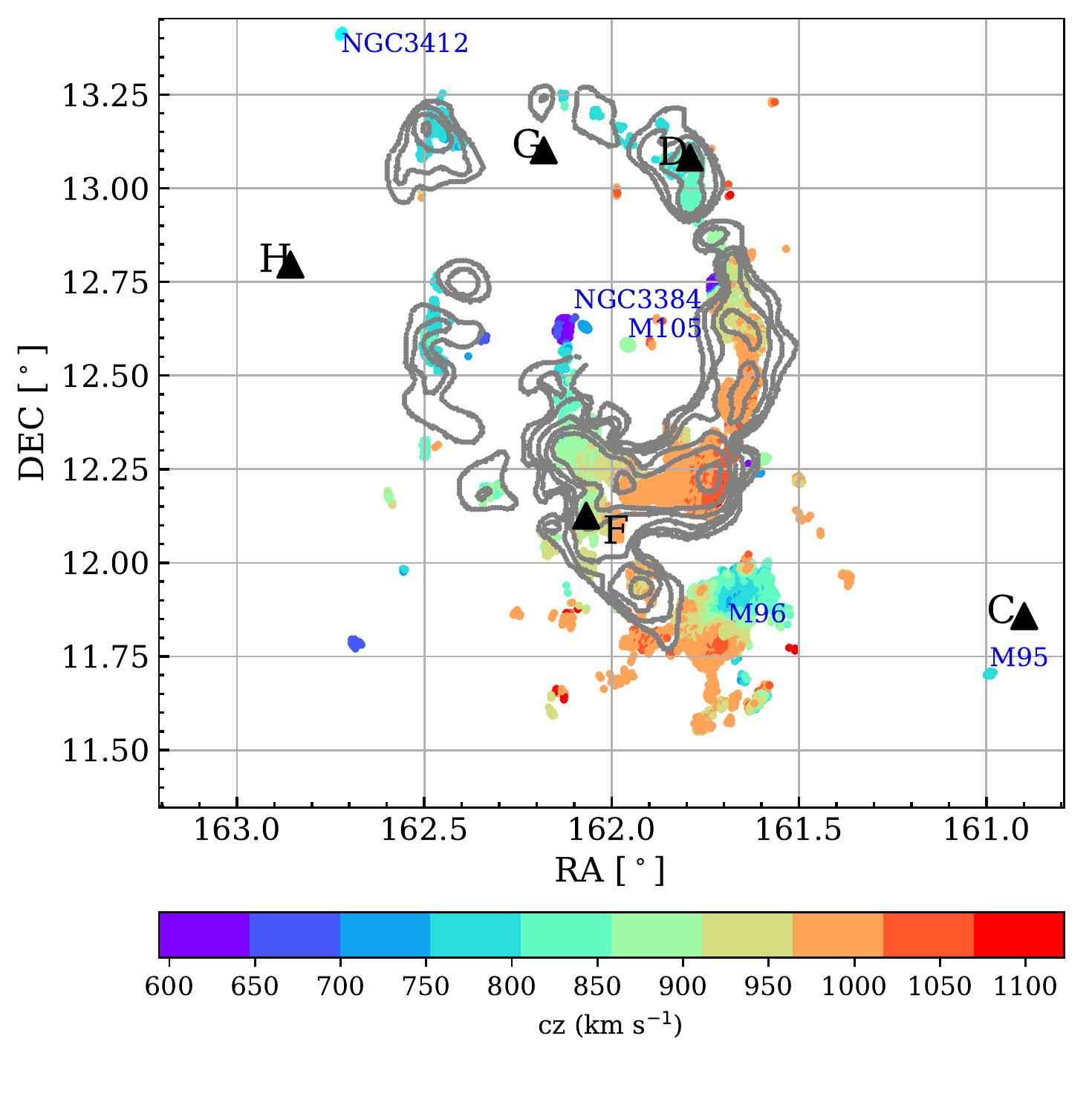} }}%
    \caption{Left - A map showing the locations of 12 sightlines (A-L) with COS observations in the $\approx$7\degree$\times$6\degree field containing the Leo Ring. We do not search for absorption towards SA because of the poor S/N ($\approx$0.1). Triangles denote the sightlines with detected absorption between 600{\kms}$< cz <$1400{\kms}, while squares denote the sightlines which do not show absorption within this velocity range. One unit length in the grid corresponds to $\approx$200 kpc at the redshift of $z$=0.0025. We indicate the heliocentric velocities of the galaxies in the field using a colormap shown below. Right - A zoom-in of the Leo Ring region shown using contours from \citet{schneider89} revealing the ringlike distribution of intergalactic {\hi}. Contours are shown at 0.1, 0.2, 0.4, 0.8, 1.6, and 3.2 Jy {\kms} per beam. 1 Jy {\kms} per beam corresponds to a column density of $\approx$ 2 $\times$ 10$^{19}$ cm$^{-2}$ for the Arecibo beam size. We also show the velocity field of the {\hi} gas mapped by Westerbork Synthesis Radio Telescope (adapted from \citealt{dansac2010}). Five sightlines C, D, F, G, and H are potentially tracing the gas associated with the Leo Ring. }%
    \label{fig:LeoRingfield}%
\end{figure*}

The background quasars in this study have UV spectra from the Cosmic Origins Spectrograph (COS) on the Hubble Space Telescope (\textit{$HST$}). Table~\ref{tab:sample} presents the details of the observations of the 12 sightlines with COS observations in the field around Leo Ring and the Leo I group. The 7\degree $\times$ 6\degree field, containing the Leo Ring along with the 12 sightlines, is shown in Figure~\ref{fig:LeoRingfield}. We do not study sightline SA because of the poor $S/N$ ratio of 0.1.

\smallskip
The UV \textit{$HST$}/COS spectra in this study are taken from the Barbara A. Mikulski Archive for Space Telescopes (MAST). The observations use the G130M and G160M gratings spanning observed wavelength ranges of 1135-1457~{\AA} and 1399-1794~{\AA}, respectively. The spectra have an average resolving power of $R \approx 20,000$ and cover a range of ions including the {\hi} Lyman series, {\cii}, {\ciii}, {\civ}, {\nv}, {\siii}, {\siiii} and {\siiv}. The spectra were reduced following the procedures in \citet{wakker2015}. In essence, spectra are processed with \textsc{CALCOS} v3.2.1 and are aligned using a crosscorrelation, and then shifted to ensure that (1) the velocities of the interstellar lines match the 21cm {\hi} profile, and (2) the velocities of the lines in a single absorption system are aligned properly. The exposures are then collectively combined by summing total counts per pixel prior to converting to flux. 


\smallskip

Along all the sightlines, on the low end of the velocity range of interest, the {\hi} {\lya} absorption could be affected by the right wing of the geocoronal airglow emission line. The strength of the airglow emission varies from observation to observation, and depends on solar activity. The shape of the emission profile, however, does not vary much. The location and amplitude change, and must be determined from a fit to the data for each sightline. We use the publicly available community generated airglow templates\footnote{\url{https://www.stsci.edu/hst/instrumentation/cos/calibration/airglow}}~\citep{bourrier2018} to fit the emission profile, and inspect whether it is affecting the {\hi} {\lya} absorption of interest. We fit the {\hi} {\lya} geocoronal emission template for each sightline and show that it does not significantly affect the {\hi} {\lya} absorption of interest. The emission line, and the corresponding template fit for sightlines SB-SL are shown in Figures~\ref{fig:SBairglow} -- \ref{fig:SLairglow}. In most of these figures a plateau can be seen on the blueward side of the Galactic {\lya} trough that is elevated about the Galactic absorption by a wing of the geocoronal airglow emission line. Such a wing is surely also elevating the flux on the redward side of Galactic {\lya} as well.  In these cases, it appears that a simple scaling of the airglow emission template to fit the observed geocoronal emission is failing. The response of the COS detector becomes non-linear when the count rate is high enough. As a result, if the model is scaled to match the counts at the peak of the {\lya} emission, the wings may be underscaled. Furthermore, because of the COS optics, the geocoronal {\lya} emission fills the COS aperture, which is difficult to model.  The excess counts from the wings of the bright geocoronal {\lya} emission could cause an incorrect flux zero point in the {\hi} {\lya} absorption profiles of our Leo I group absorbers, affecting them at velocities of 600{\kms} up to $\sim 1000$~{\kms}. To address this problem which is noticeably affecting sightlines SC, SG, and SH, we obtain 100 MC realizations, varying the flux zero level (by an amount drawn from a uniform distribution between 0 and average of affected contiguous pixels) to explore the range of parameter space that is consistent with the data in the presence of such a systematic zero-point uncertainty.

\smallskip 

In addition to the effect of {\hi} {\lya} airglow emission, the \oi~$\lambda$1302 transition is also affected by the {\oi} geocoronal emission. We minimize its effect by using observations taken during orbital night, i.e when the Sun is below the horizon. We use the \textsc{timefilter} module from the \textsc{costools} package to filter unwanted data (\hbox{SUN\_ALT $>$ 0}) and rerun the \textsc{calcos} pipeline on the filtered data to generate the airglow emission corrected spectrum. However, this step considerably reduces the signal-to-noise of the corrected region of the spectrum covering \oi~$\lambda$1302.

\smallskip

We performed continuum normalization by fitting a cubic spline to the spectrum. We estimated the statistical uncertainties in the continuum fits using ``flux randomization'' Monte Carlo simulations (e.g. \citealt{peterson1998optical}), varying the flux in each pixel of the spectrum by a random Gaussian deviate based on the spectral uncertainty. The pixel-error weighted average and standard deviation of 1000 iterations was adopted as the flux and uncertainty of the continuum fit, respectively. 

\smallskip

For galaxies with redshift measurements, we have determined their heliocentric velocities and plotted them in Figure~\ref{fig:LeoRingfield} colored by their c$z$. We adapted contours from \citet{schneider89} revealing the ringlike distribution of intergalactic {\hi}. Contours are shown at 0.1, 0.2, 0.4, 0.8, 1.6, and 3.2 Jy {\kms} per beam. 1 Jy {\kms} per beam corresponds to a column density of $\approx$ 2 $\times$ 10$^{19}$ cm$^{-2}$ for the Arecibo beam size. We also show the velocity field of the {\hi} gas mapped by Westerbork Synthesis Radio Telescope (adapted from \citealt{dansac2010}). We used the plot digitization software WebPlotDigitizer~\citep{Rohatgi2020} to extract the information from adapted figures. Five sightlines C, D, F, G, and H are potentially tracing the gas associated with the Leo Ring. 

\smallskip

We have investigated possible biases that could arise from the use of archival data, in our case, aimed at probing the general surroundings of the Leo Ring region. The sightline towards SB is from a program to investigate quasar outflows, the background quasars in this study have $z >$ 0.4 selected to study intrinsic absorption. The sightlines towards SC and SI are from a program to probe the interface between the disk and the CGM of gas-rich galaxies using a sample of 35 QSOs; the QSOs have z between 0.003--0.05. The QSO towards SC has a $z \approx$ 0.79. The sightline towards SD, SF and SG are QSOs behind the Leo Ring. The sightline towards SE is from two separate programs, investigating outskirts of galaxy clusters at z between 0.1--0.55. The sightlines towards SK, SL, and SH are from a program, to investigate baryonic cycling in galaxies with $z$ between 0.02-0.5. Since the Leo Ring absorption is expected to be below $z \approx$ 0.0035, we are certain that our sample selection is not biased in any way.

\section{METHODOLOGY}
\label{sec:methodology}

\begin{figure}
\begin{center}
\hspace*{-0.7cm}\includegraphics[scale=0.425]{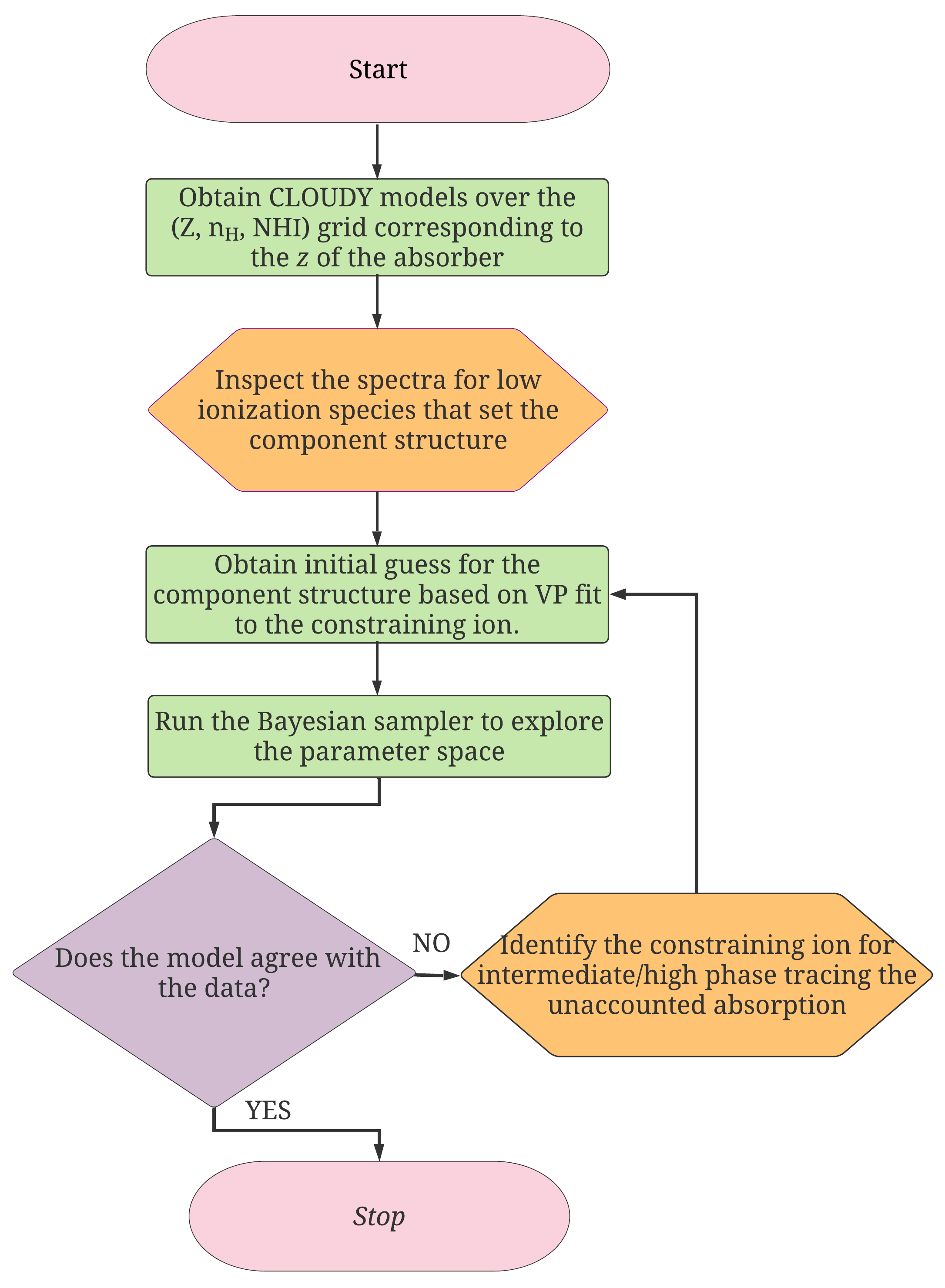}
\caption{A flowchart showing the workflow involved in obtaining the models that best describe the conditions present in an absorption system.}
\label{fig:flowchart}
\end{center}
\end{figure}

We model the observed absorption systems using the photoionization code {\CLOUDY} ver17.02~\citep{cloudy17}, and synthesize the expected absorption profiles and compare them to the observed profiles in order to infer the column densities, Doppler parameters, and physical conditions of the absorbing gas. We adopt a slightly modified approach to the CMBM method described in S21. The main difference is that, in S21, we optimized on the measured column density of an ion that was considered to trace a gas phase in the absorption system. The free parameters were metallicity and hydrogen number density. In this work, we incorporated three additional free parameters - neutral hydrogen column density, redshift of the absorber, and Doppler broadening parameter. This approach was adopted because of the comparatively lower $S/N \lesssim 10$ in the data, so as to account for the uncertainty in the precise determination of absorption centroids, Doppler parameters, and column densities. By utilizing the shapes of the absorption profiles and centering of individual components, the CMBM method allows for more rigorous constraints on model parameters than methods which average together components and utilize {\CLOUDY} modeling of total column densities derived from the data. Recent works (e.g. \citealt{zahedy2019characterizing}, S21, \citealt{Haislmaier2021}) that perform component-wise and phase-wise modelling, find that an absorption system often has a metallicity that varies by more than an order of magnitude between components and phases.

\textsc{cloudy} has many tunable parameters. The three most important tunable parameters are metallicity relative to solar abundance ($Z$), hydrogen number density ($n_{\H}$), and neutral hydrogen column density, N(\hi). We generate photoionization thermal equilibrium (PIE) models adopting the \citet{KS19}[KS19, Q18] model for the extragalactic background radiation (EBR) to determine the volume density of ionizing photons, $n_{\gamma}$. The KS19 EBR incorporates recent estimates of {\hi} distribution in the IGM~\citep{Inoue2014}, QSO emissivities, cosmic star formation rate density, and far UV dust extinction~\citep{khaire2015}. The shape and intensity of the radiation field are sources of systematic uncertainty in ionization modeling. An exploration of the systematic effects due to this choice is described in~\citet{acharya21}. Depending on whether the gas is high or low density, they find a variation of 4-6.3 and 1.6-3.2 times in the inferred density and metallicity, respectively.  The equilibrium temperature, $T$, is self-consistently determined by {\CLOUDY} by a balance between heating and cooling processes. We do not consider the systematic effects arising from the choice of EBR. We adopt a solar abundance pattern~\citep{grevesse2011chemical} when generating the {\CLOUDY} models. Though deviations from such a pattern are common~\citep{zahedy2021}, the wavelength coverage of constraining transitions and the low $S/N$ of the data make it unfeasible to constrain an additional model parameter for these data.  For similar reasons, we also do not consider depletion due to dust in our modeling.  This is a reasonable assumption as the sightlines are primarily probing the CGM of galaxies. The effects of dust depletion are expected to be more prominent for ISM gas.

\smallskip

PIE models are generated on a 3-D grid at the fiducial redshift of $z$=0.0025. The grid is $\log Z/Z_{\sun}$ $\in$ [-3.0,1.5], $\log n_{\H}$ $\in$ [-6.0, 1.0], and $\log N(\hi)$ $\in$ [11.0, 20.0], and obtained with a 0.1 dex step-size. It is a time-consuming endeavor to construct {\CLOUDY} models over such a large grid, thus to increase efficiencies we employ distributed parallel processing on the Stampede2~\footnote{\url{https://www.tacc.utexas.edu/systems/stampede2}} and the CyberLAMP~\footnote{\url{https://wikispaces.psu.edu/display/CyberLAMP/}} clusters to obtain models in a short amount of time. 

\smallskip

In certain intervening absorbers, \nv, because of its comparatively high ionization potential ($\approx$ 77.5 eV), could arise in gas with \hbox{T $>$ 10$^{5}$ K} where the ionization is regulated by collisions between energetic free electrons and metal ions. In cases, where we have strong {\nv} absorption detected, we model the absorption arising in such a phase using photo-collisional (PC) models. PC models are obtained on a 4-D grid with {\temp} $\in$ [2.0, 6.5], being the fourth dimension of the grid, in addition to [{\metallicity}, {\hden}, and {\colden}]. The temperature in such a model is no longer set by a balance between heating and cooling processes.

\smallskip

We begin by choosing a transition with unsaturated lines, with a high enough $S/N$ to resolve the component structure. We use this species to obtain an initial guess for the component structure, and refer to this as the ``constraining ion''. The constraining ion sets the location in redshift space for an absorber. The component structure is determined by fitting a Voigt profile (VP) model corresponding to the highest log-evidence and the least model complexity, evaluated using Bayes factor. We model the VPs using an analytic approximation~\citep{garcia2006voigt} for the Voigt-Hjerting function as implemented in the VoigtFit package~\citep{krogager2018voigtfit}. The properties of various atomic line transitions are adopted from linetools~\footnote{\url{https://linetools.readthedocs.io/en/latest/linelist.html}} package. If a model that fits the low-ionization transitions (e.g., \siii, \cii) cannot produce the observed amount of high-ionization absorption (e.g., \civ), or adequately match the profile of the \lya, we will use more than one constraining ion. The other constraining ions represent the different phases of gas, with distinct regions having different densities along the line-of-sight. For example, we would typically constrain on {\siii}, {\cii}, or {\siiii} for the low-ionization gas, and on {\civ} for high-ionization gas. For some components, which have only {\hi} {\lya} detected, we use the {\hi} {\lya} as the constraining ion.

\smallskip

Every cloud in a model is identified uniquely by five parameters: $\log Z$, $\log n_{\H}$, $\log N(\hi)$, $z$, $b_{nt}$. We adopt uniform priors on $\log Z$, $\log n_{\H}$, and $\log N(\hi)$ in the ranges specified above. We use a uniform prior on $b_{nt}$ (the non-thermal contribution to the Doppler parameter of a cloud) ranging between [0, 100] and defined in {\kms}. We determine the Doppler broadening parameter $b$ for all the transitions arising in a phase using the equation $b^{2}=b^{2}_{\mathrm{nt}}+b^{2}_{\mathrm{t}}$, where $b_{\mathrm{t}}=\sqrt{2kT/m}$ is the line broadening due to temperature and $b_{\mathrm{nt}}$ the line broadening due to turbulence and other non-thermal effects. The non-thermal broadening component is assumed to be the same for all transitions in the same phase. For $z$, we use a Gaussian prior with a mean value and a standard deviation based on the error estimate, both of which are obtained from the VP fit to that ion.

\smallskip

We then determine the parameters that best describe the conditions ($\log Z$, $\log n_{\H}$, $\log N(\hi)$, $z$, and $b_{nt}$) present in all the clouds by synthesizing VPs (using the compiled line profiles for the appropriate instrument/grating) and comparing to the data using PyMultinest~\citep{buchner2014x}. The exploration of the parameter space is challenging when multiple, well separated modes with similar probabilities exist. PyMultinest is a nested sampling algorithm which is adept at addressing this challenge~\citep{buchner21}. During the log-likelihood evaluation the values at the sampled points are determined by interpolation on the preconstructed 3D-grid. The interpolant is generated by triangulating the grid data with Qhull~\citep{barber1996quickhull}, and on each triangle performing linear barycentric interpolation. We ensure that the pixels which are contaminated by blending effects or are too noisy are masked during the log-likelihood evaluation.  A summary of our methodology is given in the form of a flowchart in Figure~\ref{fig:flowchart}. 

\smallskip

In all of the absorbers of interest, the {\hi} {\lya} absorption falls on the wings of the Galactic {\hi} {\lya} absorption. To determine the Galactic N({\hi}), we first fit a VP to the Galactic {\siii} absorption to determine the absorption centroid(s), depending on the number of components seen in Galactic {\siii}. These centroids are then used to fix the number of components and their locations in Galactic N({\hi}). A VPfit is then obtained by fitting the Galactic {\lya} with this information. We divide the continuum normalized spectrum with this VP fit to determine the N({\hi}) of the absorber of interest. We compare our best fit values of Galactic {\colden} with the EBHIS~\citep{winkel16} database and find agreement within 0.05 dex in all but one case - SB. \citet{wakker2011} found the ratio between $\log N$(\hi; \lya) and $\log N$(\hi; 21 cm) to be statistically consistent with 1 in their study of 59 AGN spectra. In appendix~\ref{fig:21cmvsabs}, we show the comparison between the $\log N$(\hi; \lya) and $\log N$(\hi; 21 cm) for the 11 sightlines in this study. The Galactic {\lya} is blended with the absorption from the AGN itself, preventing a reliable measurement of Galactic {\colden} towards SB. However, the absorption of interest is not affected by this measurement towards SB. The comparison is shown in Figure. The VP fits corresponding to the maximum likelihood model for the eleven sightlines are shown in Figures~\ref{fig:SBvpfitgal}--\ref{fig:SLvpfitgal}.

\section{RESULTS}
\label{sec:Results}

We discuss the findings from the application of our methodology to 11 quasar sightlines in the Leo Ring region for which we can probe absorption between heliocentric velocities of \hbox{600-1400~\kms}.  Positions of these sightlines, SB -- SL, were shown in Figure~\ref{fig:LeoRingfield} along with the galaxies in the region.  A magnified view of the Leo Ring region that could contribute to absorption in each sightline is also shown in Figure~\ref{fig:LeoRingfield}. Sightline SA is not discussed further because of the poor $S/N$ of its spectrum ($\approx 0.1$). The absorber redshift is defined based on the centroid of the strongest metal line absorption or the {\hi} {\lya} absorption when no metal-lines are observed. We present the posterior results for physical properites of the best model in Table~\ref{tab:modelparams} and show these properties using a violin plot for each sightline. We present the posterior distributions of inferred column densities and the Doppler broadening parameters for the observed transitions in all the components, and in all the model phases in Table~\ref{tab:basicinfo}.

\smallskip 

For each sightline, we also discuss the possibility of a gas cloud tracing absorption associated with particular galaxies. Since there is an ambiguity in associating a particular galaxy with an absorption line, we quantify this using a modified prescription of likelihood parameter (LP) described in \citet{french2017}. The LP is defined as follows: 
\begin{equation}
    \textit{LP} = e^{-(\frac{\rho}{R_{vir}})^2} \times e^{-(\frac{\Delta V}{V_{esc}})^2}
    \label{eqn:1}
\end{equation}
where, $\rho$ denotes the impact parameter between the galaxy and the line of sight,  $V_{esc}$ is the galaxy's escape velocity, {\rvir} is the galaxy's virial radius, and $\Delta$V denotes the difference between the absorber's velocity (V$_{abs}$) and possible velocities within the galaxy's halo,  V$_{abs}$ $-$ (V$_{gal}$ $\pm$ V$_{esc}$). We assume that all the gas associated with a galaxy is within the escape velocity of the halo; all absorption-line systems appear to be within the escape velocity of the halo~\citep{kacprzak2017}. In the case of galaxies with measured rotation curves or stellar line-of-sight velocity fields, V$_{gal}$ is determined by extrapolating the galaxy's rotation curve or velocity map in the direction of the sightline, whereas in the case of dwarfs and galaxies where this information is absent, V$_{gal}$ is set equal to the systemic velocity of the galaxy. The $\pm$ V$_{esc}$ term in equation~\ref{eqn:1} accounts for the fact that almost all absorption (even from outflows) in many transitions is bound to galaxies at or below the escape velocity~\citep{stocke2013,mathes2014,tumlinson2011large,bordoloi2014cos,Kacprzak2019ApJ}. In Table~\ref{tab:majorgalaxies}, we present the literature references for the rotation curves. The virial radius, {\rvir}, and virial mass, M$_{vir}$, of each galaxy are determined according to the combined ``halo matching'' and constant mass-to-light ratio technique described in \citet[see their Figure 1]{stocke2013}. We adopt the $B$-band magnitude from the SIMBAD database~\citep{simbad} in the determination of {\rvir} and M$_{vir}$. When the $B$-band magnitude is unavailable, we use the color transformation $B = (g + 0.21) + 0.39(g - r)$~\citep{jester2005} on SDSS model magnitudes to estimate the $B$-band magnitude. Given the low-z of our sample, we do not carry out $K$-corrections to galaxy luminosities.


\smallskip

\begin{table*}
\renewcommand\thetable{3}
\caption{\bf Gas phases in the absorbers}
\label{tab:modelparams}
 
\begin{tabular}{c@{\hspace{1.5\tabcolsep}}c@{\hspace{1.5\tabcolsep}}c@{\hspace{1.5\tabcolsep}}c@{\hspace{1.5\tabcolsep}}c@{\hspace{1.5\tabcolsep}}c@{\hspace{1.5\tabcolsep}}c@{\hspace{1.5\tabcolsep}}c@{\hspace{1.5\tabcolsep}}c@{\hspace{1.5\tabcolsep}}c@{\hspace{1.5\tabcolsep}}c@{\hspace{1.5\tabcolsep}}c@{\hspace{1.5\tabcolsep}}} \hline \hline

Sightline & Constraining & $cz$ & $b_{nt}$ & $b_{t}$ &$b_{t, \hi}$ & $b$ & $\log \frac{Z}{Z_\odot}$ & $\log \frac{n(H)}{\cc}$ & $\log \frac{T}{K} $ & $\log \frac{N(\hi)}{\cmsq}$ & $\log \frac{L}{kpc} $ \\  
& ion cloud & (\kms) & (\kms) & (\kms) & (\kms)&(\kms) & & & & & \\
(1) & (2) & (3) & (4) & (5) & (6) & (7) & (8) & (9) & (10) & (11) & (12)\\\hline

SB & \textcolor{blue}{\hi_0} & ${722.1}_{-1.6}^{+1.9}$ &	${3.8}_{-3.3}^{+12.1}$ &	${12.7}_{-4.0}^{+4.1}$ &	${12.7}_{-4.0}^{+4.1}$ & ${14.4}_{-2.8}^{+4.9}$ & $<-0.35$ &	$>-2.63$ &	${3.99}_{-0.33}^{+0.24}$ &	${15.28}_{-0.77}^{+1.02}$ &	$<-2.26$ \\

\hline

SC$^{a}$ & \textcolor{blue}{\nv_0} & ${710.0}_{-3.5}^{+2.9}$ &	${50.5}_{-4.4}^{+4.5}$ &	${6.4}_{-0.4}^{+0.4}$ &	${24.0}_{-1.6}^{+1.4}$& ${51.0}_{-4.4}^{+4.4}$ &	${0.81}_{-0.13}^{+0.15}$ &	${-4.18}_{-0.03}^{+0.03}$ &	${4.54}_{-0.06}^{+0.05}$ &	${14.62}_{-0.15}^{+0.10}$ &	${0.93}_{-0.22}^{+0.16}$    \\

& \textcolor{red}{\siiv_1} & ${716.4}_{-0.3}^{+0.2}$ &	${14.7}_{-0.6}^{+0.8}$ &	${2.9}_{-0.1}^{+0.1}$ &	${15.4}_{-0.5}^{+0.5}$ & ${14.9}_{-0.6}^{+0.8}$ &	${-0.13}_{-0.11}^{+0.08}$ &	${-2.73}_{-0.04}^{+0.03}$ &	${4.16}_{-0.03}^{+0.03}$ &	${17.97}_{-0.09}^{+0.13}$ &	${0.74}_{-0.10}^{+0.12}$  \\

& \textcolor{green}{\cii_1} & ${878.6}_{-1.0}^{+0.8}$ &	${3.7}_{-1.3}^{+1.9}$ &	${3.9}_{-0.1}^{+0.2}$ &	${13.6}_{-0.4}^{+0.5}$ & ${5.4}_{-0.8}^{+1.5}$ &	 ${-1.18}_{-0.31}^{+0.25}$ &	${-1.92}_{-0.21}^{+0.21}$ &	${4.05}_{-0.03}^{+0.03}$ &	${18.36}_{-0.04}^{+0.04}$ &	${-0.67}_{-0.32}^{+0.36}$ \\

& \textcolor{magenta}{\hi_0} & ${1035.8}_{-1.8}^{+1.8}$ &	${29.0}_{-5.4}^{+3.0}$ &	${12.7}_{-6.4}^{+3.7}$ &	${12.7}_{-6.4}^{+3.7}$ & ${31.7}_{-4.3}^{+2.5}$ &	$<-0.37$ &	$>-2.56$ &	${3.99}_{-0.62}^{+0.22}$ &	${14.12}_{-0.06}^{+0.11}$ &	$<-3.09$     \\

SC$^{b}$ & \textcolor{blue}{\nv_0} & ${706.8}_{-9.1}^{+12.3}$ &	${41.9}_{-16.9}^{+16.3}$ &	${19.8}_{-2.5}^{+3.2}$ &	${73.9}_{-9.1}^{+11.9}$ & ${46.3}_{-14.3}^{+15.1}$  &	${0.37}_{-0.65}^{+0.64}$ & 	${-1.74}_{-1.98}^{+2.49}$ &	${5.52}_{-0.11}^{+0.13}$ &	${13.10}_{-0.59}^{+0.50}$ &	${-0.59}_{-2.52}^{+2.16}$    \\

& \textcolor{red}{\siiv_1} & ${716.2}_{-0.5}^{+0.3}$ &	${16.4}_{-0.1}^{+0.1}$ &	${3.3}_{-0.1}^{+0.1}$ &	${17.3}_{-0.4}^{+0.3}$ & ${16.7}_{-0.1}^{+0.1}$ &	${-0.44}_{-0.05}^{+0.07}$ &	${-2.92}_{-0.03}^{+0.03}$ &	${4.26}_{-0.02}^{+0.02}$ &	${18.23}_{-0.07}^{+0.06}$ &	${1.30}_{-0.10}^{+0.08}$   \\

& \textcolor{green}{\cii_1} & ${878.6}_{-1.0}^{+0.8}$ &	${3.4}_{-1.2}^{+1.5}$ &	${4.0}_{-0.1}^{+0.1}$ &	${14.0}_{-0.5}^{+0.5}$ &  ${5.3}_{-0.70}^{+1.1}$ &	${-1.21}_{-0.27}^{+0.23}$ &	${-2.07}_{-0.16}^{+0.18}$ &	${4.07}_{-0.03}^{+0.03}$ &	${18.31}_{-0.05}^{+0.04}$ &	${-0.44}_{-0.32}^{+0.30}$ 	\\

& \textcolor{magenta}{\hi_0} & ${1035.8}_{-1.8}^{+1.6}$ &	${28.0}_{-4.0}^{+3.2}$ &	${13.9}_{-3.7}^{+4.7}$ &	${13.9}_{-3.7}^{+4.7}$ & ${31.4}_{-3.4}^{+2.8}$ &	$<0.12$ &	$>-2.94$ &	${4.07}_{-0.27}^{+0.25}$ &	${14.14}_{-0.06}^{+0.10}$ &	$<-2.25$      \\

\hline 

SD & \textcolor{blue}{\siiv_0} & ${844.4}_{-1.7}^{+1.6}$ &	${18.9}_{-2.8}^{+3.3}$ &	${3.2}_{-0.5}^{+0.3}$ &	${17.1}_{-2.4}^{+1.3}$ & ${19.1}_{-2.7}^{+3.2}$&	${-0.06}_{-0.41}^{+0.48}$ &	${-2.98}_{-0.12}^{+0.10}$ &	${4.25}_{-0.13}^{+0.06}$ &	${16.92}_{-0.64}^{+0.55}$ &	${0.59}_{-0.59}^{+0.51}$   \\    

& \textcolor{red}{\hi_0} & ${978.4}_{-8.0}^{+3.7}$ &	${3.0}_{-2.8}^{+3.1}$ &	${12.5}_{-8.0}^{+5.7}$ &	${12.5}_{-8.0}^{+5.7}$ &	${13.1}_{-8.0}^{+5.3}$& $<-0.65$ &	$>-2.85$ &	${3.98}_{-0.89}^{+0.32}$ &	${16.79}_{-1.52}^{+0.86}$ &	$<-1.24$ 	 \\

\hline 

SE$^{a}$  & \textcolor{blue}{\hi_0} & ${700.7}_{-0.9}^{+1.6}$ &	${14.5}_{-3.8}^{+4.5}$ &	${23.3}_{-5.5}^{+2.6}$ &	${23.3}_{-5.5}^{+2.6}$ &	${27.5}_{-2.6}^{+1.9}$ & ${-0.38}_{-0.40}^{+0.43}$ &	${-3.33}_{-0.10}^{+0.22}$ &	${4.52}_{-0.23}^{+0.09}$ &	${14.09}_{-0.06}^{+0.09}$ &	${-1.08}_{-0.80}^{+0.36}$  \\
   
& \textcolor{red}{\civ_0} & ${704.6}_{-3.3}^{+2.8}$ &	${27.0}_{-2.5}^{+2.6}$ &	${7.0}_{-0.5}^{+0.5}$ &	${24.3}_{-1.8}^{+1.9}$ &	${27.8}_{-2.4}^{+2.5}$ &${1.19}_{-0.18}^{+0.21}$ &	${-4.48}_{-0.07}^{+0.08}$ &	${4.55}_{-0.07}^{+0.06}$ &	${13.30}_{-0.23}^{+0.18}$ &	${0.16}_{-0.22}^{+0.21}$ \\

SE$^{b}$  & \textcolor{blue}{\hi_0} & ${701.7}_{-0.6}^{+0.6}$ &	${7.9}_{-3.7}^{+3.6}$ &	${15.0}_{-0.9}^{+1.6}$ &	${15.0}_{-0.9}^{+1.6}$ &	${17.1}_{-1.7}^{+2.3}$ & ${-0.04}_{-0.22}^{+0.22}$ &	${-2.71}_{-0.21}^{+0.21}$ &	${4.14}_{-0.05}^{+0.09}$ &	${15.09}_{-0.36}^{+0.41}$ &	${-1.93}_{-0.17}^{+0.28}$   \\
   
& \textcolor{red}{\civ_0} & ${706.6}_{-2.8}^{+2.2}$ &	${25.4}_{-2.6}^{+3.7}$ &	${14.3}_{-0.2}^{+0.1}$ &	${49.5}_{-0.5}^{+0.3}$ &${29.2}_{-2.3}^{+3.3}$	& ${0.59}_{-0.14}^{+0.15}$ &	${-2.19}_{-1.01}^{+0.60}$ &	${5.17}_{-0.01}^{+0.00}$ &	${12.55}_{-0.27}^{+0.23}$ &	${-1.27}_{-0.63}^{+0.94}$  \\   

\hline 

SF & \textcolor{blue}{\siii_1} & ${909.9}_{-1.3}^{+1.3}$ &	${11.9}_{-0.9}^{+0.8}$ &	${1.3}_{-0.1}^{+0.3}$ &	${7.1}_{-0.4}^{+1.4}$ & 
${12.0}_{-0.9}^{+0.8}$&	${1.01}_{-0.19}^{+0.08}$ &	${-2.75}_{-0.13}^{+0.08}$ &	${3.48}_{-0.05}^{+0.16}$ &	${17.59}_{-0.13}^{+0.16}$ &	${0.30}_{-0.21}^{+0.25}$   \\    

\hline 

SG & \textcolor{blue}{\civ_0} & ${817.2}_{-2.5}^{+1.4}$ &	${17.8}_{-4.8}^{+5.0}$ &	${4.7}_{-0.5}^{+0.5}$ &	${16.1}_{-1.5}^{+1.7}$& ${18.4}_{-4.6}^{+4.9}$	& ${1.44}_{-0.12}^{+0.05}$ &	${-4.04}_{-0.11}^{+0.11}$ &	${4.20}_{-0.09}^{+0.09}$ &	${13.61}_{-0.14}^{+0.14}$ &	${-0.67}_{-0.23}^{+0.25}$ 	    \\    

& \textcolor{red}{\siiv_0} & ${861.3}_{-3.8}^{+5.5}$ &	${10.1}_{-2.2}^{+0.9}$ &	${2.6}_{-0.4}^{+0.7}$ &	${13.5}_{-2.1}^{+3.6}$ &${10.4}_{-2.1}^{+0.9}$ &	${1.14}_{-0.47}^{+0.31}$ &	${-3.58}_{-0.17}^{+0.22}$ &	${4.04}_{-0.15}^{+0.21}$ &	${14.14}_{-0.41}^{+0.57}$ &	${-1.19}_{-0.32}^{+0.47}$ 	   \\    

& \textcolor{green}{\hi_1} & ${942.8}_{-4.5}^{+5.6}$ &	${9.2}_{-8.2}^{+10.9}$ &	${13.2}_{-3.0}^{+4.5}$ &	${13.2}_{-3.0}^{+4.5}$ & ${16.5}_{-3.8}^{+7.8}$&	$<-0.05$ &	$>-2.82$ &	${4.02}_{-0.22}^{+0.25}$ &	${15.38}_{-1.05}^{+1.05}$ &	$<-1.49$  	   \\  

& \textcolor{magenta}{\hi_2} & ${1056.7}_{-4.3}^{+4.2}$ &	${11.1}_{-9.9}^{+11.3}$ &	${12.6}_{-7.0}^{+4.5}$ &	${12.6}_{-7.0}^{+4.5}$ & ${16.9}_{-4.4}^{+9.1}$&	$<-0.01$ &	$>-2.79$ &	${3.99}_{-0.70}^{+0.26}$ &	${14.21}_{-0.40}^{+0.59}$ &	$<-2.68$ 	 	   \\  

& \textcolor{brown}{\hi_3} & ${1297.2}_{-26.3}^{+27.1}$ &	${5.3}_{-4.8}^{+5.2}$ &	${13.1}_{-7.9}^{+26.7}$ &	${13.1}_{-7.9}^{+26.7}$ & 
${14.8}_{-7.8}^{+25.5}$&	$<0.54$ &	$>-3.95$ &	${4.01}_{-0.81}^{+0.97}$ &	${13.64}_{-2.39}^{+2.98}$ &	$<1.04$  		 	   \\

\hline 

SH & \textcolor{blue}{\siiii_0} & ${766.8}_{-0.8}^{+0.3}$ &	${16.3}_{-0.7}^{+0.2}$ &
${6.7}_{-0.1}^{+0.1}$ &	${35.3}_{-0.5}^{+0.5}$ &	
${17.6}_{-0.6}^{+0.2}$ &
${-1.68}_{-0.11}^{+0.10}$ &	${-3.35}_{-0.02}^{+0.02}$ &	${4.88}_{-0.01}^{+0.01}$ &	${15.43}_{-0.09}^{+0.11}$ &	${1.68}_{-0.11}^{+0.14}$   \\

& \textcolor{red}{\siiii_1} & ${818.0}_{-0.9}^{+1.3}$ &	${1.2}_{-1.0}^{+1.0}$ &	${3.4}_{-0.3}^{+0.3}$ &	${17.8}_{-1.8}^{+1.7}$ &
${3.6}_{-0.5}^{+0.5}$ &
${1.46}_{-0.09}^{+0.03}$ &	${-4.21}_{-0.14}^{+0.15}$ &	${4.29}_{-0.09}^{+0.08}$ &	${15.45}_{-0.21}^{+0.23}$ &	${1.57}_{-0.36}^{+0.34}$	   \\  

& \textcolor{green}{\hi_2} & ${899.5}_{-2.9}^{+2.7}$ &	${4.6}_{-4.1}^{+6.6}$ &	${12.4}_{-1.8}^{+2.1}$ &	${12.4}_{-1.8}^{+2.1}$ &
${13.5}_{-1.6}^{+3.4}$&
$<-0.23$ &	$>-2.25$ &	${3.97}_{-0.14}^{+0.14}$ &	${14.52}_{-0.40}^{+0.39}$ &	$<-3.56$  	   \\  

& \textcolor{magenta}{\hi_3} &${976.8}_{-7.5}^{+5.1}$ &	${40.1}_{-11.4}^{+14.3}$ &	${36.3}_{-23.6}^{+7.2}$ &	${36.3}_{-23.6}^{+7.2}$ &
${53.9}_{-7.5}^{+9.7}$ &
$<-0.04$ &	$>-3.83$ &	${4.90}_{-0.91}^{+0.16}$ &	${14.08}_{-0.05}^{+0.05}$ &	$<1.12$     \\  

\hline 

SJ & \textcolor{red}{\hi_0} &${650.6}_{-2.4}^{+2.8}$ &	${4.0}_{-3.6}^{+6.4}$ &	${9.5}_{-4.7}^{+3.7}$ &	${9.5}_{-4.7}^{+3.7}$ & ${10.9}_{-2.9}^{+3.4}$&	$<0.51$ &	$>-2.25$ &	${3.74}_{-0.60}^{+0.29}$ &	${14.17}_{-0.30}^{+0.54}$ &	$<-3.9$    \\

& \textcolor{blue}{\civ_0} &${720.1}_{-1.5}^{+1.5}$ &	${16.4}_{-3.0}^{+3.3}$ &	${7.2}_{-0.5}^{+0.5}$ &	${24.7}_{-1.6}^{+1.7}$ &	${17.9}_{-2.7}^{+3.0}$ & ${0.55}_{-0.13}^{+0.12}$ &	${-4.04}_{-0.04}^{+0.04}$ &	${4.57}_{-0.06}^{+0.06}$ &	${14.42}_{-0.08}^{+0.09}$ &	${0.51}_{-0.17}^{+0.16}$	   \\    

\hline
\end{tabular} \\

Properties of the different gas phases contributing to the absorption towards different sightlines. Notes: (1) Sightline ID (2) Constraining ion tracing a phase. In the case of SC and SE, we explore both PIE and photo-collisional models because of the detection of {\nv} absorption, earmarked with $^{a}$ and $^{b}$, respectively; (3) Heliocentric velocity of the component; (4) Non-thermal Doppler broadening parameter of constraining ion; (5) Thermal Doppler broadening parameter of the constraining ion; (6) Thermal Doppler broadening parameter measured for {\hi}; (7) Total Doppler broadening parameter of the constraining ion, obtained from the distributions in (4) and (5); (8) log metallicity; (9) log hydrogen number density; (10) log neutral hydrogen column density; (11) log temperature in Kelvin (10) log thickness in kpc. The marginalized posterior values of model parameters are given as the median along with the upper and lower bounds associated with a 95\% credible interval. The marginalized posterior distributions for the cloud properties for columns (3-11) are presented in Figures~\ref{fig:SBHI0b} - \ref{fig:SJCIV0} respectively. We present the inferred column densities and total Doppler broadening parameter ($b$), for every ion and in each model component,  in all of the sightlines in Table \ref{tab:basicinfo}.


\end{table*}

\begin{table*}
\renewcommand\thetable{4}
\begin{center}
\caption{\bf Galaxies likely to be contributing to the absorption towards different sightlines}
\label{tab:galaxyparams}
\begin{tabular}{c@{\hspace{1.5\tabcolsep}}c@{\hspace{1.5\tabcolsep}}c@{\hspace{1.5\tabcolsep}}c@{\hspace{1.5\tabcolsep}}c@{\hspace{1.5\tabcolsep}}c@{\hspace{1.5\tabcolsep}}c@{\hspace{1.5\tabcolsep}}c@{\hspace{1.5\tabcolsep}}c@{\hspace{1.5\tabcolsep}}c@{\hspace{1.5\tabcolsep}}c@{\hspace{1.5\tabcolsep}}c@{\hspace{1.5\tabcolsep}}c@{\hspace{1.5\tabcolsep}}c@{\hspace{1.5\tabcolsep}}} \hline \hline
  Sightline     & Constraining &  V$_{abs}$ &  Galaxy &   RA  &    Dec &      V$_{gal}$ & V$_{esc}$  &    $g-r$ &   $\log \frac{M_{vir}}{M_{\sun}} $   &  {\rvir} &     $\rho$  &     $\rho$/{\rvir}  & LP\\
 ID & ion cloud & (\kms) & Name & (deg) & (deg) & (\kms) & (\kms) & (mag) & & (kpc) & (kpc) &  &  \\
(1) & (2) & (3) & (4) & (5) & (6) & (7) & (8) & (9) & (10) & (11) & (12) & (13) & (14)\\
\midrule
  SB & \textcolor{blue}{\hi_0} &   722.1 & NGC  3377 & 161.93 & 13.99 & 696.1 & 66.8 & 0.7 & 11.2 & 141.0 & 543.6 & 3.9 & ${3e-7}_{-2e-7}^{+5e-8}$ \\
&&722.1 & dw1045+16 & 161.48 & 16.92 & 695.8 & 16.5 & 0.9 & 9.4 & 37.4 & 138.9 & 3.7 & ${8e-8}_{-8e-8}^{+6e-7}$ \\

\hline

SC & \textcolor{blue}{\nv_0} &              710.0 & M  95 & 160.99 & 11.7 & 700.0 & 91.7 & 0.8 & 11.5 & 181.2 & 35.1 & 0.2 & ${0.95}_{-0.67}^{+0.01}$ \\
&& 710.0 & LeG09$^{*}$ & 160.64 & 12.15 & 695.8 & 19.8 & 0.6 & 9.6 & 44.6 & 68.9 & 1.5 & ${0.05}_{-0.05}^{+0.04}$ \\

 & \textcolor{red}{\siiv_1} &
                                     716.4 & M  95 & 160.99 & 11.7 & 700.0 & 91.7 & 0.8 & 11.5 & 181.2 & 35.1 & 0.2 & ${0.93}_{-0.69}^{+0.03}$ \\
&&716.4 & M  96 & 161.69 & 11.82 & 900.0 & 111.5 & 0.9 & 11.8 & 240.6 & 179.6 & 0.7 & ${0.04}_{-0.04}^{+0.34}$ \\

& \textcolor{green}{\cii_1}  &           878.6 & M  96 & 161.69 & 11.82 & 900.0 & 111.5 & 0.9 & 11.8 & 240.6 & 179.6 & 0.7 & ${0.55}_{-0.41}^{+0.02}$ \\
&&878.6 & NGC  3384 & 162.07 & 12.63 & 840.0 & 113.8 & 0.7 & 11.7 & 173.9 & 259.3 & 1.5 & ${0.1}_{-0.08}^{+0.01}$ \\
&& 878.6 & M 105 & 161.96 & 12.58 & 950.0 & 96.0 & 0.9 & 11.6 & 207.8 & 283.5 & 1.4 & ${0.09}_{-0.08}^{+0.07}$ \\

  & \textcolor{magenta}{\hi_0} &    1035.8 & M  96 & 161.69 & 11.82 & 900.0 & 111.5 & 0.9 & 11.8 & 240.6 & 179.6 & 0.7 & ${0.13}_{-0.13}^{+0.42}$ \\
&&1035.8 & M 105 & 161.96 & 12.58 & 950.0 & 96.0 & 0.9 & 11.6 & 207.8 & 283.5 & 1.4 & ${0.07}_{-0.07}^{+0.09}$ \\
\midrule

 SD & \textcolor{blue}{\siiv_0}     & 844.4 & M 105 & 161.96 & 12.58 & 915.0 & 96.0 & 0.9 & 11.6 & 207.8 & 118.1 & 0.6 & ${0.42}_{-0.38}^{+0.30}$ \\
&&844.4 & NGC  3384 & 162.07 & 12.63 & 740.0 & 113.8 & 0.7 & 11.7 & 173.9 & 99.2 & 0.6 & ${0.31}_{-0.29}^{+0.41}$ \\
&&844.4 & M  96 & 161.69 & 11.82 & 780.0 & 111.5 & 0.9 & 11.8 & 240.6 & 293.1 & 1.2 & ${0.16}_{-0.14}^{+0.07}$ \\
&&844.4 & NGC  3412 & 162.72 & 13.41 & 864.9 & 71.6 & 0.7 & 11.3 & 156.3 & 214.0 & 1.4 & ${0.14}_{-0.11}^{+0.01}$ \\

& \textcolor{red}{\hi_0} &   978.4 & M 105 & 161.96 & 12.58 & 915.0 & 96.0 & 0.9 & 11.6 & 207.8 & 118.1 & 0.6 & ${0.47}_{-0.42}^{+0.25}$ \\
&&978.4 & NGC  3412 & 162.72 & 13.41 & 864.9 & 71.6 & 0.7 & 11.3 & 156.3 & 214.0 & 1.4 & ${0.01}_{-0.01}^{+0.1}$ \\
\midrule
SE & \textcolor{blue}{\hi_0}      &            700.7 & NGC  3377 & 161.93 & 13.99 & 696.1 & 66.8 & 0.7 & 11.2 & 141.0 & 221.6 & 1.6 & ${0.08}_{-0.06}^{+0.00}$ \\
&&700.7 & NGC  3384 & 162.07 & 12.63 & 732.1 & 113.8 & 0.7 & 11.7 & 173.9 & 488.6 & 2.8 & ${3e-4}_{-3e-4}^{+3e-5}$ \\

& \textcolor{red}{\civ_0}      &           704.6 & NGC  3377 & 161.93 & 13.99 & 696.1 & 66.8 & 0.7 & 11.2 & 141.0 & 221.6 & 1.6 & ${0.08}_{-0.06}^{+0.00}$ \\
&&704.6 & NGC  3384 & 162.07 & 12.63 & 732.1 & 113.8 & 0.7 & 11.7 & 173.9 & 488.6 & 2.8 & ${3e-4}_{-3e-4}^{+2e-5}$ \\

    \midrule
SF & \textcolor{blue}{\siii_1}         &   909.9 & M 105 & 161.96 & 12.58 & 910.0 & 96.0 & 0.9 & 11.6 & 207.8 & 105.2 & 0.5 & ${0.77}_{-0.49}^{+0.0}$ \\
&&909.9 & M  96 & 161.69 & 11.82 & 800.0 & 111.5 & 0.9 & 11.8 & 240.6 & 111.1 & 0.5 & ${0.31}_{-0.29}^{+0.5}$ \\
&&909.9 & NGC  3384 & 162.07 & 12.63 & 800.0 & 113.8 & 0.7 & 11.7 & 173.9 & 94.5 & 0.5 & ${0.29}_{-0.27}^{+0.45}$ \\
 \midrule
 SG & \textcolor{blue}{\civ_0} &          817.2 & NGC  3384 & 162.07 & 12.63 & 720.0 & 113.8 & 0.7 & 11.7 & 173.9 & 91.0 & 0.5 & ${0.37}_{-0.35}^{+0.39}$ \\
&&817.2 & M 105 & 161.96 & 12.58 & 895.0 & 96.0 & 0.9 & 11.6 & 207.8 & 126.8 & 0.6 & ${0.36}_{-0.33}^{+0.33}$ \\
&&817.2 & NGC  3412 & 162.72 & 13.41 & 864.9 & 71.6 & 0.7 & 11.3 & 156.3 & 135.6 & 0.9 & ${0.3}_{-0.27}^{+0.17}$ \\
&&817.2 & UGC05944$^{*}$ & 162.58 & 13.27 & 810.0 & 27.5 & 0.6 & 10.0 & 61.6 & 88.0 & 1.4 & ${0.12}_{-0.09}^{+0.01}$ \\
 & \textcolor{red}{\siiv_0} &    861.3 & M 105 & 161.96 & 12.58 & 895.0 & 96.0 & 0.9 & 11.6 & 207.8 & 126.8 & 0.6 & ${0.61}_{-0.5}^{+0.08}$ \\
&&861.3 & NGC  3412 & 162.72 & 13.41 & 864.9 & 71.6 & 0.7 & 11.3 & 156.3 & 135.6 & 0.9 & ${0.47}_{-0.31}^{+0.0}$ \\
&&861.3 & NGC  3384 & 162.07 & 12.63 & 720.0 & 113.8 & 0.7 & 11.7 & 173.9 & 91.0 & 0.5 & ${0.16}_{-0.16}^{+0.56}$ \\
 & \textcolor{green}{\hi_1} &   942.8 & M 105 & 161.96 & 12.58 & 895.0 & 96.0 & 0.9 & 11.6 & 207.8 & 126.8 & 0.6 & ${0.54}_{-0.47}^{+0.15}$ \\
&&942.8 & NGC  3412 & 162.72 & 13.41 & 864.9 & 71.6 & 0.7 & 11.3 & 156.3 & 135.6 & 0.9 & ${0.14}_{-0.13}^{+0.33}$ \\
    
 & \textcolor{magenta}{\hi_2} &         1056.7 & M 105 & 161.96 & 12.58 & 895.0 & 96.0 & 0.9 & 11.6 & 207.8 & 126.8 & 0.6 & ${0.04}_{-0.04}^{+0.39}$ \\
 &&1056.7 & NGC  3412 & 162.72 & 13.41 & 864.9 & 71.6 & 0.7 & 11.3 & 156.3 & 135.6 & 0.9 & ${4e-4}_{-4e-4}^{+0.03}$ \\
          
   & \textcolor{brown}{\hi_3}  &  1297.2 &    NGC  3389 & 162.12 & 12.53 & 1298.1 & 88.4 & 0.3 & 11.4 & 130.1 & 190.1 & 1.5 & ${0.12}_{-0.08}^{+0.0}$ \\
&&1297.2 & NGC  3338 & 160.53 & 13.75 & 1301.8 & 77.5 & 0.8 & 11.4 & 168.8 & 576.5 & 3.4 & ${9e-6}_{-6e-6}^{+3e-8}$ \\
\midrule
SH & \textcolor{blue}{\siiii_0}  &         766.8 & NGC  3384 & 162.07 & 12.63 & 680.0 & 113.8 & 0.7 & 11.7 & 173.9 & 147.5 & 0.8 & ${0.27}_{-0.25}^{+0.22}$ \\
&&766.8 & M  96 & 161.69 & 11.82 & 800.0 & 111.5 & 0.9 & 11.8 & 240.6 & 347.5 & 1.4 & ${0.11}_{-0.09}^{+0.01}$ \\
 & \textcolor{red}{\siiii_1}  &             818.0 & NGC  3412 & 162.72 & 13.41 & 864.9 & 71.6 & 0.7 & 11.3 & 156.3 & 139.7 & 0.9 & ${0.29}_{-0.26}^{+0.16}$ \\
&&818.0 & M 105 & 161.96 & 12.58 & 900.0 & 96.0 & 0.9 & 11.6 & 207.8 & 203.2 & 1.0 & ${0.19}_{-0.18}^{+0.19}$ \\
&&818.0 & M  96 & 161.69 & 11.82 & 800.0 & 111.5 & 0.9 & 11.8 & 240.6 & 347.5 & 1.4 & ${0.12}_{-0.09}^{+0.0}$ \\
&&818.0 & NGC  3384 & 162.07 & 12.63 & 680.0 & 113.8 & 0.7 & 11.7 & 173.9 & 147.5 & 0.8 & ${0.11}_{-0.11}^{+0.36}$ \\

 & \textcolor{green}{\hi_2} &                899.5 & M 105 & 161.96 & 12.58 & 900.0 & 96.0 & 0.9 & 11.6 & 207.8 & 203.2 & 1.0 & ${0.38}_{-0.24}^{+0.0}$ \\
&&899.5 & NGC  3412 & 162.72 & 13.41 & 864.9 & 71.6 & 0.7 & 11.3 & 156.3 & 139.7 & 0.9 & ${0.36}_{-0.31}^{+0.09}$ \\
 & \textcolor{magenta}{\hi_3} &          976.8 & M 105 & 161.96 & 12.58 & 900.0 & 96.0 & 0.9 & 11.6 & 207.8 & 203.2 & 1.0 & ${0.2}_{-0.18}^{+0.18}$ \\
&&976.8 & NGC  3412 & 162.72 & 13.41 & 864.9 & 71.6 & 0.7 & 11.3 & 156.3 & 139.7 & 0.9 & ${0.04}_{-0.04}^{+0.29}$ \\
\midrule

\end{tabular} \\

\end{center}

\end{table*}

\begin{table*}
\renewcommand\thetable{3}
\begin{center}

\begin{tabular}{c@{\hspace{1.5\tabcolsep}}c@{\hspace{1.5\tabcolsep}}c@{\hspace{1.5\tabcolsep}}c@{\hspace{1.5\tabcolsep}}c@{\hspace{1.5\tabcolsep}}c@{\hspace{1.5\tabcolsep}}c@{\hspace{1.5\tabcolsep}}c@{\hspace{1.5\tabcolsep}}c@{\hspace{1.5\tabcolsep}}c@{\hspace{1.5\tabcolsep}}c@{\hspace{1.5\tabcolsep}}c@{\hspace{1.5\tabcolsep}}c@{\hspace{1.5\tabcolsep}}c@{\hspace{1.5\tabcolsep}}} \hline \hline
  Sightline     & Constraining &  V$_{abs}$ &  Galaxy &   RA  &    Dec &      V$_{gal}$ & V$_{esc}$  &    $g-r$ &   $\log \frac{M_{vir}}{M_{\sun}} $   &  {\rvir} &     $\rho$  &     $\rho$/{\rvir}  & LP\\
 ID & ion cloud & (\kms) & Name & (deg) & (deg) & (\kms) & (\kms) & (mag) & & (kpc) & (kpc) &  &  \\
(1) & (2) & (3) & (4) & (5) & (6) & (7) & (8) & (9) & (10) & (11) & (12) & (13) & (14)\\
\midrule

SI & No absorption  &   &              M  96 &  161.69 &  11.82 &   903.0 &  111.5 &  0.9 &      11.8 &    240.6 &           474.2 &             2.0&- \\
 &&   &                 M  95 &  160.99 &  11.70 &   767.5 &   91.7 &  0.8 &      11.5 &    181.2 &             490.7 &             2.7 &-\\
&&&                 NGC  3384 &  162.07 &  12.63 &   732.1 &  113.8 &  0.7 &      11.7 &    173.9 &             476.3&             2.7 &-\\
 && &                   M 105 &  161.96 &  12.58 &   876.0 &   96.0 &  0.9 &      11.6 &    207.8 &             570.3 &             2.7 &-\\

 \midrule

SJ & \textcolor{blue}{\hi_0}  &           650.6 & NGC  3489 & 165.08 & 13.9 & 697.0 & 68.6 & 3.8 & 11.2 & 149.8 & 144.1 & 1.0 & ${0.25}_{-0.23}^{+0.15}$ \\
&&650.6 & LSBCD640-08$^{*}$ & 165.22 & 13.88 & 656.5 & 21.6 & 0.5 & 9.7 & 48.7 & 144.5 & 3.0 & ${1e-4}_{-1e-4}^{+1e-5}$ \\

    & \textcolor{red}{\civ_0}  &              720.1 & NGC  3489 & 165.08 & 13.9 & 697.0 & 68.6 & 3.8 & 11.2 & 149.8 & 144.1 & 1.0 & ${0.35}_{-0.28}^{+0.05}$ \\
&&720.1 & LeG33 & 165.19 & 14.17 & 695.8 & 15.7 & 0.7 & 9.3 & 35.6 & 102.9 & 2.9 & ${2e-5}_{-2e-5}^{+1e-3}$ \\
        
\midrule

SK & No absorption  &    &              NGC  3489 &  165.08 &  13.90 &   697.0 &   68.6 &  3.8 &      11.2 &    149.8 &           305.2 &             2.0&- \\
    &&&             dw1101+11$^{*}$ &  165.34 &  11.75 &   695.8 &   15.3 &  0.0 &       9.3 &     34.4 &         101.2&             2.9 &-\\
    &&&             NGC  3384 &  162.07 &  12.63 &   732.1 &  113.8 &  0.7 &      11.7 &    173.9 &             540.8 &             3.1 &-\\
     &&&                M  96 &  161.69 &  11.82 &   903.0 &  111.5 &  0.9 &      11.8 &    240.6 &           751.1 &             3.1&- \\
    \midrule
    
SL & No absorption  &     &   dw1059+11$^{*}$ &  164.96 &  11.43 &   695.8 &   15.9 &  0.4 &       9.3 &     36.1 &         124.4 &             3.4&- \\
   &&&                  M  96 &  161.69 &  11.82 &   903.0 &  111.5 &  0.9 &      11.8 &    240.6 &           833.0&             3.5&- \\
  &&&               NGC  3489 &  165.08 &  13.90 &   697.0 &   68.6 &  3.8 &      11.2 &    149.8 &           560.6 &             3.7 &-\\
   &&&              NGC  3384 &  162.07 &  12.63 &   732.1 &  113.8 &  0.7 &      11.7 &    173.9 &             673.3 &             3.9&- \\ 
  \hline  
\end{tabular} 

\end{center}

Notes: (1) Sightline ID (2) Constraining ion (3) Absorption centroid (4) Galaxy name (5) Right ascension in degrees; (6) Declination in degrees; (7) Galaxy's systemic velocity. For galaxies earmarked with $\star$, we derive systemic velocity based on the adopted distance modulus of 30.06~\citep{muller2018}; (8) Escape velocity of the Galaxy halo; (9) $g-r$-band SDSS AB magnitude; (10) and (11) The virial mass and the virial radius determined according to the combined ``halo matching'' and constant mass-to-light ratio technique; (12) Impact parameter in kpc (13) Ratio of impact parameter to the virial radius; (14) Likelihood parameter determined using equation~\ref{eqn:1}. We present an estimate obtained by setting V$_{gal}$ equal to the velocity in the direction of rotation map extension, and the bounds represent the minimum and maximum values of the LP function within the range $V_{abs} - (V_{gal}+V_{esc}) <= \Delta V <= V_{abs} - (V_{gal}-V_{esc})$. We list at least two galaxies or all galaxies with LP estimate $>$0.1. Columns 4-7 are obtained from SIMBAD database~\citep{simbad}. 

\end{table*}

\begin{figure*}

\includegraphics[width=\linewidth]{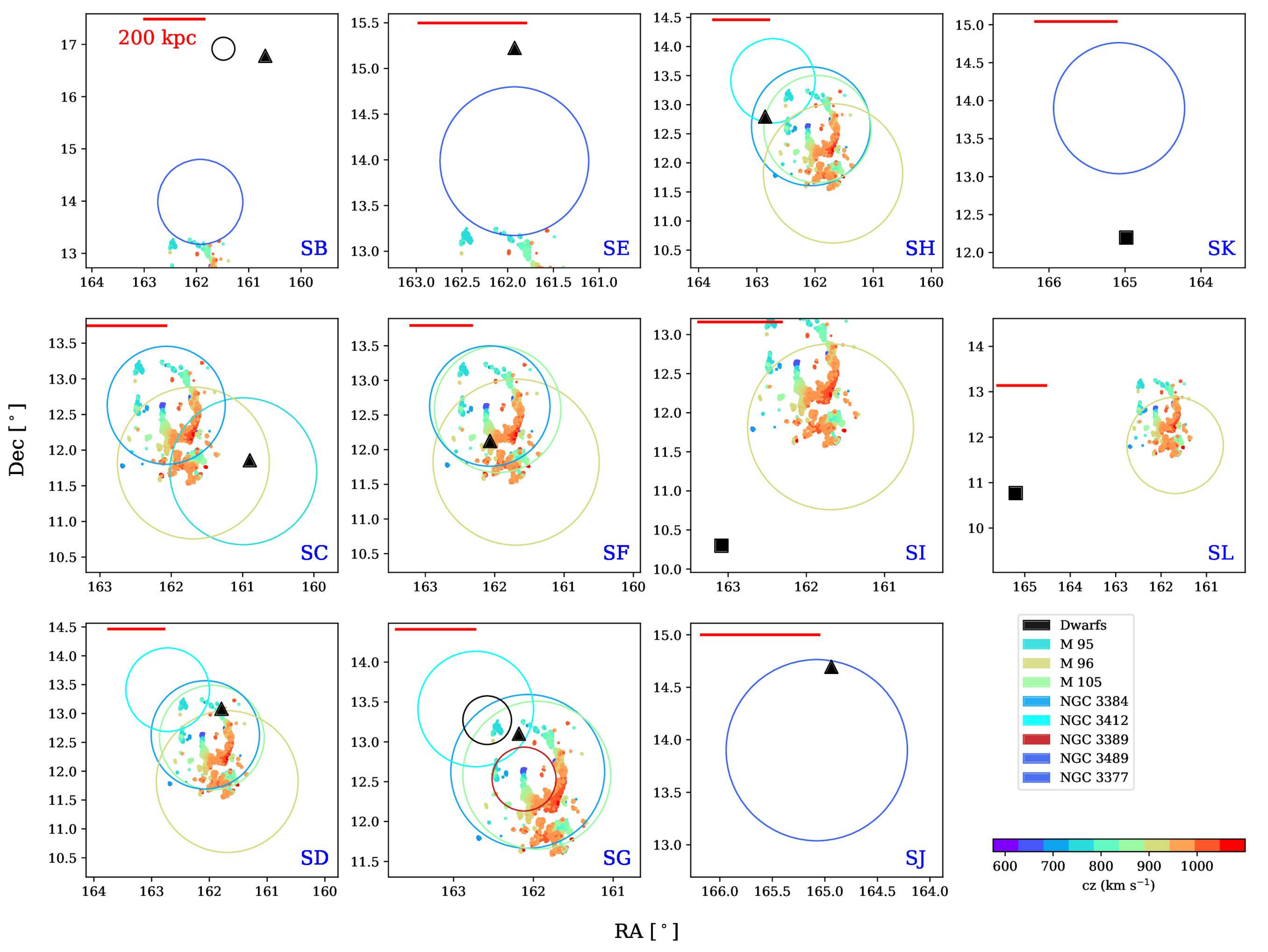}
\caption{Plots showing the galaxies that are potentially contributing to the absorption along the respective sightline indicated at the bottom-right of each panel. We show only those galaxies which have a LP $\gtrsim$ 0.1. The filled triangle denotes the location of the quasar sightline. The circle is centered at the location of the galaxy, and its radius corresponds to the virial radius of the galaxy. We indicate major galaxies in different colors corresponding to their $cz$, while the dwarfs are shown in black. We also overlay the Leo Ring as seen in Figure~\ref{fig:LeoRingfield}. The red horizontal line at the top-left in each panel corresponds to 200 kpc at the redshift of the absorber.}
\label{fig:galaxiesvirial}

\end{figure*}

\subsection{SB: The $z = 0.00240$ absorber towards the quasar J1043+1151}
\label{sec:SB}

\begin{figure*}
\begin{center}
\includegraphics[width=\linewidth]{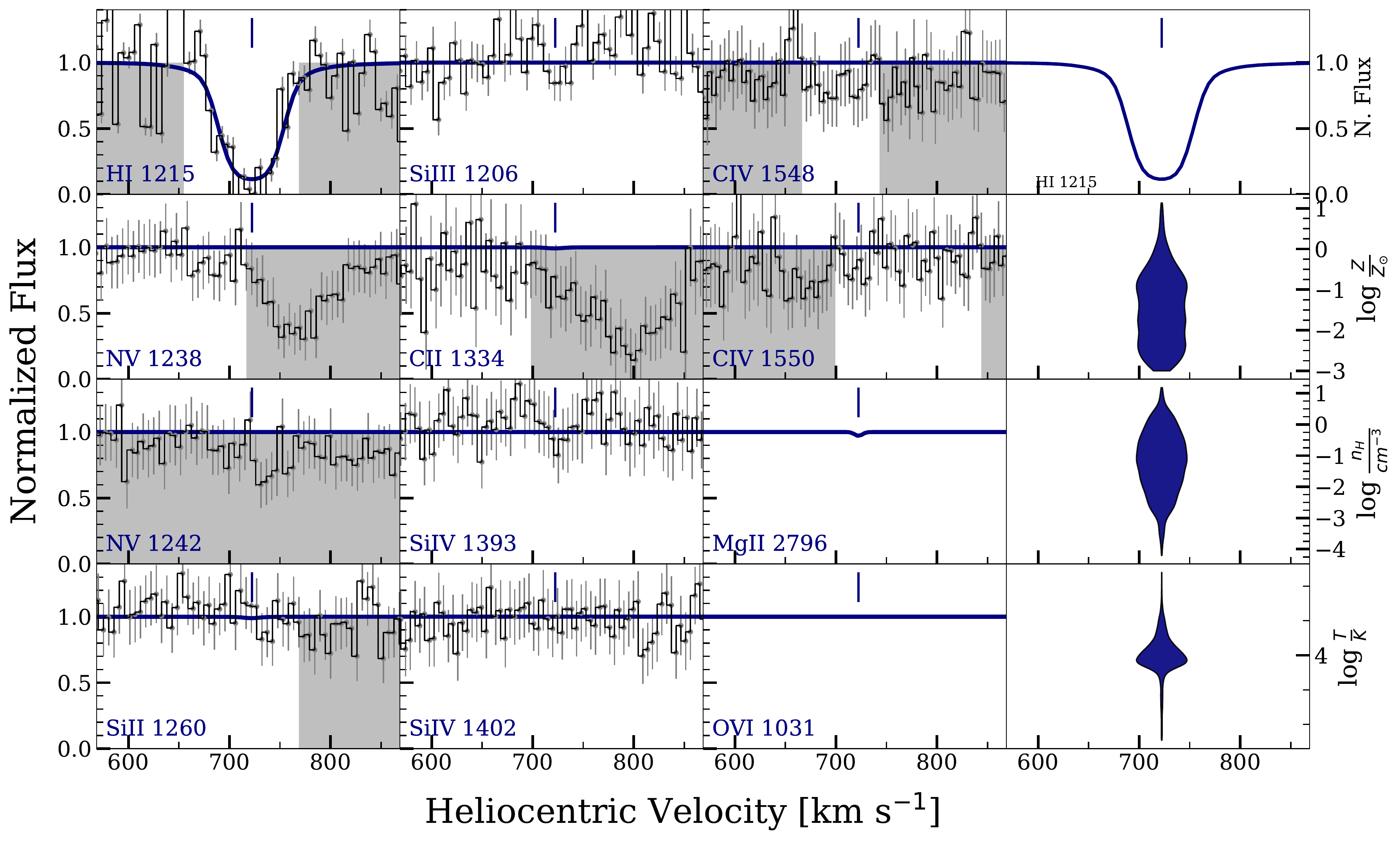}
\caption{\CLOUDY~models for the $z = 0.0024$ absorber towards SB, obtained using the MLE values of the best fit model. The models are superimposed on the observed spectra obtained with {\hst}/COS in the wavelength regions corresponding to important constraining transitions. 1$\sigma$ error bars are superimposed on the data. The centroid of the {\hi} {\lya} absorption component as determined from the VP fit is indicated by the vertical tick mark on top of each panel. The photoionized gas phase is traced only by the {\hi} cloud (shown as a blue curve). A summary of the properties of this cloud are shown in the form of violin plots using filled blue color. 
the width of the filled blue shapes to represent the relative probability densities of the values given on the vertical axes.
The regions shaded in grey show the pixels that were not used in the evaluation of the log-likelihood. We show only the transitions that might be detected for an absorber with this relatively weak {\hi} {\lya} line.}
\label{fig:SBsysplot}
\end{center}
\end{figure*}

A system plot of the $z$ = 0.0024 absorber, with {\hi} {\lya} as the constraining transition, is shown in Figure~\ref{fig:SBsysplot}. This system shows absorption in {\hi} {\lya} only. We do not see low, intermediate, or high ionization, metal line absorption. 

\smallskip

We model the absorption in {\hi} {\lya} using a single component. With access to only the {\hi} {\lya} profile among Lyman series lines, and with no detections of metal lines, the properties of this system are poorly constrained.
The {\colden} is determined to be ${15.3}_{-0.8}^{+1.0}$. Ionization modeling yields a 2$\sigma$ upper limit on the metallicity of \metallicity \hbox{$\lesssim -0.35$}, a hydrogen number density, \hden $> -2.6$, and a small line of sight thickness of $<$1 pc. The temperature is determined to be relatively well constrained at {\temp} $\approx$ 4. These limits are primarily due to the relatively narrow {\hi} {\lya} profile that does not allow for a lower density/higher temperature. 

\smallskip

The physical properties of this absorption system are summarized in Table~\ref{tab:modelparams}. The {\CLOUDY} model corresponding to the maximum likelihood estimate (MLE) is shown overlaid on the spectral data in Figure~\ref{fig:SBsysplot}. We show the properties of this {\hi} cloud graphically using violin plots to the right of the system plot. The violin plots show the probability density for different values of metallicity, hydrogen density, and temperature. In Figures \ref{fig:SBHI0b} and~\ref{fig:SBHI0}, the probability density functions (PDFs) of the properties, in the form of corner plots, for the {\hi}-only cloud are presented.

\subsubsection{Galaxy Properties and Physical Interpretation}

The closest galaxy which could contribute to the absorption at the redshift of the absorber is a dwarf galaxy, dw1045+16, consistent with a velocity of c$z$ $\approx$700 {\kms}. However, the absorber is at a large projected separation of $\approx$ 3.7 {\rvir} from that galaxy. The large separation results in a very low LP of $\sim10^{-7}$. Figure~\ref{fig:galaxiesvirial} shows the two nearest galaxies along with their virial radii relative to the position of SB. Table \ref{tab:galaxyparams} lists the properties of these galaxies. Given these large separations, the weak {\hi} {\lya} absorption ($\log N({\HI})$ $\approx$ 15.3) is as likely to be associated with a large scale filamentary structure as with any individual galaxy~\citep{penton2002,wakker2015,bouma2021}. The velocity of an extension of the Leo Ring {\hi} gas is large (~$\approx$ 850~{\kms}) compared with the velocity of this absorption at 700~{\kms},  and also SB is at an impact parameter of $\approx$750~kpc from the detected {\hi}.  It is therefore not likely to be directly related to gas released by the interaction.

\subsection{SC: The $z = 0.00238$ absorber towards the quasar J1043+1151}
\label{sec:SC}

\begin{figure*}
\begin{center}
\includegraphics[width=\linewidth]{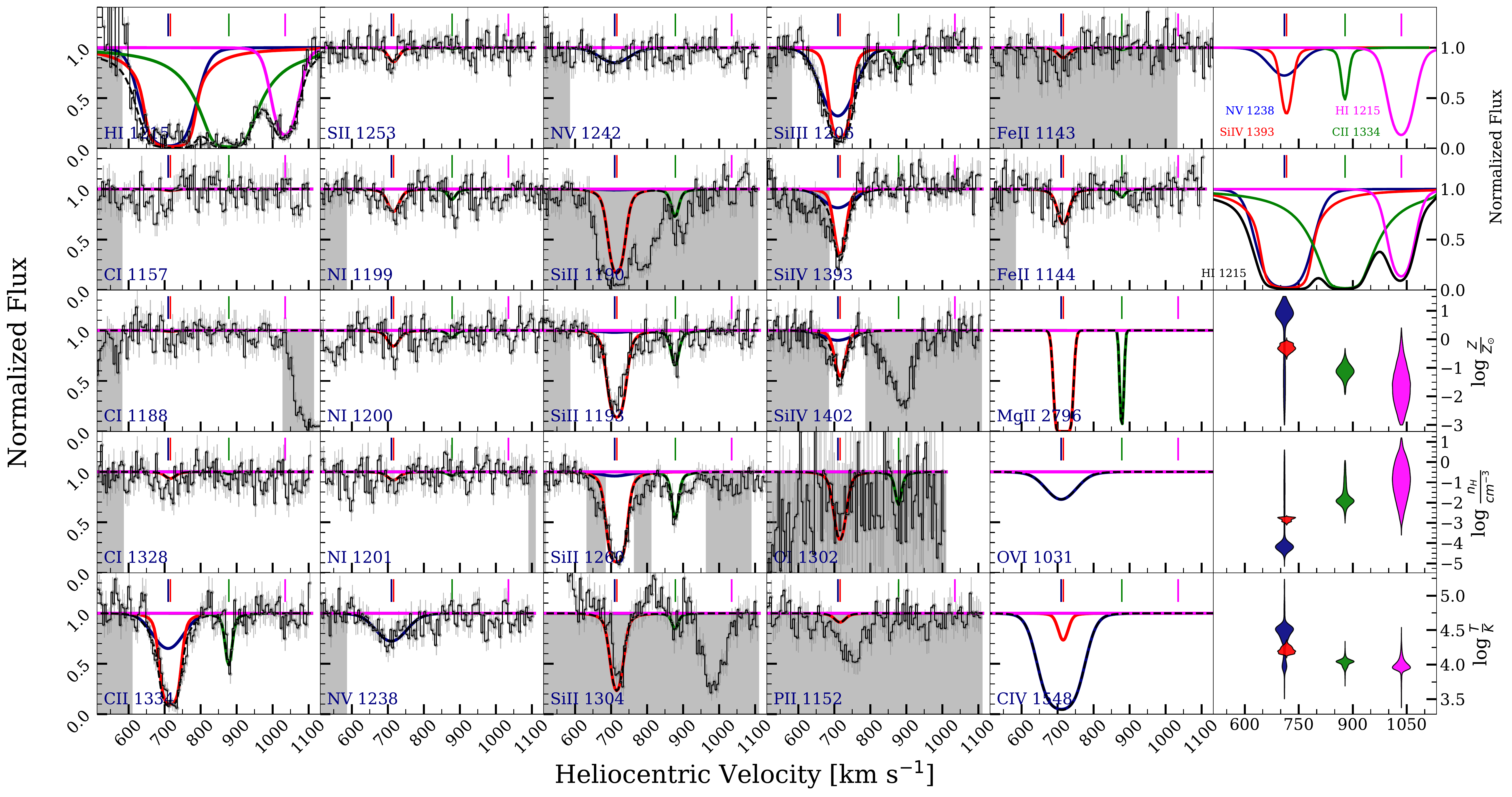}
\caption{\CLOUDY~models for the $z = 0.00238$ absorber towards SC obtained using the MLE values, superimposed on the {\hst}/COS spectra, displayed with one sigma errors. The centroids of absorption components as determined from the VP fits to constraining ions are indicated by the vertical tick marks on top of each panel. The photoionized gas phases are traced by four clouds - two coincident phases, one of which is a higher ionization phase tracing the broad {\nv} absorption (blue curve) and the other tracing the bulk of the lower ionization absorption (shown as a red curve), the offset redward cloud traced by {\cii} (shown as a green curve), and the {\hi}-only cloud (shown as a magenta curve). The superposition of these four models is shown by the black dashed curve.  A summary of the properties of these four clouds is shown in the form of violin plots using filled colors. The regions shaded in grey show the pixels that were not used in the evaluation of the log-likelihood.}
\label{fig:SCsysplot}
\end{center}
\end{figure*}

A system plot of the $z=0.00238$ absorber, with the constraining transitions, is shown in Figure~\ref{fig:SCsysplot}. This system shows absorption in the low-ionization transitions of \nitri, \cii, \Sii, \siii, and \feii, and the intermediate-ionization transitions of {\siiii} and {\siiv}. The system would classify as a strong {\mgii} absorber based on the observed strengths of {\siii} and {\cii}~\citep{narayanan2005survey}.

\smallskip

We find that a two component model is able to best explain the absorption centered at $\approx$710 {\kms}, with two nearly coincident phases of gas, a broad higher ionization component producing {\nv} absorption, and a narrow lower ionization component which even produces detected weak {\nitri} absorption. Both of these clouds contribute to the absorption detected in {\cii}, {\siiii}, and {\siiv}. We also observe absorption in {\cii} and {\siii} at $\approx$878{\kms}, which we model using a single narrow component. There is also absorption seen in {\hi} at $\approx$1032{\kms}, with no associated metals detected.

\smallskip

The absorption centered at $\approx$710 {\kms} needs a high ionization gas phase traced by {\nv} (shown as a blue curve) to explain the broad absorption associated with it. This phase of gas also explains some of the absorption seen in the wings of {\cii} and {\siiii}.  A lower ionization gas phase traced by narrow {\siiv} (shown as red curve) explains most of the absorption seen in these same transitions. The metallicity of the gas phase traced by {\nv} is \metallicity $\approx$0.8, with a low hydrogen density \hden $\approx -4.2$, and a high temperature of \temp $\approx$ 4.5. This cloud also predicts significant {\civ} absorption along with modest absorption in {\ovi}. The lower ionization gas phase traced by {\siiv} produces the dominant absorption seen in transitions of {\siii}, {\cii}, and {\siiii}, with a neutral hydrogen column density of \colden $\approx$ $18.0$, a density of \hden $\approx -$ 2.7, and a metallicity of \metallicity $\approx  -0.1$.  We also observe some {\civ} absorption arising from this cloud. The two clouds constituting this absorption have line of sight thicknesses of $\approx$ 8 kpc and 5 kpc, respectively. 

\smallskip

We also explore the possibility of {\nv} arising in a collisionally ionized phase by using PC models to explain the {\nv} absorption. The metallicities are quite similar to the values obtained in the photoionized case. The temperature for the {\nv} phase is, however, high with {\temp} $\approx$5.5. We present the results for both the PIE and the PC model in Table~\ref{tab:modelparams}, and the models superimposed on the data are presented in Figures~\ref{fig:SCsysplot} and \ref{fig:SCpici}, respectively. The most plausible origin of the high ionization {\nv} phase could be better determined if the {\civ} and {\ovi} were covered. We cannot discriminate between the two models due to the lack of coverage of the other higher ionization transitions.  The constraints on the {\siiv} cloud at $\approx$716~{\kms} are affected by which model for the origin of the {\nv} absorption applies, however, the difference is only a $0.3$ dex reduction in \metallicity and a $0.2$ dex reduction in \hden when the {\nv} is collisionally ionized.

\smallskip

Interestingly, the lower ionization gas phase at $\approx$ 716 {\kms} shows appreciable detections in {\nitri} and {\oi}. For the assumed solar abundance pattern, the inferred metallicity of this phase is consistent with solar metallicity within the uncertainties.

\smallskip

The absorption centered at $\approx$878 {\kms} produces damping wings in {\hi} {\lya} with a column density of {\colden} $\approx$ 18.4. This gas phase explains the narrow absorption profiles seen in {\cii} and \siii. The metallicity is constrained to be \metallicity $\approx -1.0$. Only low ionization and intermediate ionization absorption is observed to arise in this gas phase with a density of \hden $\approx -1.9$. The line of sight thickness is constrained to be $\approx$ 0.2 kpc.

\smallskip

The absorption centered at $\approx$1032{\kms} does not show a clear detection of absorption in metals. There is a possible detection of {\siii}$\lambda$1260 at 2$\sigma$ level at this velocity. The absence of absorption in {\siii}$\lambda\lambda$1190, 1193, and in {\cii}, however, argues against this identification. The background quasar has a relatively high redshift, so its spectrum has a higher density of lines and this candidate {\siii} is ruled as a blend, and the spectral region is masked out when modelling. We obtain an upper limit on the metallicity of $\lesssim -0.4$. The relatively narrow {\hi} absorption constrains the temperature to be {\temp} $\approx$ 4.

\smallskip

In Table~\ref{tab:modelparams} we present the posterior results for the four clouds identified in this absorption system. Figure~\ref{fig:SCsysplot} shows {\CLOUDY} models, using the MLE values superimposed on the data. The probability density distributions of metallicity, hydrogen number density, and temperature corresponding to each one of the constraining ions are summarized in the form of violin plots. The PDFs of the properties of the different clouds are presented using corner plots in Figures~\ref{fig:SCNV0b} -- \ref{fig:SCHI0}.  

\smallskip

We explore the consequences of zero-point uncertainty of the {\hi} {\lya} flux for the components between 710--720 {\kms}. This uncertainty could arise through a contribution to the observed flux from wings to the geocoronal airglow emission as we described in \S~\ref{sec:observations}. The violin plots showing the distribution of parameters for 100 MC realizations (with different zero point values) are presented in Figure~\ref{fig:MCSC}. We find that the posteriors for various phases are still in agreement within uncertainties of the values presented in Table~\ref{tab:modelparams} for SC$^{b}$.

\subsubsection{Galaxy Properties and Physical Interpretation}

The nearest galaxy which could be responsible for the dominant absorption at $\approx$710 {\kms} is M95, which is at an impact parameter of 35kpc/0.2\rvir. An extension of its rotation curve in the direction of SC would be consistent with the absorption at $\approx$710 {\kms}. The LP for M95 to explain these components is determined to be $\sim 0.9$. There is also a dwarf galaxy candidate LeG09, and it has a LP of $\approx 0.04$, so it could also contribute. However, given the predicted strength of {\civ} absorption from this phase, it is unlikely to have a significant contribution from a dwarf galaxy at a separation of 1.5\rvir~\citep{bordoloi2014cos}. Moreover, the Lyman Limit system, with {\colden} $\approx$ 18 and solar metallicity, in the low ionization phase at this velocity is consistent with gas arising in {\hii} regions of low-$z$ star forming-galaxies~\citep{pilyugin2014}, and not with dwarf galaxies (see Figure 10 in \citealt{zahedy2021}). M96 could also contribute, and shows an LP of $\approx$ 0.04, however the absorption is much beyond the escape velocity of halo of M96. Hence these high metallicity components, \metallicity $\approx$ 0.8 and $\approx -0.1$ are highly likely to be associated with M95.

\smallskip

The absorption at $\approx$879 {\kms} has a LP of $\approx 0.5$ for association with M96. M96 is at an impact parameter of 0.7\rvir from SC and an extension of its rotation curve would be consistent with producing absorption at $\approx$900 {\kms}. The next likely galaxy candidate is NGC 3384 which is at an impact parameter of 1.5\rvir and has a LP of $\sim$ 0.1. M105 is at an impact parameter 1.4\rvir and it also has a LP of $\lesssim$0.1. These galaxies along with their virial radii are shown in Figure~\ref{fig:galaxiesvirial}. Based on the predicted EW$_{\mgii, \lambda2796}$ $\approx$ 0.12\AA, we do not expect contribution to absorption from NGC 3384 or M 105, both of which are at $>$ 1.4\rvir~\citep{Nielsen2013}. We suggest that this sub-solar metallicity component, \metallicity$\approx-1.0$, is associated with M96. 

\smallskip

The Leo Ring itself is thought to have formed from a head on collision between M96 and NGC 3384 (now at the center of the ring)~\citep{dansac2010}. The gas at the lower right of the ring (Fig. ~\ref{fig:LeoRingfield}), closest to M96, is at a higher velocity than the two galaxies, $\approx$ 1050 {\kms}, and so cannot arise directly from the galaxies themselves. The absorption that we see only in {\hi} {\lya} at 1035 {\kms} could share the kinematics with the lower right of the ring, though the sightline is at an impact parameter of $\approx100$~kpc from that location. It is also possible that the absorption is unrelated to the interaction that created the ring.

\subsection{SD: The $z = 0.00279$ absorber towards the quasar J1047+1304}
\label{sec:SD}

\begin{figure*}
\begin{center}
\includegraphics[width=\linewidth]{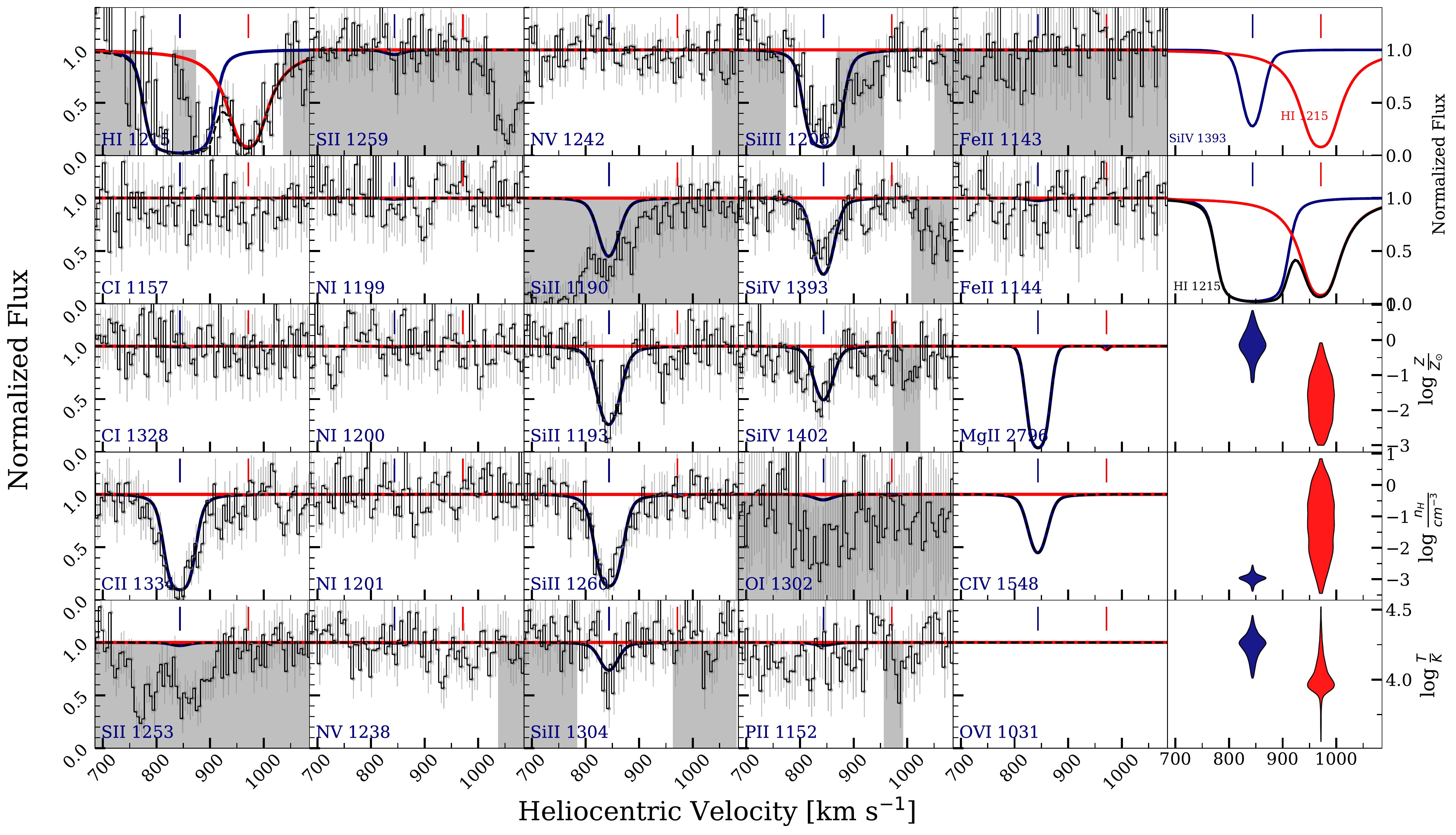}
\caption{\CLOUDY~models for the $z = 0.00279$ absorber towards SD obtained using the MLE values, superimposed on the {\hst}/COS spectra, displayed with their $1\sigma$ errors. The centroids of absorption components as determined from the VP fits to constraining ions are indicated by the vertical tick marks on top of each panel. The photoionized gas phases are traced by two clouds - the blueward cloud (shown as blue curve), and the offset redward cloud (shown as a red curve). The superposition of these two models is shown by the black dashed curve. A summary of the properties of these two clouds is shown in the form of violin plots using filled colors. The region shaded in grey shows the pixels that were not used in the evaluation of the log-likelihood.}
\label{fig:SDsysplot}
\end{center}
\end{figure*}

A system plot of the $z = 0.00279$ absorber with the constraining transitions is shown in Figure~\ref{fig:SDsysplot}. This system shows absorption in the low ionization transitions of {\cii} and {\siii}, and the intermediate ionization transitions of {\siiii} and {\siiv}. Absorption is not detected in the lower ionization (neutral) transitions. The higher ionization transitions, {\civ} and {\ovi} are not covered by the observations, but {\nv} is not detected. We model the absorption as two components, one component tracing the absorption seen in low and intermediate ionization transitions, at $\approx$845 {\kms}, and one component in {\hi}, without any metals,  at $\approx$978 {\kms}. Table~\ref{tab:modelparams} summarizes the posterior results for these two clouds. We observe a peculiarity in the {\hi} {\lya} profile at $\approx$ 845 {\kms}; the flux shows a sharp discontinuity from the saturation level at this velocity which is unexpected based on the observed metal absorption. The profile at this velocity is therefore assumed to be saturated.

\smallskip

The blueward cloud at $\approx$845 {\kms} is consistent with solar metallicity, has a density of \hden $\approx -3.0$, and a temperature of \temp $\approx 4.25$. The line of sight thickness is $L \approx 4$ kpc. The {\hi} cloud at 978 {\kms} is unconstrained. We determine an upper limit on the metallicity of \metallicity $\lesssim - 0.65$. The density is determined to be $\gtrsim -2.85$, and the line of sight thickness is an upper limit, $L \lesssim 60$pc. The properties of the two clouds are sketched in Figure~\ref{fig:SDsysplot}. The PDFs of the properties of the two clouds are presented using corner plots in Figures~\ref{fig:SDSiIV0b} -- \ref{fig:SDHI2b}. 

\smallskip

Previously, the absorption at $z=$ 0.00279 towards this quasar was studied by \citet{rosenberg2014}. They determine \colden=19.1 from the Westerbork 21cm emission map, and obtain \colden = 18.8 $\pm$ 0.3 from a nominal fit to the {\hi} {\lya}. We obtain a \colden $\approx$ 17.2 when we add the absorption in the two components. \citet{rosenberg2014} obtain a lower metallicity of \metallicity $\approx -1$ because of the larger adopted column density and also because all the {\hi} is associated with only one component which is all assumed to be related to the metals.

\subsubsection{Galaxy Properties and Physical Interpretation}

The absorption component at $\approx$845 {\kms} has solar metallicity and is kinematically consistent with arising in the Leo Ring at the location of sightline SD. The galaxies NGC3384 and M105 are also within an impact parameter of 0.6{\rvir} of the sightline. M96 and NGC 3412 are within an impact parameter of 1.4{\rvir}. These four galaxies have LP between 0.1--0.4, but  we do not anticipate significant contribution to {\mgii} absorption (with predicted EW$_{\mgii, \lambda2796}$ $\approx$ 0.4) from M96 or NGC 3412 at their large impact parameters. Table \ref{tab:galaxyparams} lists galaxies with LP $>$ 0.1. The Leo Ring and/or the galaxies, M 105 and NGC 3384, could potentially explain the solar metallicity component along this sightline.

\smallskip

The {\hi}-only absorption at 978 {\kms} is most likely to be related to M105, because it is relatively close to an extension of the kinematics based on its velocity map, and it is at an impact parameter of just 0.6{\rvir}. The LP for M105 is determined to be 0.5. The next likely candidate, NGC 3412, has a lower LP of 0.01 and the absorption is much beyond the escape velocity of halo of NGC 3412. This absorption is unlikely to be associated with the Leo Ring as we do not see gas consistent with the observed velocity at this location. Alternatively, rather than being associated with a specific galaxy, this absorption could be related to loose material at large in this interacting group environment.

\subsection{SE: The $z = 0.0023$ absorber towards the quasar J1047+1513}
\label{sec:SE}

\begin{figure*}
\begin{center}
\includegraphics[width=\linewidth]{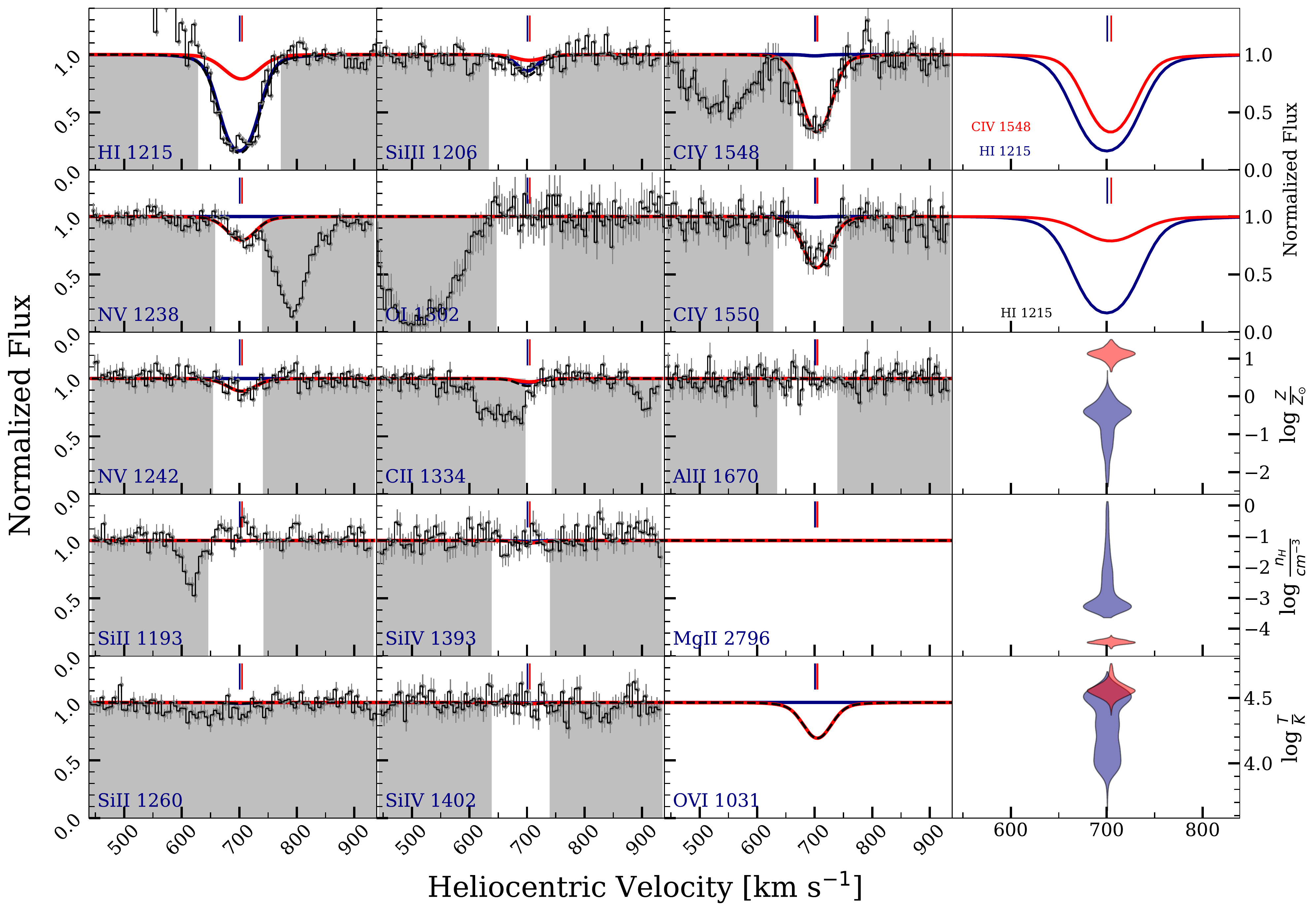}
\caption{\CLOUDY~models for the $z = 0.0023$ absorber towards SE obtained using the MLE values. The spectral data are shown in gray with 1$\sigma$~errors. The centroids of absorption components as determined from the VP fits to constraining ions are indicated by the vertical tick marks on top of each panel. The blue-colored curve traces the dominant absorption seen in {\hi}. The red-colored curve traces the absorption seen in the higher ionization {\civ} and {\nv} transitions.} The violin plots show the full posterior distribution of the physical properties of the cloud. The region shaded in grey shows the pixels that were not used in the evaluation of the log-likelihood.
\label{fig:SEsysplot}
\end{center}
\end{figure*}

A system plot of the $z = 0.0023$ absorber is shown in Figure~\ref{fig:SEsysplot}. This system shows strong {\civ} and {\nv} absorption, and modest absorption in {\siiii}. We find that a single phase model does not satisfactorily explain the {\hi} profile shape, and also does not produce the observed amount of {\nv} absorption in a phase that would agree both with {\siiii} and {\civ}. We, therefore, model the absorption using two phases, a low ionization phase centered at $\approx$ 701 {\kms} that produces the modest absorption seen in {\siiii}, and the bulk of the absorption in {\hi}, and a high ionization phase centered at $\approx$ 705 {\kms} that produces absorption seen in {\civ} and {\nv}. 

\smallskip

The properties of the low ionization cloud are determined to be {\metallicity} $\approx -0.4$, {\hden} $\approx$ $-$3.3, and \temp $\approx$ 4.5. The high ionization phase is determined to have a {\metallicity} $\approx$ 1, a low density of {\hden} $-4.5$, and a temperature of {\temp} $\approx$ 4.5. The PDFs of the properties of the two clouds are presented using corner plots in Figures~\ref{fig:SEHI0b} -- \ref{fig:SECIV1}. 

\smallskip

We also obtain a model in which the high ionization phase is collisionally ionized. We obtain similar properties for the metallicities of the two clouds compared with the PIE case. The temperature, however, for the high phase is higher {\temp} $\approx$5.2 compared to {\temp} $\approx$4.5 for the photoionized case. We also see from Figure~\ref{fig:SEvpfitgal} that possible zero-point uncertainties are negligible for $v>600$~{\kms}. A system plot showing the PC model is shown in Figure~\ref{fig:SEpici}. Other physical properties of these clouds are summarized in Table~\ref{tab:modelparams}.

\subsubsection{Galaxy Properties and Physical Interpretation}

Only the galaxy NGC 3377 is within 2{\rvir} of sightline SE, and it is at the fairly large separation of 1.6\rvir. The LP is found to be $\lesssim$0.1 for the absorption to be associated with this galaxy. The absorption is consistent with being within the escape velocity of the halo of NGC 3377. The Leo Ring gas would also potentially match kinematically, within $50$--$100$~{\kms} of this absorber, if it extended to lower {\HI} column densities on the topside~(see Fig.~\ref{fig:LeoRingfield}). The relatively strong {\civ} absorption in the high ionization phase, EW$_{\civ, \lambda1548}$ $\approx$ 0.23\AA, at this rather large distance ($\rho$ $\approx$ 220 kpc) is unexpected from an elliptical galaxy like NGC 3377~\citep{liangchen2014}. We attribute this high metallicity absorption to be associated with the Leo Ring gas, and the bulk of {\hi} absorption that also produces the modest absorption in {\siiii} could either be from the Leo Ring gas or from NGC 3377.

\subsection{SF: The $z = 0.0030$ absorber towards the quasar J1047+1513}
\label{sec:SF}

\begin{figure*}
\begin{center}
\includegraphics[width=\linewidth]{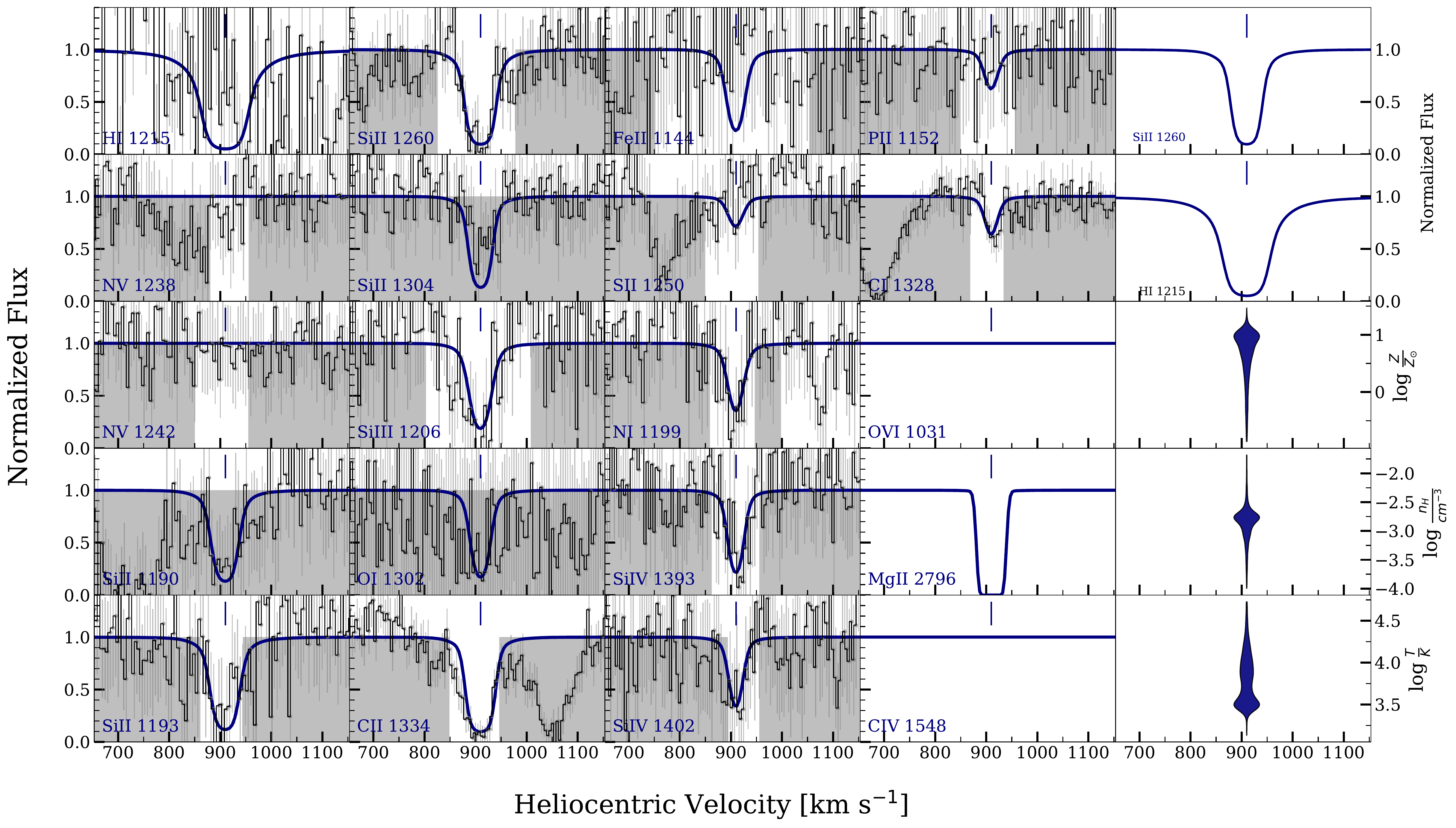}
\caption{\CLOUDY~models for the $z = 0.0030$ absorber towards SF obtained using the MLE values. The spectral data are shown in gray with 1$\sigma$~errors. The centroids of absorption components as determined from the VP fits to constraining ions are indicated by the vertical tick marks on top of each panel. One phase of photoionized gas if found to explain the absorption seen in the data. A summary of the properties of this cloud is shown using the violin plots. The region shaded in grey shows the pixels that were not used in the evaluation of the log-likelihood.}
\label{fig:SFsysplot}
\end{center}
\end{figure*}

A system plot of the $z = 0.0030$ absorber, with the constraining transitions, is shown in Figure~\ref{fig:SFsysplot}. This system shows absorption in many low ionization transitions, including, {\ci},  {\nitri}, {\siii}, and {\cii}. We also observe absorption in the intermediate ionization transitions, {\siiii} and {\siiv}. The data are noisy for this quasar and that leads to some uncertainty in the shape of the {\hi} {\lya} profile.  Nonetheless, we are able to derive a one component model, that is consistent with the data. In addition to this metal-line absorption, the {\hi} {\lya} profile extends up to $\approx$ 1400 {\kms}.  Only the component with metal lines detected can be constrained by models.

\smallskip

The physical parameters describing this absorption system are summarized using violin plots in Figure~\ref{fig:SFsysplot}, and also presented in Table~\ref{tab:modelparams}. The metallicity of the cloud centered at $\approx$910 {\kms} is \metallicity $\approx 1.0$ , while its density is \hden $\approx -2.7$. The size of the cloud along the line of sight is $\approx$2 kpc. The PDFs of the properties of this cloud are presented using corner plots in Figures~\ref{fig:SFSiIV0b}--\ref{fig:SFSiIV0}. 

\smallskip

We performed two experiments to account for any systematic uncertainty in correctly modeling the Galactic {\lya} absorption and the geocoronal emission due to the noisy nature of the {\hi} profile.  The main goal is to assess whether the metallicity of the metal line absorber is indeed super--solar.  In the first experiment, our prior included only the information from metals, the observed {\hi} profile was masked when determining the constraints. Even considering just the metal lines, we inferred a metallicity of {\metallicity} = 0.60 $\pm$ 0.15 for the cloud at $\approx$910 {\kms}. The synthetic profiles based on this model are shown in Figure~\ref{fig:SFextreme1}. It may seem counterintuitive that it is possible to infer a metallicity without using the hydrogen lines.  However, the strong detected {\siiv} absorption must arise in an ionized zone of absorption, which has only $\log N(\rm H)$ $\sim$ 19 regardless of the size of any shielded neutral layers, and thus must have high metallicity to produce the strong {\siiv}. In the second experiment, we forced the prior on {\colden} to only take values greater than 19.5.  The system plot showing the model profiles for this experiment is shown in Figure~\ref{fig:SFextreme2}. It would appear that the solution is not an acceptable fit to the {\lya}, but it is hard to tell for certain.  Regardless of this, even for this much larger $N({\hi})$ the metallicity that we obtained was {\metallicity} = 0.21 $\pm$ 0.08. Both these extreme experiments reveal that the ``true'' metallicity of this system has to be super-solar. 

\smallskip

This absorber was previously studied by \citet{rosenberg2014}. They used a column density, {\colden}, of 19.5 as measured from the 21cm emission map. By associating all of the {\hi} absorption to the observed metal line absorption, they determined a metallicity consistent with {\metallicity} $\approx -1$. However, from our modeling we find that a one component gas-phase with {\colden} of $\approx$17.6, centered on the metal absorption, explains all the observed metal absorption.

\subsubsection{Galaxy Properties and Physical Interpretation}

Three luminous galaxies lie within a virial radius of sightline SF, as shown in Figure~\ref{fig:galaxiesvirial}. M105 is at an impact parameter of 0.5\rvir, and the extension of its velocity map in the direction of SF could explain the absorption at $\approx 900$ {\kms}. The LP is determined to be $\approx 0.8$. The other galaxies, M96 and NGC 3384, are also within 0.5{\rvir} of this sightline, and the absorption centroid is consistent within the escape velocity of halo of these galaxies. The LP for these two galaxies is $\approx 0.3$.

\smallskip

The Leo Ring gas, however, would also be at a velocity consistent with the metal line absorption in this region.  The gas could be more directly related to the ring itself if the interaction that produced the ring has distributed the gas that was connected to the individual galaxies. 


\smallskip

Note also the extension of {\hi} {\lya} absorption (without associated metals) out to velocities as high as 1400 {\kms}, indicating a kinematically complex structure along this sightline.

\subsection{SG: The $z = 0.0027$ absorber towards the quasar J1048+1306}
\label{sec:SG}

\begin{figure*}
\begin{center}
\includegraphics[width=\linewidth]{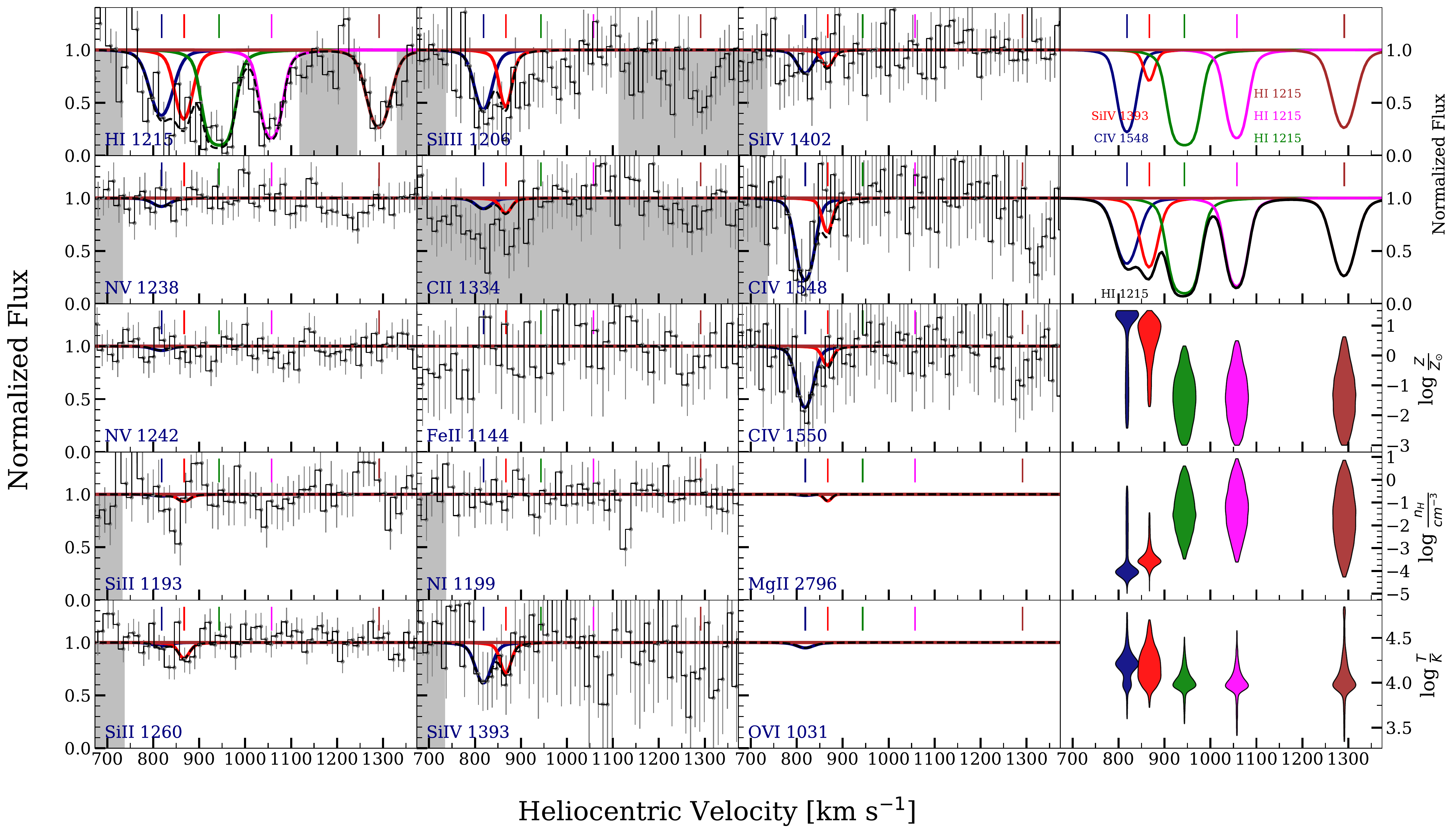}
\caption{\CLOUDY~models for the $z = 0.0027$ absorber towards SG obtained using the MLE values. The spectral data are shown in gray with 1$\sigma$~errors. The centroids of absorption components as determined from the VP fits to constraining ions are indicated by the vertical tick marks on top of each panel. The data has been binned by 2 pixels for display purposes. The photoionized gas phases are traced by five clouds - the blueward cloud constrained on {\siiii} (shown as a blue curve), and the blended redward cloud constrained on {\siiii} (shown as red curve), the three offset {\hi}-only clouds shown as green ,magenta, and brown curves. The superposition of these three models is shown by the black dashed curve. A summary of the properties of these five clouds is shown using violin plots. The region shaded in grey shows the pixels that were not used in the evaluation of the log-likelihood.}
\label{fig:SGsysplot}
\end{center}
\end{figure*}

A system plot of the $z = 0.0027$ absorber, with the constraining transitions, is shown in Figure~\ref{fig:SGsysplot}. This system shows strong metal line absorption only at two velocities, $\approx$817 {\kms} and $\approx$861 {\kms}, detected in {\siiii}, {\siiv}, and {\civ}. Five distinct components are apparent in {\hi} {\lya}, centered at velocities of 817, 861, 943, 1057, and 1297 {\kms}. 

\smallskip

We find that the {\siiii}, {\siiv}, and {\civ} arise in the same gas phase for the gas clouds at $\approx$817 and $\approx$861{\kms}. These two clouds have similar properties, high metallicities of \metallicity $\approx$1.4, temperatures of \temp $\approx$4.2, and low densities of \hden $\approx-4.0$ and $-3.6$, respectively. The line of sight cloud thicknesses are $L \approx$ 200 pc and 600 pc, respectively. The physical parameters describing this absorption system are summarized using violin plots in Figure~\ref{fig:SGsysplot} and in Table~\ref{tab:modelparams}.  The three components detected in {\hi} {\lya}, but not in metal lines, are poorly constrained. The PDFs of the properties of these clouds are presented using corner plots in Figures~\ref{fig:SGCIV0b} -- \ref{fig:SGHI3}. 

\smallskip

This absorber has been previously studied by \citet{rosenberg2014}. Based on the absence of radio emission at the location of the QSO, they place an upper limit on the {\colden} of 18.73, and do not derive a metallicity because of the noisy data.

\smallskip

We explore the consequences of zero-point uncertainty of the {\hi} {\lya} flux for the components between 815--865 {\kms}. The violin plots showing the distribution of parameters for 100 MC realizations are presented in Figure~\ref{fig:MCSG}. We find that the posteriors for various phases are still in agreement within uncertainties of the values presented in Table~\ref{tab:modelparams} for SG. The metallicity of the {\civ} phase at $815$~{\kms} is still well constrained by the rapid decline on the blueward side of the {\lya} profile, and not heavily dependent on the zero-point.  The {\siiv} phase at $\approx$861 {\kms}, however, is more significantly affected by the presence of a zero-point uncertainty, since its {\lya} profile is blended with others on both sides.

\subsubsection{Galaxy Properties and Physical Interpretation}

The metal-line components at $817$ and $861$ {\kms} are consistent with the Leo Ring kinematics at the position of SG. The galaxy M105, with an impact parameter of 0.6 {\rvir} and a velocity of 880-900 {\kms} in this direction, could also contribute to the absorption. It has an LP of $\approx$0.35. NGC 3412 with an impact parameter 0.9 {\rvir} and velocity $\approx$ 860 {\kms} could also contribute (a LP of $\approx$0.3). As with SF, it is not straightforward to distinguish between absorption arising in individual galaxies as compared to the ring because the gas in the ring likely originated in the surrounding galaxies.

\smallskip

The {\hi}-only component at $943${\kms} is likely to be associated with M105, and it has a LP of $\approx$0.6. The next likely candidate is NGC 3412 with a LP of $\approx$0.2. The {\hi}-only component at $1057${\kms} could also be associated with M105, but it has a very low LP of $\approx$0.04. The {\hi}-only component at $1314${\kms} shows a possible association with NGC 3389 with a LP of 0.1, however, it is at an impact parameter of 1.5{\rvir} from the galaxy. These {\hi}-only components are unlikely to be associated with the Leo Ring at the location of SG because of the mismatch in kinematics, but they could be from other gas "at large" in the group.

\subsection{SH: The $z = 0.0028$ absorber towards the quasar J1051+1247}
\label{sec:SH}

\begin{figure*}
\begin{center}
\includegraphics[width=\linewidth]{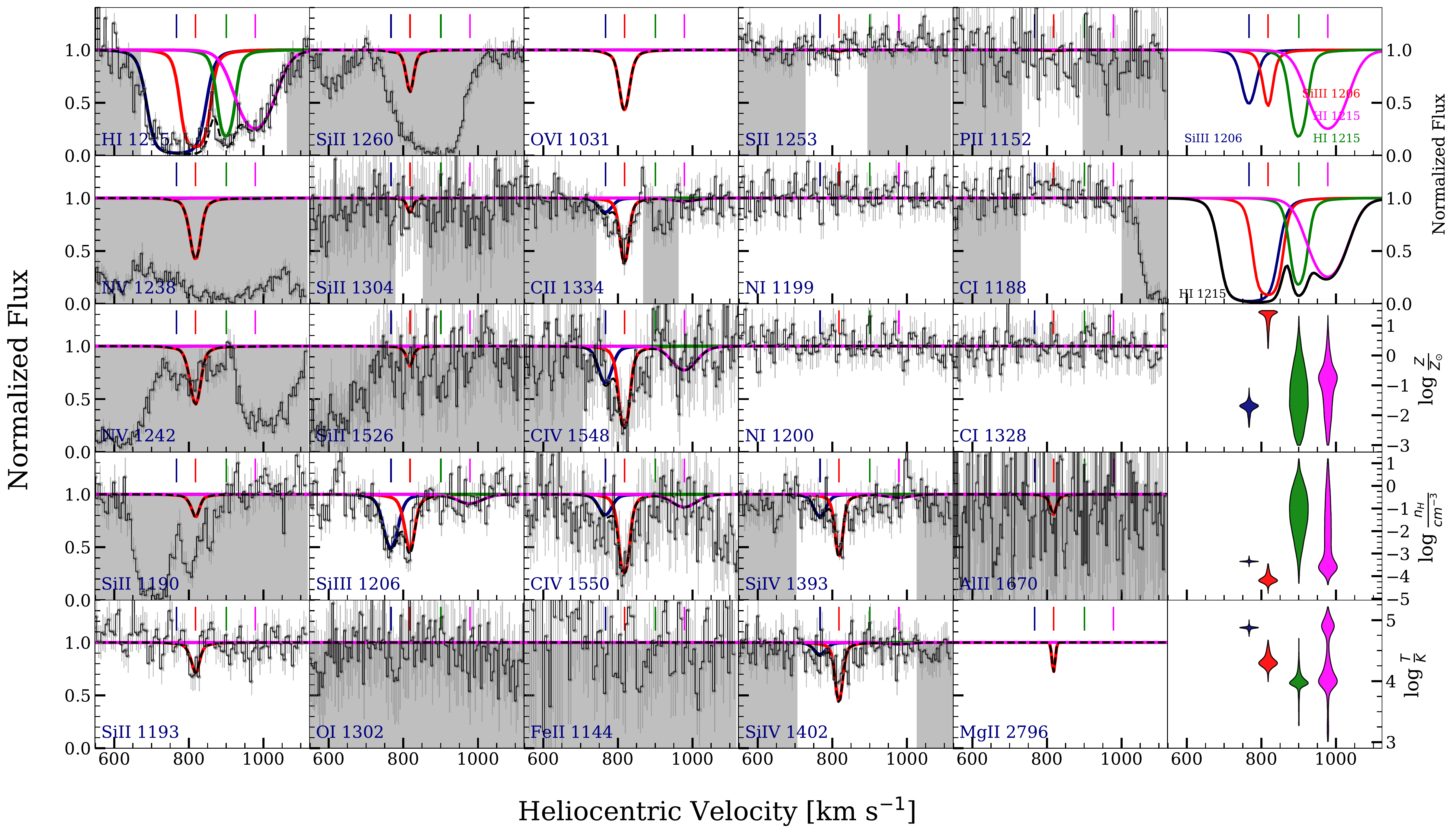}
\caption{\CLOUDY~models for the $z = 0.0028$ absorber towards SH obtained using the MLE values. The spectral data are shown in gray with 1$\sigma$~errorss. The centroids of absorption components as determined from the VP fits to constraining ions are indicated by the vertical tick marks on top of each panel. The photoionized gas phases are traced by four clouds - the blueward cloud optimized on {\siiv} (shown as a blue curve), and the blended redward clouds optimized on {\siiv} (shown as red and green curves), and the offset redward cloud optimized on {\cii} (shown as a magenta curve). The superposition of these three models is shown by the black dashed curve. A summary of the properties of these four clouds is shown in violin plots. The region shaded in grey shows the pixels that were not used in the evaluation of the log-likelihood.}
\label{fig:SHsysplot}
\end{center}
\end{figure*}

A system plot of the $z = 0.0028$ absorber, with the constraining transitions, is shown in Figure~\ref{fig:SHsysplot}. This system shows metal line absorption at $\approx$767 {\kms} and $\approx$818 {\kms} in {\cii} and {\siiii}, and in the {\siiv} and {\civ} doublets. The G160M observation of the {\civ} is quite noisy to place a meaningful constraint on the presence/absence of {\civ} for the $\approx$818 {\kms} component. Two additional, distinct {\hi} components, without detected metals, are also seen, centered at $\approx$900 and $\approx$977 {\kms}. 

\smallskip

Table~\ref{tab:modelparams} lists constraints on the physical parameters describing this absorption system, which are summarized using violin plots in Figure~\ref{fig:SHsysplot}. The component at 767 {\kms} has a low metallicity of {\metallicity} $\approx -1.7$, a low density of {\hden} $\approx-3.4$, a high temperature of {\temp} $\approx4.9$, and a line of sight thickness of $\approx$50 kpc. The stronger metal line component at 818 {\kms} has a supersolar metallicity of \metallicity $\approx$ 1.4, hydrogen density \hden $\approx -4.2$, temperature of \temp $\approx 4.3$, and line of sight thickness of $\approx$40 kpc. The {\hi} clouds at $\approx$900 {\kms} and $\approx$977 {\kms} are less certain in their properties. The metallicity is determined to be {\metallicity} $<-0.2$ and $<0$, and {\hden} $>-2.25$ and $>-3.8$. The PDFs of the properties of these clouds are presented using corner plots in Figures~\ref{fig:SHSiIII0b} -- \ref{fig:SHHI3}.

\smallskip

We explore the consequences of zero-point uncertainty of the {\hi} {\lya} flux for the components between 765--820 {\kms}. The violin plots showing the distribution of parameters for 100 MC realizations with different zero-point values are presented in Figure~\ref{fig:MCSH}. We find that the posteriors for various phases are still in agreement within uncertainties of the values presented in Table~\ref{tab:modelparams} for SH. Our conclusions about the nature of the absorption system are not affected by the presence of a zero-point uncertainty.

\subsubsection{Galaxy Properties and Physical Interpretation}

The galaxies NGC 3384 and M96 which are at 0.8{\rvir} and 1.4{\rvir} could contribute to the low metallicity absorption at 767 {\kms}. They have LPs of 0.3 and 0.1, respectively. With only {\metallicity} $\approx -1.7$, the gas would have to be inflowing toward one/both of these galaxies.  Several galaxies could contribute to the contrasting, supersolar metallicity absorption at 818 {\kms}, however their impact parameters are large to produce such strong absorption, with NGC 3412 being the closest at 0.9{\rvir}. The other galaxies that could contribute are listed in Table~\ref{tab:galaxyparams}. The velocity of an extension of the Leo Ring would also be kinematically consistent with the super-solar metallicity component at 818 {\kms}, so we favor an interpretation with a contribution from the ring, with possible contributions from the other galaxies and intragroup environment.

\smallskip

Among the sightlines studied here, SH is unusual in the contrasting metallicities of components at similar velocities.  However, such examples are seen in the literature~\citep{zahedy2019characterizing, jackson2021}
where they can be interpreted as a combination of pristine inflow and metal-enriched outflows or recycled accretion.

\subsection{SI: Lack of absorption towards the quasar J1052+1017}
\label{sec:SI}

\begin{figure}
\begin{center}
\includegraphics[scale=0.75]{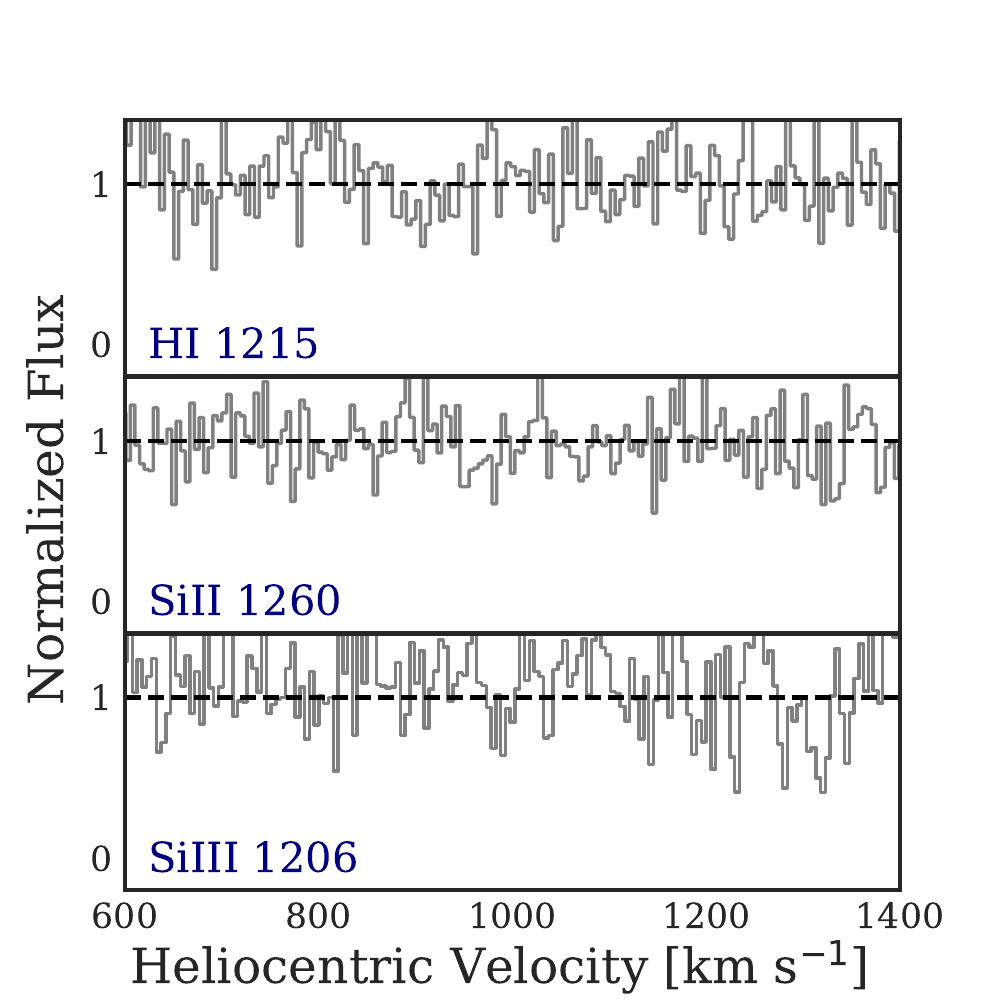}
\caption{The spectral data for SI showing lack of absorption in {\hi} {\lya}, {\siii}, and {\siiii} within the heliocentric velocity range of 600-1400 {\kms}.}
\label{fig:SIsysplot}
\end{center}
\end{figure}

We do not see any absorption in {\hi} {\lya}, the low ionization transition of {\siii}, and the intermediate ionization transition of {\siiii} towards this quasar within the 
heliocentric velocity range of 600-1400 {\kms}. Figure~\ref{fig:SIsysplot} shows the non-detections in these transitions.

\subsubsection{Galaxy Properties and Physical Interpretation}

The nearest galaxy to this sightline is M96 at a projected separation of $\approx$2.0 {\rvir}. The galaxies within a projected separation of 3{\rvir} are listed in Table~\ref{tab:galaxyparams}.

\subsection{SJ: The $z = 0.0024$ absorber towards the quasar J1059+1441}
\label{sec:SJ}

\begin{figure*}
\begin{center}
\includegraphics[width=\linewidth]{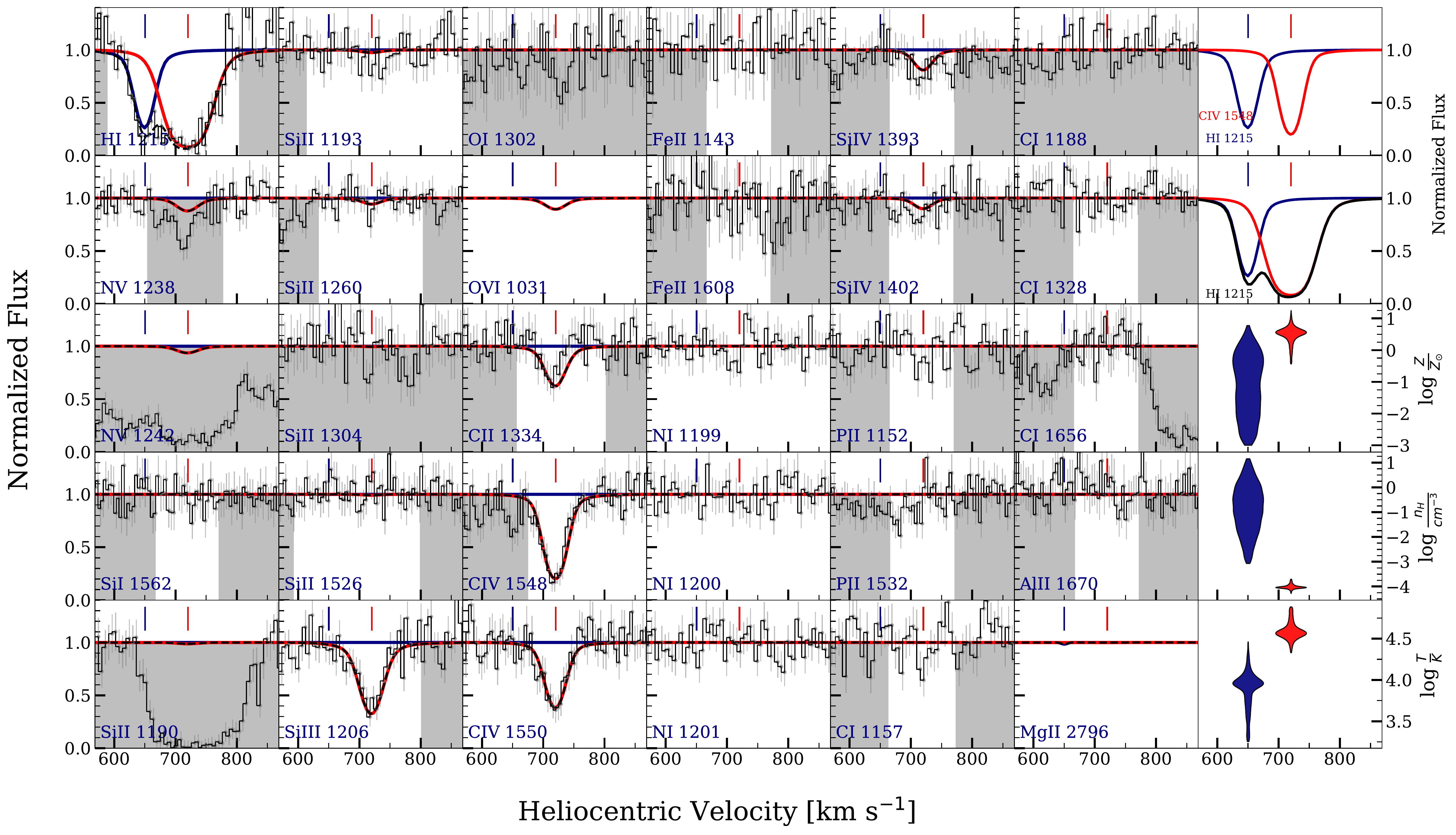}
\caption{\CLOUDY~models for the $z = 0.0028$ absorber towards SJ obtained using the MLE values. The spectral data are shown in gray with 1$\sigma$~errors. The centroids of absorption components as determined from the VP fits to constraining ions are indicated by the vertical tick marks on top of each panel. The photoionized gas phases are traced by two clouds - the blueward cloud constrained on {\hi} (shown as a blue curve), and the redward cloud constrained on {\civ} (shown as red curve). The superposition of these two models is shown by the black dashed curve. A summary of the properties of these two clouds is shown using the violin plots. The region shaded in grey shows the pixels that were not used in the evaluation of the log-likelihood. The {\nv}~$\lambda$1238 line is contaminated by {\oiv}$\lambda$787 absorption from a z = 0.5765 absorber, hence it is masked out.}
\label{fig:SJsysplot}
\end{center}
\end{figure*}

A system plot of the $z = 0.0027$ absorber, with the constraining transitions is shown in Figure~\ref{fig:SJsysplot}. This system shows absorption in two components, a {\hi}-only component without any metals at $\approx$ 651 {\kms} (shown as a blue curve), and another component at $\approx$ 720 {\kms} (shown as a red curve) which shows absorption in the low ionization transition of {\cii}, but only a 3$\sigma$ detection in {\siii}, we also observe absorption in the intermediate ionization transitions of {\siiii} and {\siiv}, and in the higher ionization transition of {\civ}. We do not observe absorption in any of the neutral transitions.

The blueward component is not well constrained. We set an upper limit on its metallicity of {\metallicity} $<0.5$ and the hydrogen density is a lower limit with {\hden} $>-2.25$. The redward component is well constrained to have $\log n_H \approx -4.0$, while its metallicity would be {\metallicity} $\approx$ 0.5 in order to explain all of the {\hi} {\lya} absorption centered at 720 {\kms} and produce the strong absorption seen in {\civ}. The constraints are illustrated in the violin plots included in Figure~\ref{fig:SJsysplot}. The PDFs of the properties of this cloud are presented using corner plots in Figures~\ref{fig:SJHI0b} -- \ref{fig:SJCIV0}.

\subsubsection{Galaxy Properties and Physical Interpretation}

Sightline J is relatively isolated from galaxies as compared to most of the others, with only one galaxy, NGC 3489, at 1{\rvir} that could be responsible for the absorption, with an LP of $\sim$ 0.25. There are also a couple of dwarf galaxies, however, they are at large separations $\gtrsim$ 2.9{\rvir}, and unlikely to contribute to the absorption with an LP of $\sim$ 10$^{-4}$. The component at $\approx$ 650 {\kms} is likely to be associated with NGC 3489. An extension of the Leo Ring in this direction would have a relatively low velocity, consistent with the absorption at 720 {\kms}, although it is at a relatively large distance of 550 kpc from the part of the ring detected in 21-cm. The observed {\civ} absorption is quite strong compared to expectations for such  galaxies~\citep{Burchett2016}. Given the strong {\civ} absorption, we favor an origin of the metal-rich material in gas related to the interaction that produced the Leo Ring.

\subsection{SK: Lack of absorption towards the quasar J1059+1211}
\label{sec:SK}

\begin{figure}
\begin{center}
\includegraphics[scale=0.75]{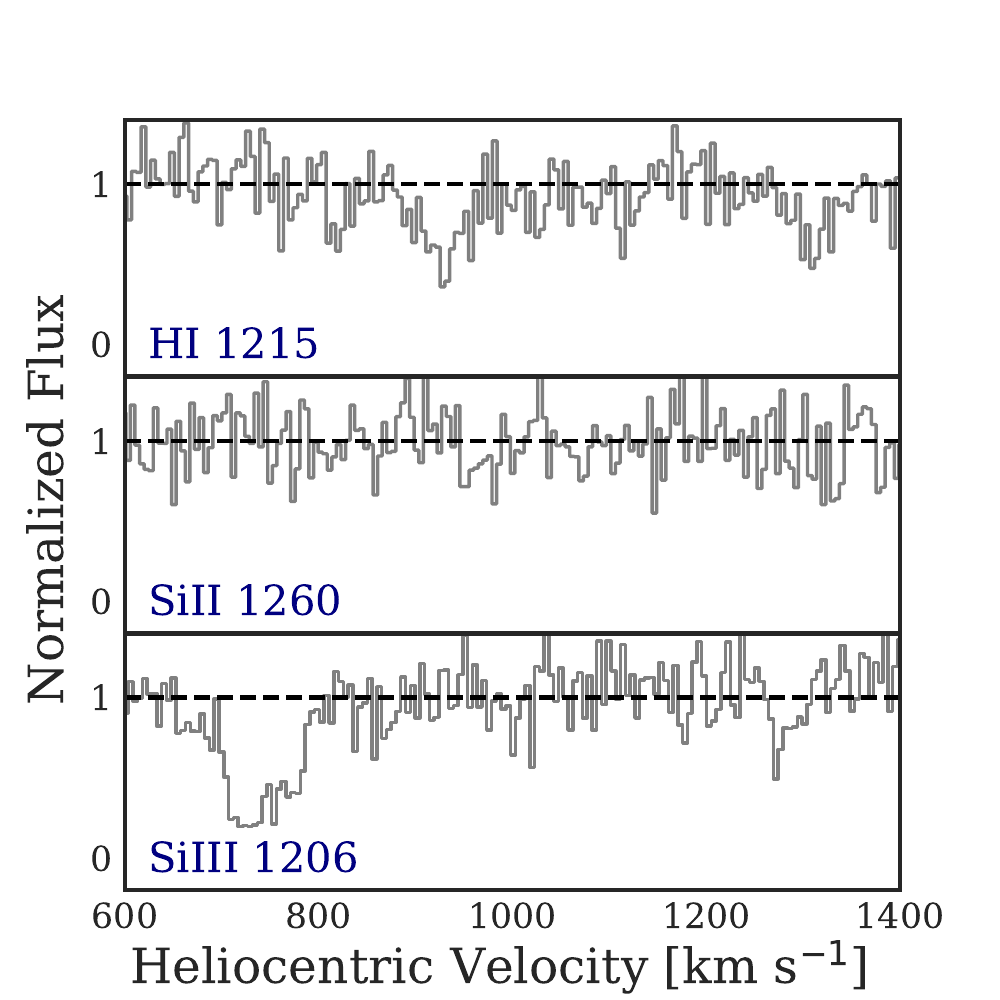}
\caption{The spectral data for SK showing lack of absorption in {\hi} {\lya}, {\siii}, and {\siiii} within the heliocentric velocity range of 600-1400 {\kms}.}
\label{fig:SKsysplot}
\end{center}
\end{figure}

We do not see any absorption in {\hi} {\lya}, the low ionization transition of {\siii}, and the intermediate ionization transition of {\siiii} towards this quasar within the 
heliocentric velocity range of 600-1400 {\kms}. Figure~\ref{fig:SKsysplot} shows the non-detections in these transitions.

\subsubsection{Galaxy Properties and Physical Interpretation}

The nearest galaxies to this sightline is NGC 3489 at a projected separation of $\approx$2.0 {\rvir}.  It is reasonable that no absorption is detected in SK given such a large galaxy separation. Galaxies within $\approx$3.0 {\rvir} to this sightline are shown in Table~\ref{tab:galaxyparams}.

\subsection{SL: Lack of absorption towards the quasar J1100+1046}
\label{sec:SL}

\begin{figure}
\begin{center}
\includegraphics[scale=0.75]{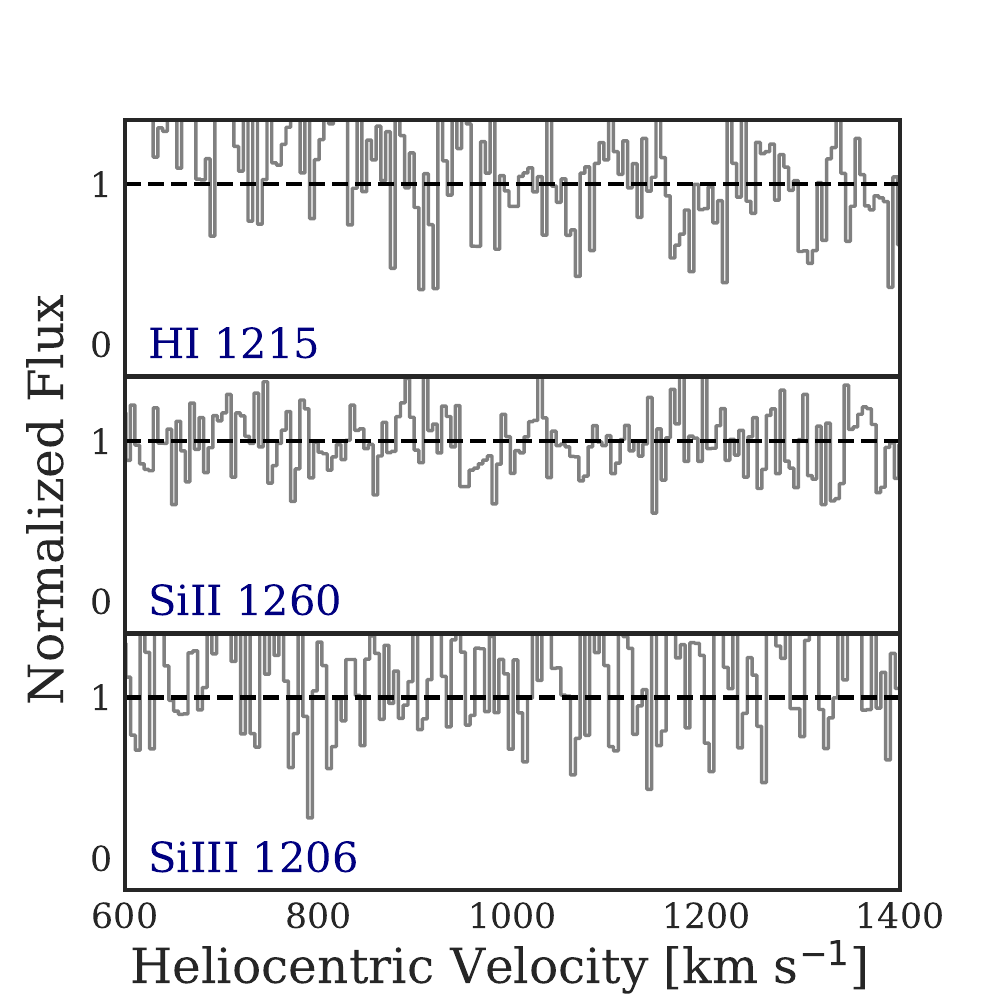}
\caption{The spectral data for SL showing lack of absorption in {\hi} {\lya}, {\siii}, and {\siiii} within the heliocentric velocity range of 600-1400 {\kms}.}
\label{fig:SLsysplot}
\end{center}
\end{figure}

We do not see any absorption in {\hi} {\lya}, the low ionization transition of {\siii}, and the intermediate ionization transition of {\siiii} towards this quasar within the 
heliocentric velocity range of 600-1400 {\kms}. Figure~\ref{fig:SLsysplot} shows the non-detections in these transitions.

\subsubsection{Galaxy Properties and Physical Interpretation}

The nearest galaxy to this sightline is a dwarf galaxy, dw1059+11,  at a separation of 3.4 {\rvir}. The large separation is consistent with the lack of absorption along SL.

\section{SUMMARY AND DISCUSSION}
\label{sec:discussion}
A general picture of the gaseous content of the Universe is emerging from studies of quasar sightlines through regions at various impact parameters from galaxies of different types (masses, morphologies, star formation rates, etc.). From these we learn in a statistical sense the covering factors and radial distributions of gas around galaxies in {\hi}, low, intermediate, and high ionization gas. The dependencies of these properties on the environment (isolated, group, or cluster) of the galaxies also provide a measure of the contributing processes and also on the larger-scale gas distribution. In addition to the statistical studies, there have been several cases of multiple lines of sight through individual galaxies (e.g. \citealt{sandhya2013,lehner2015,lehner2020}) that yield information about the variations of properties within a single galaxy. 

\smallskip

With this study of the Leo Ring/group we aimed to make connections between these different methods by studying 11 lines of sight through a 7$\degree$ $\times$ 6$\degree$ region.  In each case where absorption is detected, there is potential contribution to the absorption from one or more luminous galaxies (within one or two virial radii), from dwarf galaxies, from an intragroup medium, from the specific interaction debris known as the Leo Ring, and from the general larger scale filamentary structure.  We benefit from the close proximity of the group because it is possible to detect even faint galaxies in the region, and thus know a great deal about the galaxy and {\hi} distribution. It is so nearby, however, at 600--1400~{\kms}, that we had to overcome the challenge of fitting Galactic {\lya} and correcting for geocoronal emission in order to analyze the {\lya} absorption from our Leo Ring/Group sightlines. With our component-by-component, Bayesian methods, and careful treatment for Galactic \lya, we are able to consider which objects/processes are shaping the multiphase absorption along each sightline. This allows us to consider whether there is a coherency in properties across the region. We can also infer whether similar complexities, which may be unseen, are potentially influencing the general absorption properties along sightlines in surveys of galaxies at higher redshifts.

\begin{figure}
\begin{center}
\includegraphics[scale=0.75]{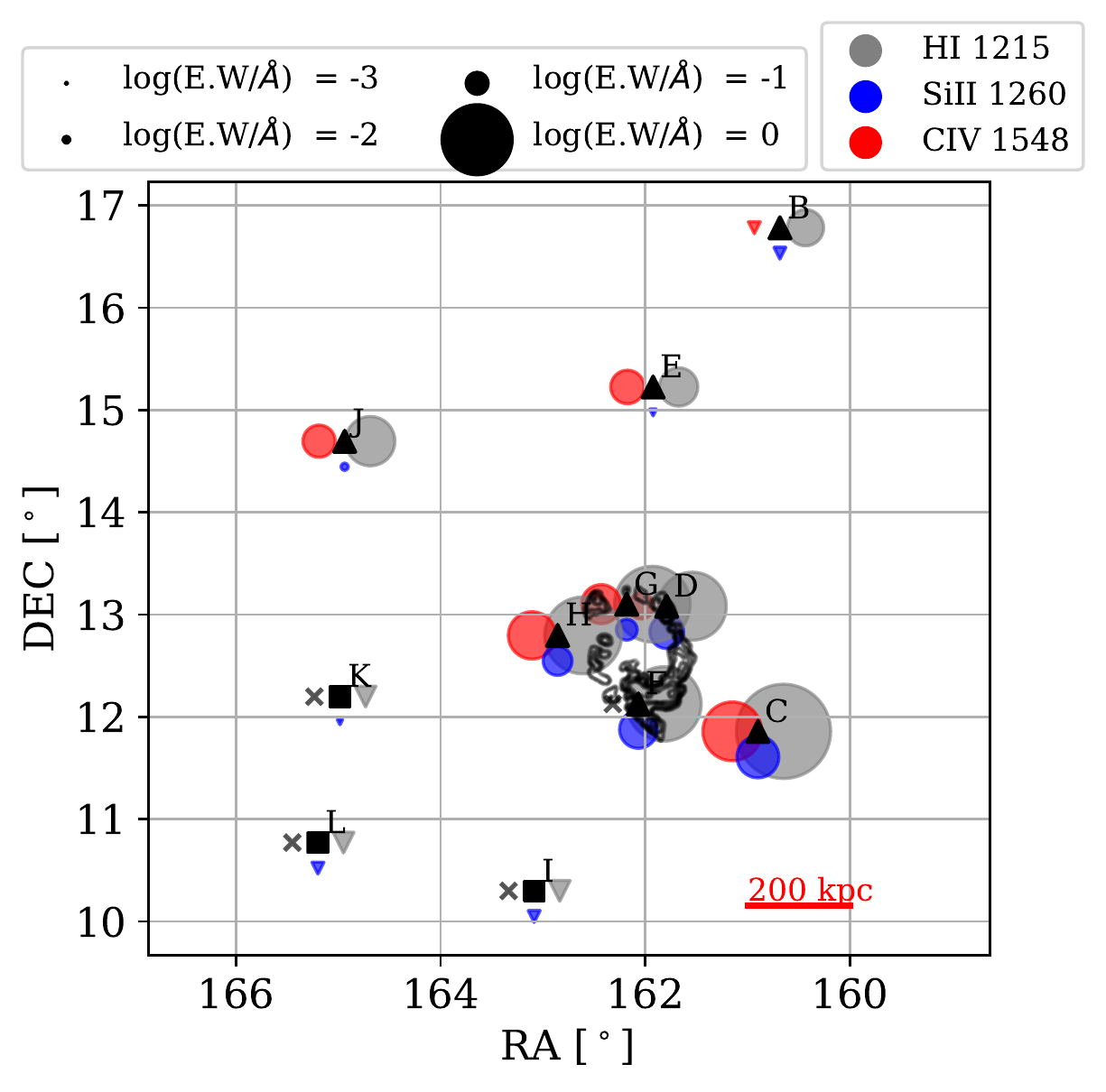}
\caption{A map of the Leo Ring field showing the 11 sightlines and the rest-frame EW measurement for absorption seen in {\hi} {\lya}, {\siii}, and {\civ}. Sightlines with {\hi} absorption detected in the range between 600 and 1400 {\kms} are shown as triangles. Sightlines with {\hi} non-detections between 600 and 1400 {\kms} are shown as squares. The EWs of absorption in several transitions are indicated by the size of the colored circles (gray for {\hi} {\lya}, blue for {\siii}, and red for {\civ}). We overlay the contours from \citet{schneider89} to indicate the distribution of intergalactic {\hi} which is potentially traced by some of the sightlines.}
\label{fig:Leo RingEWS}
\end{center}
\end{figure}

Figure~\ref{fig:Leo RingEWS} shows a map of the Leo Ring region, with the equivalent widths (EWs) of absorption (between 600 and 1400 \kms) indicated by the size of the colored circles for {\hi} {\lya},  {\siii}, and  {\civ}.  We can see that the absorption is stronger close to the 200 $\times$ 200 kpc$^{2}$ region of the {\hi} ring.  The EW of {\hi} {\lya}, in these cases, is tied to the kinematic spread of the {\hi} more so than to the column density of {\hi}.  Besides in the sightlines close to the ring (SC, SD, SF, SG, and SH within 250~kpc), metals are also detected in SE and SJ. Sightline SB has only a {\hi} {\lya} detection, while SI, SK, and SL do not have any detected absorption in the relevant velocity range. 

\begin{figure*}
\begin{center}
\includegraphics[scale=0.70]{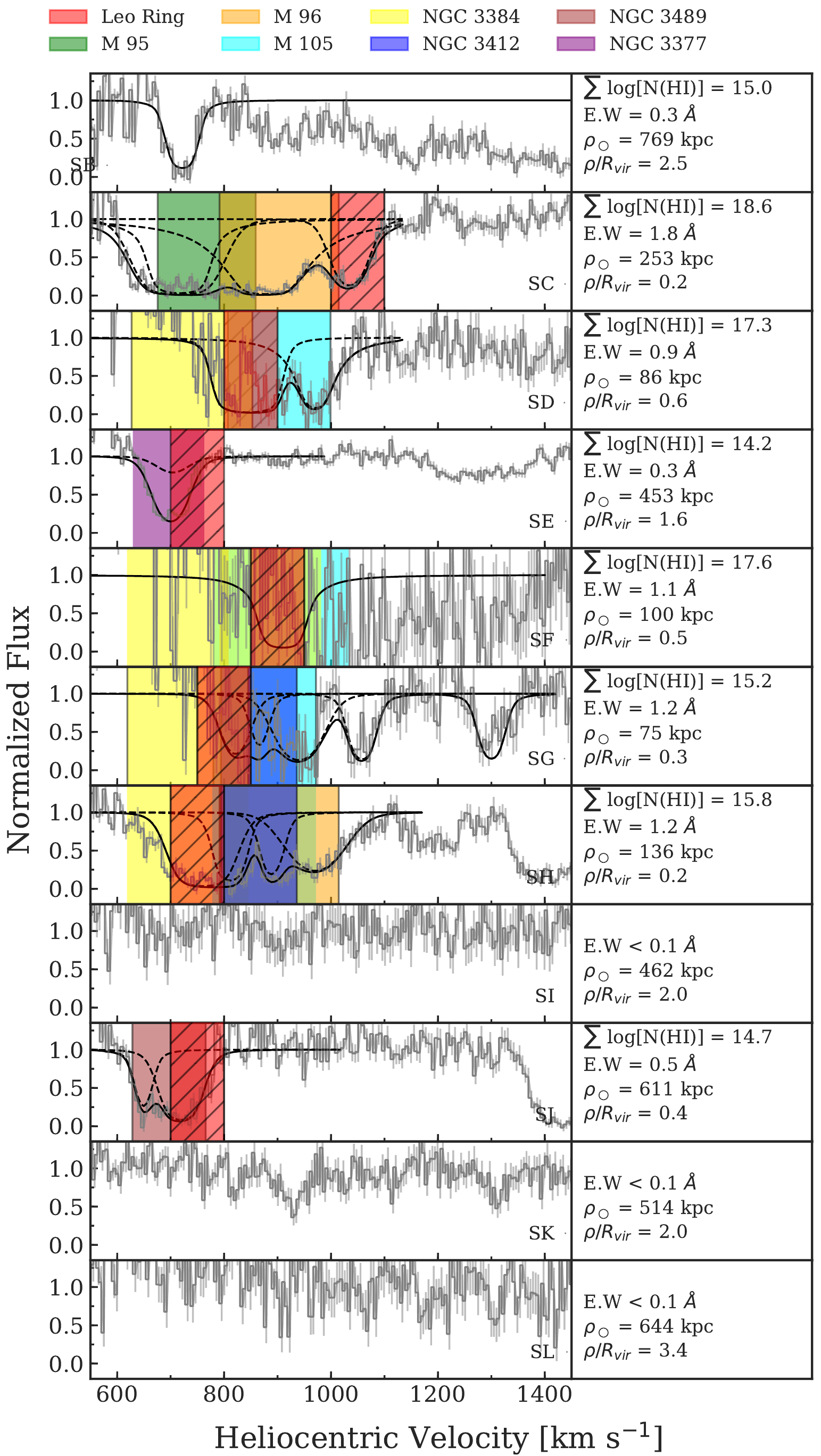}
\caption{A plot showing the {\hi} {\lya} profiles for the 11 sightlines in this study, with the model profiles superimposed. The expected velocities for extensions of Leo Ring and other large nearby galaxies are shown as shaded regions, in colors according to the key. The Leo Ring association is shaded in red and hatched. The width of the shaded region accounts for expected velocities within the escape velocity of a galaxy's halo.} To the right of the spectra, we list the sum of the N({\hi}) values from our model, the total EW measured from the solid black model profile (except for SF, where we give the EW measured for the entire absorption trough), the impact parameter to the center of the Leo Ring, and the ratio of that impact parameter to the virial radius of the nearest galaxy.
\label{fig:galaxysights}
\end{center}
\end{figure*}

Figure~\ref{fig:galaxysights} shows the absorption profiles of {\hi} {\lya} for all of the sightlines, with the model profiles superimposed. The expected velocities for extensions of large nearby galaxies (M95, M96, M105, NGC3384, and NGC3412) in the direction of the quasar sightlines are shown as shaded regions superimposed on the profiles. The expected velocities of the Leo Ring absorption at the position of the sightline are also shaded in pink for sightlines SC, SD, SE, SF, SG, SH, and SJ for which some of the {\hi} {\lya} absorption is seen coincident with that velocity. The width of the shaded regions account for expected velocities within the escape velocity of a galaxy's halo. For sightline SB, the absorption does not appear to be related to specific galaxies. For each sightline, the panel also shows the EW of {\hi} {\lya}, the column density of {\hi} given by the sum of the model components, the distance to the center of the Leo Ring ({$\rho_{o}$}), and the $\rho$/{\rvir} value to the nearest galaxy. Note that the $N({\hi})$ value need not be large for a system with large EW, if it is composed of several components.  In that case the EW is large due to the kinematic spread of these components.  We see that sightlines with the strongest {\hi} {\lya} profiles are both close to the Leo Ring, and close to their nearest galaxy.  This is not surprising given that the ring formed in a galaxy group environment, but it does lead to some ambiguity in distinguishing between the two origins.  As we have discussed, multiple origins for the absorption components at different velocities along such single sightlines is likely.

\begin{figure*}
\begin{center}
\includegraphics[scale=0.70]{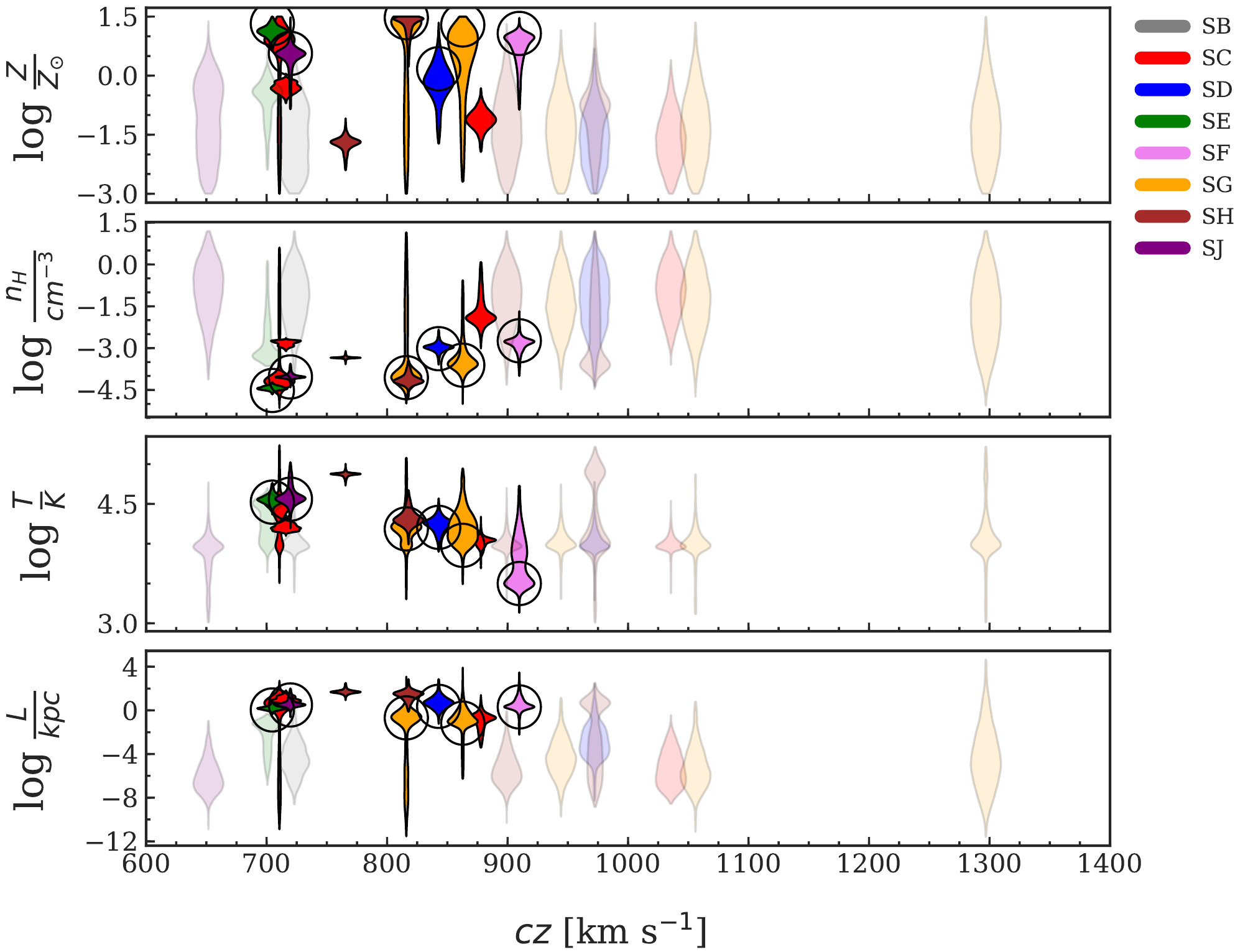}
\caption{Constraints on metallicity, density, temperature, and line of sight thickness of the gas have been derived for each absorption component using {\CLOUDY} modeling. The different sightlines are denoted by different colors as indicated in the key. Symbols circled in black are components with velocities consistent with those expected for the Leo {\hi} ring in that direction. Very light shading is used to designate violin plots corresponding to components with {\hi}-only absorption. These light shaded violin plots are presented as limits in Table~\ref{tab:modelparams}.}
\label{fig:params}
\end{center}
\end{figure*}

Constraints on metallicity, density, temperature, and line of sight thickness of the gas have been derived for each absorption component using {\CLOUDY} modeling as described in section~\ref{sec:Results}. Figure~\ref{fig:params} summarizes these results. Most violin plots have large uncertainties because there are no detections in the component for any of the metal line transitions.  We note that most components, and particularly all the ones that are plausibly kinematically associated with the Leo Ring, have metallicities of greater than the solar value (up to several times Z$_{\sun}$).  Densities are typically low, between $10^{-4}$ and $10^{-3}$ \cc. Most absorption components are produced by gas with a line of sight thickness between 1 and 10kpc.  Those components with a velocity consistent with an extension of the Leo Ring (as detected in 21cm) in the direction of the sightline are circled in black.   We find that all sightlines with detections except SB have one or more components likely to be associated with the Leo Ring or extended gas that resulted from the same interaction.  The properties of this subset of well-constrained components are uniform, with supersolar metallicities and low densities.

\subsection{LEO RING ABSORPTION}

The similarity in properties of the subset of components identified in Figure~\ref{fig:params} suggests that there is a coherent structure of tidal debris in this region, consistent with the origin of the Leo Ring proposed in other previous investigations~\citep{dansac2010}. \citet{corbelli2021} also found metallicities close to the solar value for three {\hi} clumps near SF using nebular emission lines, and concluded that the metals arose in the nearby large galaxy, M105, and were removed by tidal forces.  The metallicities and kinematic spreads of the absorbers that we confirm throughout the entire Leo Ring region are consistent with this, given that the origin of the debris is from giant galaxies that themselves have solar or supersolar metallicity, based on the expectations from the mass-metallicity relationship~\citep{tremonti2004}.  The mass of the ring, from 21-cm studies, is about two billion solar masses \citep{schneider1985}. Here, we estimate the
mass of the ring: $\pi \frac{d^{2}}{4}Ln_{H}\mu m_{H}$

\begin{equation}
    M \approxeq \left[\frac{d/2}{kpc}\right]^{2}\left[\frac{L}{kpc}\right]\left[\frac{n_{H}}{cm^{-3}}\right]\times1.1\times10^{8}M_{\sun}
\end{equation}

Assuming the gas is spread over a circular region with diameter 200 kpc, and that the thickness of the region is 10 kpc, the average gas density is 10$^{-3.5}$cm$^{-3}$, and $\mu$=1.4 accounts for the presence of helium and heavy elements in the gas, we determine the mass to be 3.5 billion solar masses. 

\smallskip

It is reassuring that the basic properties that we have inferred for our quasar sightlines yield a mass roughly consistent with 21-cm observations for the Leo Ring.  However, it is also clear that the region is not as homogeneous in density as we have assumed. Evidence for a clumpy medium is provided by the fact that the {\hi} column densities from 21-cm emission observations (over a large beam size of about $\approx$1.75\arcmin $\times$ 0.65\arcmin) are an order of magnitude or more larger than those from our models along the pencil beams we are studying.  For example, \citet{rosenberg2014} found from Westerbork observations that \colden=19.1 for SD compared to our model sum of components of only 17.2.  For SF, \citet{rosenberg2014} find {\colden}=19.5 and obtain a metallicity of {\metallicity} $\approx$ -1 by associating all the metal absorption to a phase with {\colden} = 19.5, compared to {\colden} = 17.6 and a metallicity of $\approx$ 1 for our models.  For SG, our model values only sum to \colden=15.4 and the sightline is just outside of the contours for detected 21-cm emission which implies an {\colden} value greater than 18 is unlikely.  In a clumpy medium we expect that small regions inside the beam might have N(\hi) values one or more orders of magnitude bigger than others. These small regions may occupy a very small fraction of the beam but would average out to a much larger value than typical for a pencil beam measurements. We thus believe that there is no inconsistency, and that the tidal debris has structure on small scales.

\begin{figure*}
\begin{center}
\includegraphics[scale=0.70]{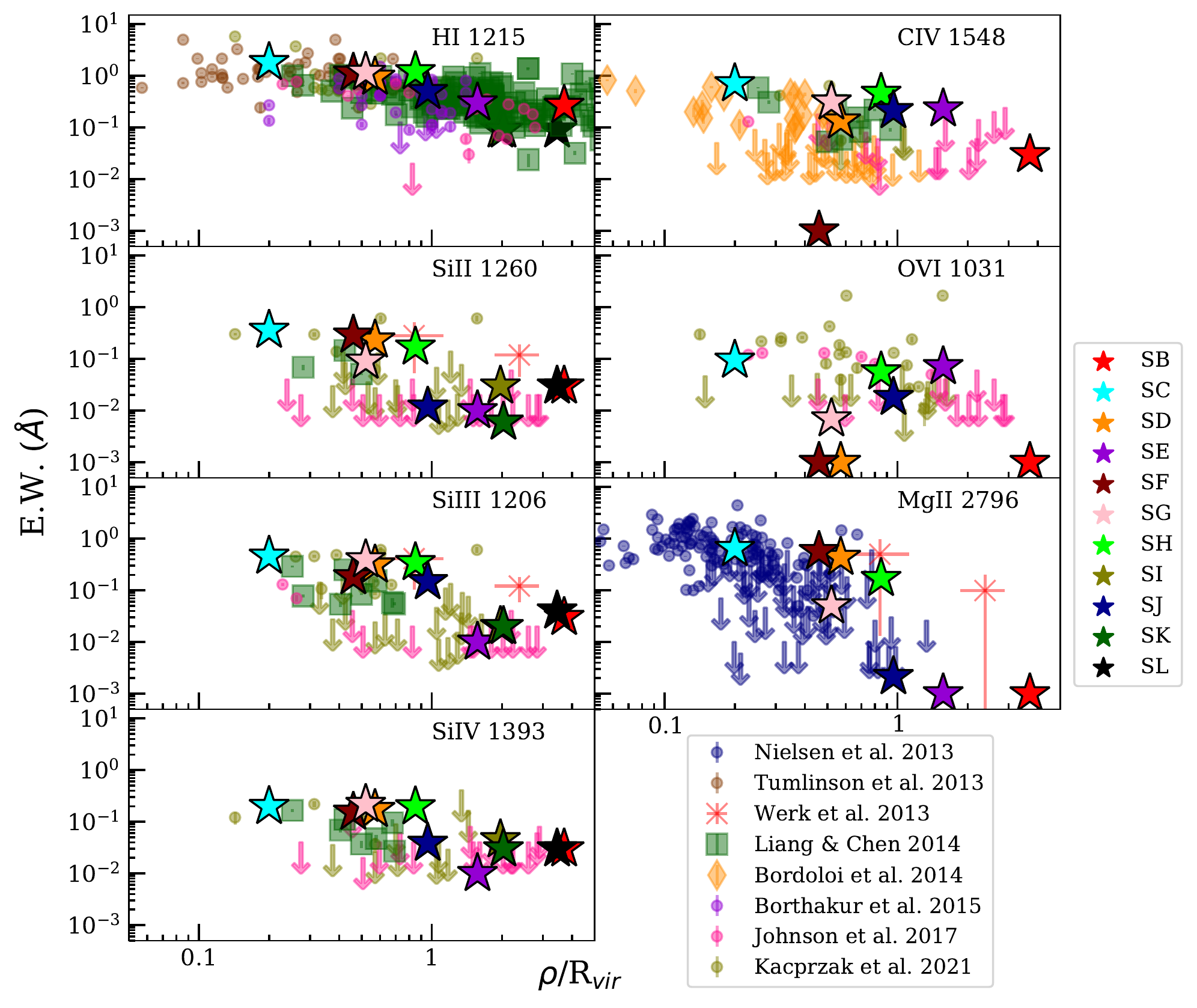}
\caption{Rest frame EWs for key transitions are plotted against the impact parameter to the nearest galaxy, in units of its virial radius. The EWs are determined based on our best model fits to the observed data. For {\mgii}, {\ovi}, and in some cases {\civ} (for SC, SD and SF), the measurements are based on predicted absorption profiles from our {\CLOUDY} models, since the spectra did not cover these transitions. In these cases, an additional hidden phase could produce absorption in these transitions, which would not be shown here. These values are compared between our sightlines through the Leo Group, shown as different colored stars, and observations of various other low redshift samples of absorption line systems, noted by the colors/types of the smaller symbols.}
\label{fig:ewcompare}
\end{center}
\end{figure*}

\subsection{COMPARISON OF SIGHTLINES TO ABSORPTION SIGNATURES OF GALAXIES AND GALAXY GROUPS}

It is clear that lines of sight through the Leo Ring environment produce relatively strong absorption from gas with high metallicity, and with complex kinematics. However, sightlines close to the ring are also close to the galaxy group that is related to the ring. In order to better understand the major contributing factors to the observed absorption signatures, we consider the EWs of several key transitions as a function of impact parameter from the nearest galaxy and perform as simple benchmarking with CGM absorbers in the literature. Through this, we explore the impact on observed signatures due to proximity to a gas rich region, i.e. the Leo Ring. To consider how to interpret absorption as related to individual galaxies, to the group gas, and/or to the Leo ring itself, we compare to studies that probe the CGM of a wide range of galaxies. We have plotted the observed equivalent widths as a function of impact parameter for star forming and for quenched galaxies~\citep{tumlinson2013cos,Kacprzak2021}, for dwarfs up to massive ellipticals~\citep{bordoloi2014cos,johnson2017,zahedy2019characterizing}, from surveys probing galaxy disk-CGM interace~\citep{Borthakur2015} to galaxy halos~\citep{werk2013cos}, and from surveys indifferent to galaxy properties~\citep{Nielsen2013,liangchen2014}. Our goal is to consider whether the equivalent widths are typical at that impact parameter from the individual galaxies, or whether the complex group environment is leading to larger values.


\smallskip

Figure~\ref{fig:ewcompare} shows this comparison for several key transitions. The EWs plotted are the values for the entire system, integrating over the full velocity range, and they are plotted against the $\rho$/{\rvir} value for the nearest galaxy. For {\mgii}, {\civ} and {\ovi} these are model predictions, since these transitions are not covered.  The actual values could be larger if hidden phases (e.g. collisionally ionized gas) are contributing~\citep{narayanan2021}.  In general, the five sightlines closest to the ring center (SC, SD, SF, SG, and SH) have {\hi} {\lya}, low ionization ({\siii} and \mgii), intermediate ionization ({\siiii} and \siiv), and {\civ} EWs that are among the largest of absorbers at the same $\rho$/{\rvir} from their nearest galaxies. These strong absorption lines in our sample, compared to respective ones in relatively isolated environments, could be potentially due to contributions from multiple galaxies and/or from the gaseous Leo Ring.  At larger distances from the Leo Ring, we note that SE ($\approx$ 400 kpc) and SJ ($\approx$ 550 kpc) have unusually large EWs of {\siiii} and {\civ}.  The even more distant sightline SB (at $\approx$825~kpc), on the other hand, does not have metal absorption detected, and has only typical absorption in {\lya}. 
In summary, however, it is clear that there is a lot of gas spread out throughout this entire Leo Ring/Group region. Although, this analysis invokes various assumptions and simplifications, it serves as an illustrative tool to highlight the impact of complex environments on the observables in our pencil beam probes.

\smallskip

The Leo Ring region presents an ideal location where we can consider the differences in the absorption properties for sightlines near isolated galaxies and those through galaxy groups. \citet{nielsen2018magiicat} found that sightlines through galaxy groups have stronger {\mgii} absorption than isolated galaxies, with a median EW of 0.65~{\AA} in groups compared to 0.41~{\AA} for isolated galaxies, within an impact parameter of 200 kpc from the nearest galaxy. A similar finding was reported by \citet{dutta2021} who found $\approx$2-3 times stronger {\mgii} absorption in higher overdensity group environments. Group sightlines also have more absorption components at larger velocities, though they do not have a large enough kinematic spread to be consistent with a superposition of two or more galaxies contributing to the sightline. This leads \citet{nielsen2018magiicat} to conclude that the gas from the individual galaxies is spread out in an intragroup medium through outflows and tidal stripping.  In this nearby Leo Ring environment we see a distribution of intragroup gas consistent with such a scenario.  The collision between NGC3384 and M96 is thought to have occurred 1.2Gyr ago, based on simulations~\citep{dansac2010}, and we would expect that in another couple of billion years the gas will to some extent disperse.  The cross section for metal-enriched, absorbing gas could remain substantial for a significant amount of time.

\smallskip

Absorption in {\civ} is commonly found to occur in a variety of environments, from isolated galaxies to galaxy groups, mainly detected within the virial radii of the galaxies~\citep{Burchett2016,jayadev19,manuwal2021}. In the Leo Ring region there is detected {\civ} in 4/5 of the sightlines for which there is {\civ} coverage.  The one sightline, SB, which does not have detected {\civ} has its closest galaxy at 3.7\rvir. The equivalent widths in Fig.~\ref{fig:ewcompare} of {\civ} along our sightlines are at the high end of the distributions from other samples at the same $\rho/R_{vir}$. The high {\civ} equivalent widths are suggestive that more than just a single undisturbed galaxy is contributing along these sightlines.

\smallskip

The {\ovi} EWs in the Leo Ring region, predicted by our models, are low compared to the other systems.  This is consistent with the results of \citet{Pointon2017} that absorbers through galaxy groups have weaker {\ovi} absorption than isolated galaxies.  We note, however, that in our Leo Ring sightlines we would not be aware of collisionally ionized {\ovi} gas, because the {\ovi} is not covered by the observations and we are only predicting it based on the phases of gas that we can observe.

\smallskip

Outside of the region that is likely to show absorption due to the Leo Ring itself, we also see {\lya} absorption towards SB. The {\hi} {\lya} EW is 0.2~{\AA}, and the N(\hi) value from our model/fit is 14.7.  Sightline SB has no known galaxies within 3.5{\rvir}, however, absorption of this strength is not unusual due to larger scale structures like filaments \citep{wakker2015,bouma2021,wilde2021}.
It is also not unusual to have sightlines 1-3{\rvir} from the nearest galaxy without detected {\hi} {\lya} absorption, like sightlines SI, SK, and SL.  In fact, using $dN/dz=30$ for $\log N({\hi}) > 14$~\citep{kim2021}, we estimate there is only an 8\% chance of detecting such absorption along a random sightline between 600 and 1400 {\kms}. The small probability of seeing {\hi} {\lya} absorption along a random sightline highlights the fact that SI, SK, and SL are not unusual.

\smallskip

\subsection{GALAXY CENTERED CONSIDERATIONS}

In this discussion the focus has so far been on understanding the contributions from the Leo Ring, galaxies, and large scale structure to the absorption along the various sightlines.  We have concluded that the Leo Ring gas itself is responsible for significant absorption along sightlines within $\approx$500kpc of its center, and that this high metallicity gas is the result of interactions between high metallicity galaxies which spreads it out over the region.  Now, rather than considering the origin of the absorbing gas and its relationship to the galaxies, we also want to reverse this and consider what absorption the large galaxies in the field are contributing at the positions of the sightlines, and whether this is reasonable based on their properties.  Note that in cases like SC, SG, and SH, with several absorbing components along the same sightline, it is likely that both galaxies and the Leo Ring itself are contributing absorption at different velocities.

\smallskip

The SB(r)b spiral M95 is only 0.2\rvir away from SC.  It would be very unusual not to see metal-line absorption at that impact parameter, and thus we associate the two components at $\approx 710$~{\kms} directly with this galaxy.  They are each $\sim$ 10 kpc in dimension and separated by few~{\kms} in velocity, consistent with arising in the same halo.  The next closest sightline to M95 is SF, at an impact parameter of 1.2\rvir for which we do not necessarily expect metal line absorption, but would probably detect {\hi} {\lya}.  SF is, however, is clearly in a region disturbed by the interaction so there may not be an identifiable contribution from M95 along the SF sightline.

\smallskip

The SAB(rs)ab galaxy, M96, is likely to be responsible for the 879{\kms} component in SC, given its impact parameter of 0.7\rvir. The EWs of the metal lines for just this component are {\hi} = 0.97\AA, {\siii} = 0.06\AA, {\siiii} = 0.03\AA, {\siiv}$<0.01$\AA, with {\civ} and {\ovi} not detected, and this would be consistent with expectations at that impact parameter. The metallicity of {\metallicity} $-1$, however, is low, and possibly suggestive of a contribution from infalling material. There is an even closer sightline to M96, sightline SF at 0.5\rvir.  It would not be unusual that we don't see detected metal line absorption at the expected velocity of $\approx 800$~{\kms}, however, we see from Fig. \ref{fig:ewcompare} that {\hi} {\lya} is almost always detected at this small impact parameter.  The absence of {\lya} absorption from M96 along the SF sightline could simply be a reflection of the noisy spectrum, or it is possible  that since M96 was involved in the interaction that produced the ring, that its {\hi} distribution is disturbed.

\smallskip

M105 is an early type E1 galaxy at $\approx$0.5\rvir from both SD and SF. For both sightlines we would expect absorption between 900 and 950~{\kms} based on extensions of the {\hi} emission map of the galaxy~\citep{Silchenko2003}.  From \citet{johnson2017}, we would expect to detect {\hi} {\lya} with an EW between 0.3 and  2~{\AA}, but we may or may not see metal lines detected.  Along sightline SD we find {\hi} {\lya} absorption at this velocity and the EW in the aligned component is at the lower end of that range typical for early type galaxies.  It is likely that the metals along sightline SD, at lower velocity, are related to the Leo Ring.  For SF, because of its location so close to the ring, it is difficult to separate contributions from individual galaxies and those from the debris.

\smallskip

The SB type galaxy, NGC3412 is probed by sightlines SG and SH within a virial radius (at 0.9\rvir) for both. It could contribute to the absorption seen in metals and the {\hi} component at $\approx$943{\kms}, however, NGC3384 and M105 can also contribute to absorption at these velocities, as could the Leo Ring.  Similarly, in SH,  NGC3412 could contribute at 818~{\kms} and 900~{\kms}, but other galaxies and the Leo Ring are equally or more likely to contribute.

\smallskip

NGC 3384, an SB(s) galaxy, is close to sightlines SD ($0.6R_{vir}$), SF ($0.5R_{vir}$), SG ($0.5R_{vir}$), and SH ($0.8R_{vir}$), for which absorption would be expected at $\approx 700$--$800$~{\kms}.  For all four of these sightlines we would expect {\hi} {\lya} absorption with EW between 0.1 and 2~{\AA} based on Figure~\ref{fig:ewcompare}. We may or may not typically see metal lines at these impact parameters.  In Figure~\ref{fig:galaxysights} we see that we do not detect {\hi} {\lya} absorption at the predicted velocity towards any of the four sightlines.  This implies that NGC 3384 is unusual, and we propose that this is because it specifically was the victim of a head on collision with M96, which left it with little gas of its own \citep{dansac2010}. 

\smallskip

There are many dwarf galaxies in the Leo Ring region, cataloged by~\citet{muller2018}, however only a few are at small impact parameters from our sightlines.  It is possible to detect fainter dwarfs than usual for absorption lines studies because of the proximity of this Leo Group region.  Nonetheless, we do not see a contribution from dwarfs to the absorption profiles along the sightlines we have studied.

\section{CONCLUSION: LESSONS FOR HIGHER REDSHIFT}
\label{sec:conclusion}
We have determined that the relatively large, Leo Ring region including the sightlines SC-SH and SJ, has a covering factor of nearly 1 of supersolar metallicity, low density gas.  Though it fills the region, the distribution of the gas is clumpy and the kinematics are complex.  Some of the absorption components are associated with specific galaxies in the Leo Group, while others are generally associated with gas in the region which has been liberated from individual galaxies.  The absorption strengths in both {\hi} {\lya} and metal lines are large, suggesting the need for contributions from the gas from more than one galaxy.  Much of the absorption that we see within the region is thus likely to be due to tidal debris and the intragroup medium.

\smallskip

The specifics of the distribution of debris in the region is surely related to the nature of the galaxy encounter, which is believed to be a head on collision between NGC 3384 and M96 \citep{dansac2010}. The geometry of the encounter is important for creating a ring without other obvious signs of interaction such as tidal tails and shells.  Though there is some star formation seen in the ring, the high metallicities, inferred through both our absorption studies and by \citet{corbelli2021} in emission, must be due to pre-enriched gas with its origin in the galaxies in the region.  The Leo Ring is a structure that is visible in {\hi} emission because it is in the local Universe, but there is presumably hidden debris in other more distant galaxy groups, which would be detected through absorption studies using background quasars.  Indeed \citet{nielsen2018magiicat} concluded that differences between {\mgii} absorption in isolated galaxies and in groups could be explained by intragroup gas.  The importance of such enriched gas with tidal origins in groups will depend on its covering factor and its longevity.  Certain types of interactions will distribute larger amounts of gas, and tidal features may last 1-2 billion years, but afterwards the dispersed gas will continue to have absorption signatures.

\smallskip

Most of the absorption that we are observing from the Leo Group intragroup medium arises in a single phase (with similar density and temperature over a large region) though it does have clumpiness and substructures that we may not be probing with our random sightlines. It is important to note that gas with the same density will yield different absorption signatures at different redshifts.  The gas we are probing in the Leo Ring region now produces {\civ} absorption due to photoionization because of its density as compared to the density of ionizing photons in the extragalactic background.  At a higher redshift, it is likely that significant photoionized {\ovi} absorption would be observed from gas with these same conditions. We also cannot address whether an additional warm/hot phase is present without coverage of higher ionization transitions (e.g. {\ovi} and \neviii).

\smallskip

Observations of lines of sight through other known interacting galaxies in the local universe, and through a variety of galaxy groups will be important diagnostics of the conditions of such regions.
For example, the M81 triplet \citep{Chynoweth2008,deblok2018}, has a very extended distribution of {\hi} debris detected through 21-cm emission.
Looking at conditions component by component is very important in such analysis.  For example, in this work, if all of the {\hi} was incorrectly associated with the single components that show metals in most sightlines, the inferred metallicity would be much lower, and one might mistake the material to be of primordial origin. As we look back to earlier epochs in the Universe, where there is no hope of detecting tidal debris through any other method but quasar absorption lines,
the ability to recognize the absorption signatures of tidal material will be very important.

\section*{ACKNOWLEDGEMENTS}
We thank the anonymous reviewer for their constructive feedback which helped improve the quality of manuscript. Computations for this research were performed on the Pennsylvania State University's Institute for Computational and Data Sciences' Roar supercomputer. The authors also acknowledge the Texas Advanced Computing Center (TACC) at The University of Texas at Austin for providing HPC resources that have contributed to the research results reported within this paper. G.G.K. and N.M.N.~acknowledge the support of the Australian Research Council through {\it Discovery Project} grant DP170103470. Parts of this research were supported by the Australian Research Council Centre of Excellence for All Sky Astrophysics in 3 Dimensions (ASTRO 3D), through project number CE170100013. A.N acknowledges the support for this work through grant number EMR/2017/002531 from the Department of Science and Technology, Government of India. This research made use of Astropy, a community-developed core
Python package for Astronomy~\citep{astropy2013}, NumPy~\citep{harris2020},\textsc{matplotlib}~\citep{Hunter2007}, and SciPy~\citep{virtanen2020}. This research has made use of
the HSLA database, developed and maintained at STScI, Baltimore,
USA. This research has made use of the NASA/IPAC Extragalactic
Database (NED), which is operated by the Jet Propulsion Laboratory,
California Institute of Technology, under contract with NASA. We acknowledge the work of people involved in the design, construction and deployment of the COS on-board the Hubble Space Telescope, and thank all those who obtained data for the sight-lines studied in this paper. We thank KITP for hosting the Fundamentals of
Gaseous Halos workshop (supported by the NSF under Grant No. NSF PHY-1748958) during which some of the ideas presented in this work were formulated.

\section*{Data availability}
The data underlying this article will be shared on reasonable request to the corresponding author.

\bibliographystyle{mnras}
\bibliography{references}
\bsp
\clearpage
\appendix
\label{appendix}

\section{Comparison of N(\hi; 21 cm) with N(\hi; {\lya})}
In this section we present a comparison between measurement of {\colden} from  EBHIS~\citep{winkel16} database and the {\colden} measured from a VP fit to the Galactic {\lya}.
\label{appendix:SA}
\begin{figure*}
\begin{center}
\includegraphics[scale=0.75]{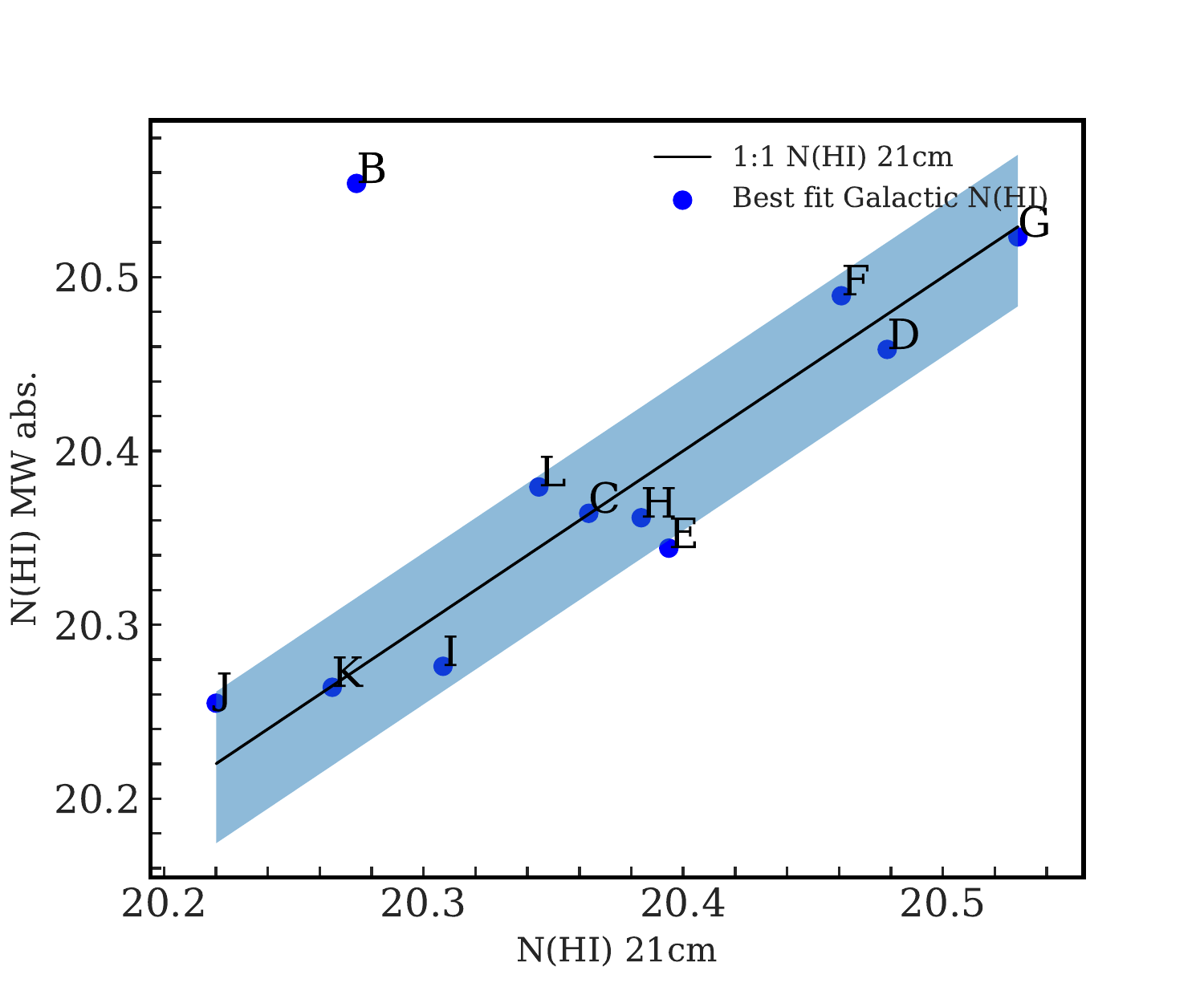}
\caption{Comparison between $\log N$(\hi; \lya) and $\log N$(\hi; 21 cm) for the 11 sightlines in this study. The blue shaded region shows the 1$\sigma$ uncertainty range for the ratio between $\log N$(\hi; \lya) and $\log N$(\hi; 21 cm) as determined in \citet{wakker2011}. All the values, except SB, are in good agreement. The Galactic {\lya} in the case of SB is blended with very strong intrinsic absorption lines of {\mgx}$\lambda$624, {\cavii}$\lambda$624, {\oxyv}$\lambda$630 which are associated with outflows from the background AGN~\citep{xu2020}, preventing an accurate modeling of the absorption. However, the uncertainty in the {\colden} of Galactic {\lya} does not significantly affect the {\lya} profile of the absorber of interest.}
\label{fig:21cmvsabs}
\end{center}
\end{figure*}
\clearpage

\section{Plots for SB}
\label{appendix:SB}

\subsection{Airglow template fit towards SB}

In this section we present the airglow template fit to the geocoronal {\lya} emission towards SB. The templates were obtained from the publically available community generated airglow templates\footnote{\url{https://www.stsci.edu/hst/instrumentation/cos/calibration/airglow}}~\citep{bourrier2018} to fit the emission profile.

\begin{figure*}
\begin{center}
\includegraphics[width=0.75\linewidth]{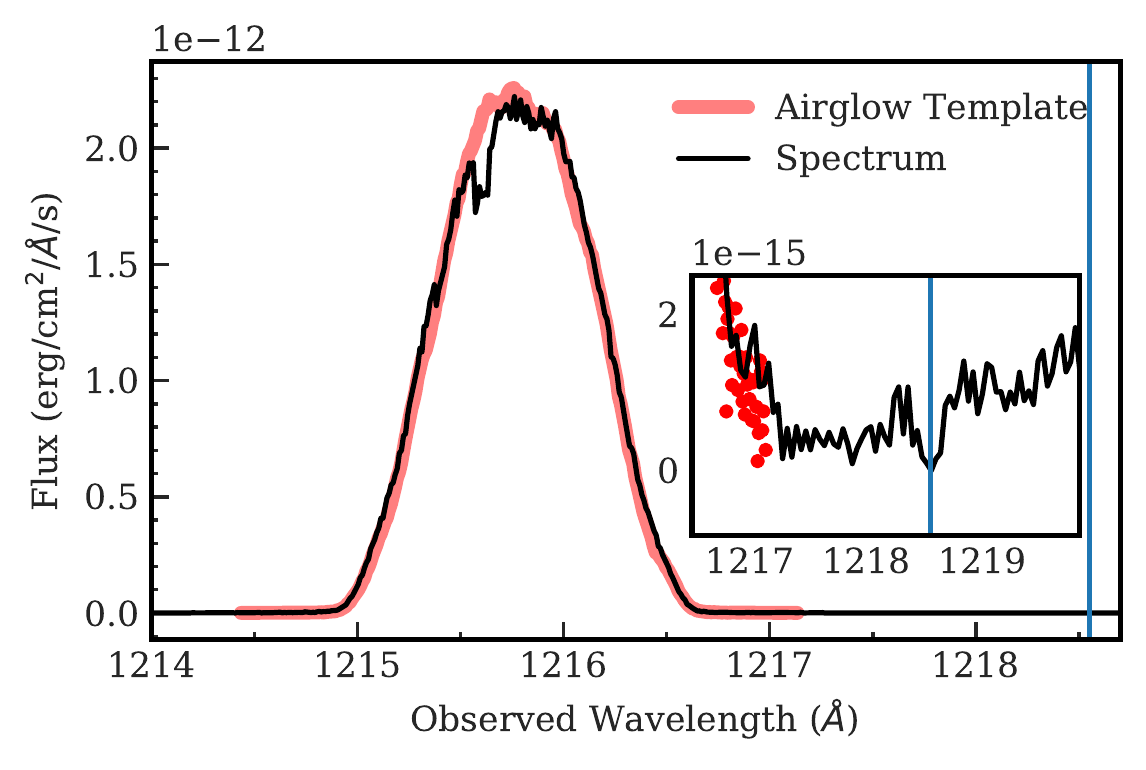}
\caption{The airglow template fit to the spectrum. Inset shows the zoom of the region containing the absorption of interest; the vertical blue line is placed at the absorption centroid}. There is no contribution of the {\hi} {\lya} geocoronal emission to the {\hi} {\lya} absorption of interest.
\label{fig:SBairglow}
\end{center}
\end{figure*}

\subsection{Best VP fit to the Galactic {\lya} towards SB}

In this section we show the VP fit to the Galactic {\lya}. We determine the unresolved component structure of Galactic {\lya} using the component structure determined from a VP fit to the Galactic {\siii} lines. The best VP fit to Galactic {\lya} is divided with the normalized spectrum towards a sightline to obtain the {\hi} profile of interest.

\begin{figure*}
\begin{center}
\includegraphics[width=\linewidth]{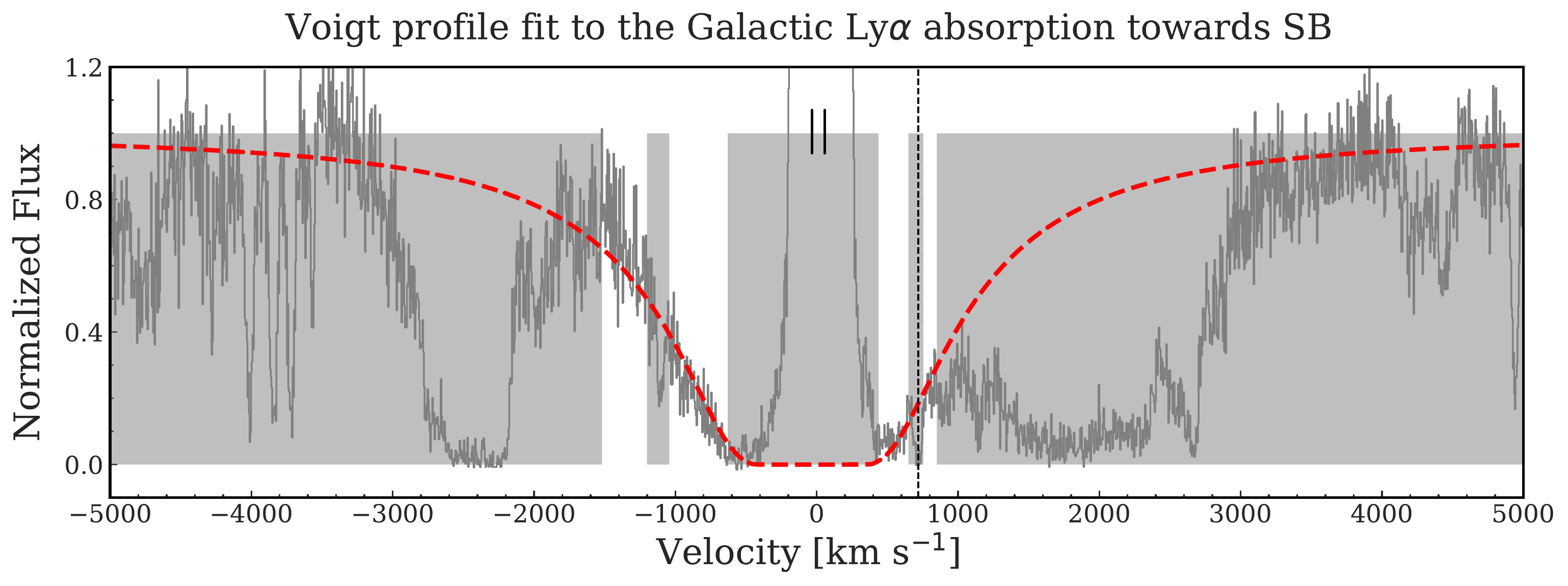}
\caption{The VP fit to the MW {\lya} absorption towards SB shown as a red curve. The two small thick vertical lines denote the absorption centroids determined from a VPfit to the Galacitc {\siii} line. The absorption of interest is at $\approx$ 711 {\kms}, and identified with a vertical dashed line. The absorption between $\approx$1700{\kms} -- 1800{\kms} is intrinsic {\oxyv} $\lambda$630 absorption associated with an outflow from the background quasar at $z=$ 0.978.}
\label{fig:SBvpfitgal}
\end{center}
\end{figure*}

\subsection{Posterior distributions for the absorber properties towards SB}
\label{appendix:SBparams}

In this section we present the posteriors summarizing the parameter distributions obtained from our Bayesian analysis. The prior distribution is defined in section 3 and the likelihood function is defined in S21.

\begin{figure*}
\begin{center}
\includegraphics[width=\linewidth]{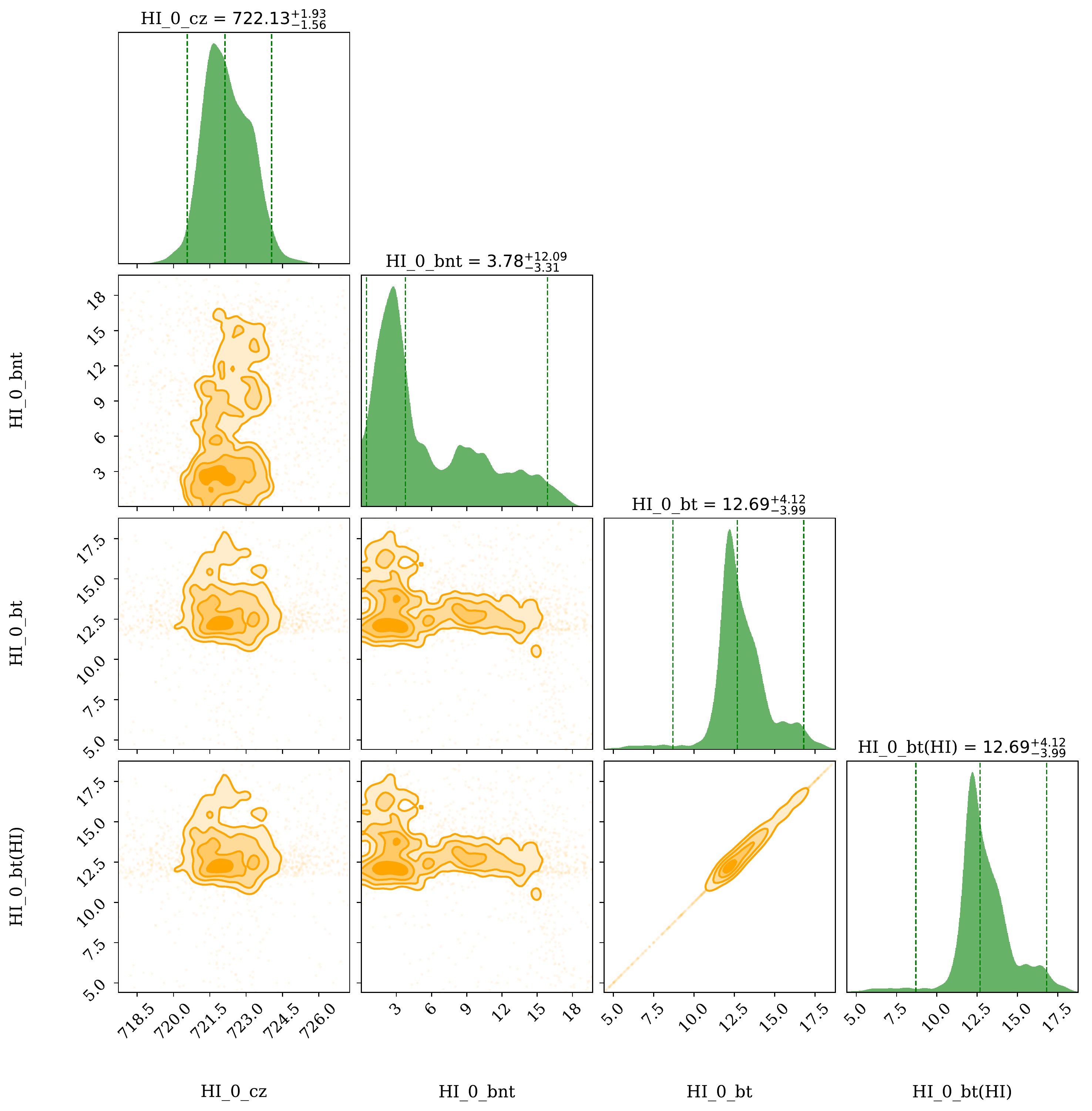}
\caption{The corner plot showing the marginalized posterior distributions for the absorption centroid ($z$), non-thermal Doppler broadening ($b_{nt}$), thermal Doppler broadening ($b_{t}$), and total Doppler broadening ($b$), of the phase traced by the {\hi} cloud of the $z=0.0024$ absorber towards SB. The over-plotted vertical lines in the posterior distribution span the 95\% credible interval. The contours indicate 0.5$\sigma$, 1$\sigma$, 1.5$\sigma$, and 2$\sigma$ levels. The model results are summarised in Table~\ref{tab:modelparams}, and the synthetic profiles based on these models are shown in Figure~\ref{fig:SBsysplot}.}
\label{fig:SBHI0b}
\end{center}
\end{figure*}

\begin{figure*}
\begin{center}
\includegraphics[width=\linewidth]{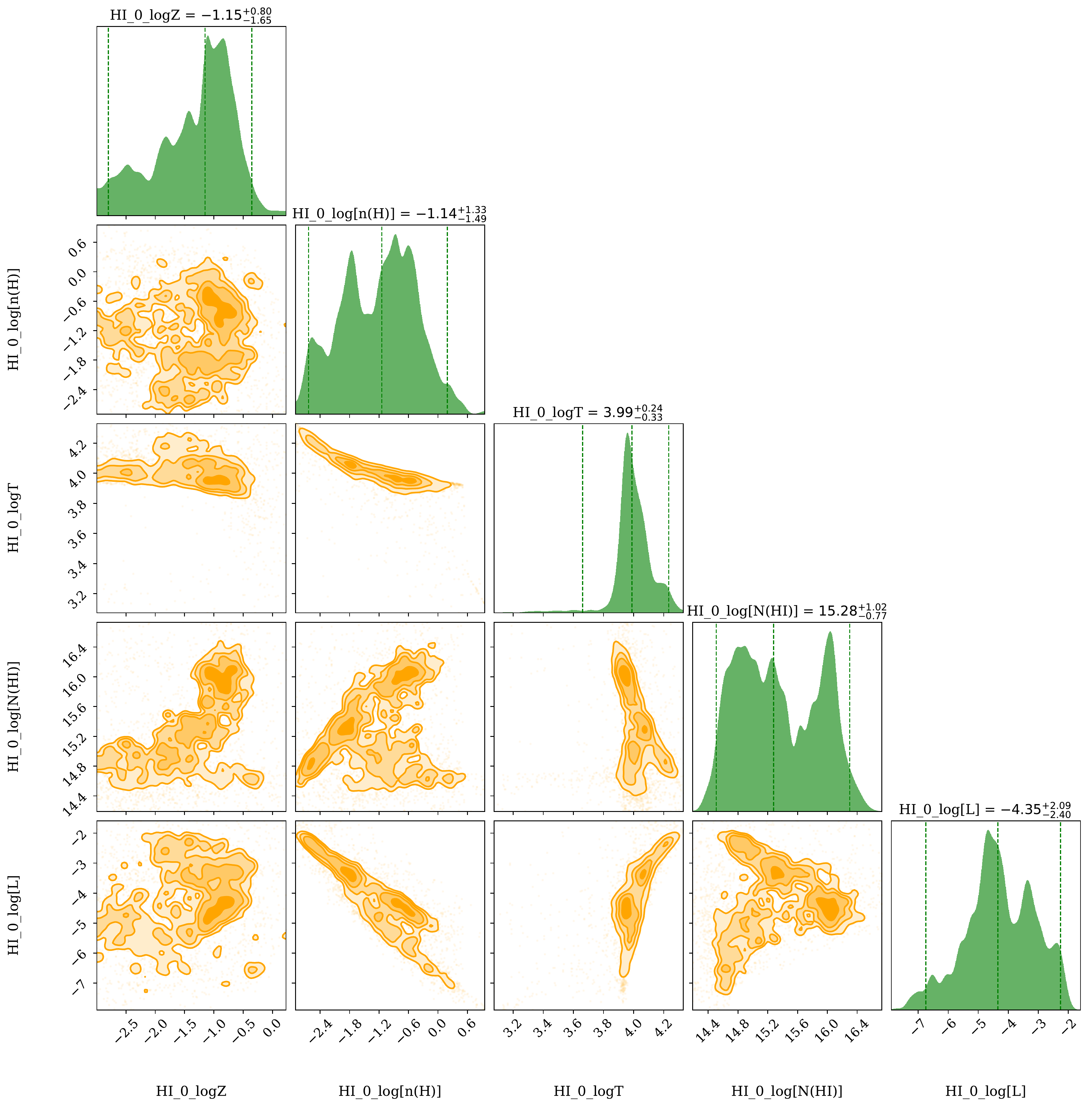}
\caption{The corner plot showing the marginalized posterior distributions for the metallicity ($\log Z$), hydrogen number density ($\log n_{H}$), temperature ($\log T$), neutral hydrogen column density ($\log N(\hi)$), and the line of sight thickness ($\log L$), of the phase traced by the {\hi} cloud of the $z=0.0024$ absorber towards SB. The over-plotted vertical lines in the posterior distribution span the 95\% credible interval. The contours indicate 0.5$\sigma$, 1$\sigma$, 1.5$\sigma$, and 2$\sigma$ levels. The model results are summarised in Table~\ref{tab:modelparams}, and the synthetic profiles based on these models are shown in Figure~\ref{fig:SBsysplot}.}
\label{fig:SBHI0}
\end{center}
\end{figure*}

\clearpage

\section{Plots for SC}
\label{appendix:SC}

\subsection{Airglow template fit towards SC}
\begin{figure*}
\begin{center}
\includegraphics[width=0.75\linewidth]{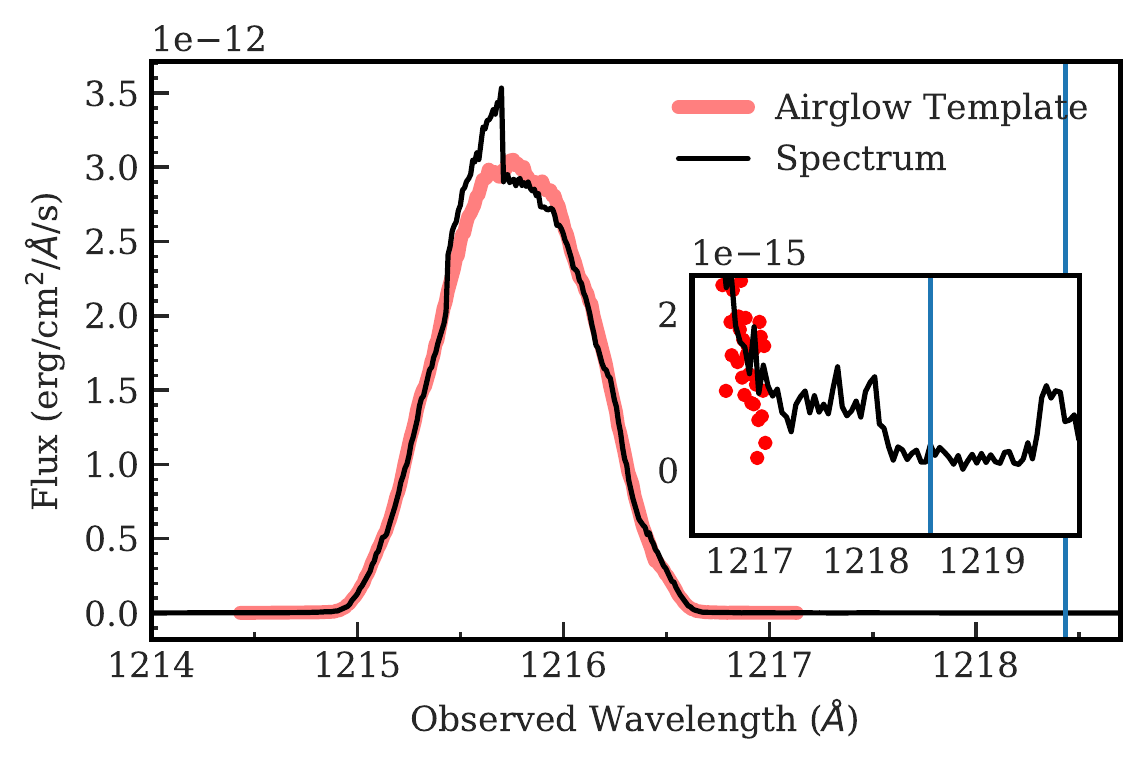}
\caption{Same as in Figure~\ref{fig:SBairglow}, but for SC.}
\label{fig:SCairglow}
\end{center}
\end{figure*}

\subsection{Best VP fit to the Galactic {\lya} towards SC}
\begin{figure*}
\begin{center}
\includegraphics[width=\linewidth]{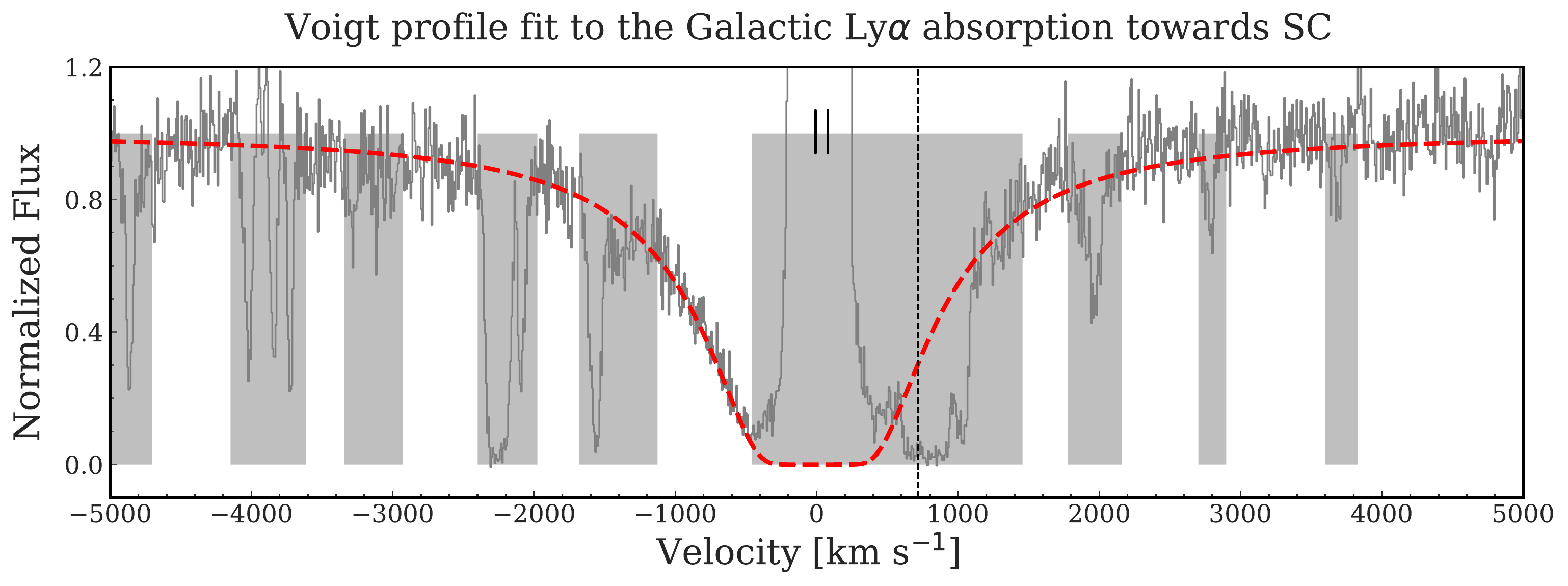}
\caption{Same as in Figure~\ref{fig:SBvpfitgal}, but for SC. Towards SC, we fit 3 components to the absorption profile of Galactic {\hi} {\lya}, as we identify 3 components in the Galactic {\siii} absorption profile. }
\label{fig:SCvpfit}
\end{center}
\end{figure*}

\subsection{PC model for the high-ionization phase towards SC}
\begin{figure*}
\begin{center}
\includegraphics[width=\linewidth]{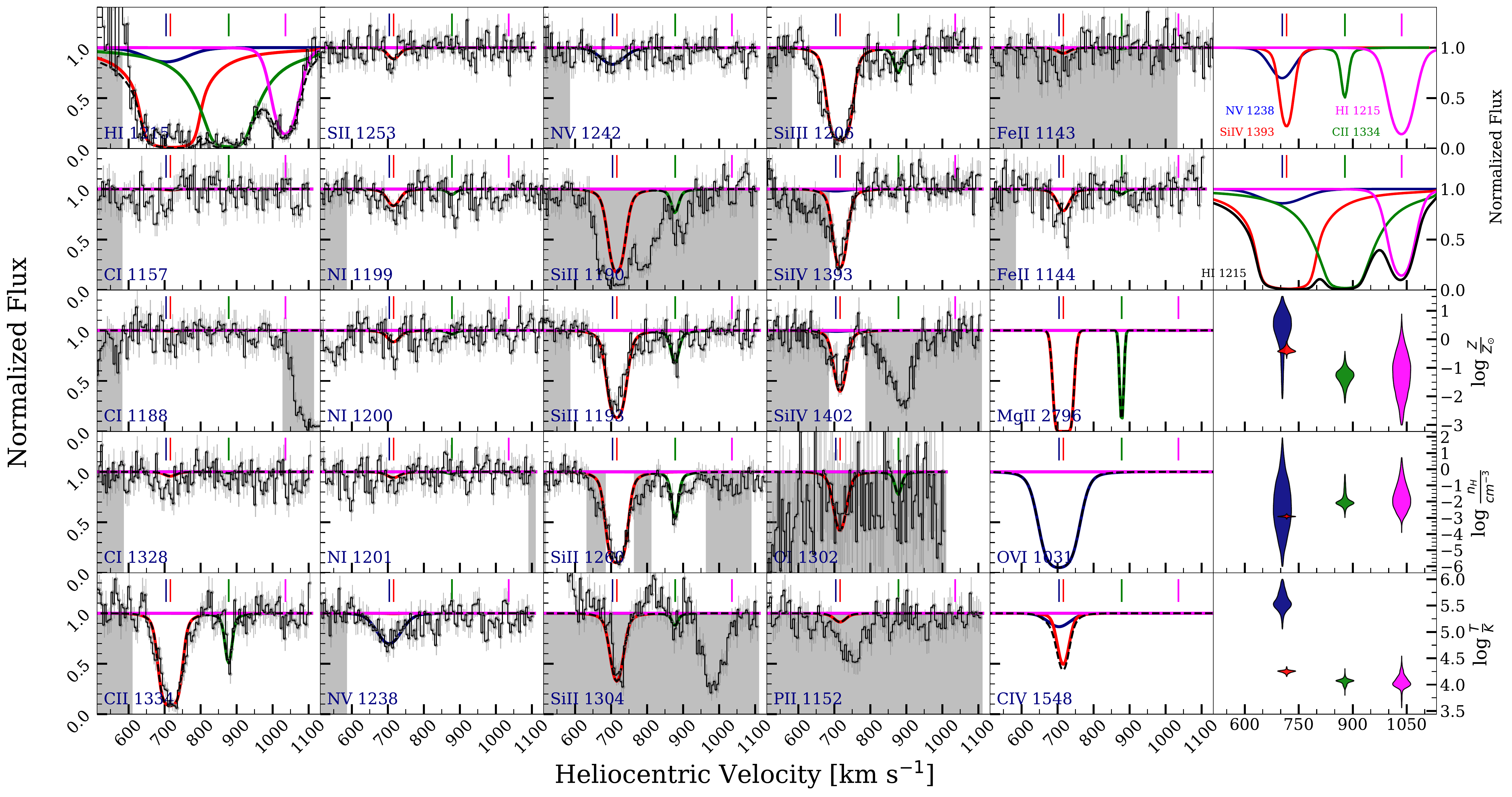}
\caption{A PC model for the absorption towards SC. Only {\nv} absorption (shown as a blue curve) is modeled as a collisionally ionized phase. The other phases are still modeled as photoionized only.}
\label{fig:SCpici}
\end{center}
\end{figure*}

\subsection{MC exploration of the zero-point uncertainty for SC}

\begin{figure*}
\begin{center}
\includegraphics[scale=0.5]{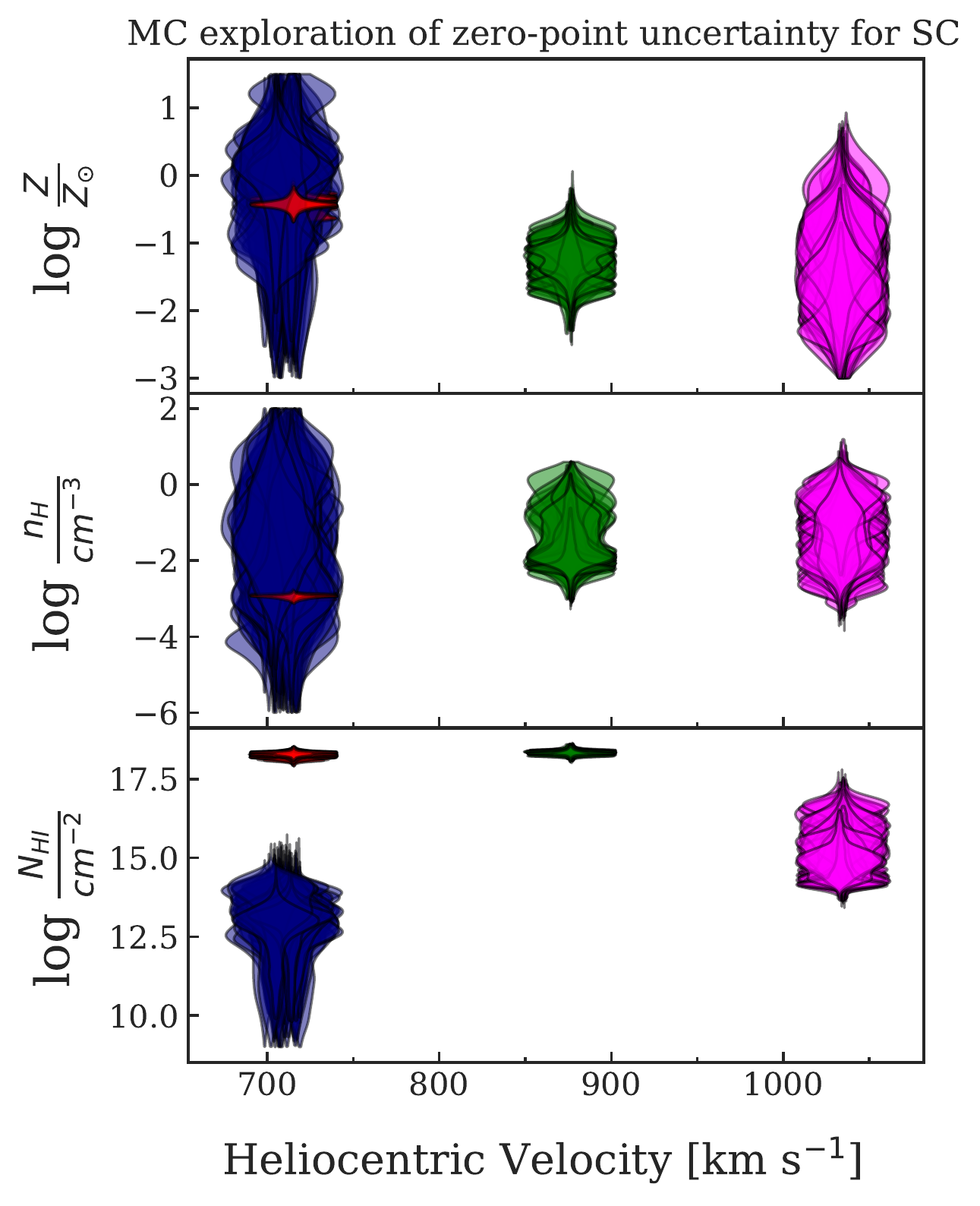}
\caption{Violin plots showing the parameter distributions for 100 different realizations of the {\hi} {\lya} profile modified between 710--720 {\kms} to account for the zero-point uncertainty. The blue violins show the parameters of {\nv}, modeled as a collisionally ionized phase. The red violins correspond to the {\siiv} phase, the green violins correspond to the {\cii} traced phase, and the magenta violins correspond to the {\hi} only phase; all of which are PIE models. }
\label{fig:MCSC}
\end{center}
\end{figure*}

\subsection{Posterior distributions for the absorber properties towards SC}
\label{appendix:SCparams}
\begin{figure*}
\begin{center}
\includegraphics[width=\linewidth]{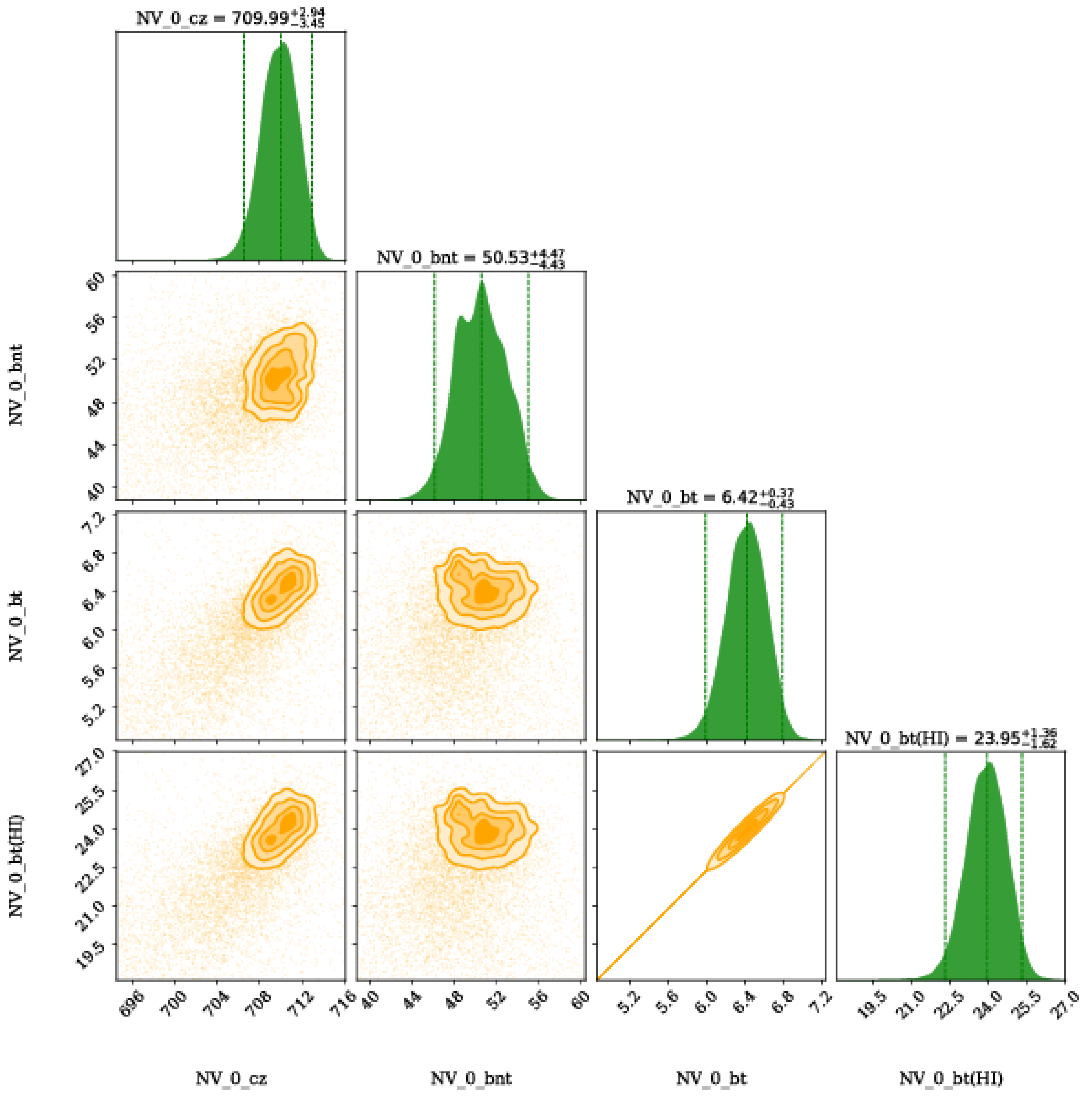}
\caption{The corner plot showing the marginalized posterior distributions for the absorption centroid ($z$), non-thermal Doppler broadening ($b_{nt}$), thermal Doppler broadening ($b_{t}$), total Doppler broadening ($b$), of the phase traced by the broad {\nv} cloud of the $z=0.00238$ absorber towards SC. The over-plotted vertical lines in the posterior distribution span the 95\% credible interval. The contours indicate 0.5$\sigma$, 1$\sigma$, 1.5$\sigma$, and 2$\sigma$ levels. The model results are summarised in Table~\ref{tab:modelparams}, and the synthetic profiles based on these models are shown in Figure~\ref{fig:SCsysplot}.}
\label{fig:SCNV0b}
\end{center}
\end{figure*}

\begin{figure*}
\begin{center}
\includegraphics[width=\linewidth]{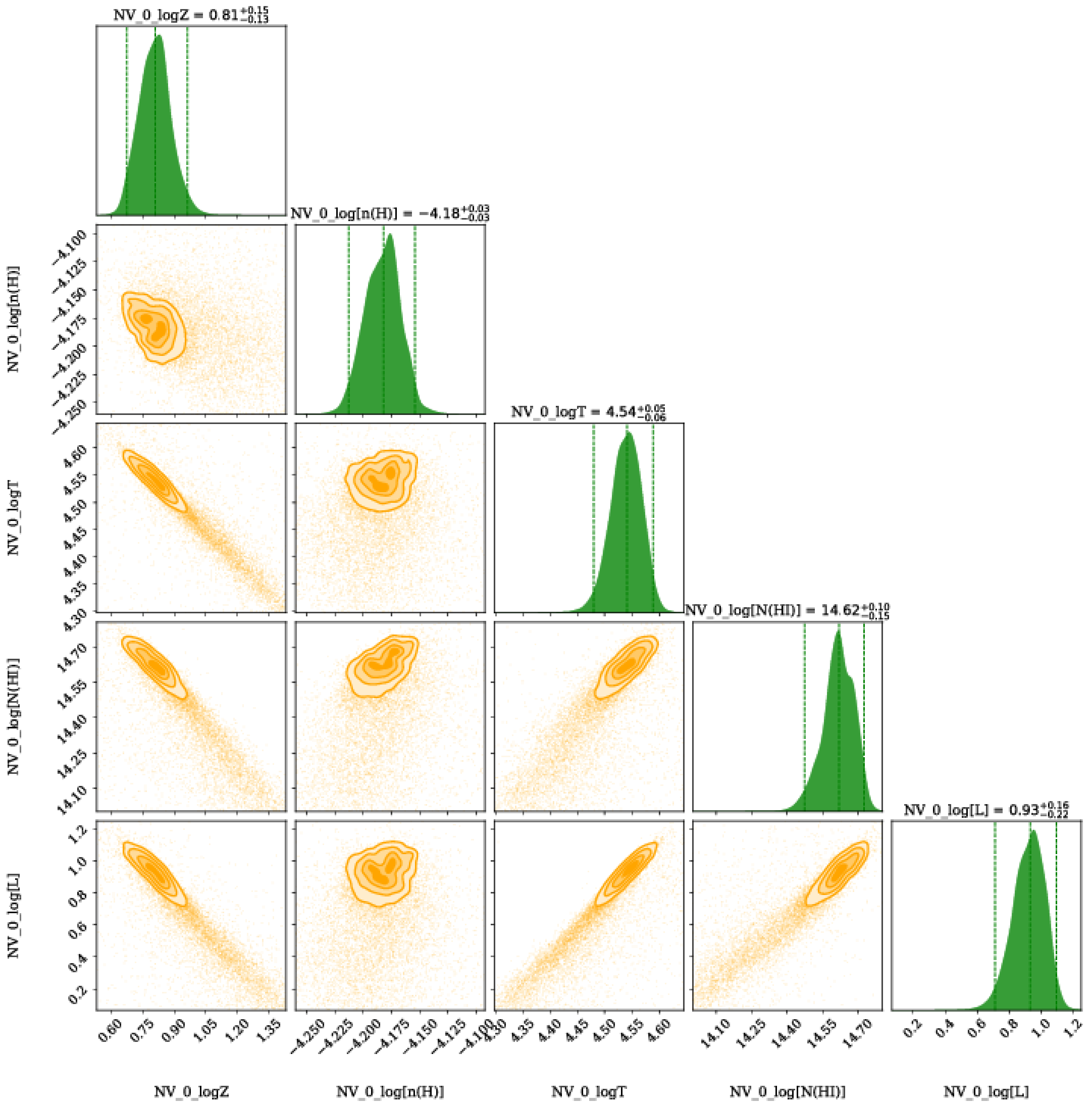}
\caption{The corner plot showing the marginalized posterior distributions for the metallicity ($\log Z$), hydrogen number density ($\log n_{H}$), and other physical properties of the broad higher ionization phase gas traced by the {\nv} cloud of the $z=0.00238$ absorber towards SC. The over-plotted vertical lines in the posterior distribution span the 95\% credible interval. The contours indicate 0.5$\sigma$, 1$\sigma$, 1.5$\sigma$, and 2$\sigma$ levels. The model results are summarised in Table~\ref{tab:modelparams}, and the synthetic profiles based on these models are shown in Figure~\ref{fig:SCsysplot}.}
\label{fig:SCNV0}
\end{center}
\end{figure*}
\begin{figure*}
\begin{center}

\includegraphics[width=\linewidth]{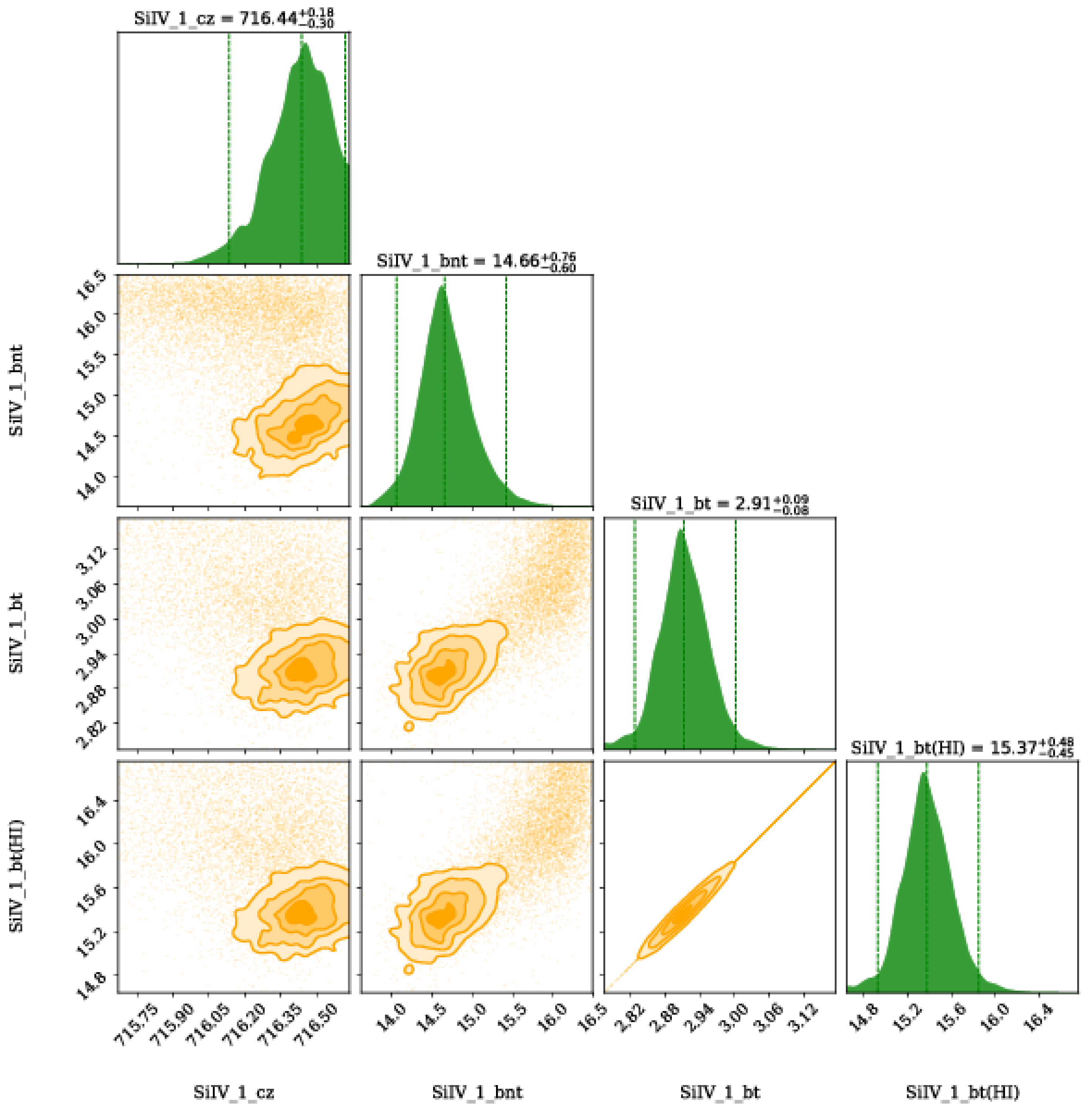}
\caption{The corner plot showing the marginalized posterior distributions for the absorption centroid ($z$), non-thermal Doppler broadening ($b_{nt}$), thermal Doppler broadening ($b_{t}$), total Doppler broadening ($b$), of the lower ionization phase traced by the {\siiv} cloud of the $z=0.00238$ absorber towards SC. The over-plotted vertical lines in the posterior distribution span the 95\% credible interval. The contours indicate 0.5$\sigma$, 1$\sigma$, 1.5$\sigma$, and 2$\sigma$ levels. The model results are summarised in Table~\ref{tab:modelparams}, and the synthetic profiles based on these models are shown in Figure~\ref{fig:SCsysplot}.}
\label{fig:SCSiIV1b}
\end{center}
\end{figure*}

\begin{figure*}
\begin{center}
\includegraphics[width=\linewidth]{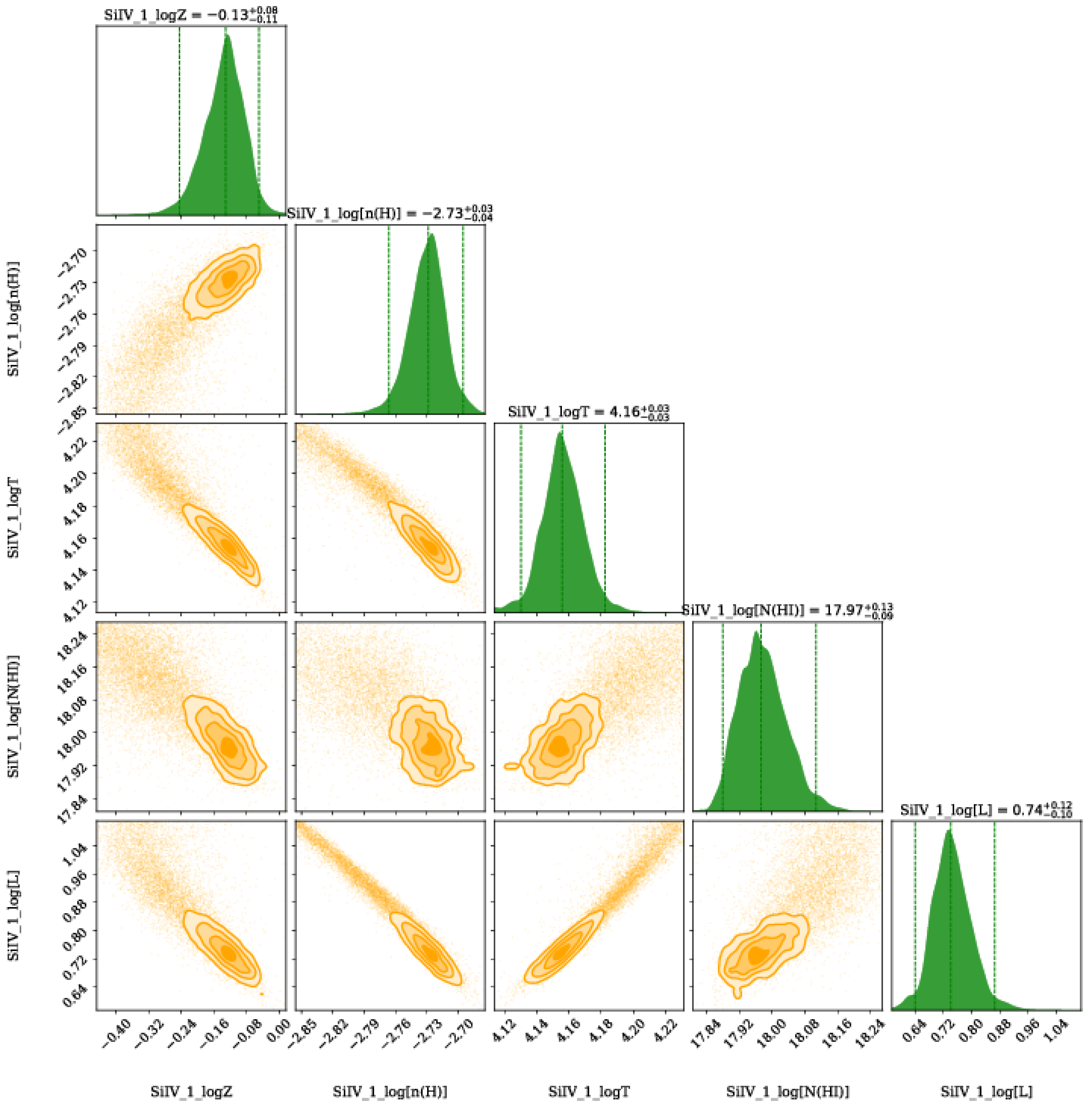}
\caption{The corner plot showing the marginalized posterior distributions for the metallicity ($\log Z$), hydrogen number density ($\log n_{H}$), and other physical properties of the low ionization phase traced by the {\siiv} cloud of the $z=0.00238$ absorber towards SC. The over-plotted vertical lines in the posterior distribution span the 95\% credible interval. The contours indicate 0.5$\sigma$, 1$\sigma$, 1.5$\sigma$, and 2$\sigma$ levels. The model results are summarised in Table~\ref{tab:modelparams}, and the synthetic profiles based on these models are shown in Figure~\ref{fig:SCsysplot}.}
\label{fig:SCSiIV1}
\end{center}
\end{figure*}

\begin{figure*}
\begin{center}
\includegraphics[width=\linewidth]{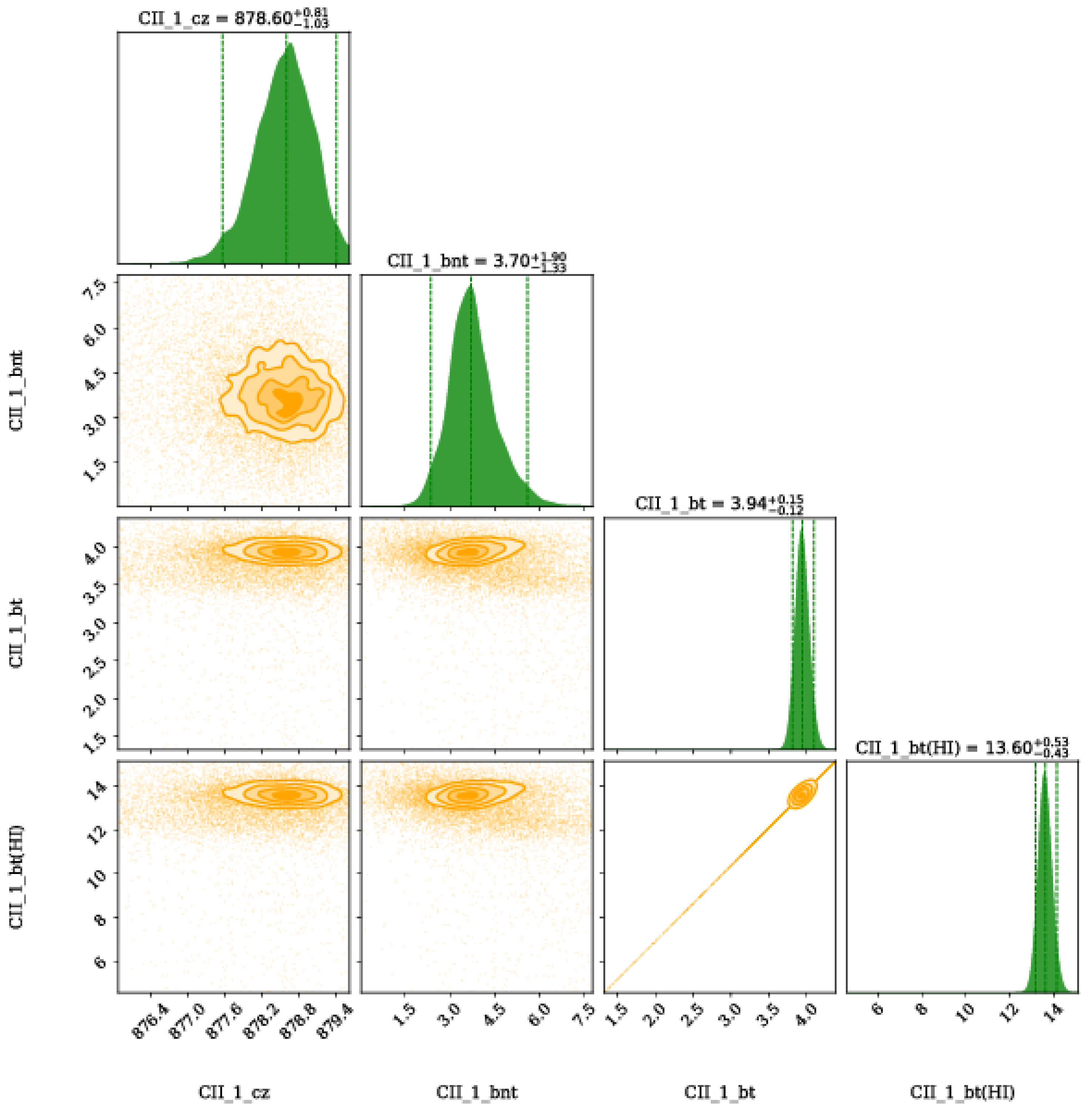}
\caption{The corner plot showing the marginalized posterior distributions for the absorption centroid ($z$), non-thermal Doppler broadening ($b_{nt}$), thermal Doppler broadening ($b_{t}$), total Doppler broadening ($b$), of the phase traced by the {\cii} cloud of the $z=0.00238$ absorber towards SC. The over-plotted vertical lines in the posterior distribution span the 95\% credible interval. The contours indicate 0.5$\sigma$, 1$\sigma$, 1.5$\sigma$, and 2$\sigma$ levels. The model results are summarised in Table~\ref{tab:modelparams}, and the synthetic profiles based on these models are shown in Figure~\ref{fig:SCsysplot}.}
\label{fig:SCCII1b}
\end{center}
\end{figure*}

\begin{figure*}
\begin{center}
\includegraphics[width=\linewidth]{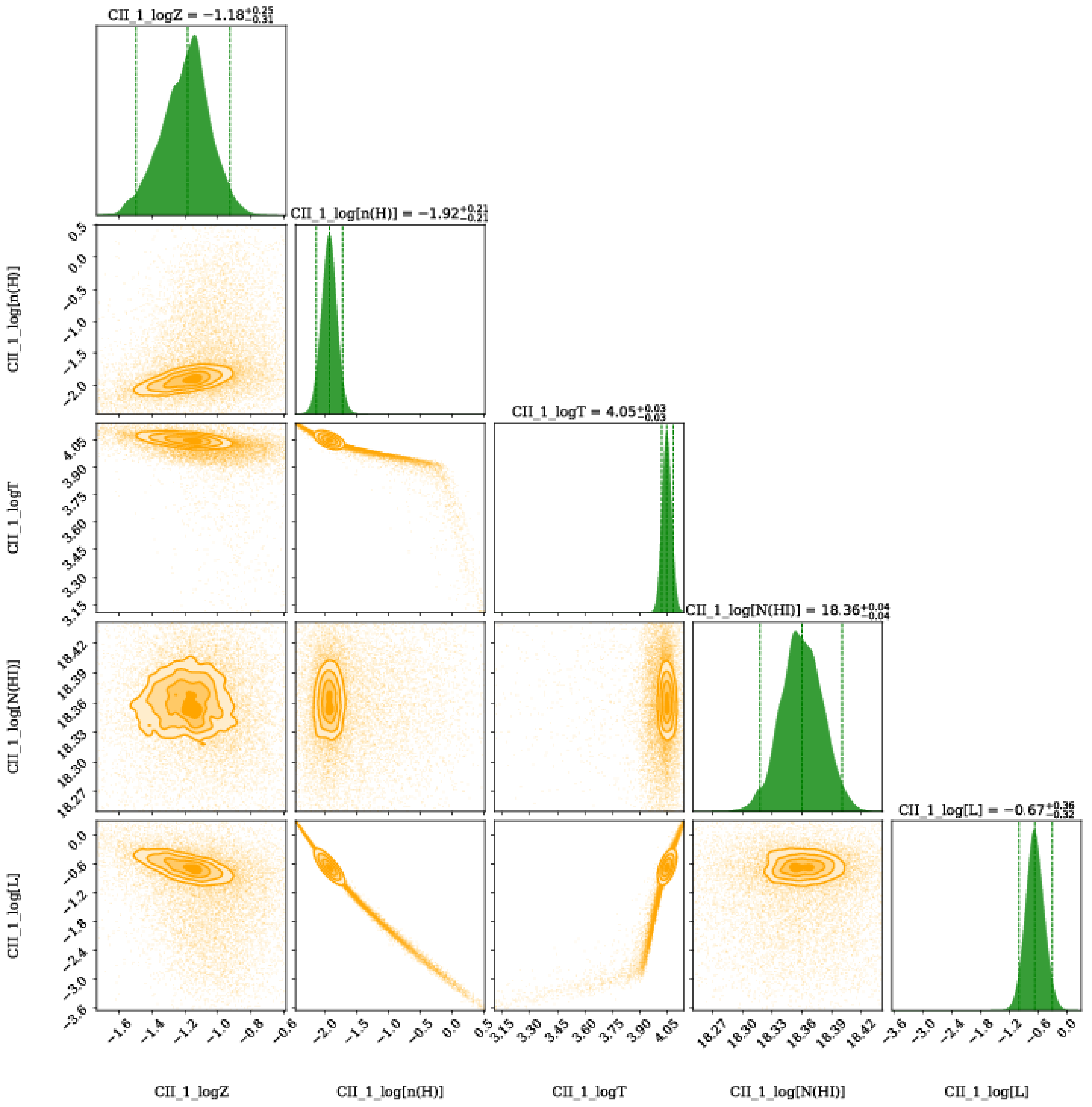}
\caption{The corner plot showing the marginalized posterior distributions for the metallicity ($\log Z$), hydrogen number density ($\log n_{H}$), and other physical properties of the low ionization phase traced by the {\cii} cloud of the $z=0.00238$ absorber towards SC. The over-plotted vertical lines in the posterior distribution span the 95\% credible interval. The contours indicate 0.5$\sigma$, 1$\sigma$, 1.5$\sigma$, and 2$\sigma$ levels. The model results are summarised in Table~\ref{tab:modelparams}, and the synthetic profiles based on these models are shown in Figure~\ref{fig:SCsysplot}.}
\label{fig:SCCII1}
\end{center}
\end{figure*}

\begin{figure*}
\begin{center}
\includegraphics[width=\linewidth]{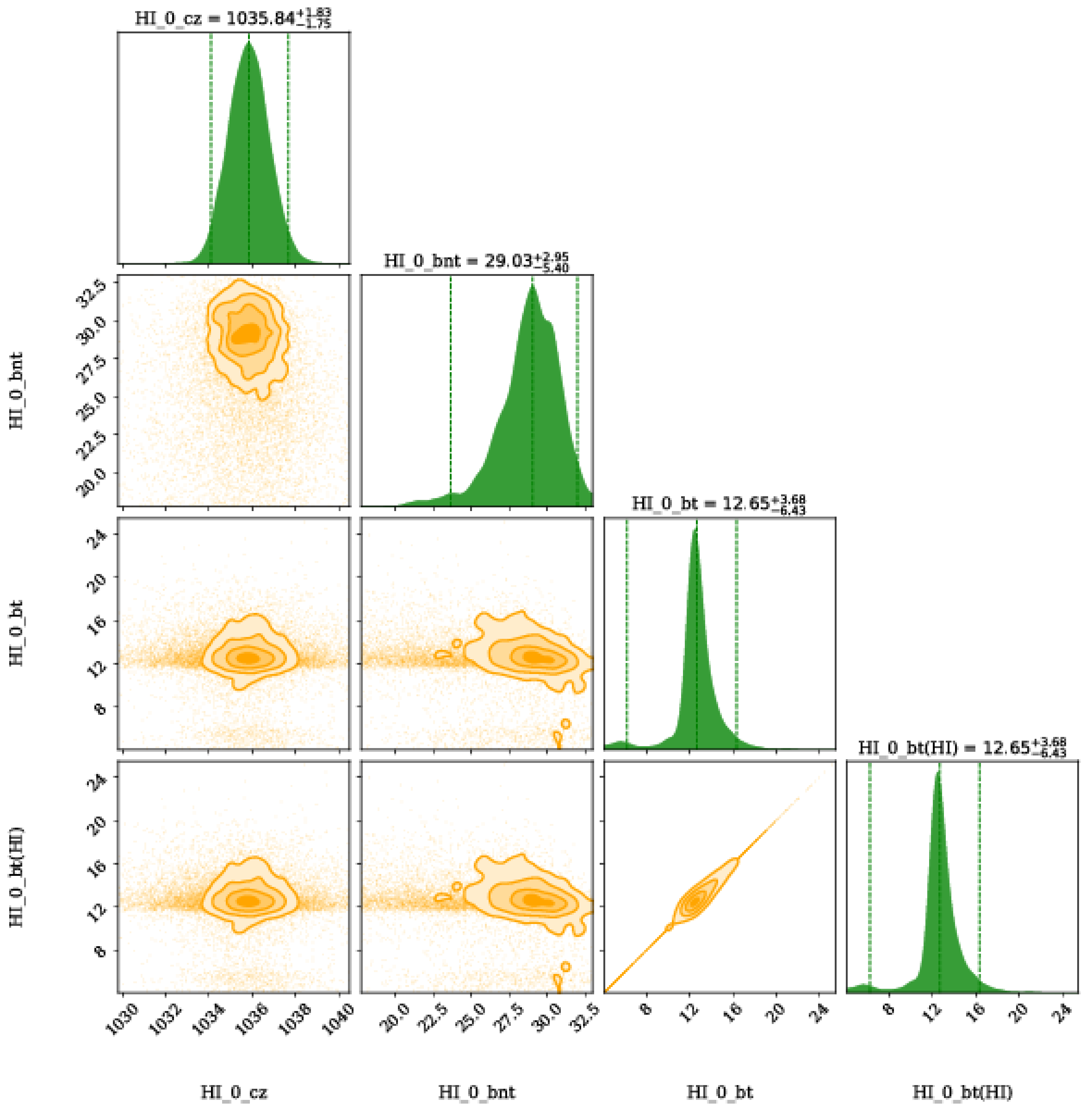}
\caption{The corner plot showing the marginalized posterior distributions for the absorption centroid ($z$), non-thermal Doppler broadening ($b_{nt}$), thermal Doppler broadening ($b_{t}$), total Doppler broadening ($b$), of the phase traced by the {\hi} cloud of the $z=0.00238$ absorber towards SC. The over-plotted vertical lines in the posterior distribution span the 95\% credible interval. The contours indicate 0.5$\sigma$, 1$\sigma$, 1.5$\sigma$, and 2$\sigma$ levels. The model results are summarised in Table~\ref{tab:modelparams}, and the synthetic profiles based on these models are shown in Figure~\ref{fig:SCsysplot}.}
\label{fig:SCHI0b}
\end{center}
\end{figure*}

\begin{figure*}
\begin{center}
\includegraphics[width=\linewidth]{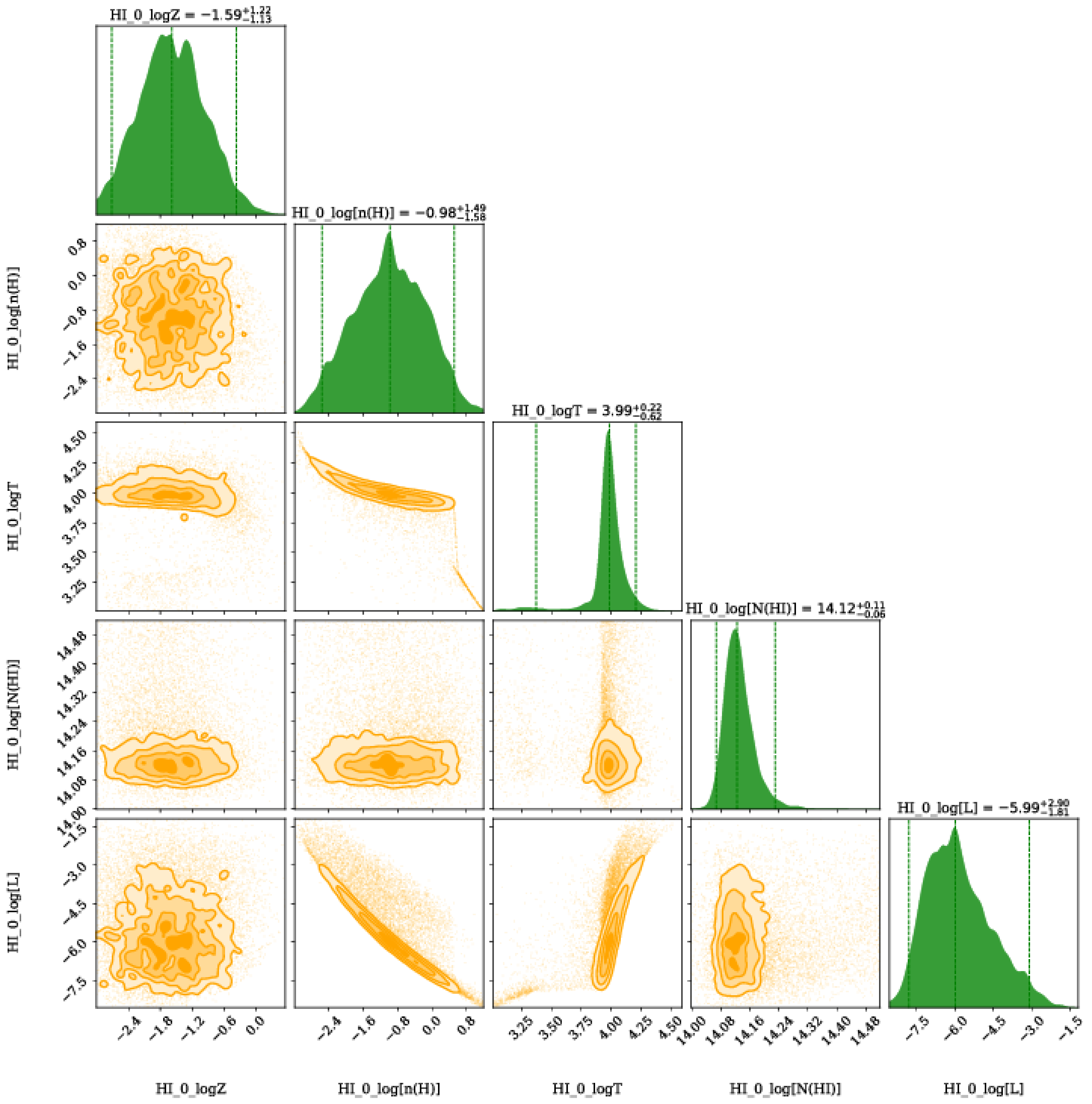}
\caption{The corner plot showing the marginalized posterior distributions for the metallicity ($\log Z$), hydrogen number density ($\log n_{H}$), and other physical properties of the low ionization phase traced by the {\hi} cloud of the $z=0.00238$ absorber towards SC. The over-plotted vertical lines in the posterior distribution span the 95\% credible interval. The contours indicate 0.5$\sigma$, 1$\sigma$, 1.5$\sigma$, and 2$\sigma$ levels. The model results are summarised in Table~\ref{tab:modelparams}, and the synthetic profiles based on these models are shown in Figure~\ref{fig:SCsysplot}.}
\label{fig:SCHI0}
\end{center}
\end{figure*}
\clearpage
\section{Plots for SD}
\label{appendix:SD}
\subsection{Airglow template fit towards SD}
\begin{figure*}
\begin{center}
\includegraphics[width=0.75\linewidth]{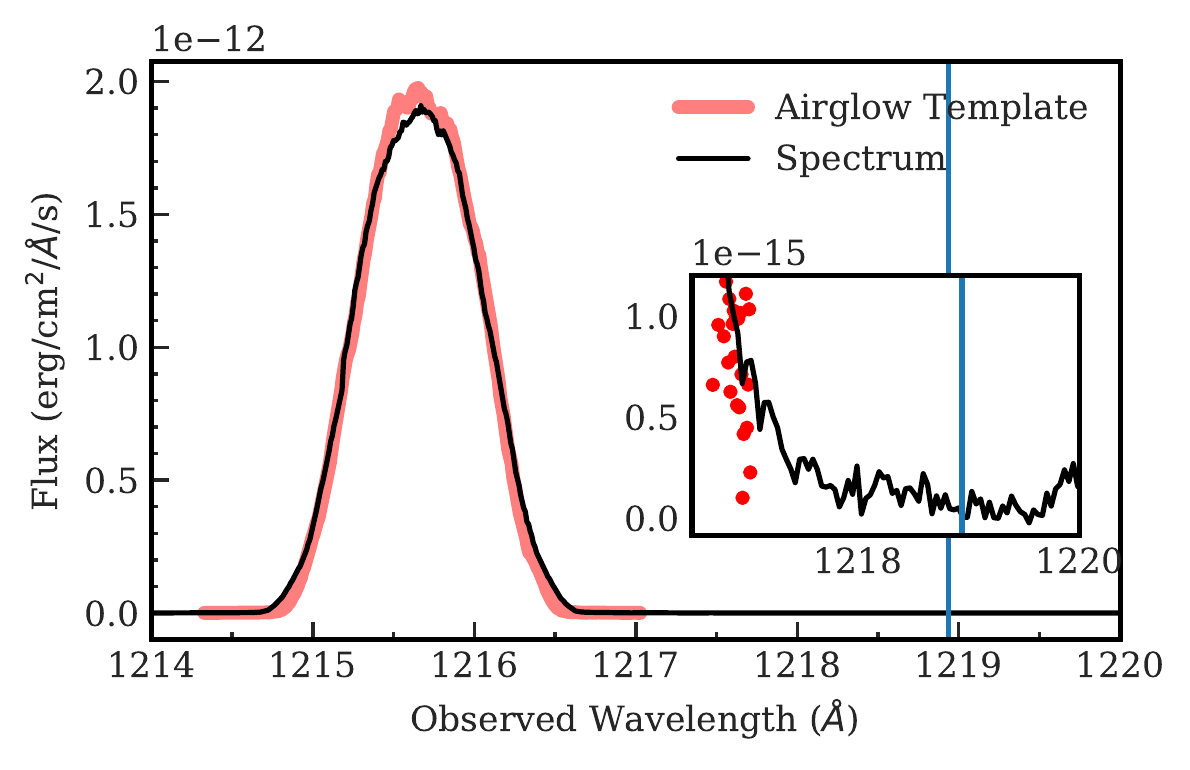}
\caption{Same as in Figure~\ref{fig:SBairglow}, but for SD.}
\label{fig:SDairglow}
\end{center}
\end{figure*}
\subsection{Airglow template fit towards SD}

\subsection{Best VP fit to the Galactic {\lya} towards SD}
\begin{figure*}
\begin{center}
\includegraphics[width=\linewidth]{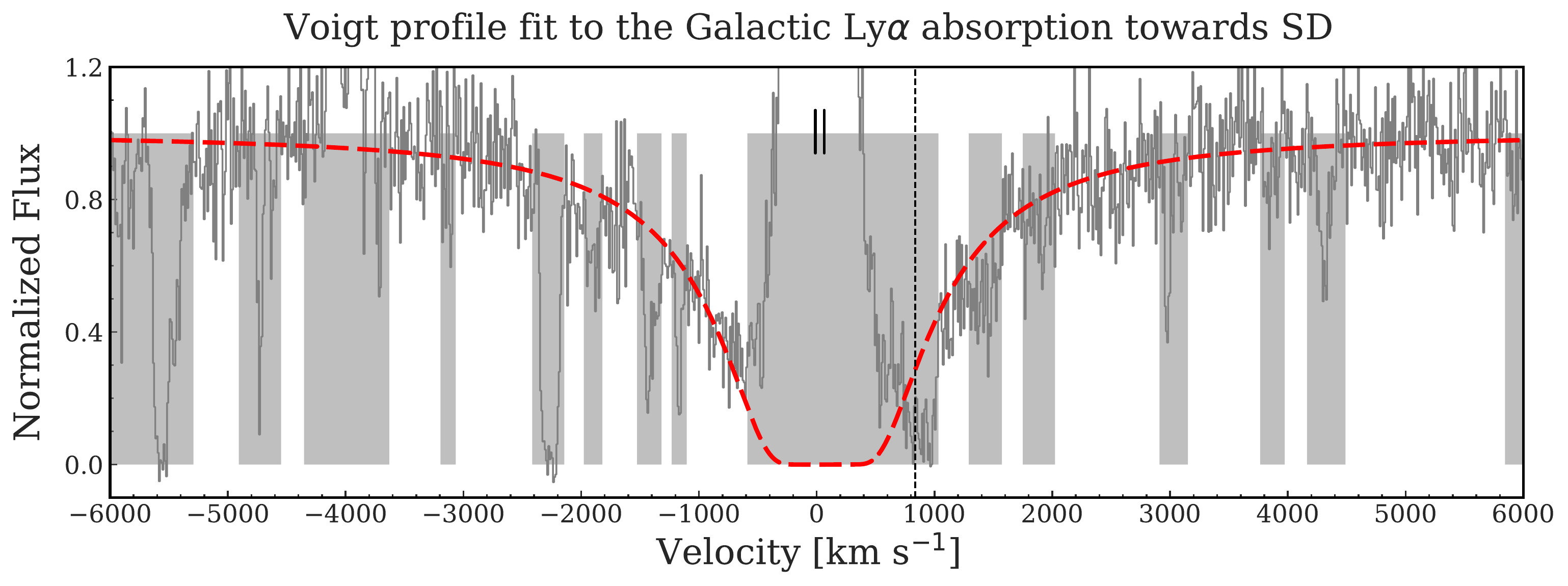}
\caption{Same as in Figure~\ref{fig:SBvpfitgal}, but for SD.}
\label{fig:SDvpfitgal}
\end{center}
\end{figure*}

\subsection{Posterior distributions for the absorber properties towards SD}
\label{appendix:SDparams}
\begin{figure*}
\begin{center}
\includegraphics[width=\linewidth]{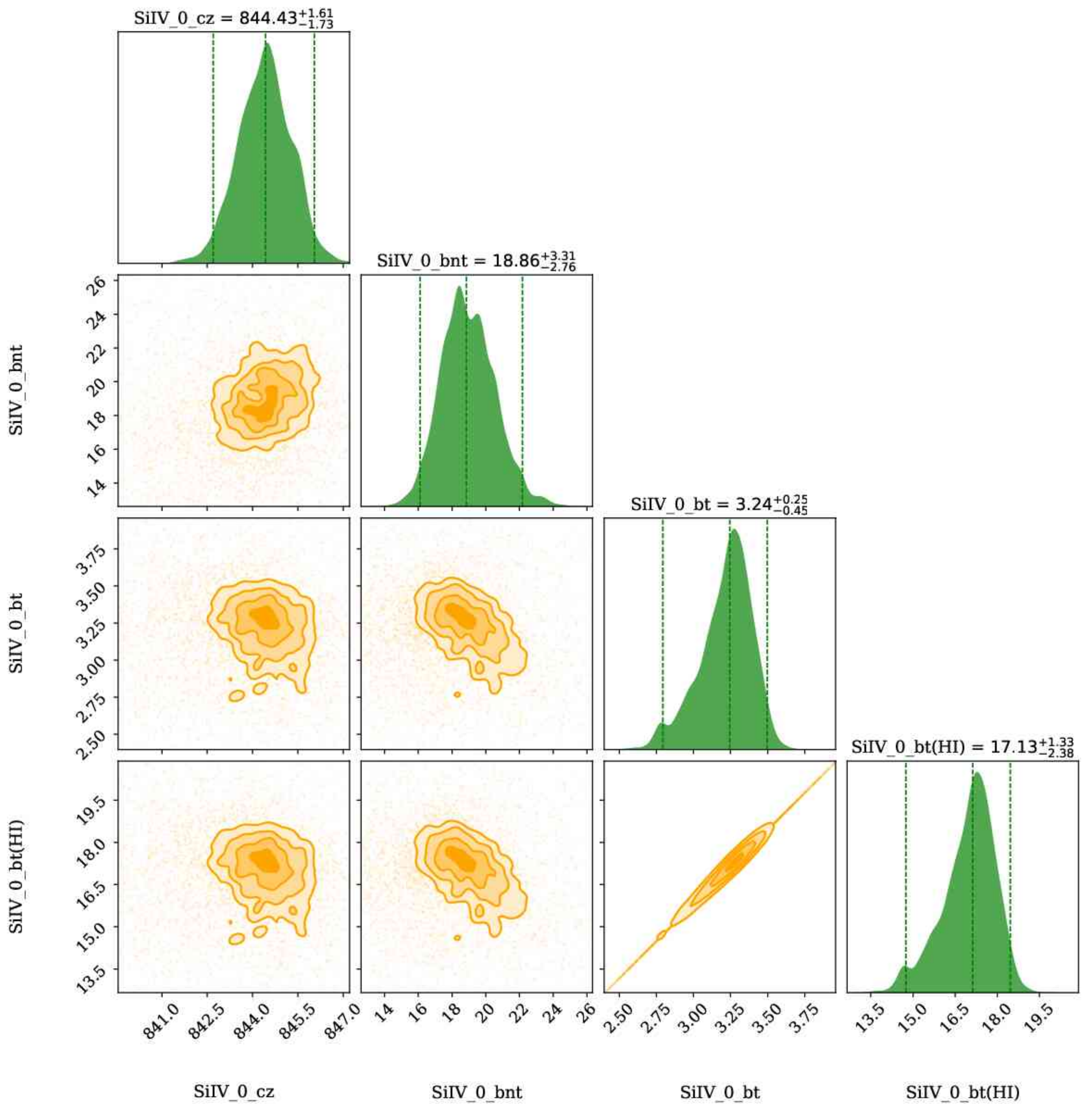}
\caption{The corner plot showing the marginalized posterior distributions for the absorption centroid ($z$), non-thermal Doppler broadening ($b_{nt}$), thermal Doppler broadening ($b_{t}$), total Doppler broadening ($b$), of the phase traced by the blueward {\siiv} cloud of the $z=0.00279$ absorber towards SD. The over-plotted vertical lines in the posterior distribution span the 95\% credible interval. The contours indicate 0.5$\sigma$, 1$\sigma$, 1.5$\sigma$, and 2$\sigma$ levels. The model results are summarised in Table~\ref{tab:modelparams}, and the synthetic profiles based on these models are shown in Figure~\ref{fig:SDsysplot}.}
\label{fig:SDSiIV0b}
\end{center}
\end{figure*}

\begin{figure*}
\begin{center}
\includegraphics[width=\linewidth]{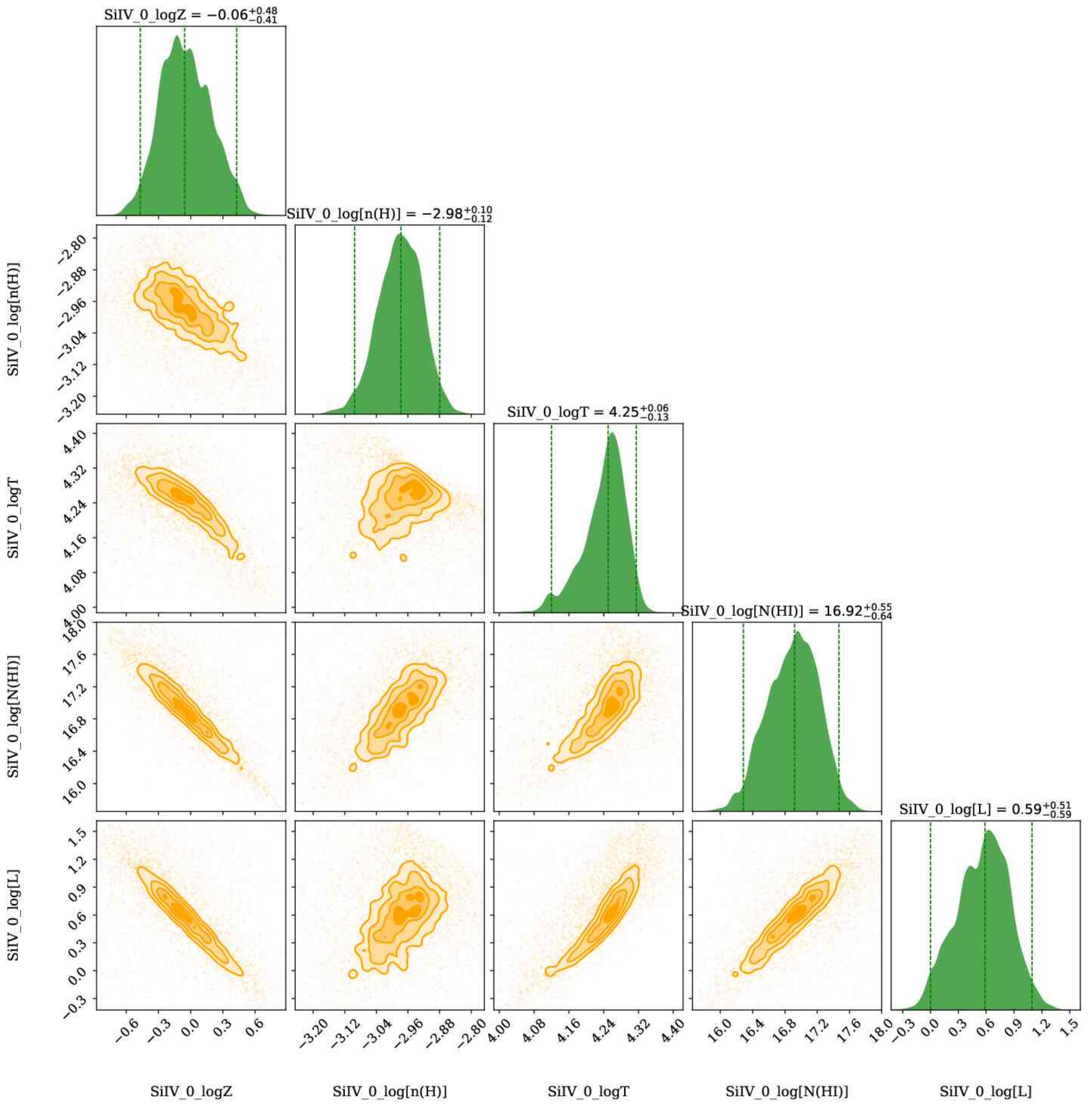}
\caption{The corner plot showing the marginalized posterior distributions for the metallicity ($\log Z$), hydrogen number density ($\log n_{H}$), temperature ($\log T$), neutral hydrogen column density ($\log N(\hi)$), and the line of sight thickness ($\log L$), of the low ionization phase traced by the blueward {\siiv} cloud of the $z=0.00279$ absorber towards SD. The over-plotted vertical lines in the posterior distribution span the 95\% credible interval. The contours indicate 0.5$\sigma$, 1$\sigma$, 1.5$\sigma$, and 2$\sigma$ levels. The model results are summarised in Table~\ref{tab:modelparams}, and the synthetic profiles based on these models are shown in Figure~\ref{fig:SDsysplot}.}
\label{fig:SDSiIV0}
\end{center}
\end{figure*}

\begin{figure*}
\begin{center}
\includegraphics[width=\linewidth]{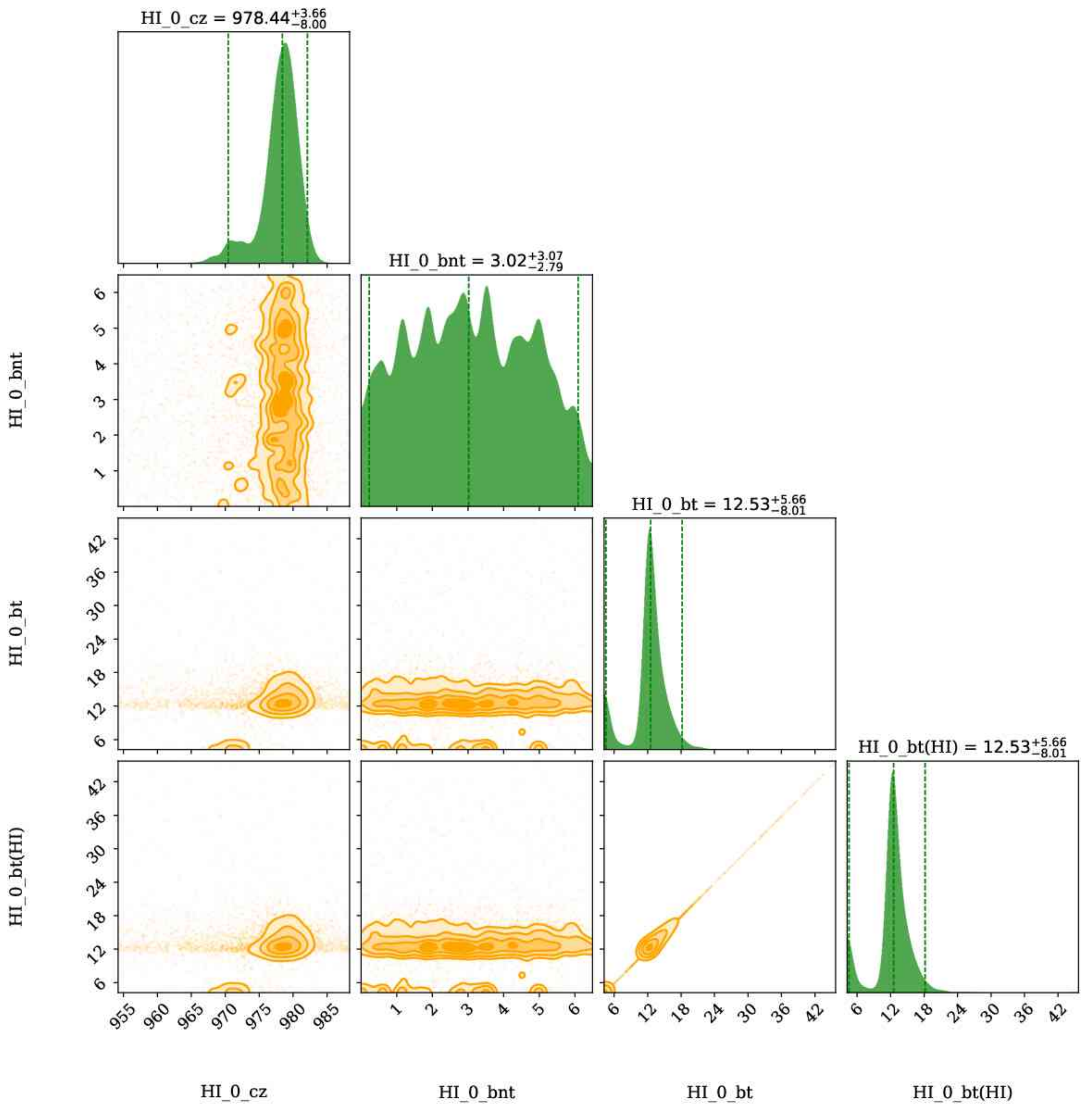}
\caption{The corner plot showing the marginalized posterior distributions for the absorption centroid ($z$), non-thermal Doppler broadening ($b_{nt}$), thermal Doppler broadening ($b_{t}$), total Doppler broadening ($b$), of the phase traced by the redward {\hi} cloud of the $z=0.00279$ absorber towards SD. The over-plotted vertical lines in the posterior distribution span the 95\% credible interval. The contours indicate 0.5$\sigma$, 1$\sigma$, 1.5$\sigma$, and 2$\sigma$ levels. The model results are summarised in Table~\ref{tab:modelparams}, and the synthetic profiles based on these models are shown in Figure~\ref{fig:SDsysplot}.}
\label{fig:SDHI2b}
\end{center}
\end{figure*}

\begin{figure*}
\begin{center}
\includegraphics[width=\linewidth]{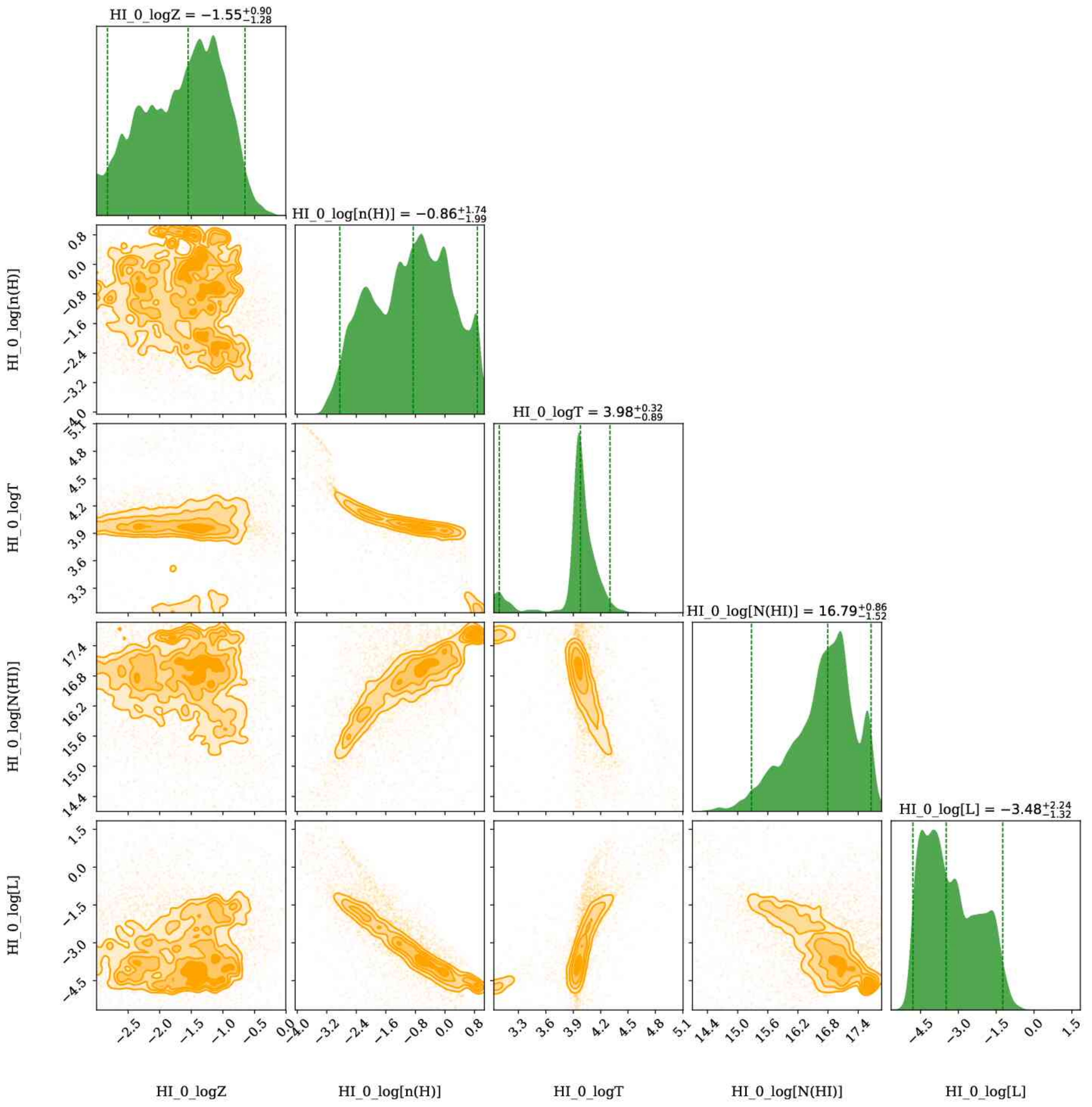}
\caption{The corner plot showing the marginalized posterior distributions for the metallicity ($\log Z$), hydrogen number density ($\log n_{H}$), and other physical properties of the low ionization phase traced by the redward {\hi} cloud of the $z=0.00279$ absorber towards SD. The over-plotted vertical lines in the posterior distribution span the 95\% credible interval. The contours indicate 0.5$\sigma$, 1$\sigma$, 1.5$\sigma$, and 2$\sigma$ levels. The model results are summarised in Table~\ref{tab:modelparams}, and the synthetic profiles based on these models are shown in Figure~\ref{fig:SDsysplot}.}
\label{fig:SDHI2}
\end{center}
\end{figure*}

\clearpage
\section{Plots for SE}
\label{appendix:SE}

\subsection{Airglow template fit towards SE}

\begin{figure*}
\begin{center}
\includegraphics[width=0.75\linewidth]{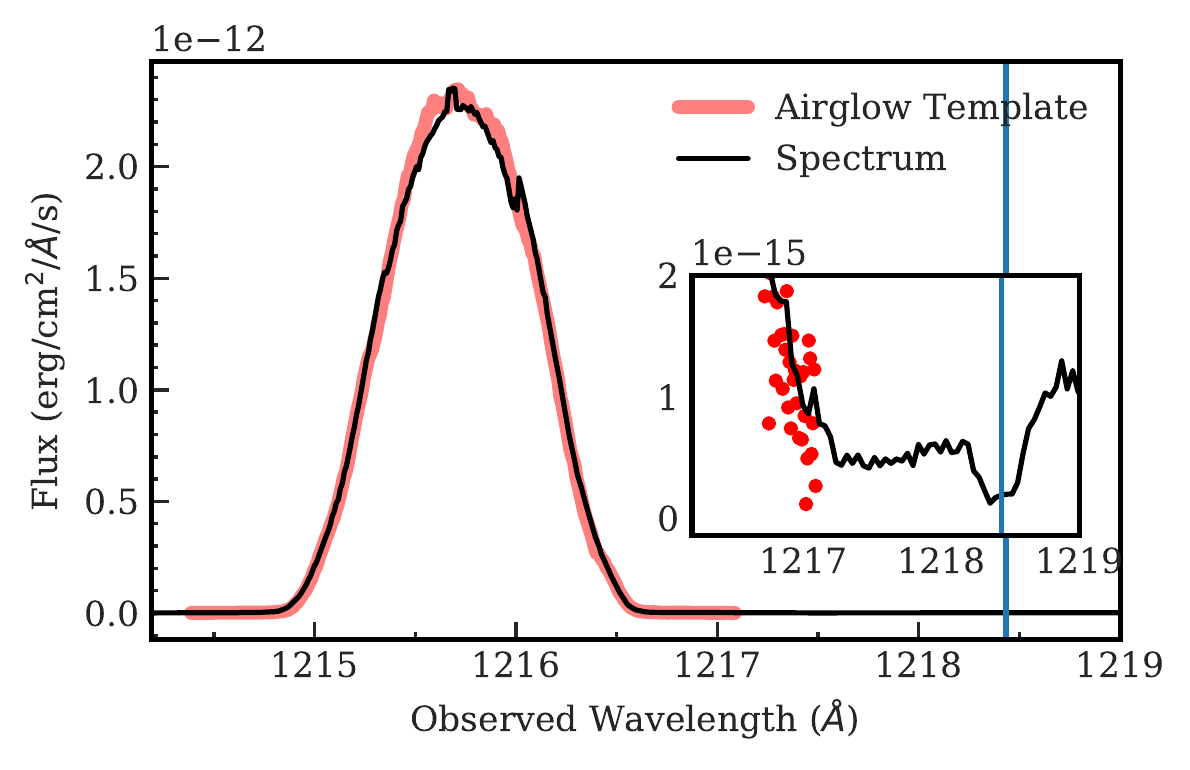}
\caption{Same as in Figure~\ref{fig:SBairglow}, but for SE.}
\label{fig:SEairglow}
\end{center}
\end{figure*}

\subsection{Best VP fit to the Galactic {\lya} towards SE}
\begin{figure*}
\begin{center}
\includegraphics[width=\linewidth]{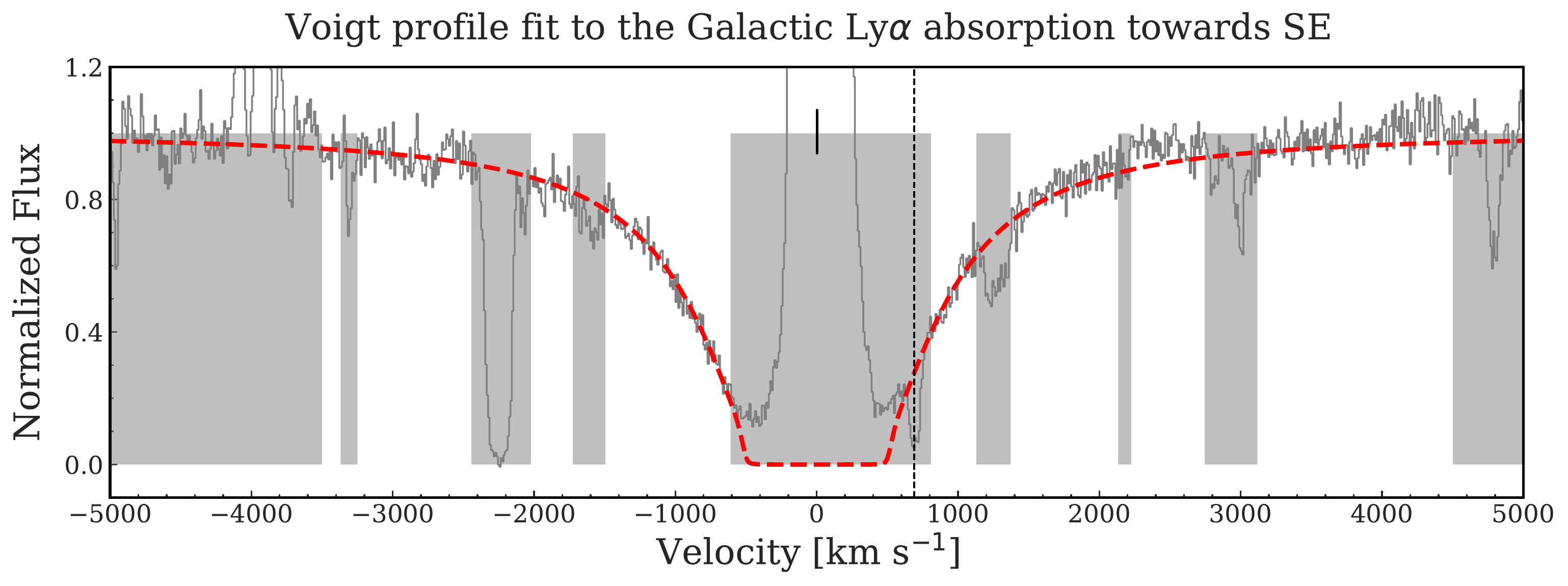}
\caption{Same as in Figure~\ref{fig:SBvpfitgal}, but for SE.}
\label{fig:SEvpfitgal}
\end{center}
\end{figure*}

\subsection{PC model for the high-ionization phase towards SE}

\begin{figure*}
\begin{center}
\includegraphics[width=\linewidth]{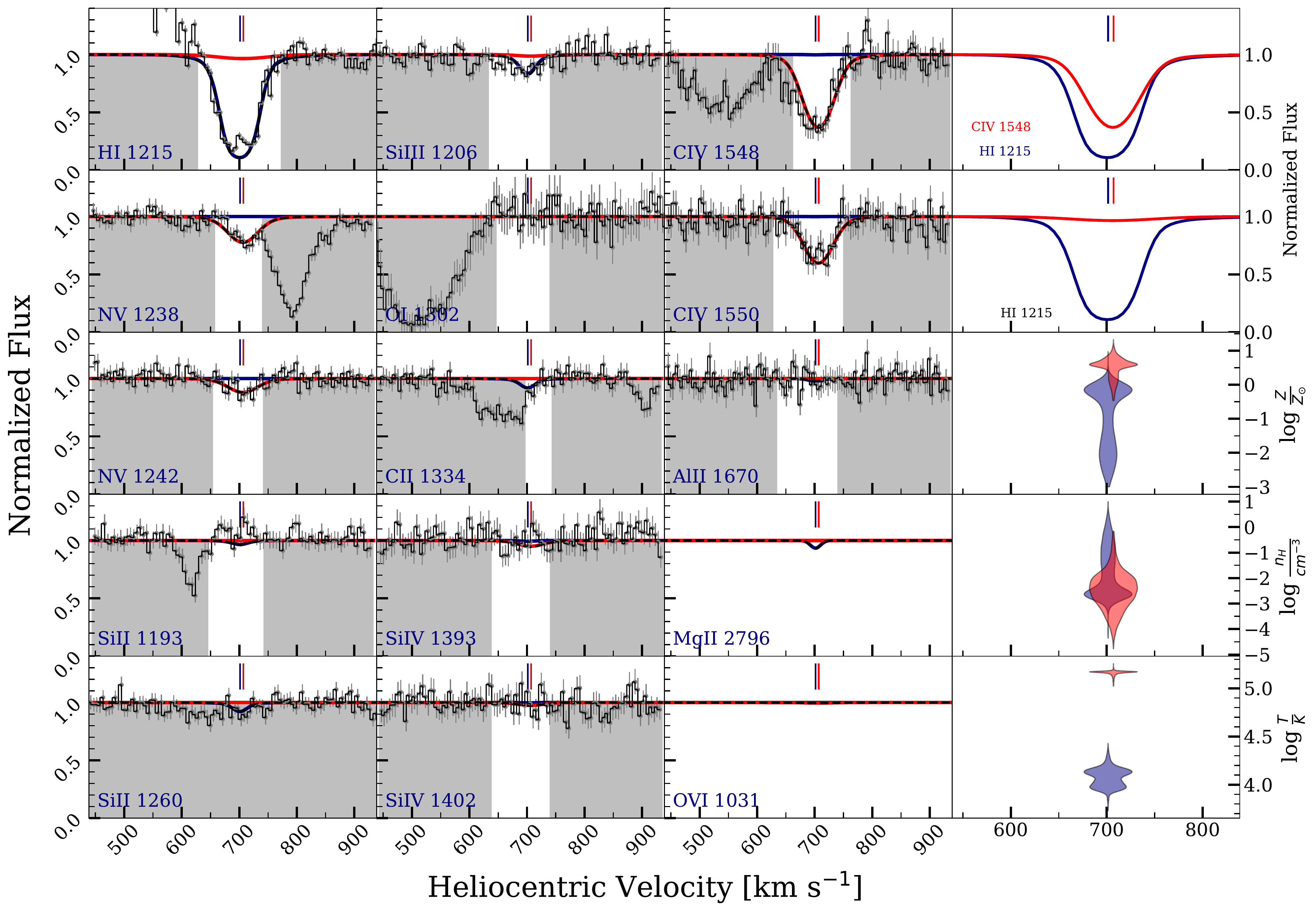}
\caption{A PC model for the absorption towards SE. The {\civ} absorption is modelled as a collisionally ionized phase, and {\nv} is also found to arise in the same phase with a temperature of {\temp} $\approx$ 5.2. The dominant {\hi} absorption, which also produces modest absorption in {\siiii} is still modelled as a photoionized phase only.}
\label{fig:SEpici}
\end{center}
\end{figure*}

\subsection{Posterior distributions for the absorber properties towards SE}

\label{appendix:SEparams}

\begin{figure*}
\begin{center}
\includegraphics[width=\linewidth]{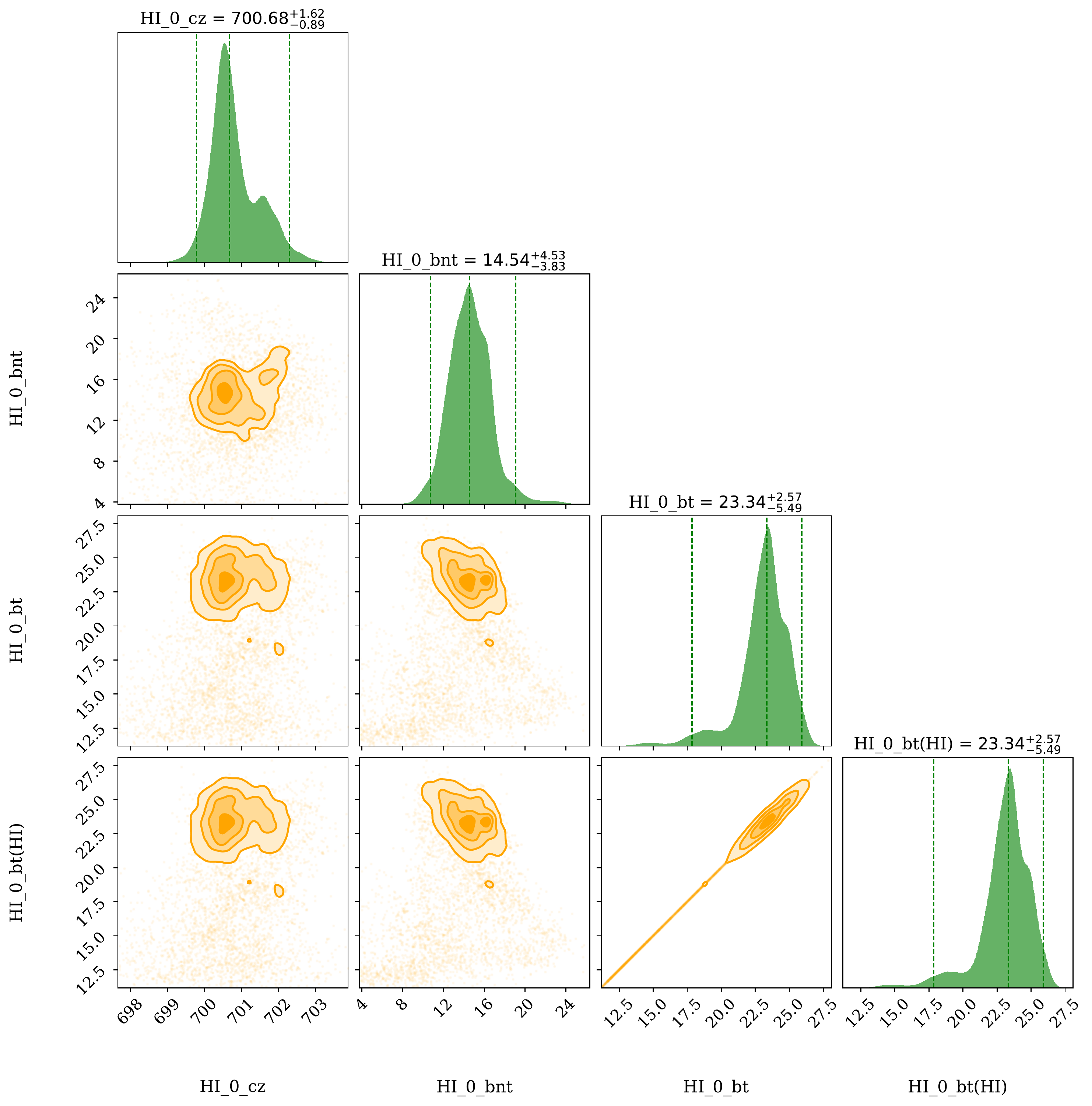}
\caption{The corner plot showing the marginalized posterior distributions for the absorption centroid ($z$), non-thermal Doppler broadening ($b_{nt}$), thermal Doppler broadening ($b_{t}$), total Doppler broadening ($b$), of the phase traced by the dominant {\hi}-producing absorption cloud of the $z=0.0023$ absorber towards SE. The over-plotted vertical lines in the posterior distribution span the 95\% credible interval. The contours indicate 0.5$\sigma$, 1$\sigma$, 1.5$\sigma$, and 2$\sigma$ levels. The model results are summarised in Table~\ref{tab:modelparams}, and the synthetic profiles based on these models are shown in Figure~\ref{fig:SEsysplot}.}
\label{fig:SEHI0b}
\end{center}
\end{figure*}

\begin{figure*}
\begin{center}
\includegraphics[width=\linewidth]{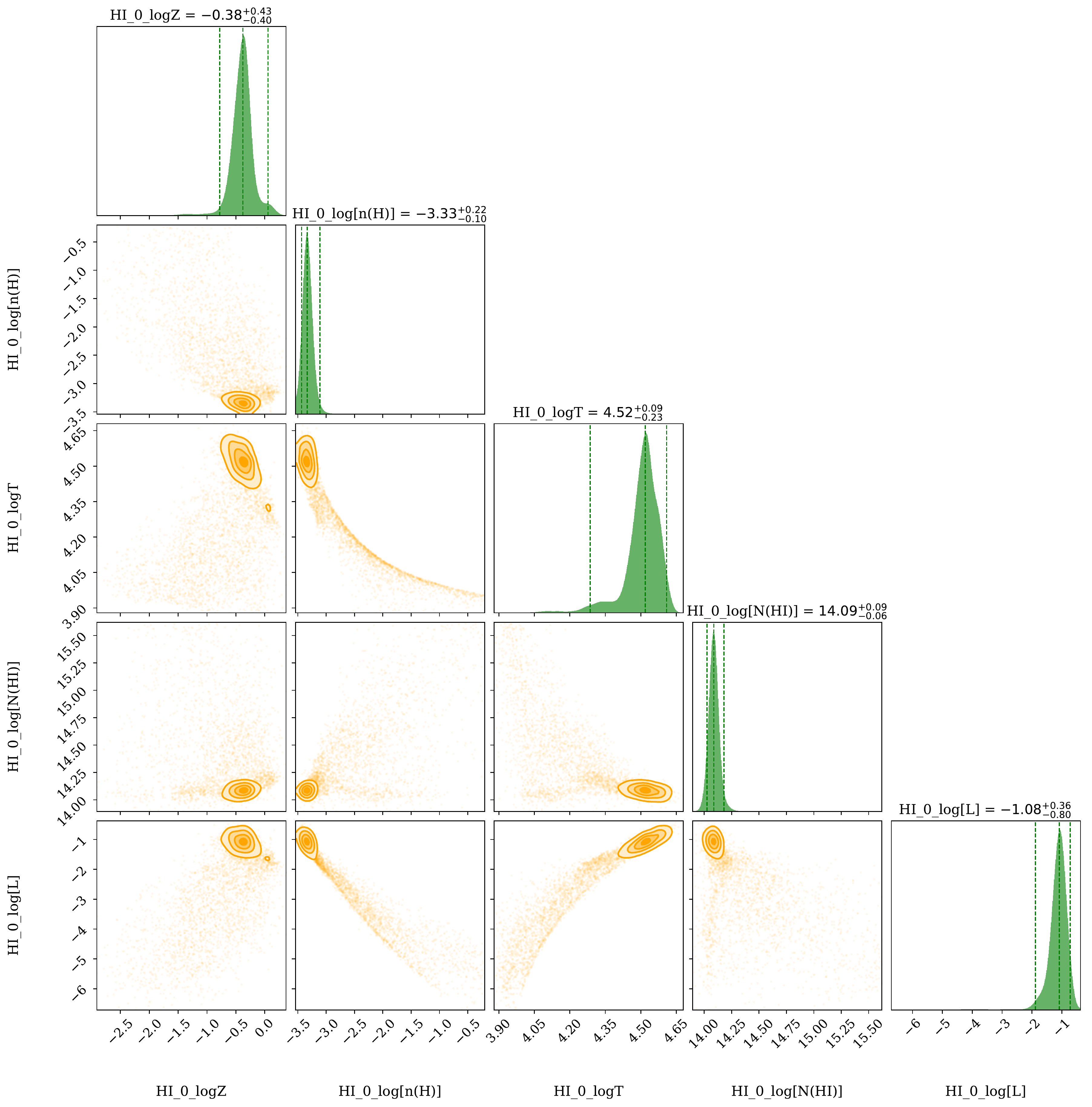}
\caption{The corner plot showing the marginalized posterior distributions for the metallicity ($\log Z$), hydrogen number density ($\log n_{H}$), temperature ($\log T$), neutral hydrogen column density ($\log N(\hi)$), and the line of sight thickness ($\log L$), of the phase traced by the dominant {\hi}-producing absorption cloud of the $z=0.0023$ absorber towards SE. The over-plotted vertical lines in the posterior distribution span the 95\% credible interval. The contours indicate 0.5$\sigma$, 1$\sigma$, 1.5$\sigma$, and 2$\sigma$ levels. The model results are summarised in Table~\ref{tab:modelparams}, and the synthetic profiles based on these models are shown in Figure~\ref{fig:SEsysplot}.}
\label{fig:SEHI0}
\end{center}
\end{figure*}

\begin{figure*}
\begin{center}
\includegraphics[width=\linewidth]{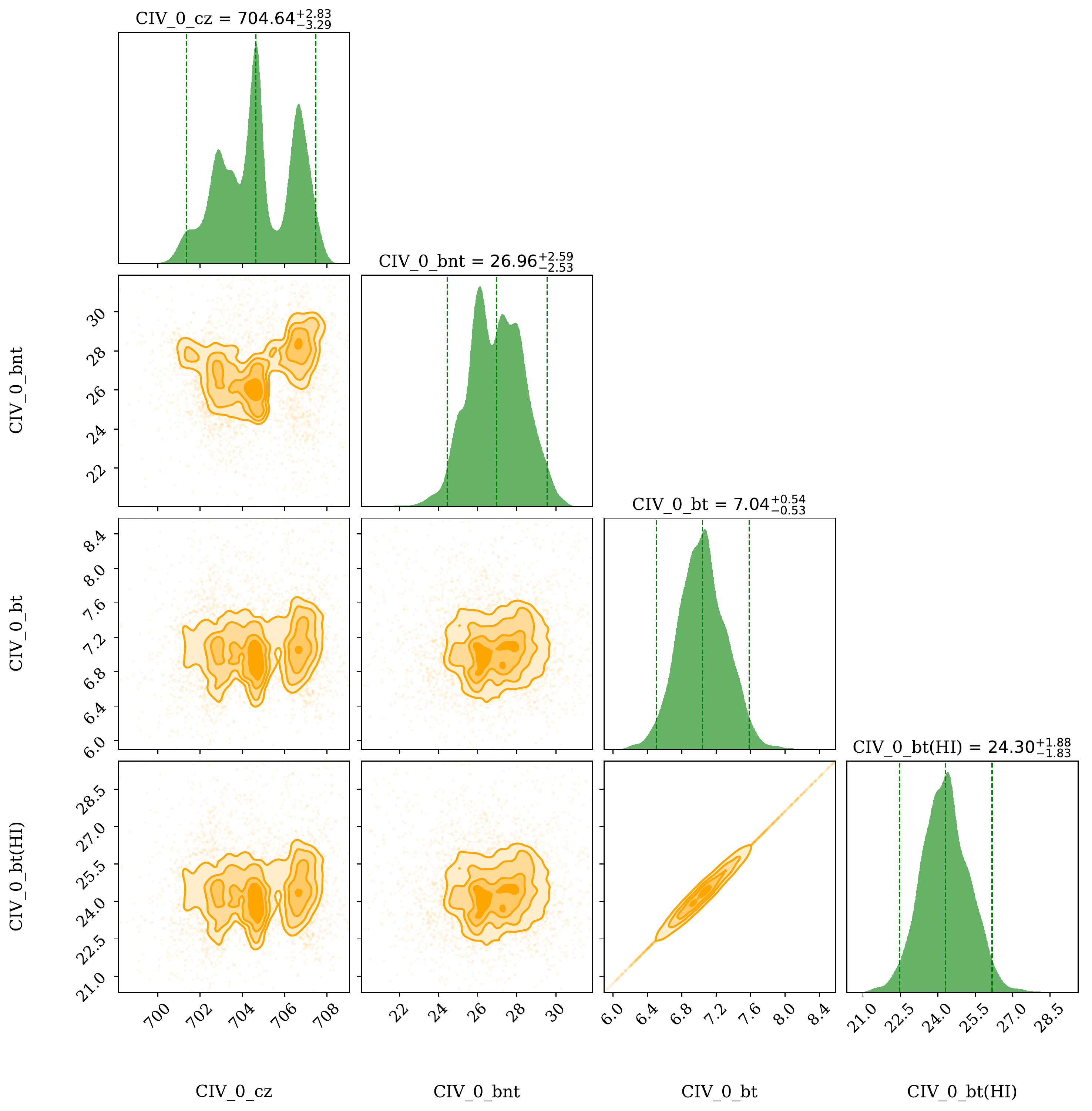}
\caption{The corner plot showing the marginalized posterior distributions for the absorption centroid ($z$), non-thermal Doppler broadening ($b_{nt}$), thermal Doppler broadening ($b_{t}$), total Doppler broadening ($b$), of the phase traced by the {\civ} cloud of the $z=0.0023$ absorber towards SE. The over-plotted vertical lines in the posterior distribution span the 95\% credible interval. The contours indicate 0.5$\sigma$, 1$\sigma$, 1.5$\sigma$, and 2$\sigma$ levels. The model results are summarised in Table~\ref{tab:modelparams}, and the synthetic profiles based on these models are shown in Figure~\ref{fig:SEsysplot}.}
\label{fig:SECIV1b}
\end{center}
\end{figure*}

\begin{figure*}
\begin{center}
\includegraphics[width=\linewidth]{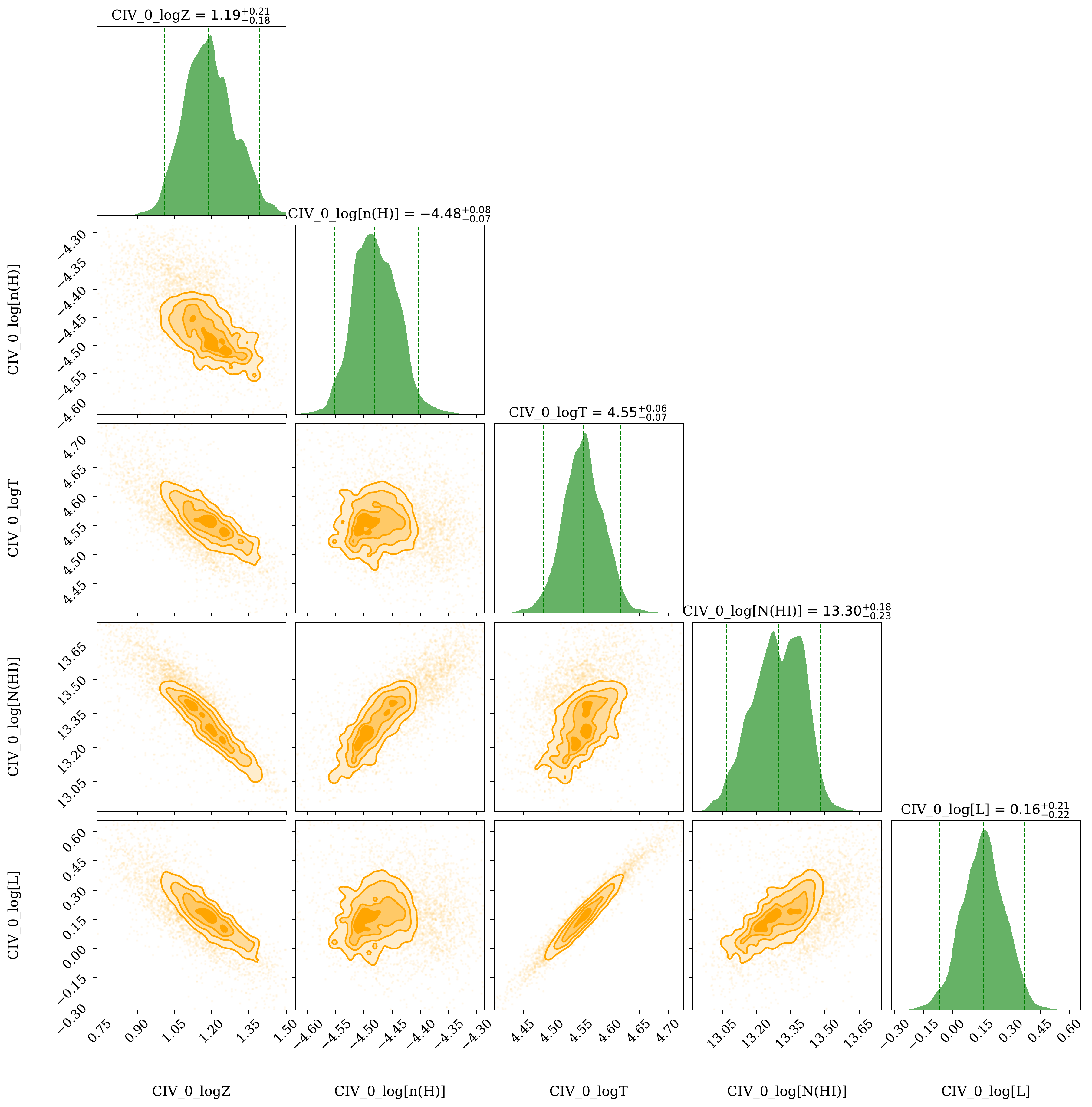}
\caption{The corner plot showing the marginalized posterior distributions for the metallicity ($\log Z$), hydrogen number density ($\log n_{H}$), temperature ($\log T$), neutral hydrogen column density ($\log N(\hi)$), and the line of sight thickness ($\log L$), of the phase traced by the {\civ} cloud of the $z=0.0023$ absorber towards SE. The over-plotted vertical lines in the posterior distribution span the 95\% credible interval. The contours indicate 0.5$\sigma$, 1$\sigma$, 1.5$\sigma$, and 2$\sigma$ levels. The model results are summarised in Table~\ref{tab:modelparams}, and the synthetic profiles based on these models are shown in Figure~\ref{fig:SEsysplot}.}
\label{fig:SECIV1}
\end{center}
\end{figure*}

\clearpage
\section{Plots for SF}
\label{appendix:SF}

\subsection{Airglow template fit towards SF}

\begin{figure*}
\begin{center}
\includegraphics[width=0.75\linewidth]{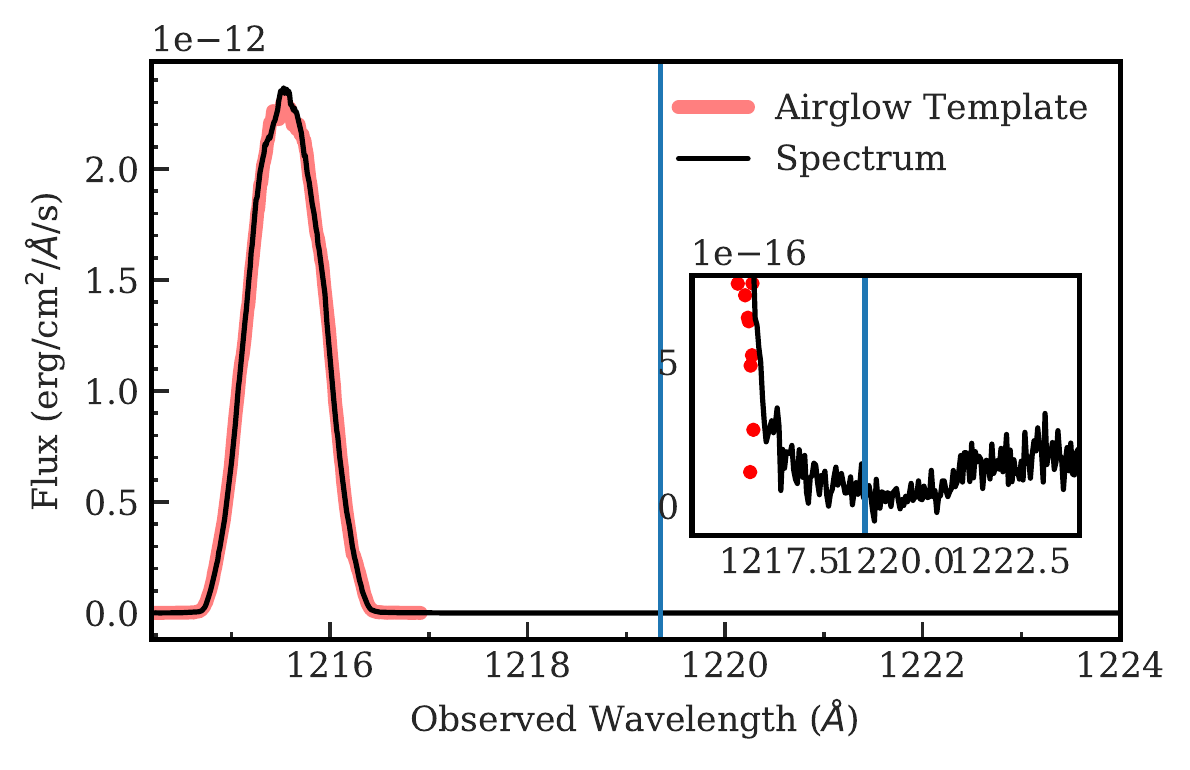}
\caption{Same as in Figure~\ref{fig:SBairglow}, but for SF.}
\label{fig:SFairglow}
\end{center}
\end{figure*}
\subsection{Best VP fit to the Galactic {\lya} towards SF}

\begin{figure*}
\begin{center}
\includegraphics[width=\linewidth]{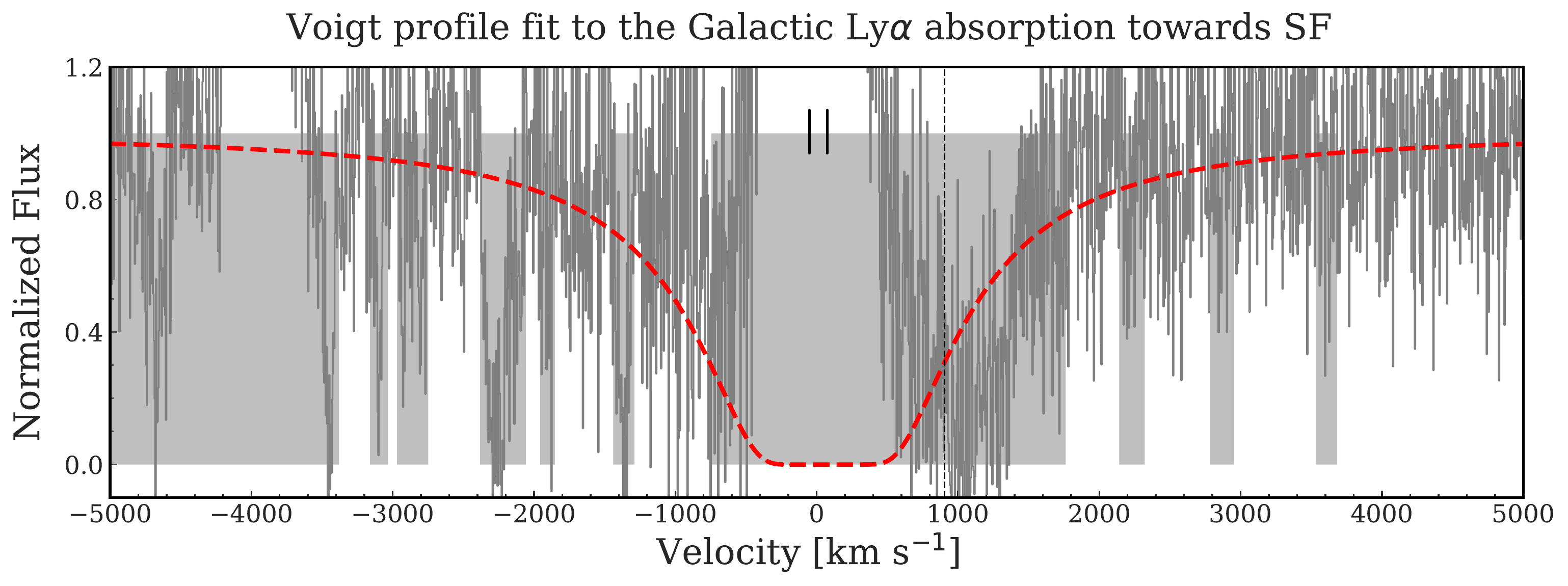}
\caption{Same as in Figure~\ref{fig:SBvpfitgal}, but for SF.}
\label{fig:SFvpfitgal}
\end{center}
\end{figure*}

\subsection{System plot for sightline SF with {\hi} data masked}

\begin{figure*}
\begin{center}
\includegraphics[width=\linewidth]{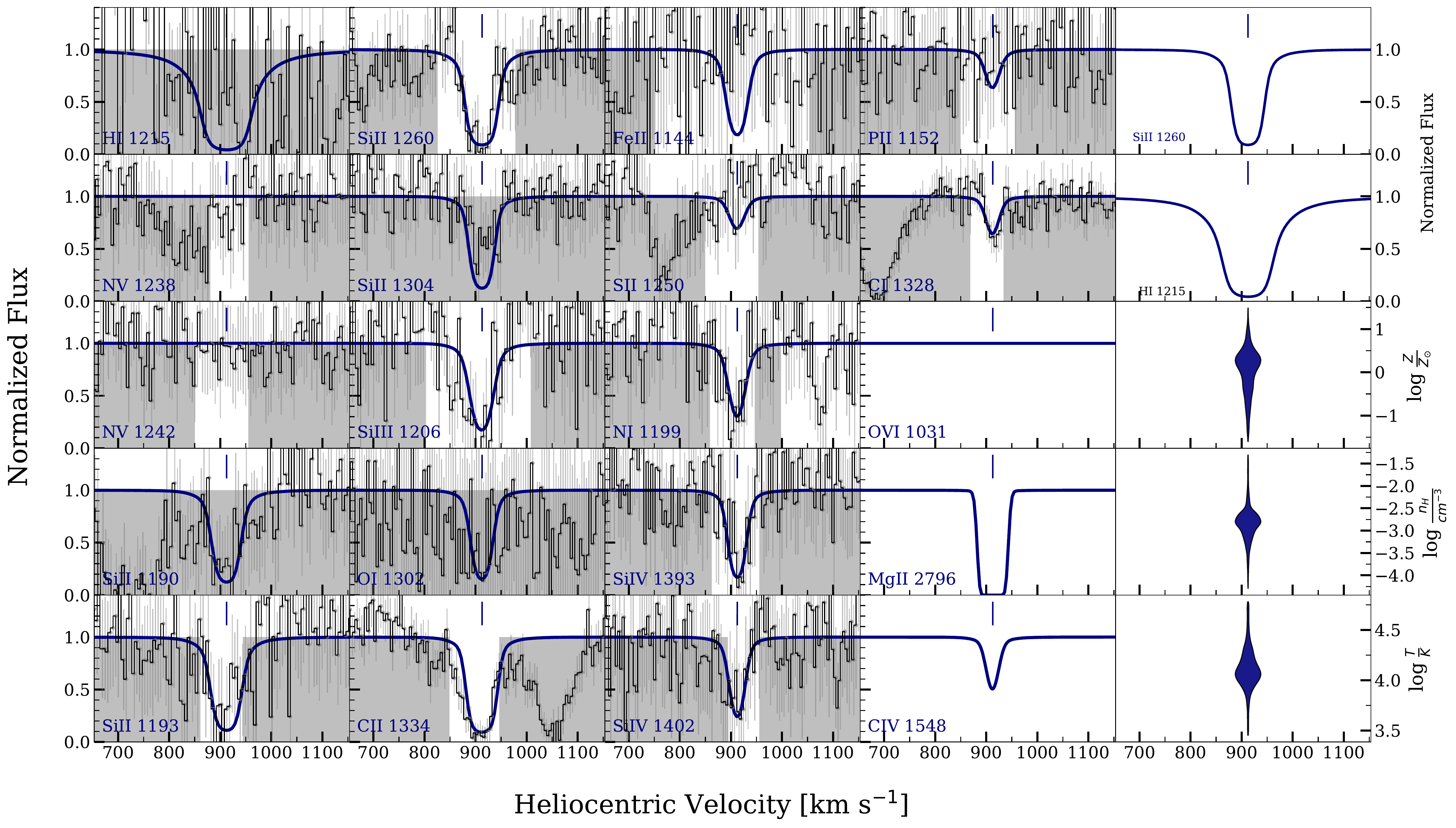}
\caption{The system plots showing the model in which the {\hi} data is masked. The detection of neutral species, {\ci} and {\nitri}, provides a constraint on the metallicity and density of the phase.}
\label{fig:SFextreme1}
\end{center}
\end{figure*}

\subsection{System plot for sightline SF with a restricted prior on \colden}

\begin{figure*}
\begin{center}
\includegraphics[width=\linewidth]{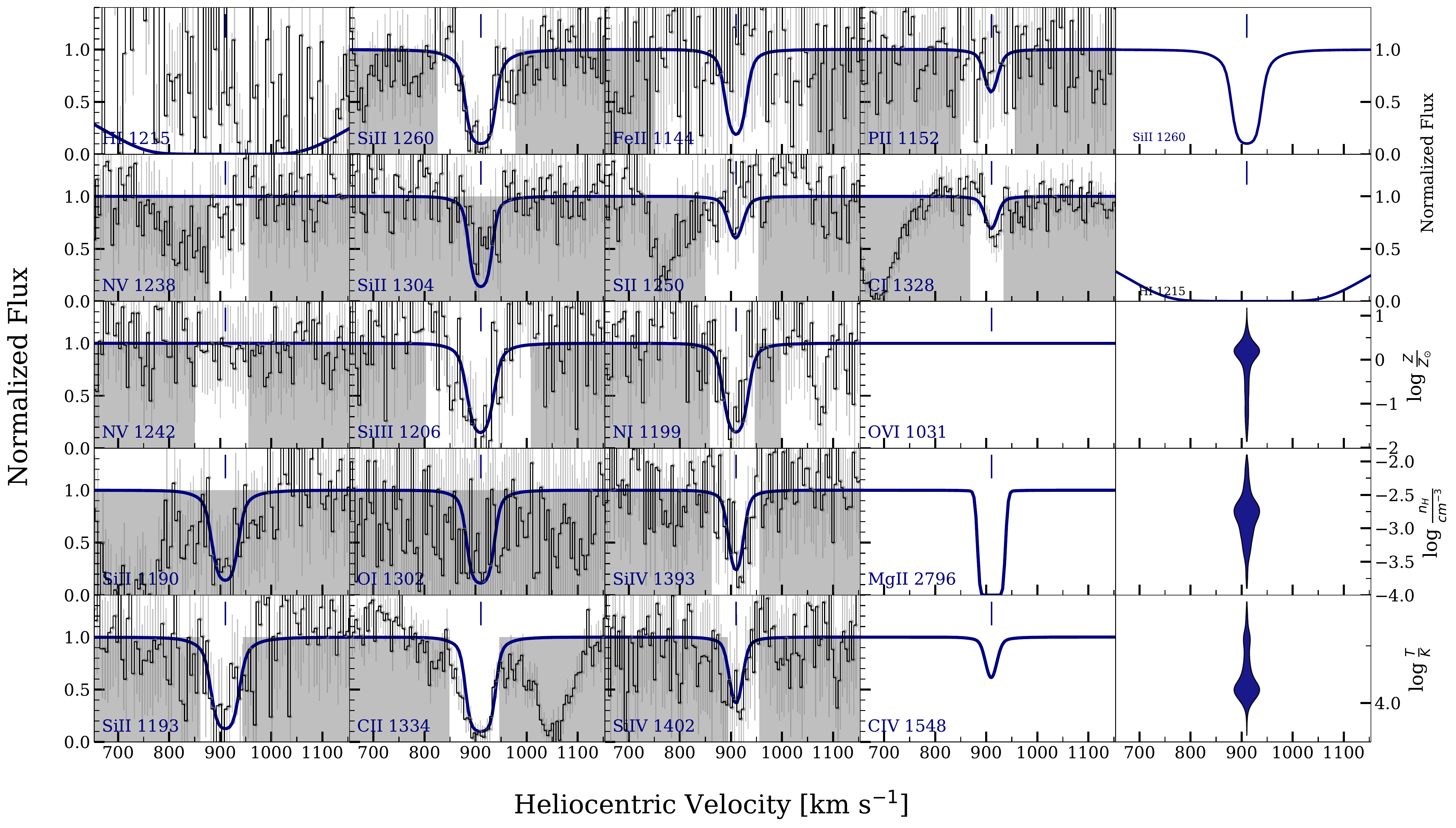}
\caption{The system plots showing the model in which the prior on {\colden} is restricted to be greater than 19.5.   Even for this larger value of {\colden}, we find a high value of {\metallicity} = 0.21 $\pm$ 0.08.}
\label{fig:SFextreme2}
\end{center}
\end{figure*}
\subsection{Posterior distributions for the absorber properties towards SF}

\label{appendix:SFparams}

\begin{figure*}
\begin{center}
\includegraphics[width=\linewidth]{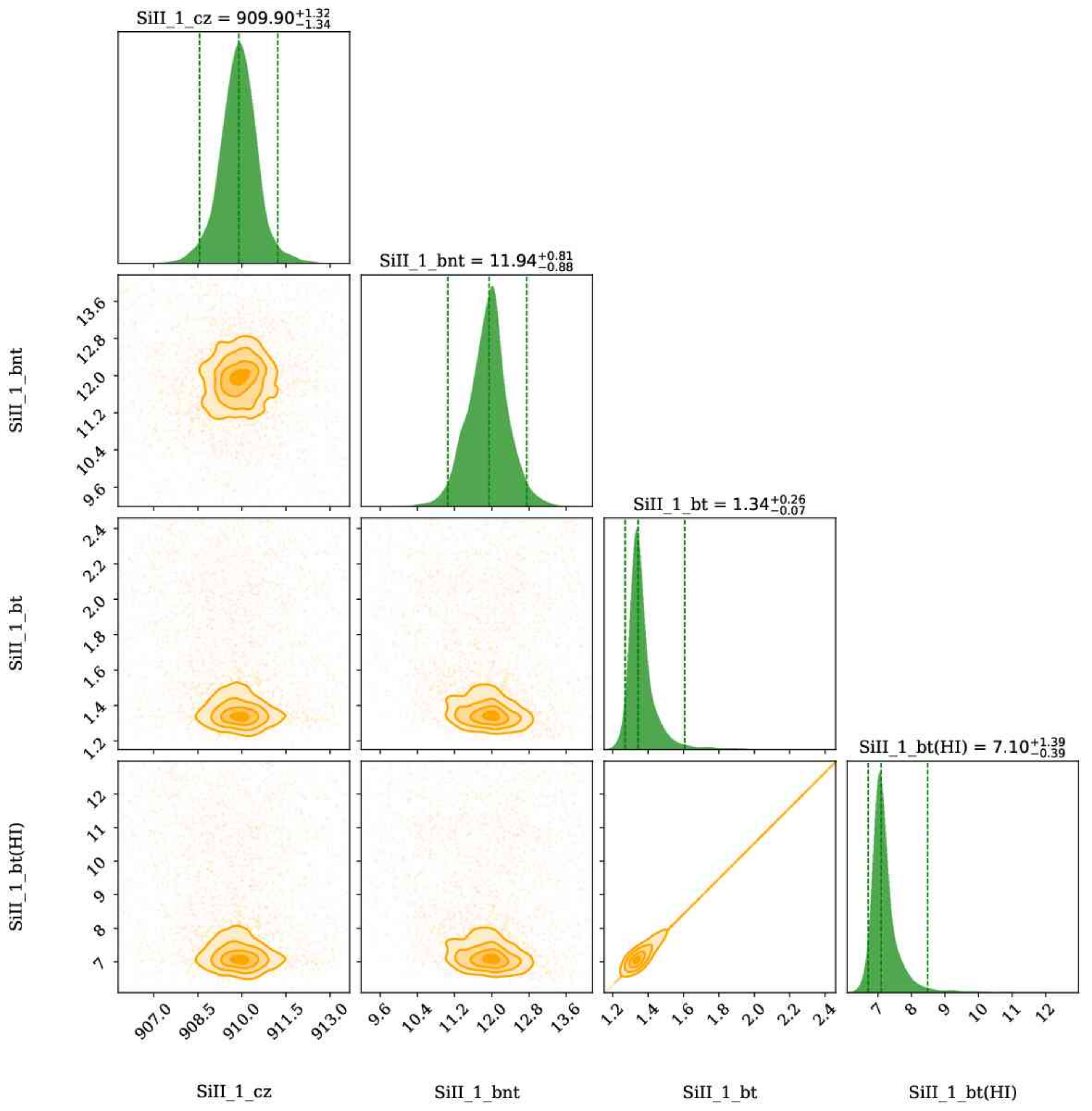}
\caption{The corner plot showing the marginalized posterior distributions for the absorption centroid ($z$), non-thermal Doppler broadening ($b_{nt}$), thermal Doppler broadening ($b_{t}$), total Doppler broadening ($b$), of the phase traced by the {\siii} cloud of the $z=0.0030$ absorber towards SF. The over-plotted vertical lines in the posterior distribution span the 95\% credible interval. The contours indicate 0.5$\sigma$, 1$\sigma$, 1.5$\sigma$, and 2$\sigma$ levels. The model results are summarised in Table~\ref{tab:modelparams}, and the synthetic profiles based on these models are shown in Figure~\ref{fig:SFsysplot}.}
\label{fig:SFSiIV0b}
\end{center}
\end{figure*}

\begin{figure*}
\begin{center}
\includegraphics[width=\linewidth]{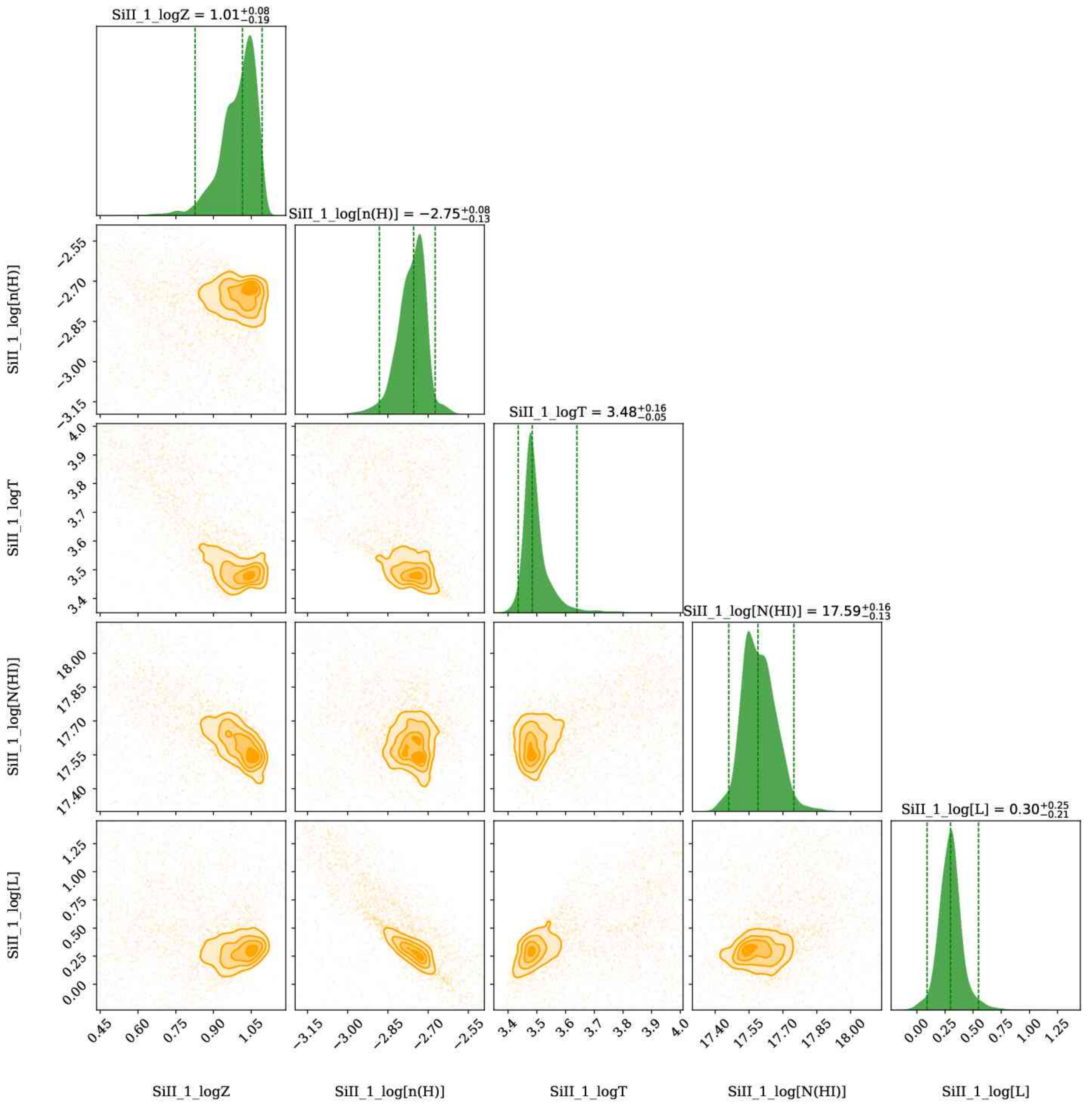}
\caption{The corner plot showing the marginalized posterior distributions for the metallicity ($\log Z$), hydrogen number density ($\log n_{H}$), temperature ($\log T$), neutral hydrogen column density ($\log N(\hi)$), and the line of sight thickness ($\log L$), of the phase traced by the {\siii} cloud of the $z=0.0030$ absorber towards SF. The over-plotted vertical lines in the posterior distribution span the 95\% credible interval. The contours indicate 0.5$\sigma$, 1$\sigma$, 1.5$\sigma$, and 2$\sigma$ levels. The model results are summarised in Table~\ref{tab:modelparams}, and the synthetic profiles based on these models are shown in Figure~\ref{fig:SFsysplot}.}
\label{fig:SFSiIV0}
\end{center}
\end{figure*}

\clearpage
\section{Plots for SG}
\label{appendix:SG}
\subsection{Airglow template fit towards SG}

\begin{figure*}
\begin{center}
\includegraphics[width=0.75\linewidth]{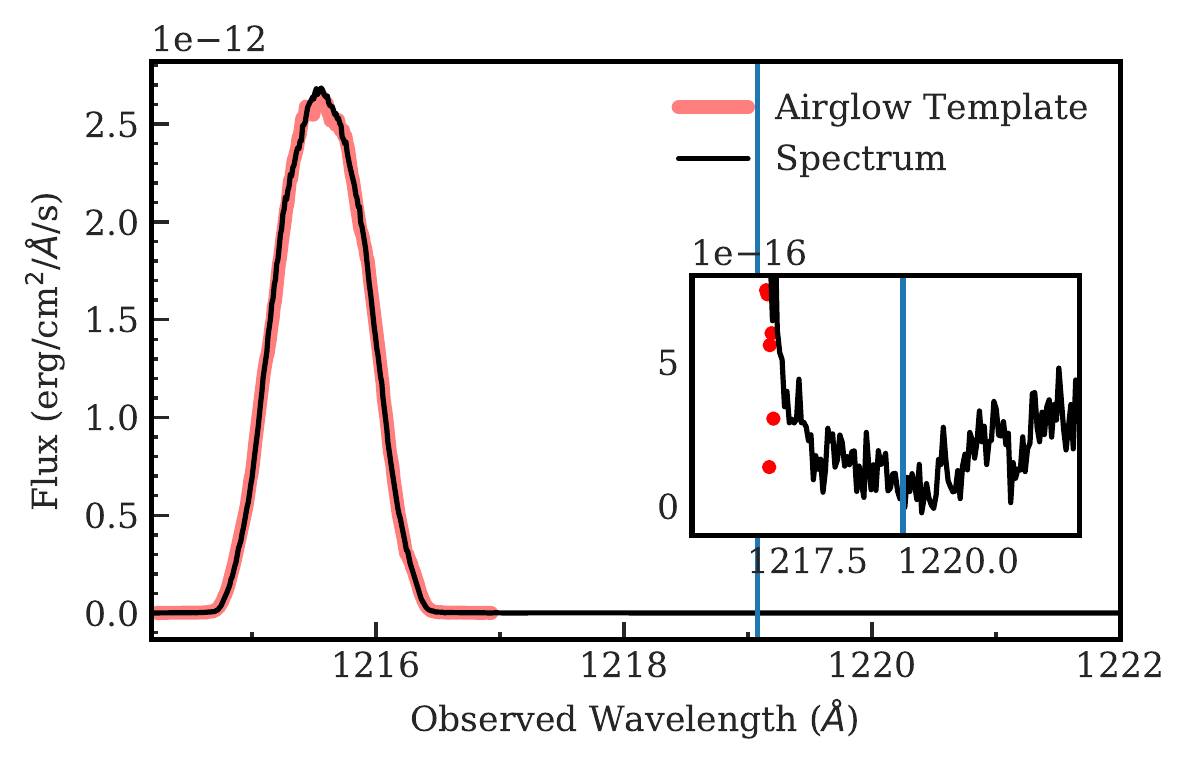}
\caption{Same as in Figure~\ref{fig:SBairglow}, but for SG.}
\label{fig:SGairglow}
\end{center}
\end{figure*}

\subsection{Best VP fit to the Galactic {\lya} towards SG}

\begin{figure*}
\begin{center}
\includegraphics[width=\linewidth]{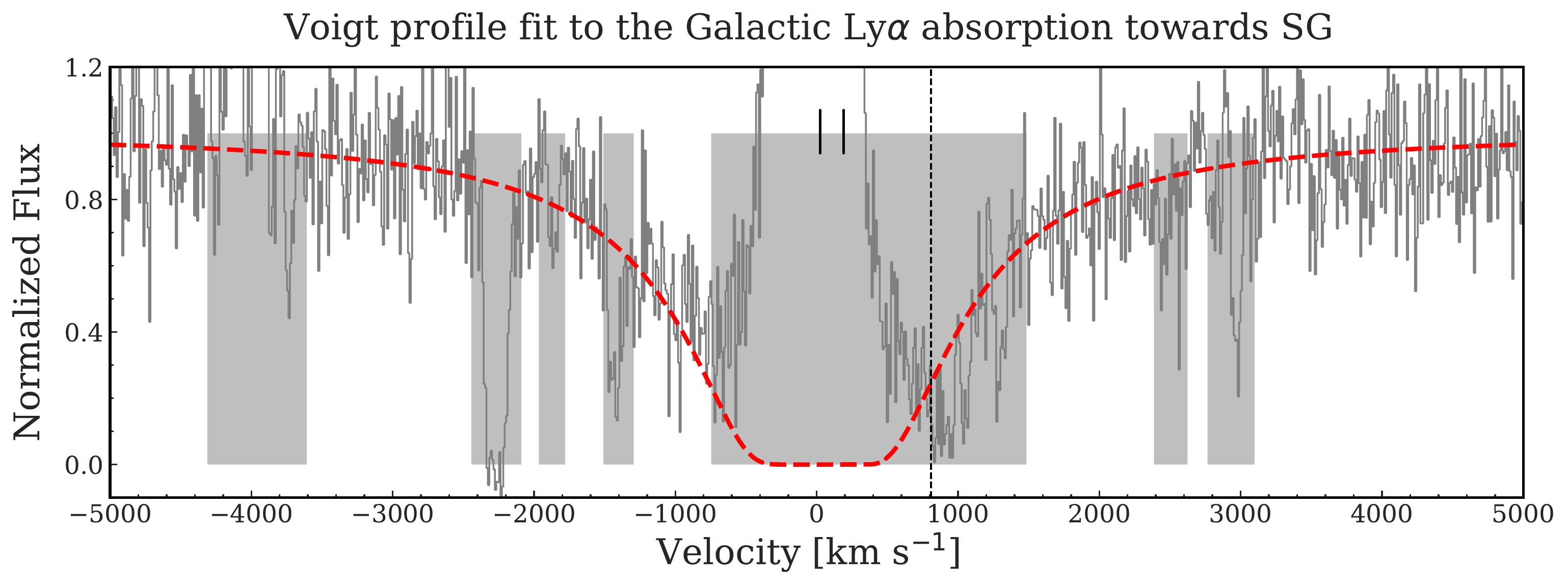}
\caption{Same as in Figure~\ref{fig:SBvpfitgal}, but for SG.}
\label{fig:SGvpfitgal}
\end{center}
\end{figure*}

\subsection{MC exploration of the zero-point uncertainty for SG}

\begin{figure*}
\begin{center}
\includegraphics[scale=0.5]{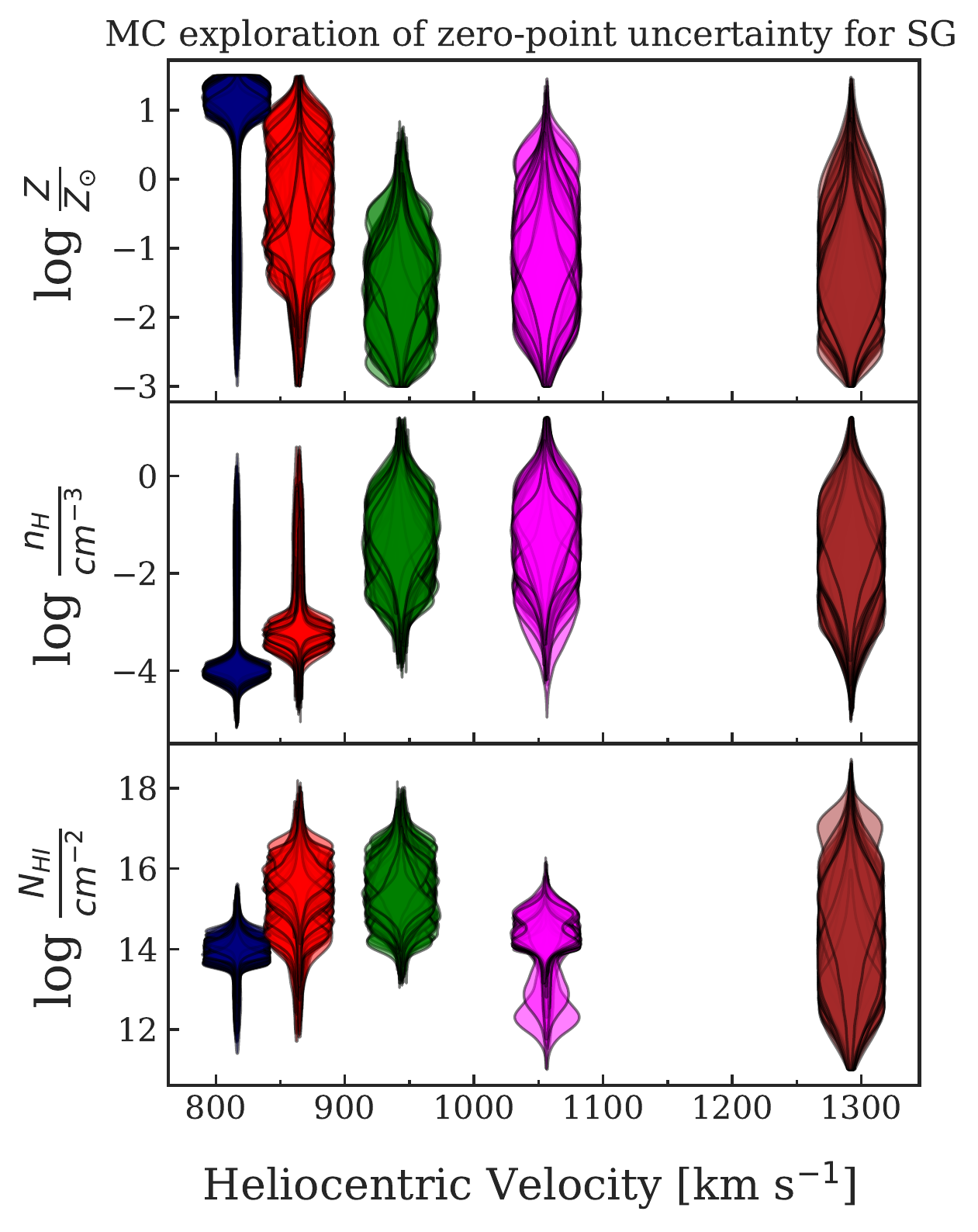}
\caption{Violin plots showing the parameter distributions for 100 different realizations of the {\hi} {\lya} profile modified  between 815--865 {\kms} to account for the zero-point uncertainty. The blue violins show the parameters of {\civ} phase. The red violins correspond to the {\siiv} phase, the green, magenta, and brown violins correspond to the {\hi} only phases.}
\label{fig:MCSG}
\end{center}
\end{figure*}

\subsection{Posterior distributions for the absorber properties towards SG}

\label{appendix:SGparams}

\begin{figure*}
\begin{center}
\includegraphics[width=\linewidth]{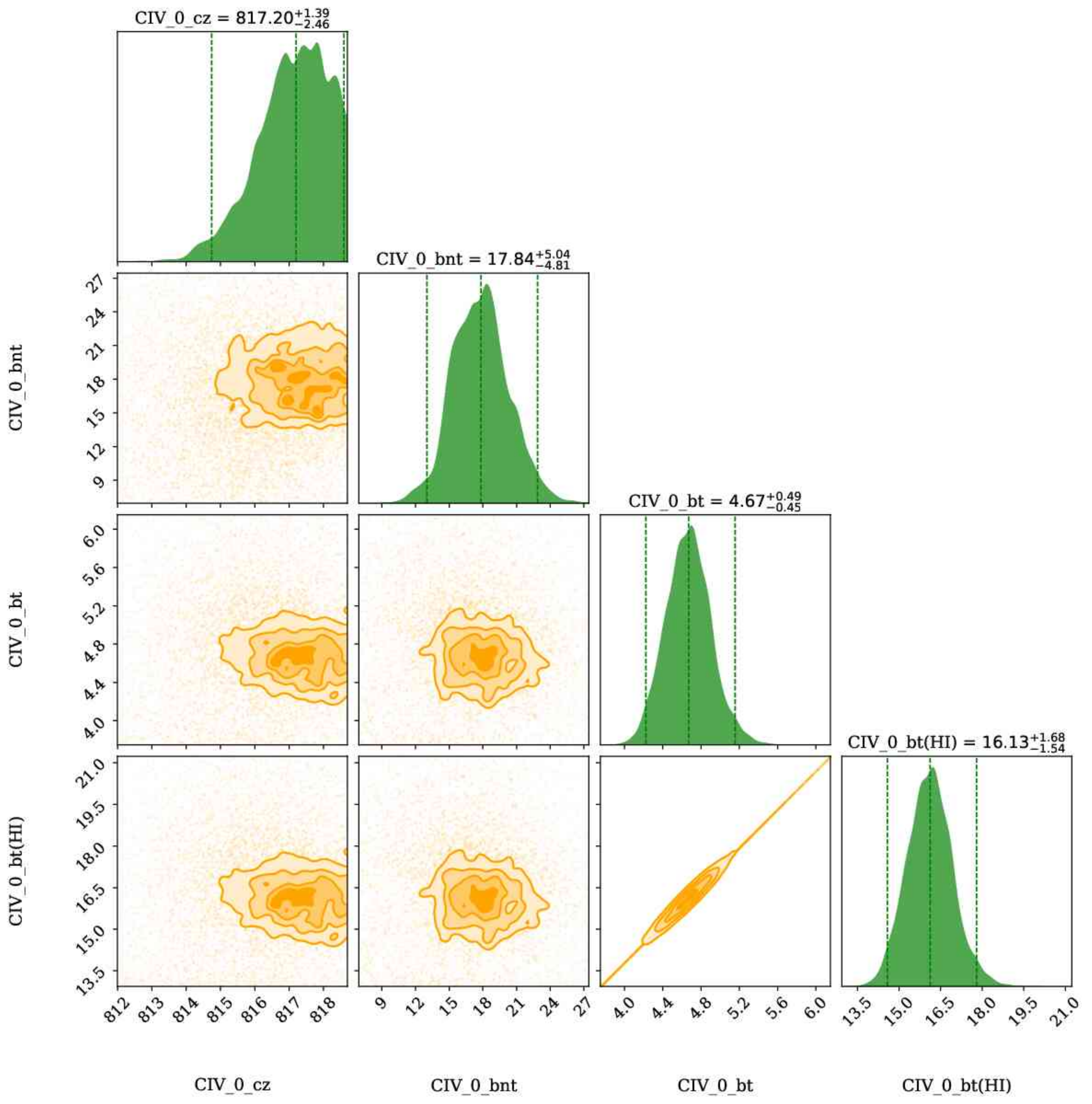}
\caption{The corner plot showing the marginalized posterior distributions for the absorption centroid ($z$), non-thermal Doppler broadening ($b_{nt}$), thermal Doppler broadening ($b_{t}$), total Doppler broadening ($b$), of the phase traced by the {\civ} cloud of the $z=0.0027$ absorber towards SG. The over-plotted vertical lines in the posterior distribution span the 95\% credible interval. The contours indicate 0.5$\sigma$, 1$\sigma$, 1.5$\sigma$, and 2$\sigma$ levels. The model results are summarised in Table~\ref{tab:modelparams}, and the synthetic profiles based on these models are shown in Figure~\ref{fig:SGsysplot}.}
\label{fig:SGCIV0b}
\end{center}
\end{figure*}

\begin{figure*}
\begin{center}
\includegraphics[width=\linewidth]{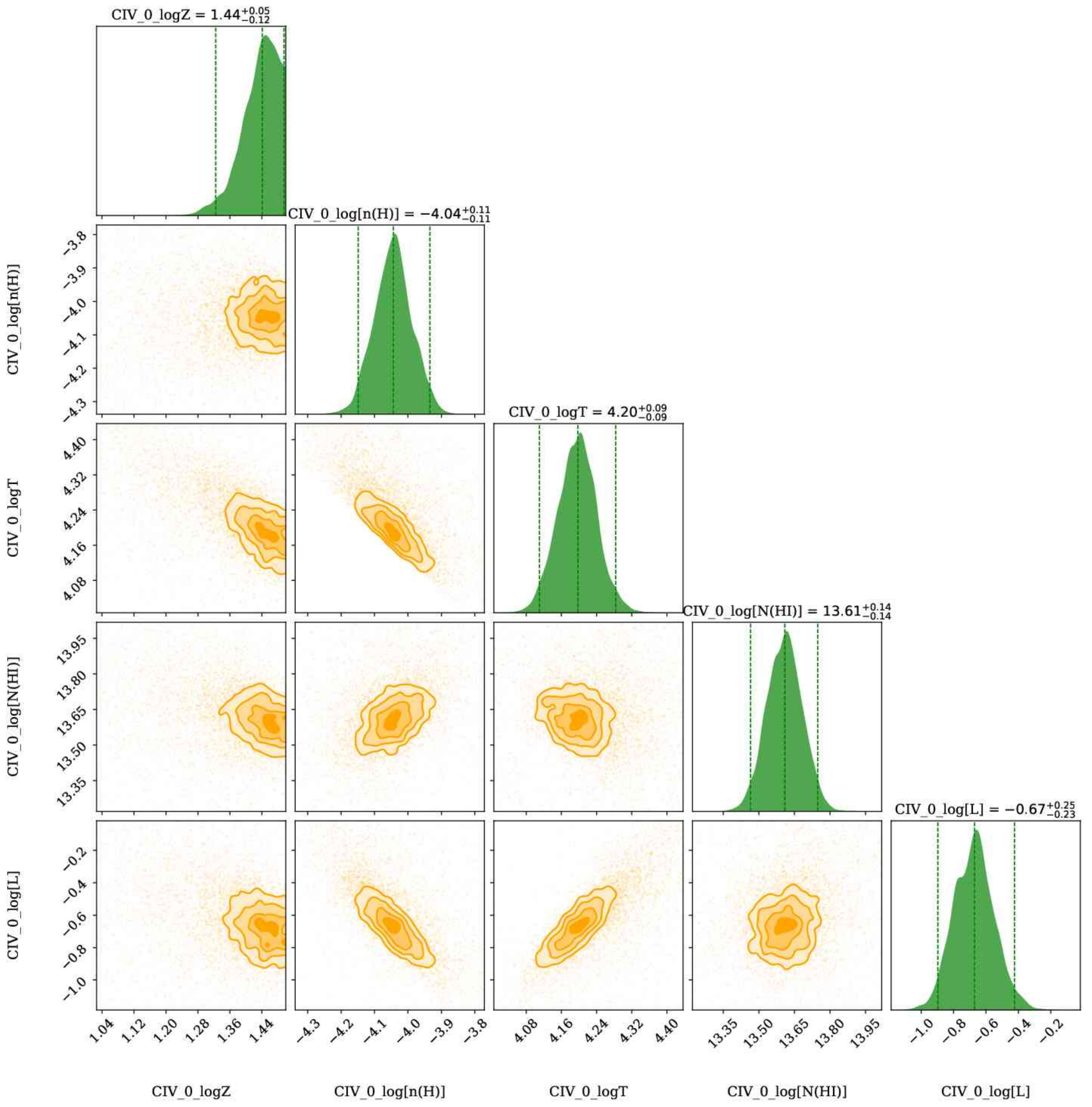}
\caption{The corner plot showing the marginalized posterior distributions for the metallicity ($\log Z$), hydrogen number density ($\log n_{H}$), temperature ($\log T$), neutral hydrogen column density ($\log N(\hi)$), and the line of sight thickness ($\log L$), of the low ionization phase traced by the {\civ} cloud of the $z=0.0027$ absorber towards SG. The over-plotted vertical lines in the posterior distribution span the 95\% credible interval. The contours indicate 0.5$\sigma$, 1$\sigma$, 1.5$\sigma$, and 2$\sigma$ levels. The model results are summarised in Table~\ref{tab:modelparams}, and the synthetic profiles based on these models are shown in Figure~\ref{fig:SGsysplot}.}
\label{fig:SGCIV0}
\end{center}
\end{figure*}

\begin{figure*}
\begin{center}
\includegraphics[width=\linewidth]{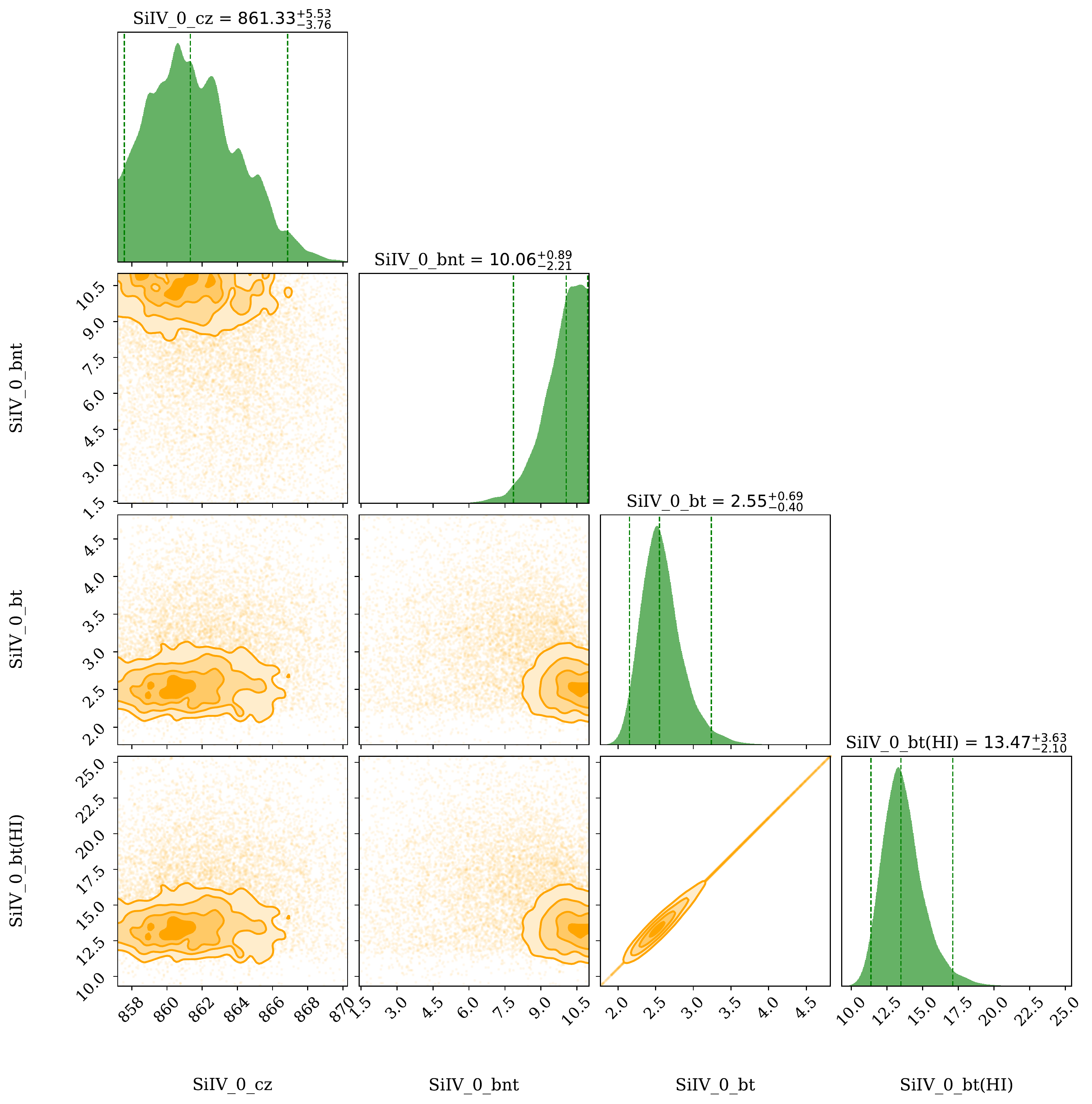}
\caption{The corner plot showing the marginalized posterior distributions for the absorption centroid ($z$), non-thermal Doppler broadening ($b_{nt}$), thermal Doppler broadening ($b_{t}$), total Doppler broadening ($b$), of the phase traced by the {\siiv} cloud of the $z=0.0027$ absorber towards SG. The over-plotted vertical lines in the posterior distribution span the 95\% credible interval. The contours indicate 0.5$\sigma$, 1$\sigma$, 1.5$\sigma$, and 2$\sigma$ levels. The model results are summarised in Table~\ref{tab:modelparams}, and the synthetic profiles based on these models are shown in Figure~\ref{fig:SGsysplot}.}
\label{fig:SGSiIV0b}
\end{center}
\end{figure*}

\begin{figure*}
\begin{center}
\includegraphics[width=\linewidth]{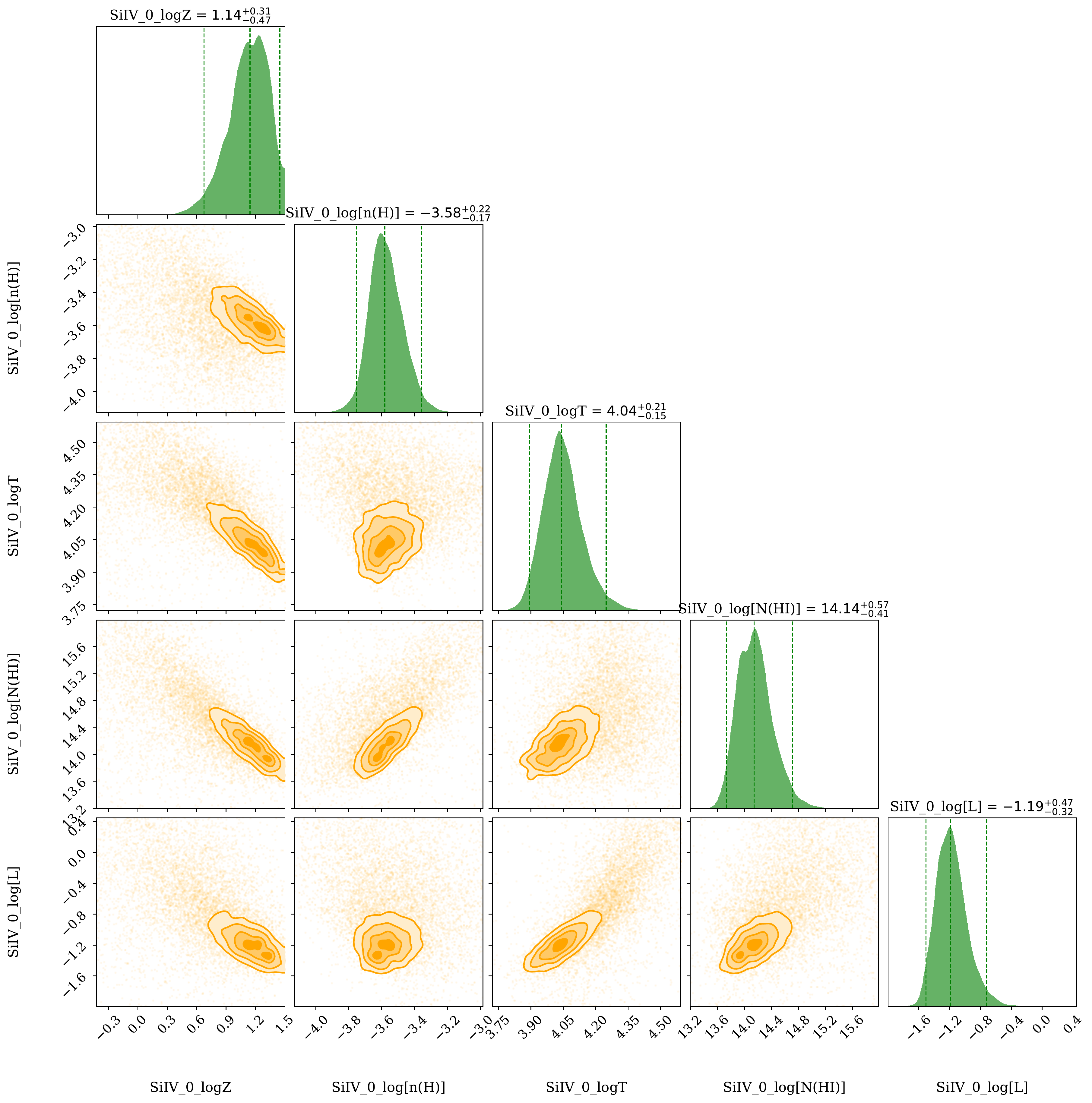}
\caption{The corner plot showing the marginalized posterior distributions for the metallicity ($\log Z$), hydrogen number density ($\log n_{H}$), temperature ($\log T$), neutral hydrogen column density ($\log N(\hi)$), and the line of sight thickness ($\log L$), of the low ionization phase traced by the {\siiv} cloud of the $z=0.0027$ absorber towards SG. The over-plotted vertical lines in the posterior distribution span the 95\% credible interval. The contours indicate 0.5$\sigma$, 1$\sigma$, 1.5$\sigma$, and 2$\sigma$ levels. The model results are summarised in Table~\ref{tab:modelparams}, and the synthetic profiles based on these models are shown in Figure~\ref{fig:SGsysplot}.}
\label{fig:SGSiIV0}
\end{center}
\end{figure*}

\begin{figure*}
\begin{center}
\includegraphics[width=\linewidth]{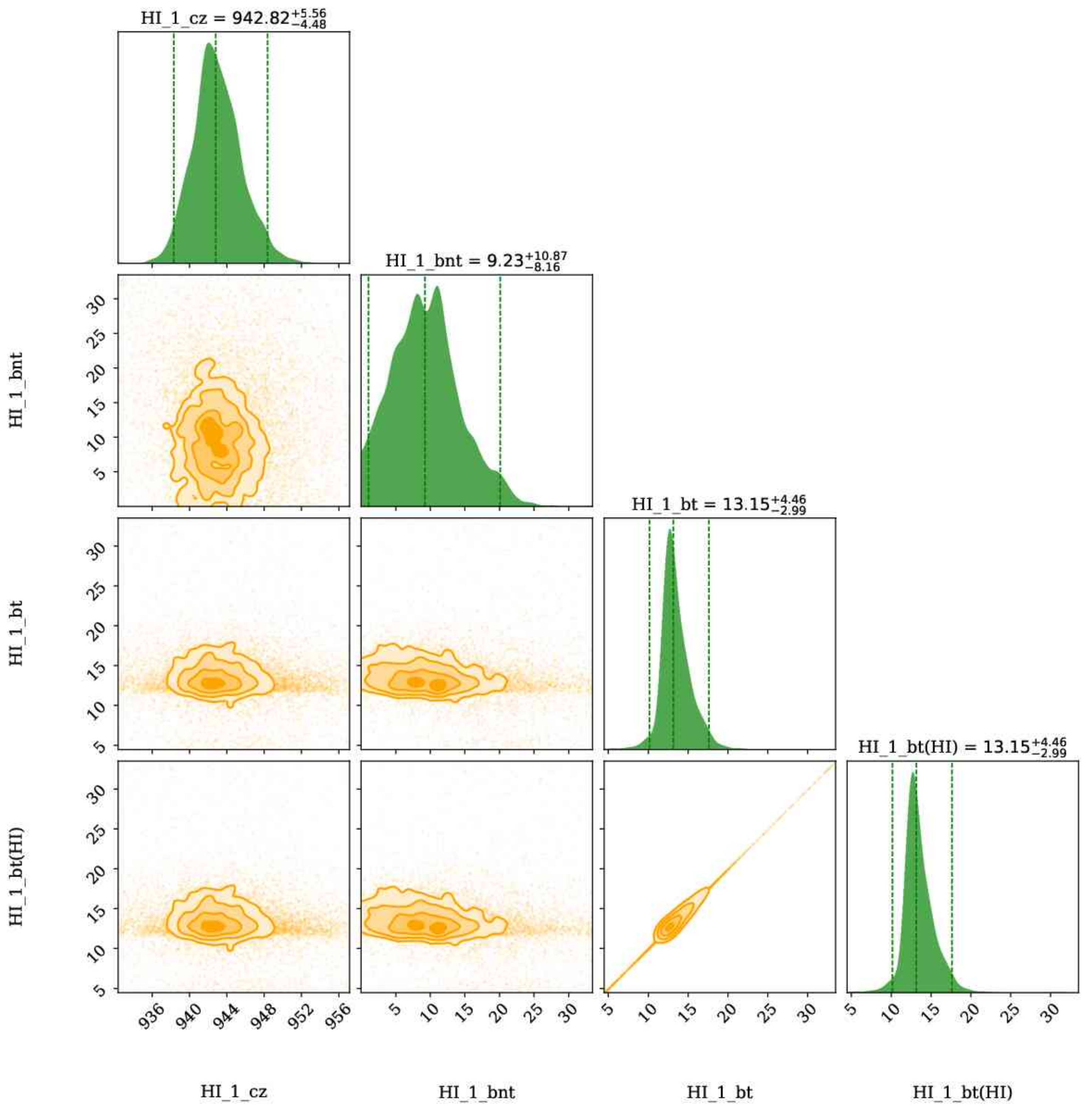}
\caption{The corner plot showing the marginalized posterior distributions for the absorption centroid ($z$), non-thermal Doppler broadening ($b_{nt}$), thermal Doppler broadening ($b_{t}$), total Doppler broadening ($b$), of the phase traced by the blueward {\hi} cloud of the $z=0.0027$ absorber towards SG. The over-plotted vertical lines in the posterior distribution span the 95\% credible interval. The contours indicate 0.5$\sigma$, 1$\sigma$, 1.5$\sigma$, and 2$\sigma$ levels. The model results are summarised in Table~\ref{tab:modelparams}, and the synthetic profiles based on these models are shown in Figure~\ref{fig:SGsysplot}.}
\label{fig:SGHI1b}
\end{center}
\end{figure*}

\begin{figure*}
\begin{center}
\includegraphics[width=\linewidth]{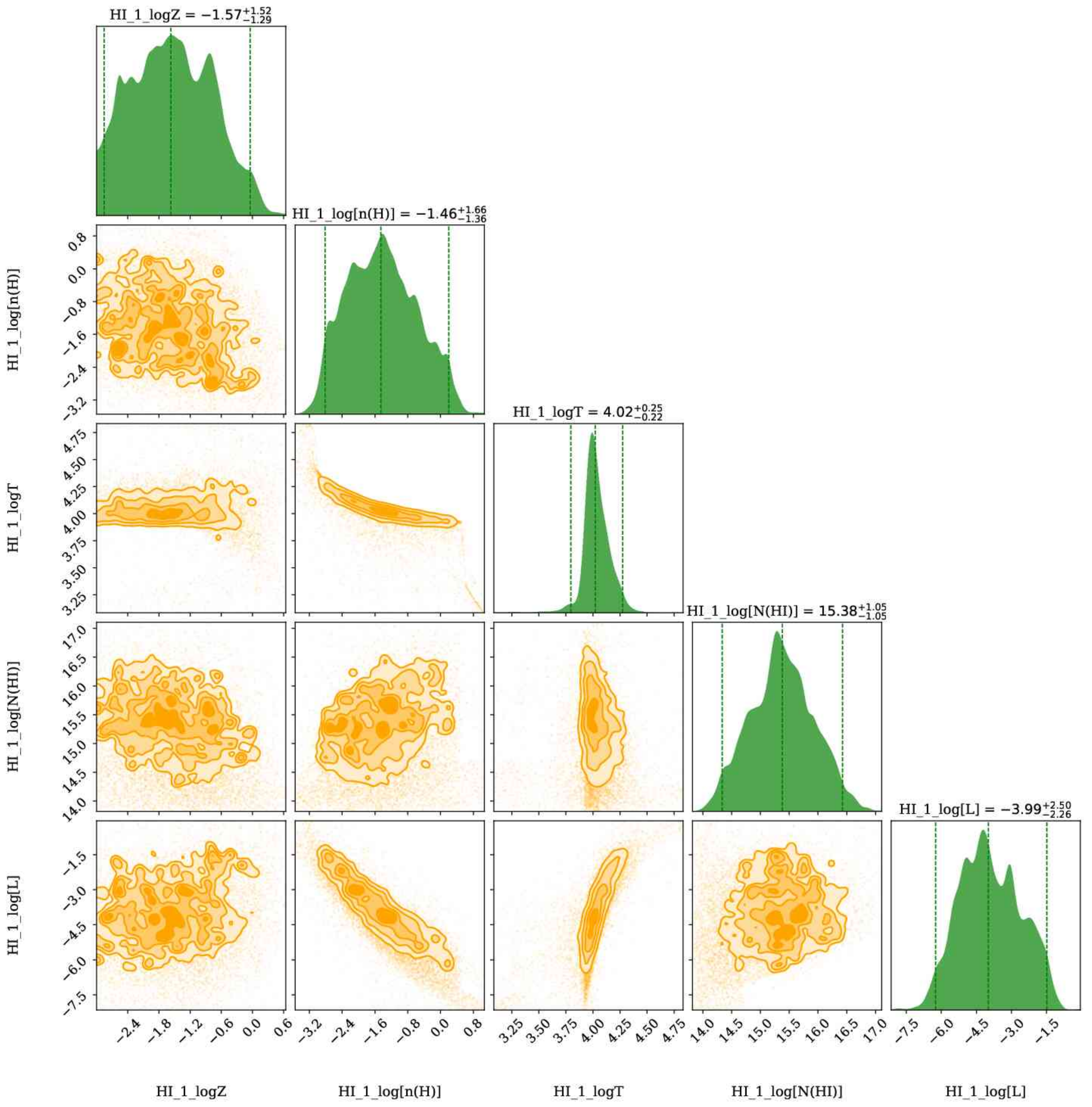}
\caption{The corner plot showing the marginalized posterior distributions for the metallicity ($\log Z$), hydrogen number density ($\log n_{H}$), temperature ($\log T$), neutral hydrogen column density ($\log N(\hi)$), and the line of sight thickness ($\log L$), of the low ionization phase traced by the blueward {\hi} cloud of the $z=0.0027$ absorber towards SG. The over-plotted vertical lines in the posterior distribution span the 95\% credible interval. The contours indicate 0.5$\sigma$, 1$\sigma$, 1.5$\sigma$, and 2$\sigma$ levels. The model results are summarised in Table~\ref{tab:modelparams}, and the synthetic profiles based on these models are shown in Figure~\ref{fig:SGsysplot}.}
\label{fig:SGHI1}
\end{center}
\end{figure*}

\begin{figure*}
\begin{center}
\includegraphics[width=\linewidth]{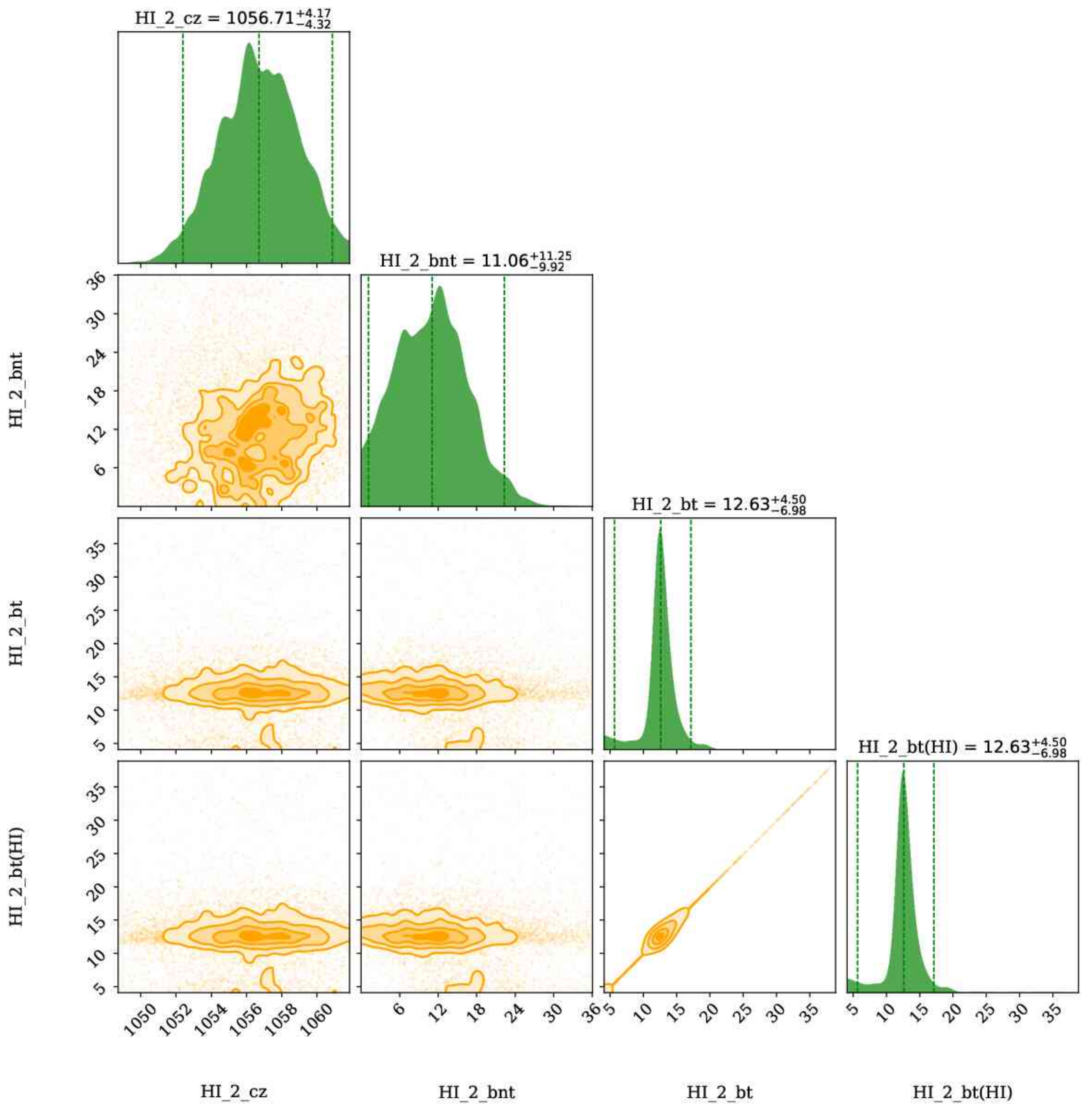}
\caption{The corner plot showing the marginalized posterior distributions for the absorption centroid ($z$), non-thermal Doppler broadening ($b_{nt}$), thermal Doppler broadening ($b_{t}$), total Doppler broadening ($b$), of the phase traced by the central {\hi} cloud of the $z=0.0027$ absorber towards SG. The over-plotted vertical lines in the posterior distribution span the 95\% credible interval. The contours indicate 0.5$\sigma$, 1$\sigma$, 1.5$\sigma$, and 2$\sigma$ levels. The model results are summarised in Table~\ref{tab:modelparams}, and the synthetic profiles based on these models are shown in Figure~\ref{fig:SGsysplot}.}
\label{fig:SGHI2b}
\end{center}
\end{figure*}

\begin{figure*}
\begin{center}
\includegraphics[width=\linewidth]{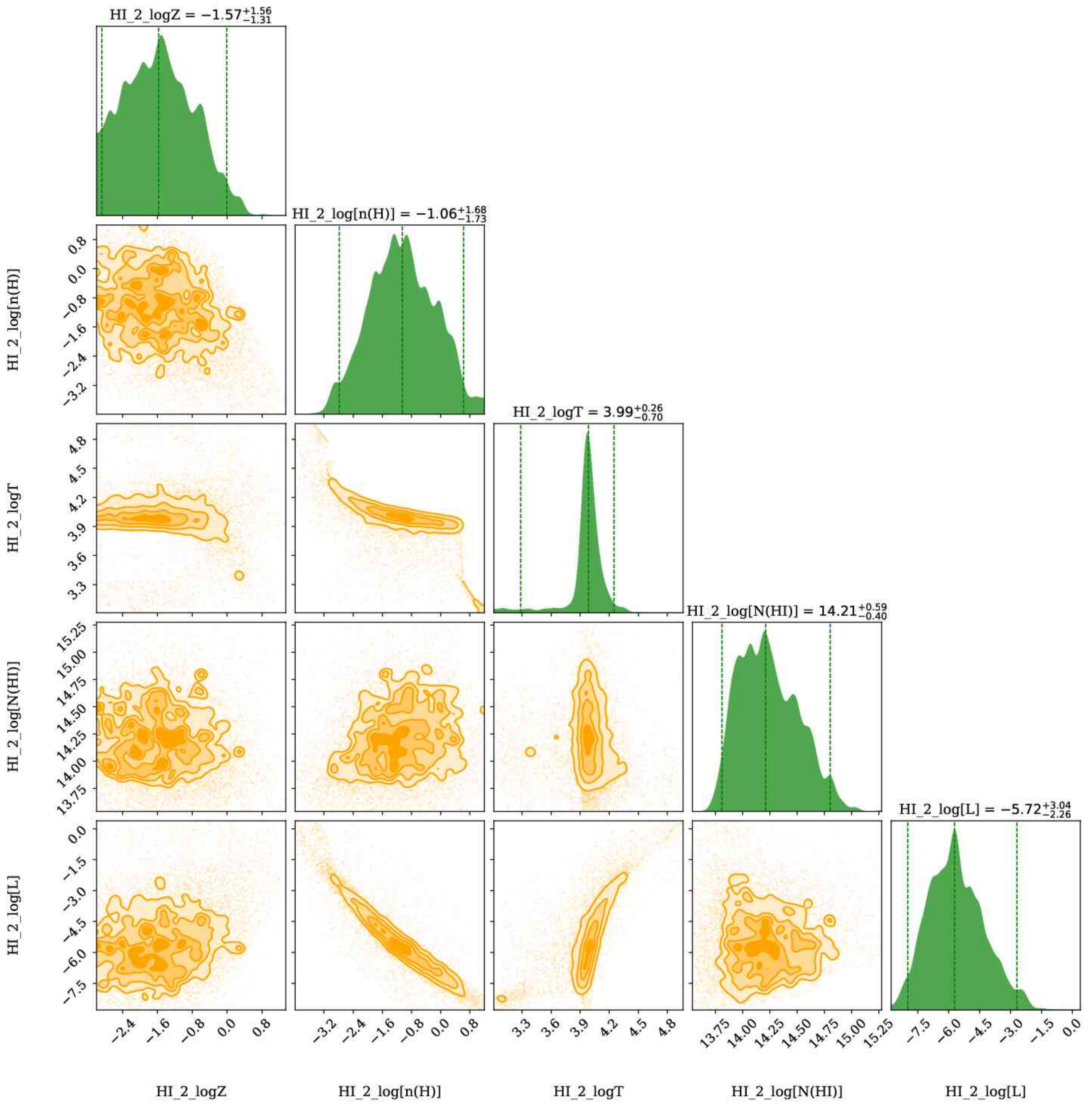}
\caption{The corner plot showing the marginalized posterior distributions for the metallicity ($\log Z$), hydrogen number density ($\log n_{H}$), temperature ($\log T$), neutral hydrogen column density ($\log N(\hi)$), and the line of sight thickness ($\log L$), of the low ionization phase traced by the central {\hi} cloud of the $z=0.0027$ absorber towards SG. The over-plotted vertical lines in the posterior distribution span the 95\% credible interval. The contours indicate 0.5$\sigma$, 1$\sigma$, 1.5$\sigma$, and 2$\sigma$ levels. The model results are summarised in Table~\ref{tab:modelparams}, and the synthetic profiles based on these models are shown in Figure~\ref{fig:SGsysplot}.}
\label{fig:SGHI2}
\end{center}
\end{figure*}

\begin{figure*}
\begin{center}
\includegraphics[width=\linewidth]{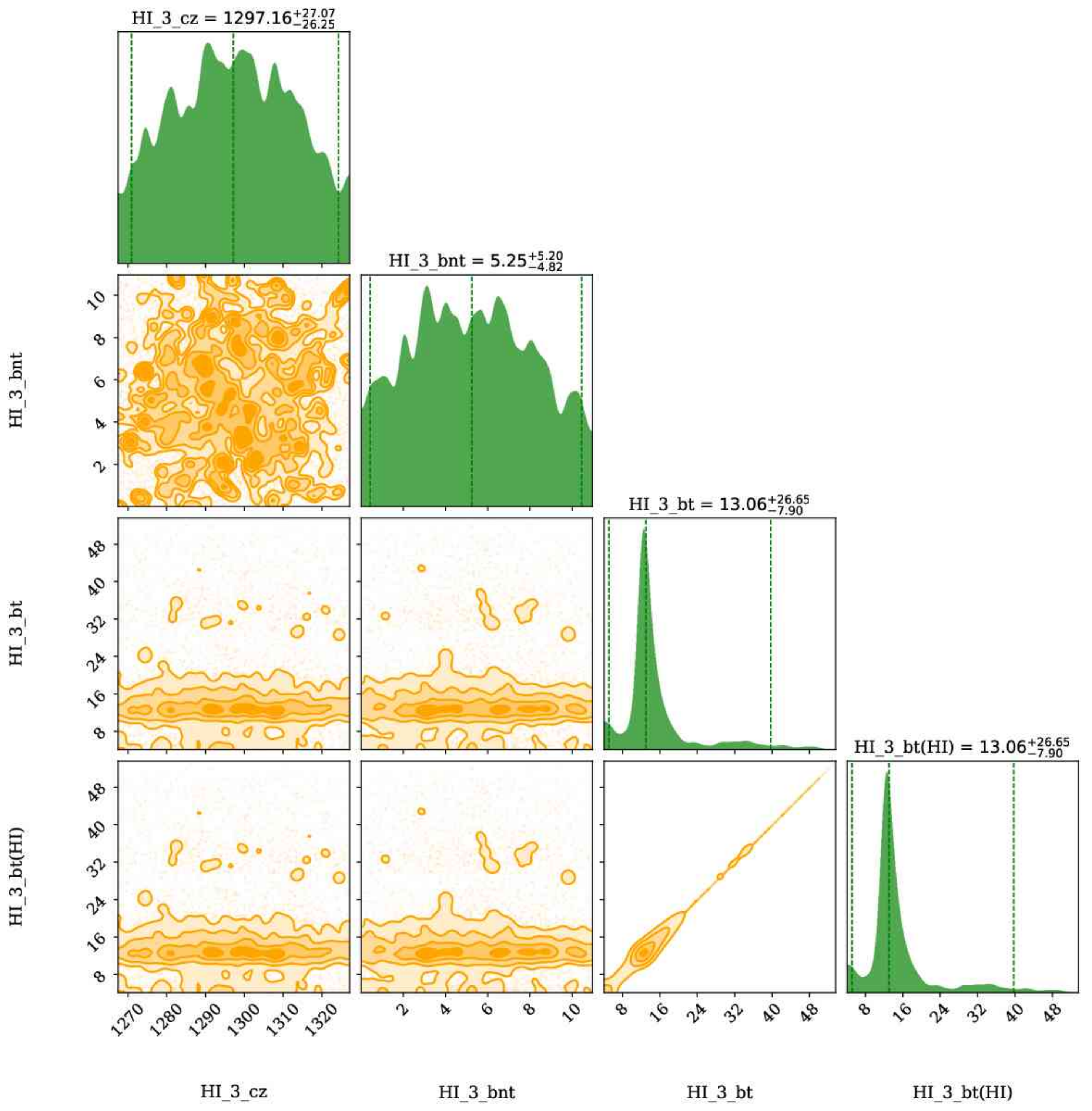}
\caption{The corner plot showing the marginalized posterior distributions for the absorption centroid ($z$), non-thermal Doppler broadening ($b_{nt}$), thermal Doppler broadening ($b_{t}$), total Doppler broadening ($b$), of the phase traced by the redward {\hi} cloud of the $z=0.0027$ absorber towards SG. The over-plotted vertical lines in the posterior distribution span the 95\% credible interval. The contours indicate 0.5$\sigma$, 1$\sigma$, 1.5$\sigma$, and 2$\sigma$ levels. The model results are summarised in Table~\ref{tab:modelparams}, and the synthetic profiles based on these models are shown in Figure~\ref{fig:SGsysplot}.}
\label{fig:SGHI3b}
\end{center}
\end{figure*}

\begin{figure*}
\begin{center}
\includegraphics[width=\linewidth]{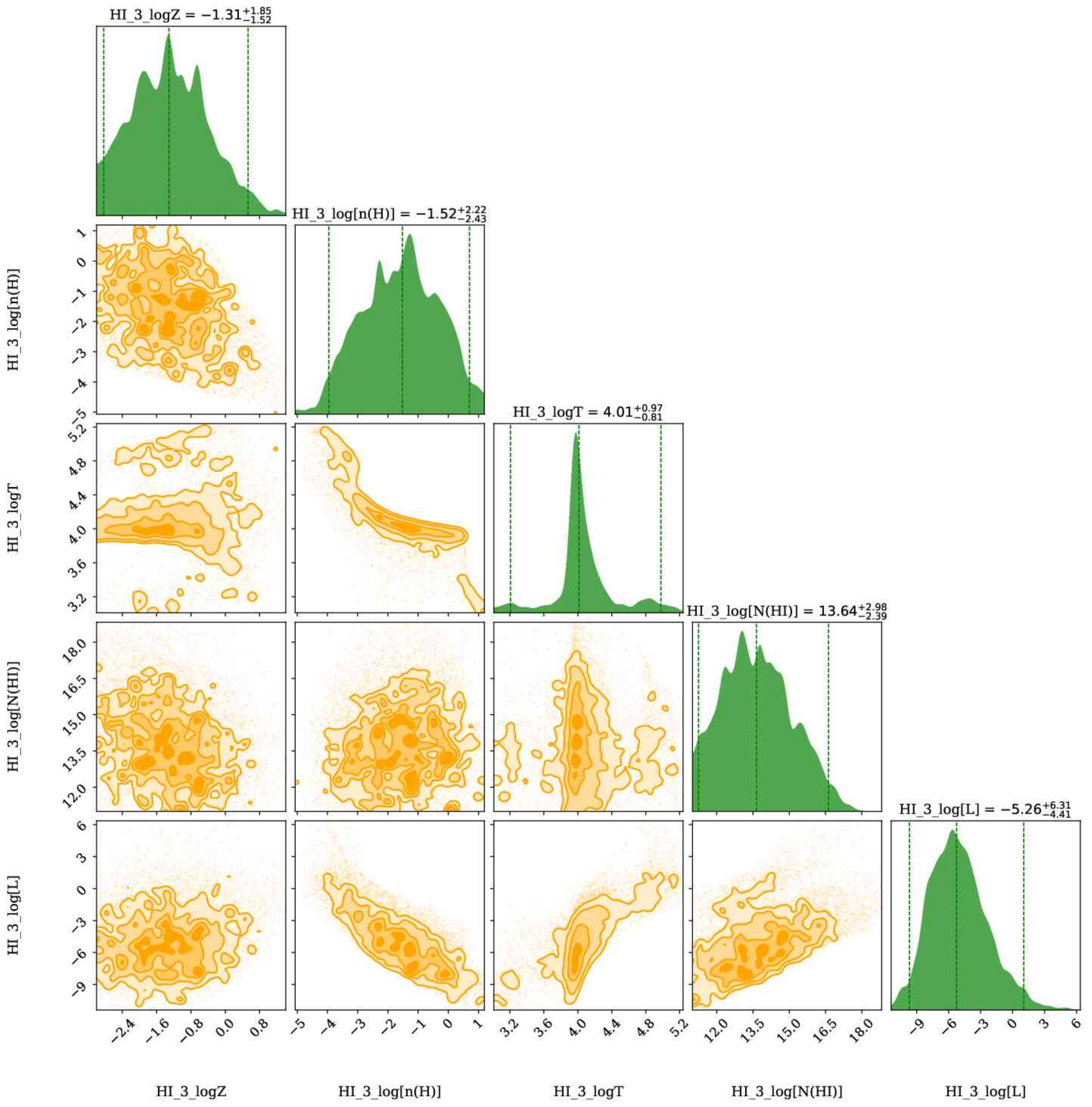}
\caption{The corner plot showing the marginalized posterior distributions for the metallicity ($\log Z$), hydrogen number density ($\log n_{H}$), temperature ($\log T$), neutral hydrogen column density ($\log N(\hi)$), and the line of sight thickness ($\log L$), of the low ionization phase traced by the redward {\hi} cloud of the $z=0.0027$ absorber towards SG. The over-plotted vertical lines in the posterior distribution span the 95\% credible interval. The contours indicate 0.5$\sigma$, 1$\sigma$, 1.5$\sigma$, and 2$\sigma$ levels. The model results are summarised in Table~\ref{tab:modelparams}, and the synthetic profiles based on these models are shown in Figure~\ref{fig:SGsysplot}.}
\label{fig:SGHI3}
\end{center}
\end{figure*}

\clearpage
\section{Plots for SH}
\label{appendix:SH}

\subsection{Airglow template fit towards SH}

\begin{figure*}
\begin{center}
\includegraphics[width=0.75\linewidth]{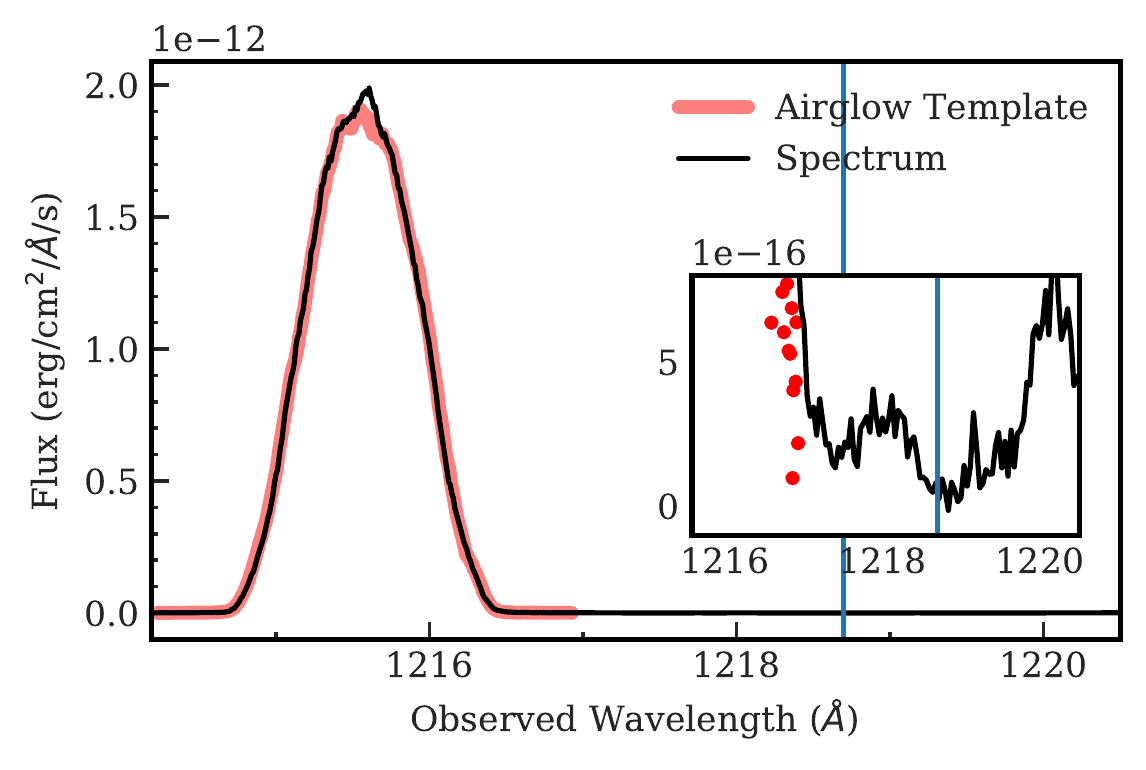}
\caption{Same as in Figure~\ref{fig:SBairglow}, but for SH.}
\label{fig:SHairglow}
\end{center}
\end{figure*}

\subsection{Best VP fit to the Galactic {\lya} towards SH}

\begin{figure*}
\begin{center}
\includegraphics[width=\linewidth]{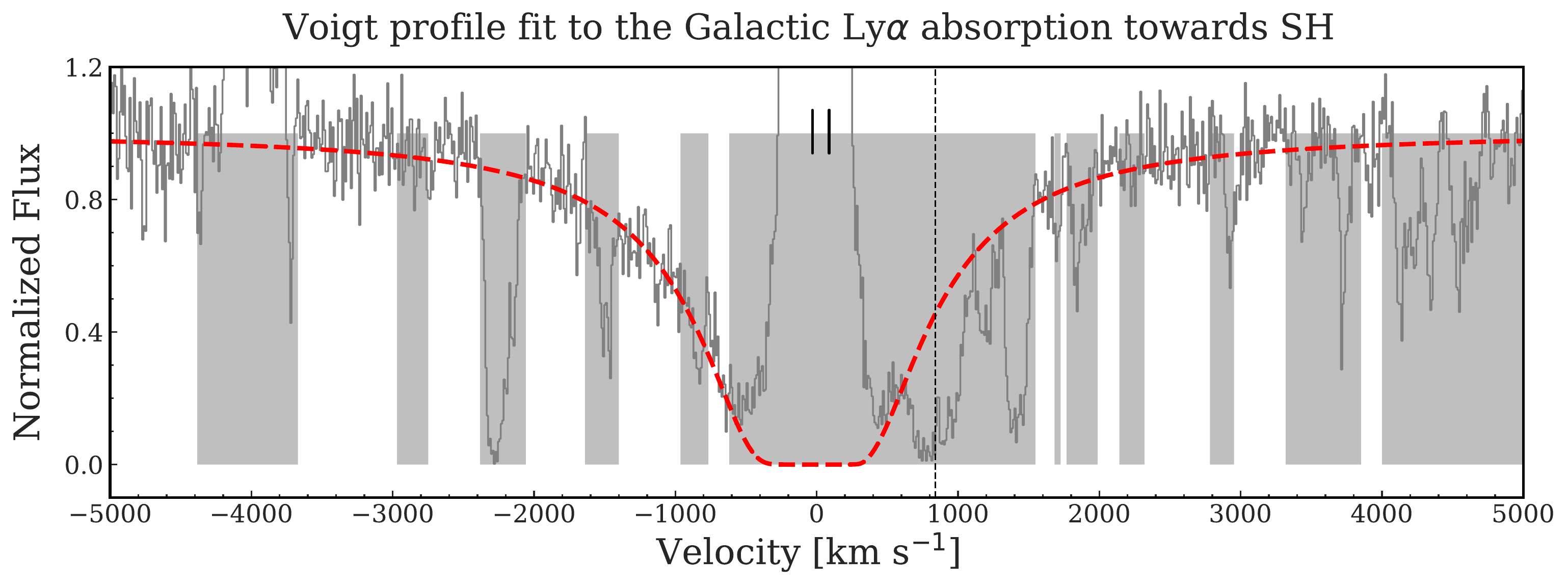}
\caption{Same as in Figure~\ref{fig:SBvpfitgal}, but for SH.}
\label{fig:SHvpfitgal}
\end{center}
\end{figure*}

\subsection{MC exploration of the zero-point uncertainty for SH}

\begin{figure*}
\begin{center}
\includegraphics[scale=0.5]{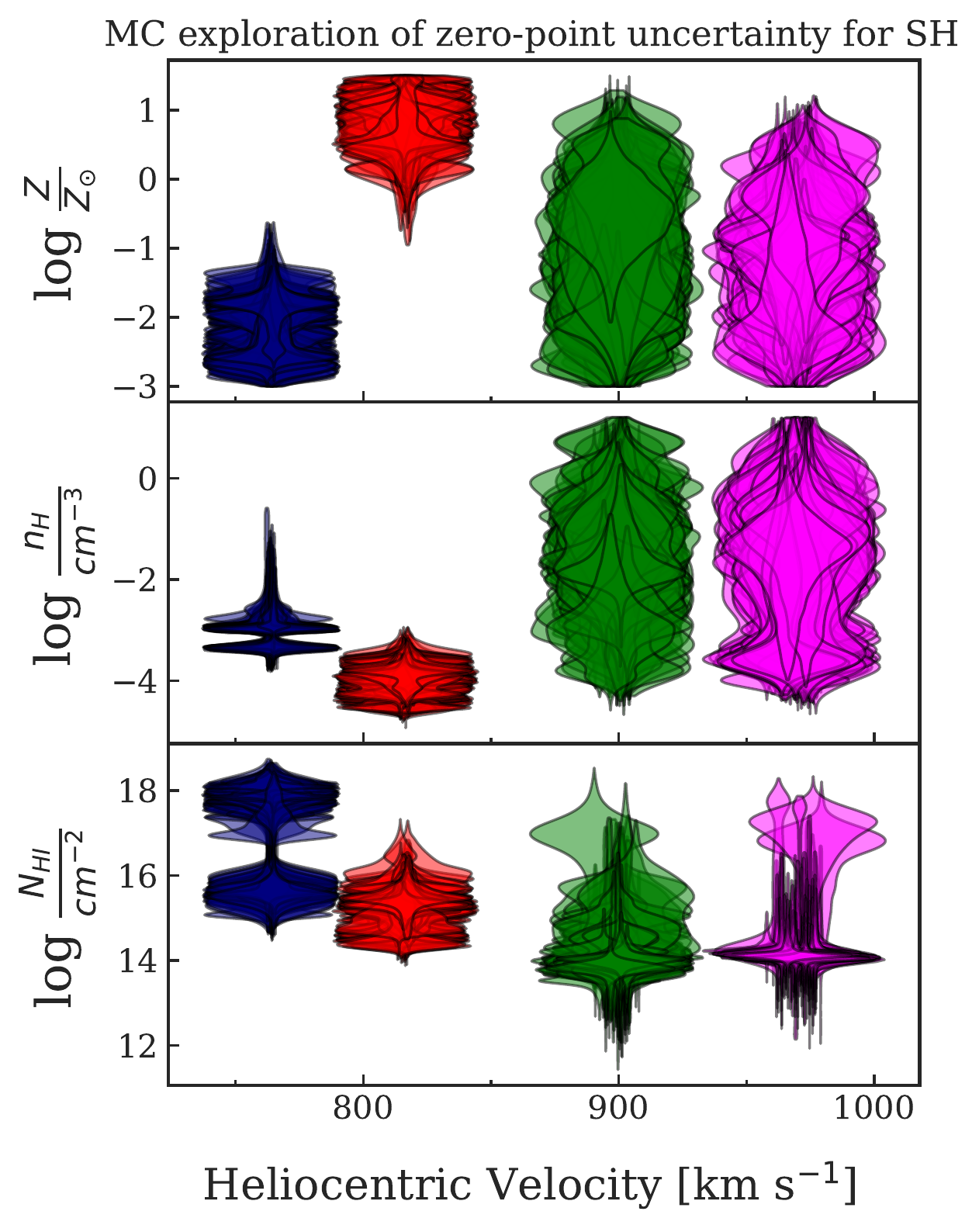}
\caption{Violin plots showing the parameter distributions for 100 different realizations of the {\hi} {\lya} profile modified between 765--820 {\kms} to account for the zero-point uncertainty. The blue violins show the parameters of blueward {\siiii} phase. The red violins correspond to the redward {\siiii} phase, the green and magenta violins correspond to the {\hi} only phases. }
\label{fig:MCSH}
\end{center}
\end{figure*}

\subsection{Posterior distributions for the absorber properties towards SH}

\label{appendix:SHparams}

\begin{figure*}
\begin{center}
\includegraphics[width=\linewidth]{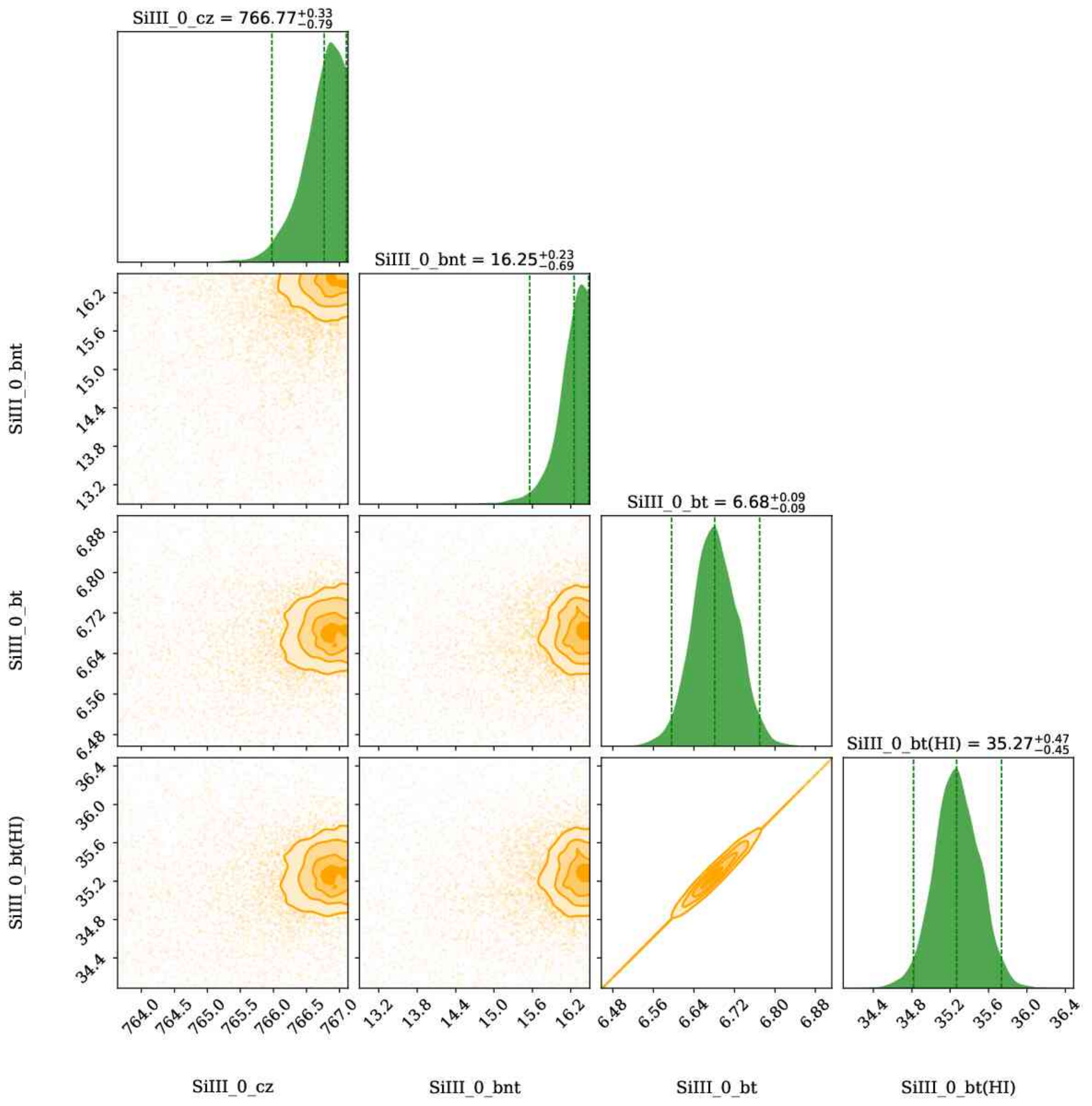}
\caption{The corner plot showing the marginalized posterior distributions for the absorption centroid ($z$), non-thermal Doppler broadening ($b_{nt}$), thermal Doppler broadening ($b_{t}$), total Doppler broadening ($b$), of the phase traced by the blueward {\hi} cloud of the $z=0.0028$ absorber towards SH. The over-plotted vertical lines in the posterior distribution span the 95\% credible interval. The contours indicate 0.5$\sigma$, 1$\sigma$, 1.5$\sigma$, and 2$\sigma$ levels. The model results are summarised in Table~\ref{tab:modelparams}, and the synthetic profiles based on these models are shown in Figure~\ref{fig:SHsysplot}.}
\label{fig:SHSiIII0b}
\end{center}
\end{figure*}

\begin{figure*}
\begin{center}
\includegraphics[width=\linewidth]{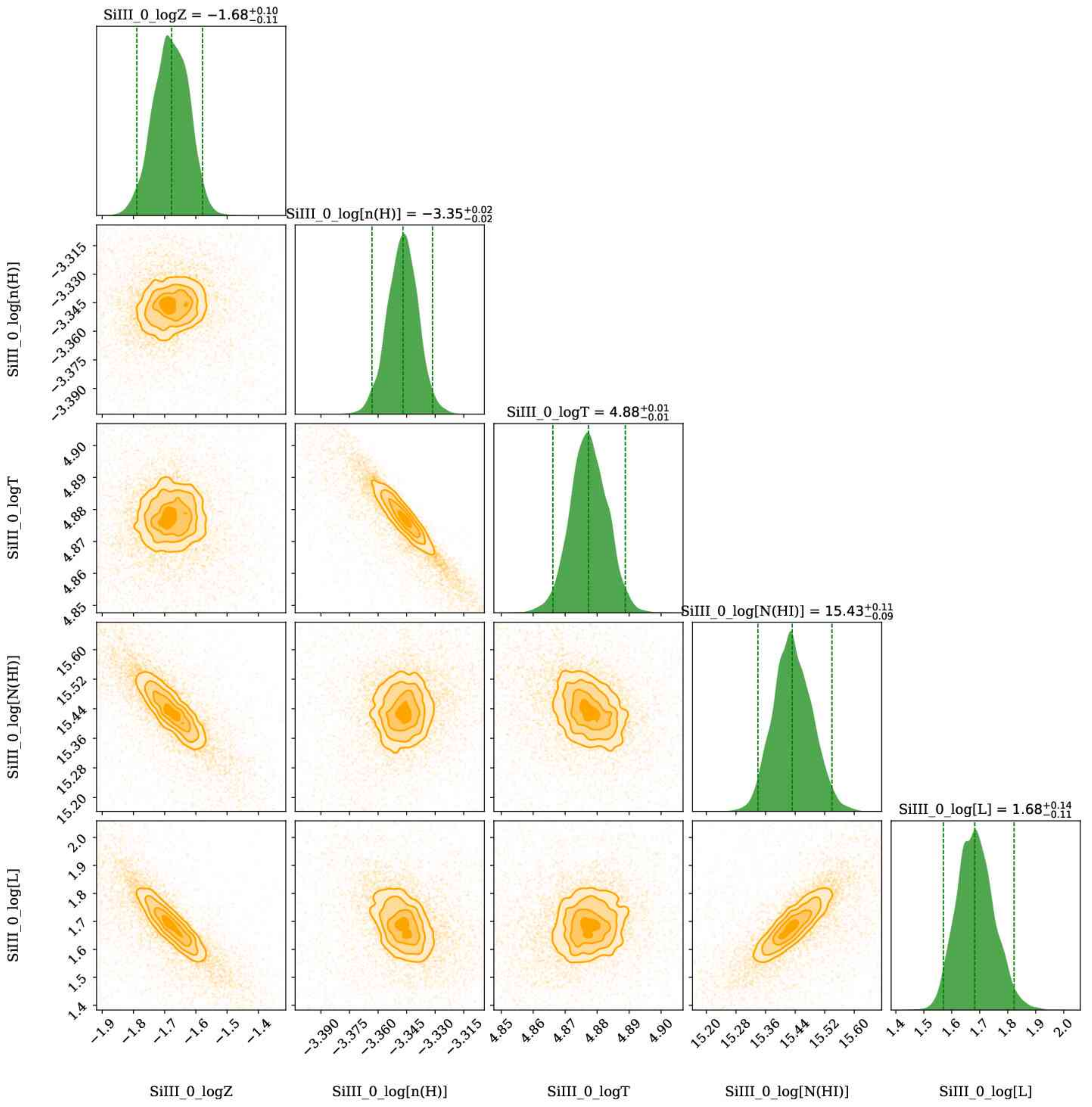}
\caption{The corner plot showing the marginalized posterior distributions for the metallicity ($\log Z$), hydrogen number density ($\log n_{H}$), temperature ($\log T$), neutral hydrogen column density ($\log N(\hi)$), and the line of sight thickness ($\log L$), of the low ionization phase traced by the {\hi} cloud of the $z=0.0028$ absorber towards SH. The over-plotted vertical lines in the posterior distribution span the 95\% credible interval. The contours indicate 0.5$\sigma$, 1$\sigma$, 1.5$\sigma$, and 2$\sigma$ levels. The model results are summarised in Table~\ref{tab:modelparams}, and the synthetic profiles based on these models are shown in Figure~\ref{fig:SHsysplot}.}
\label{fig:SHSiIII0}
\end{center}
\end{figure*}

\begin{figure*}
\begin{center}
\includegraphics[width=\linewidth]{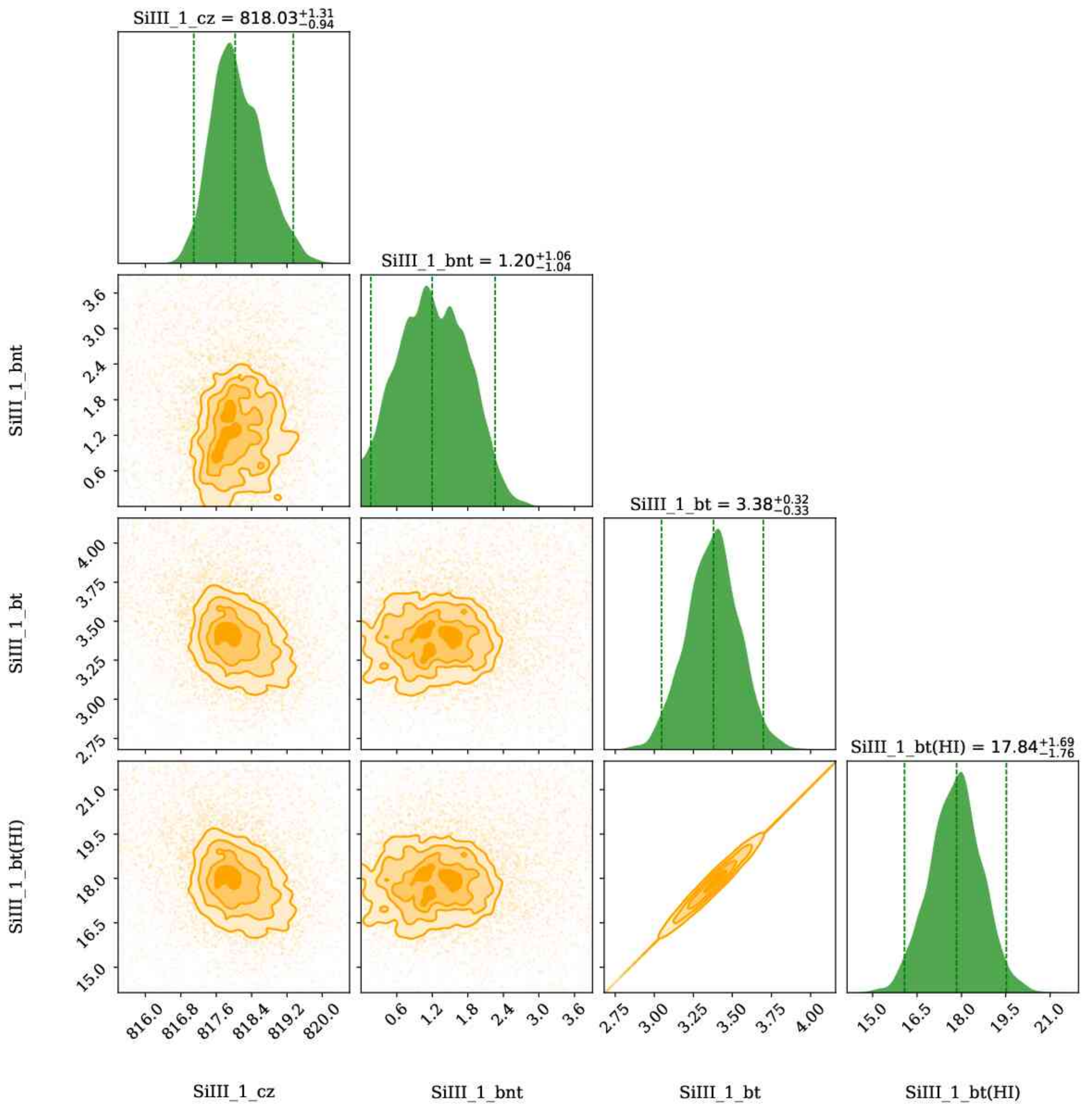}
\caption{The corner plot showing the marginalized posterior distributions for the absorption centroid ($z$), non-thermal Doppler broadening ($b_{nt}$), thermal Doppler broadening ($b_{t}$), total Doppler broadening ($b$), of the phase traced by the blueward {\hi} cloud of the $z=0.0028$ absorber towards SH. The over-plotted vertical lines in the posterior distribution span the 95\% credible interval. The contours indicate 0.5$\sigma$, 1$\sigma$, 1.5$\sigma$, and 2$\sigma$ levels. The model results are summarised in Table~\ref{tab:modelparams}, and the synthetic profiles based on these models are shown in Figure~\ref{fig:SHsysplot}.}
\label{fig:SHSiIII1b}
\end{center}
\end{figure*}

\begin{figure*}
\begin{center}
\includegraphics[width=\linewidth]{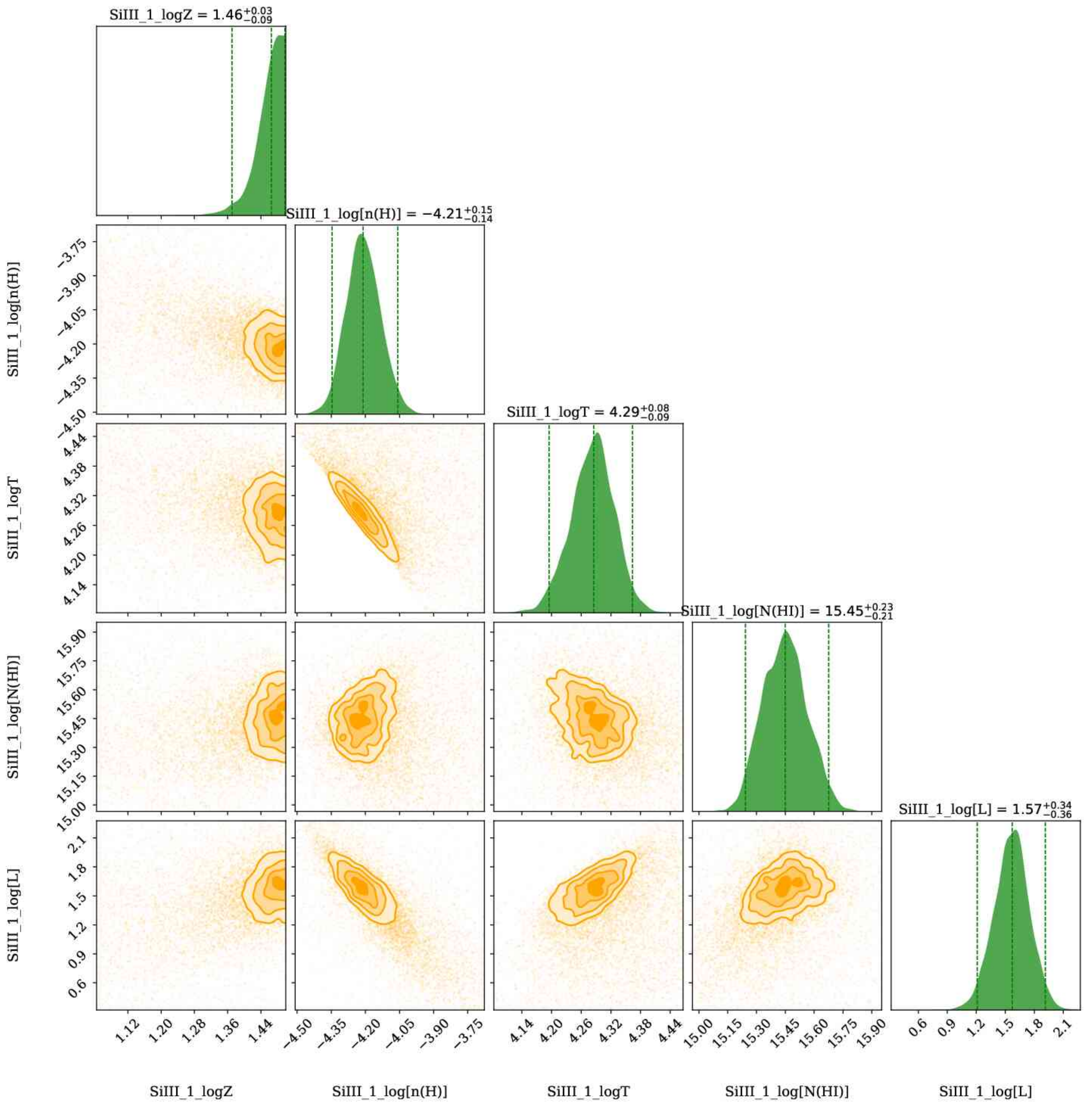}
\caption{The corner plot showing the marginalized posterior distributions for the metallicity ($\log Z$), hydrogen number density ($\log n_{H}$), temperature ($\log T$), neutral hydrogen column density ($\log N(\hi)$), and the line of sight thickness ($\log L$), of the low ionization phase traced by the {\hi} cloud of the $z=0.0028$ absorber towards SH. The over-plotted vertical lines in the posterior distribution span the 95\% credible interval. The contours indicate 0.5$\sigma$, 1$\sigma$, 1.5$\sigma$, and 2$\sigma$ levels. The model results are summarised in Table~\ref{tab:modelparams}, and the synthetic profiles based on these models are shown in Figure~\ref{fig:SHsysplot}.}
\label{fig:SHSiIII1}
\end{center}
\end{figure*}

\begin{figure*}
\begin{center}
\includegraphics[width=\linewidth]{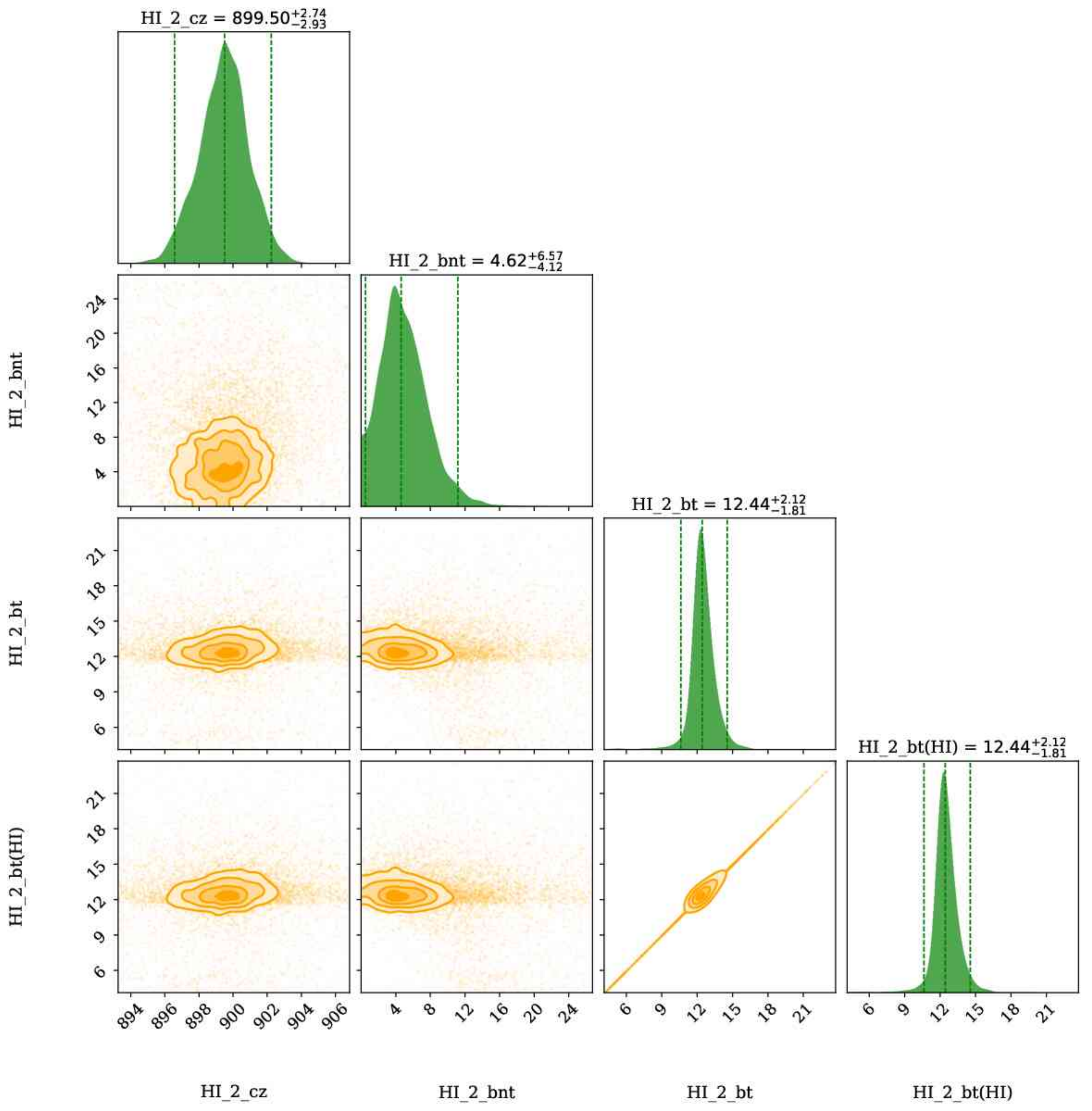}
\caption{The corner plot showing the marginalized posterior distributions for the absorption centroid ($z$), non-thermal Doppler broadening ($b_{nt}$), thermal Doppler broadening ($b_{t}$), total Doppler broadening ($b$), of the phase traced by the middle redward {\hi} cloud of the $z=0.0028$ absorber towards SH. The over-plotted vertical lines in the posterior distribution span the 95\% credible interval. The contours indicate 0.5$\sigma$, 1$\sigma$, 1.5$\sigma$, and 2$\sigma$ levels. The model results are summarised in Table~\ref{tab:modelparams}, and the synthetic profiles based on these models are shown in Figure~\ref{fig:SHsysplot}.}
\label{fig:SHHI2b}
\end{center}
\end{figure*}

\begin{figure*}
\begin{center}
\includegraphics[width=\linewidth]{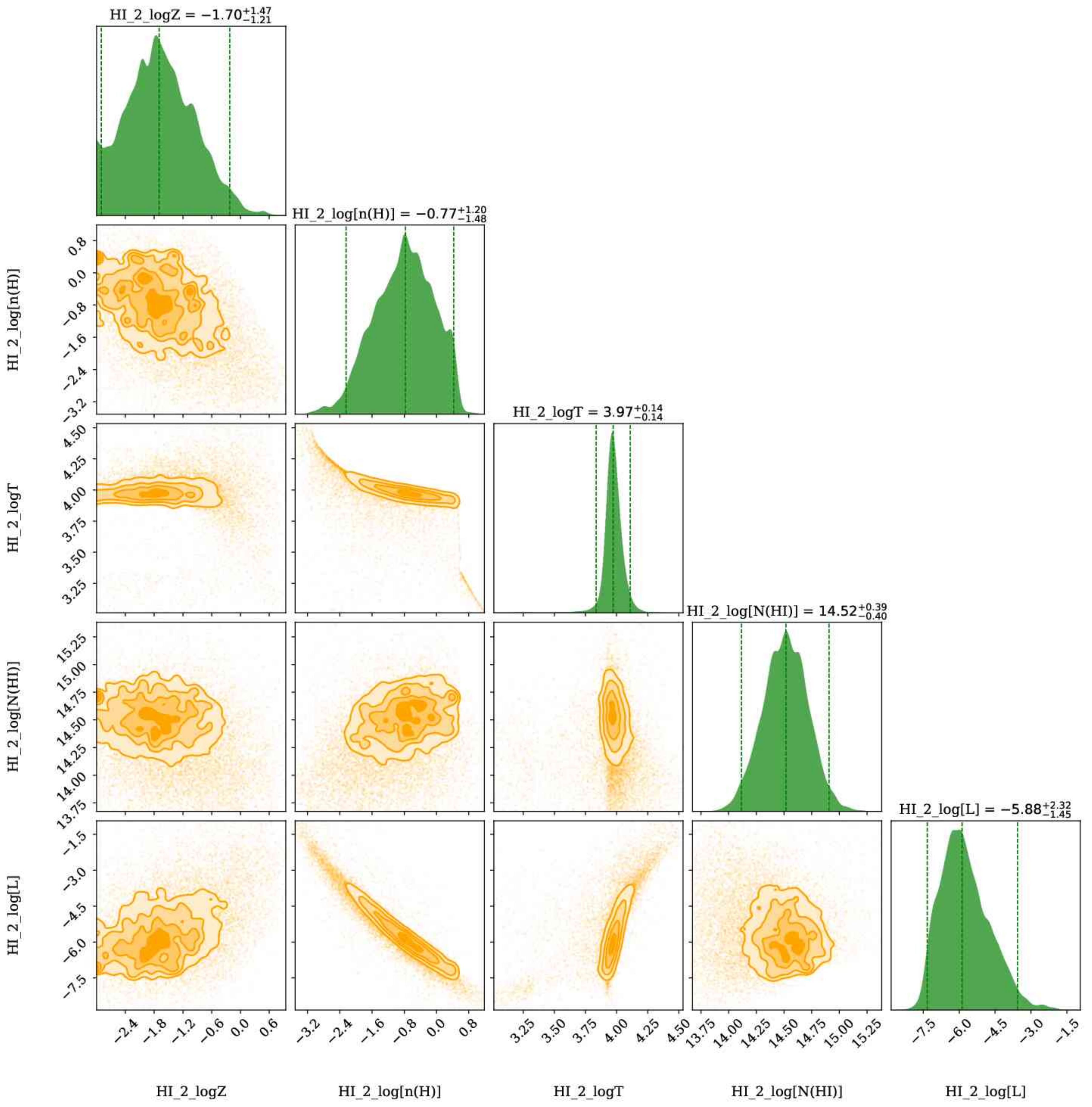}
\caption{The corner plot showing the marginalized posterior distributions for the metallicity ($\log Z$), hydrogen number density ($\log n_{H}$), temperature ($\log T$), neutral hydrogen column density ($\log N(\hi)$), and the line of sight thickness ($\log L$), of the low ionization phase traced by the middle redward {\hi} cloud of the $z=0.0028$ absorber towards SH. The over-plotted vertical lines in the posterior distribution span the 95\% credible interval. The contours indicate 0.5$\sigma$, 1$\sigma$, 1.5$\sigma$, and 2$\sigma$ levels. The model results are summarised in Table~\ref{tab:modelparams}, and the synthetic profiles based on these models are shown in Figure~\ref{fig:SHsysplot}.}
\label{fig:SHHI2}
\end{center}
\end{figure*}

\begin{figure*}
\begin{center}

\includegraphics[width=\linewidth]{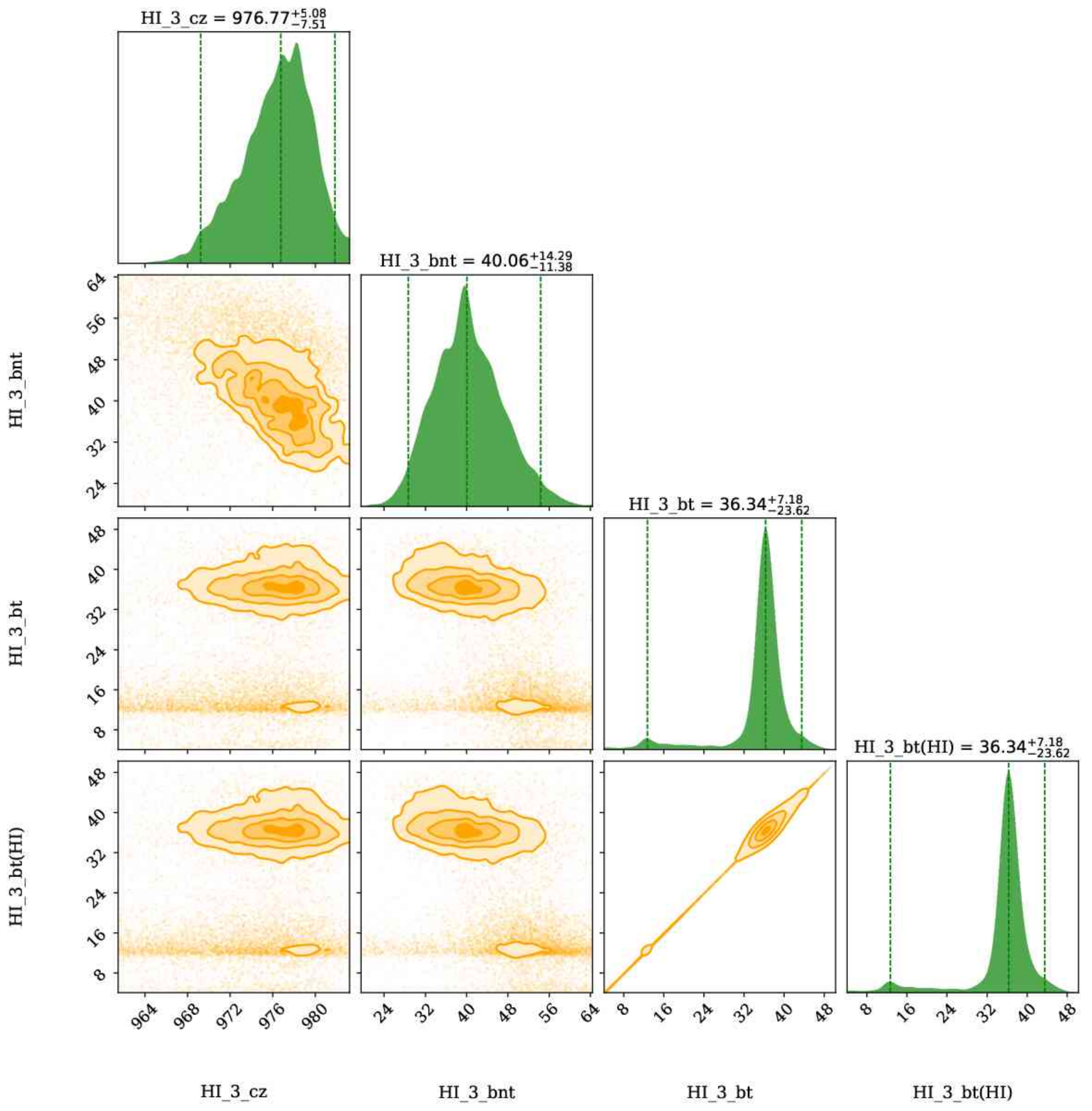}
\caption{The corner plot showing the marginalized posterior distributions for the absorption centroid ($z$), non-thermal Doppler broadening ($b_{nt}$), thermal Doppler broadening ($b_{t}$), total Doppler broadening ($b$), of the phase traced by the redward {\hi} cloud of the $z=0.0028$ absorber towards SH. The over-plotted vertical lines in the posterior distribution span the 95\% credible interval. The contours indicate 0.5$\sigma$, 1$\sigma$, 1.5$\sigma$, and 2$\sigma$ levels. The model results are summarised in Table~\ref{tab:modelparams}, and the synthetic profiles based on these models are shown in Figure~\ref{fig:SHsysplot}.}
\label{fig:SHHI3b}
\end{center}
\end{figure*}

\begin{figure*}
\begin{center}
\includegraphics[width=\linewidth]{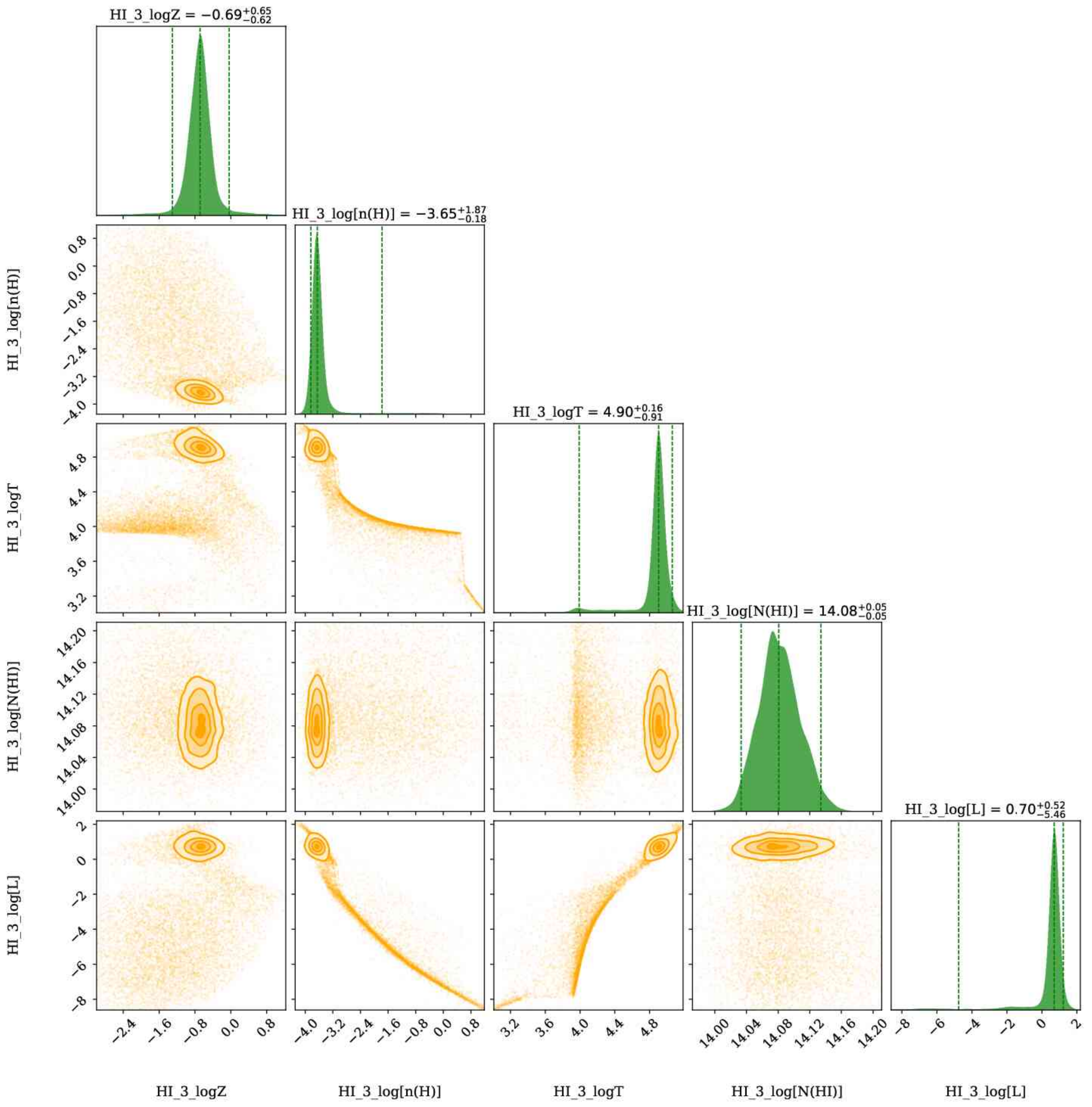}
\caption{The corner plot showing the marginalized posterior distributions for the metallicity ($\log Z$), hydrogen number density ($\log n_{H}$), temperature ($\log T$), neutral hydrogen column density ($\log N(\hi)$), and the line of sight thickness ($\log L$), of the low ionization phase traced by the redward {\hi} cloud of the $z=0.0028$ absorber towards SH. The over-plotted vertical lines in the posterior distribution span the 95\% credible interval. The contours indicate 0.5$\sigma$, 1$\sigma$, 1.5$\sigma$, and 2$\sigma$ levels. The model results are summarised in Table~\ref{tab:modelparams}, and the synthetic profiles based on these models are shown in Figure~\ref{fig:SHsysplot}.}
\label{fig:SHHI3}
\end{center}
\end{figure*}

\clearpage
\section{Plots for SI}
\label{appendix:SI}

\subsection{Airglow template fit towards SI}

\begin{figure*}
\begin{center}
\includegraphics[width=0.75\linewidth]{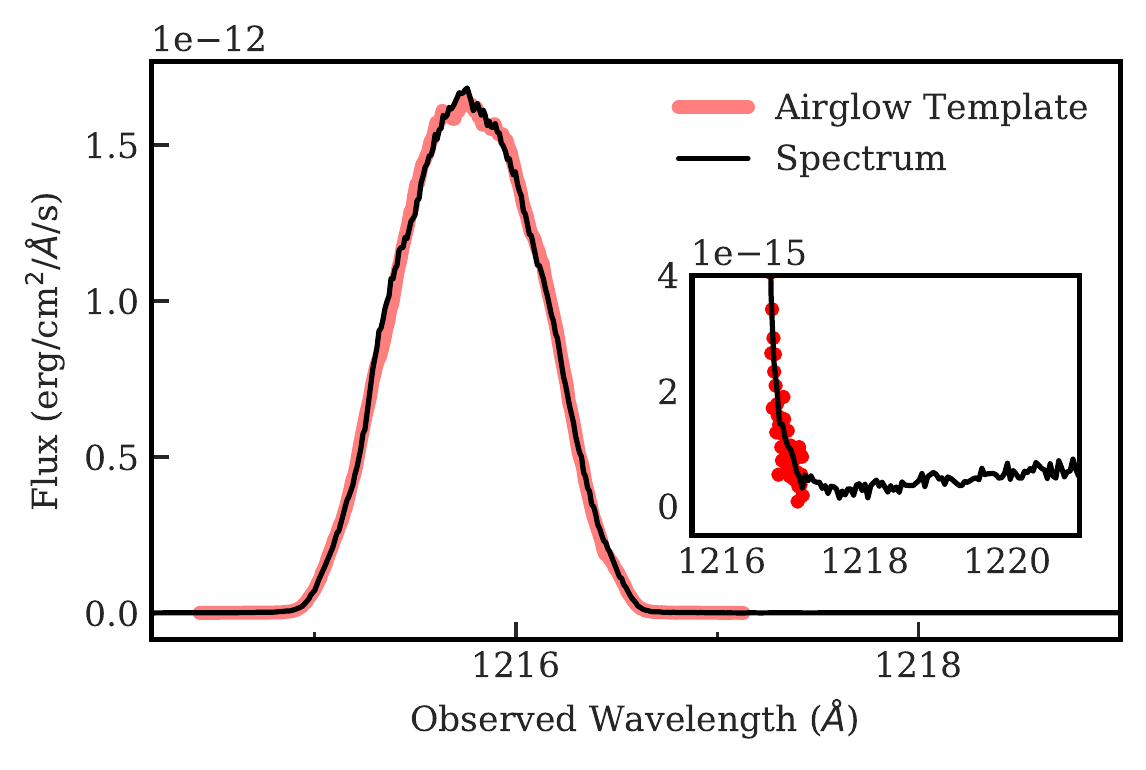}
\caption{Same as in Figure~\ref{fig:SBairglow}, but for SI.}
\label{fig:SIairglow}
\end{center}
\end{figure*}

\subsection{Best VP fit to the Galactic {\lya} towards SI}

\begin{figure*}
\begin{center}
\includegraphics[width=\linewidth]{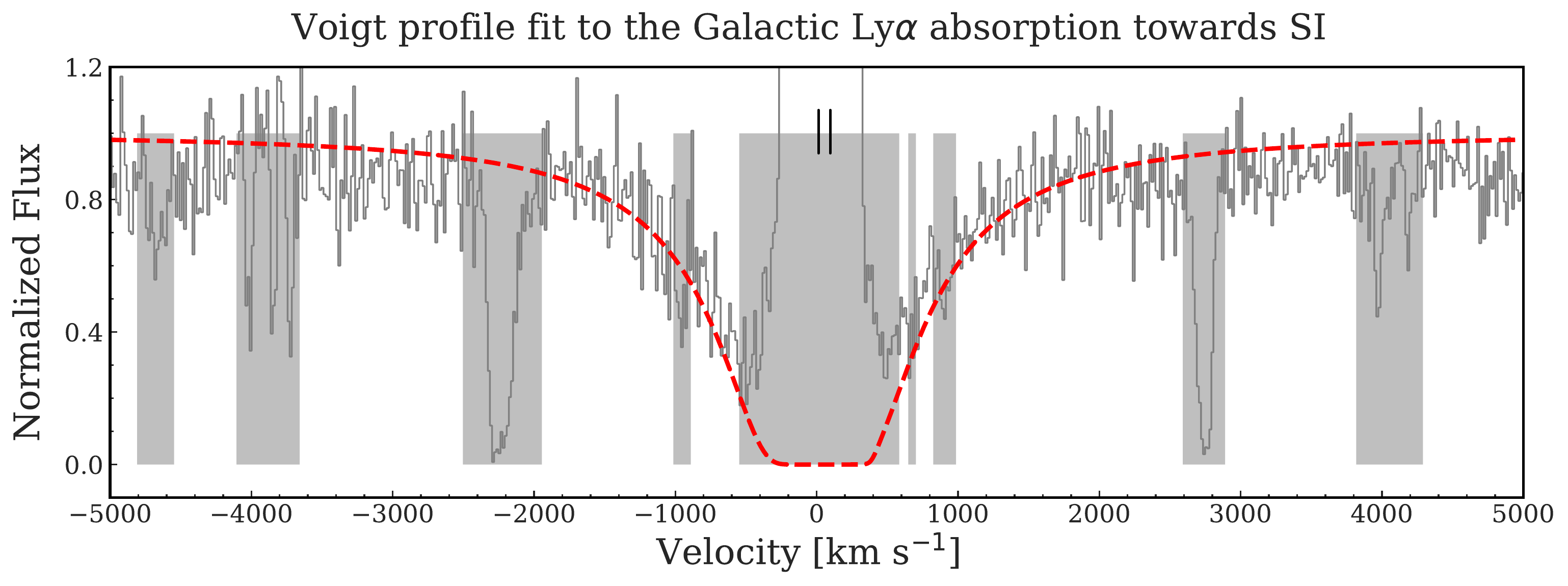}
\caption{Same as in Figure~\ref{fig:SBvpfitgal}, but for SI.}
\label{fig:SIvpfitgal}
\end{center}
\end{figure*}

\clearpage
\section{Plots for SJ}
\label{appendix:SJ}

\subsection{Airglow template fit towards SJ}

\begin{figure*}
\begin{center}
\includegraphics[width=0.75\linewidth]{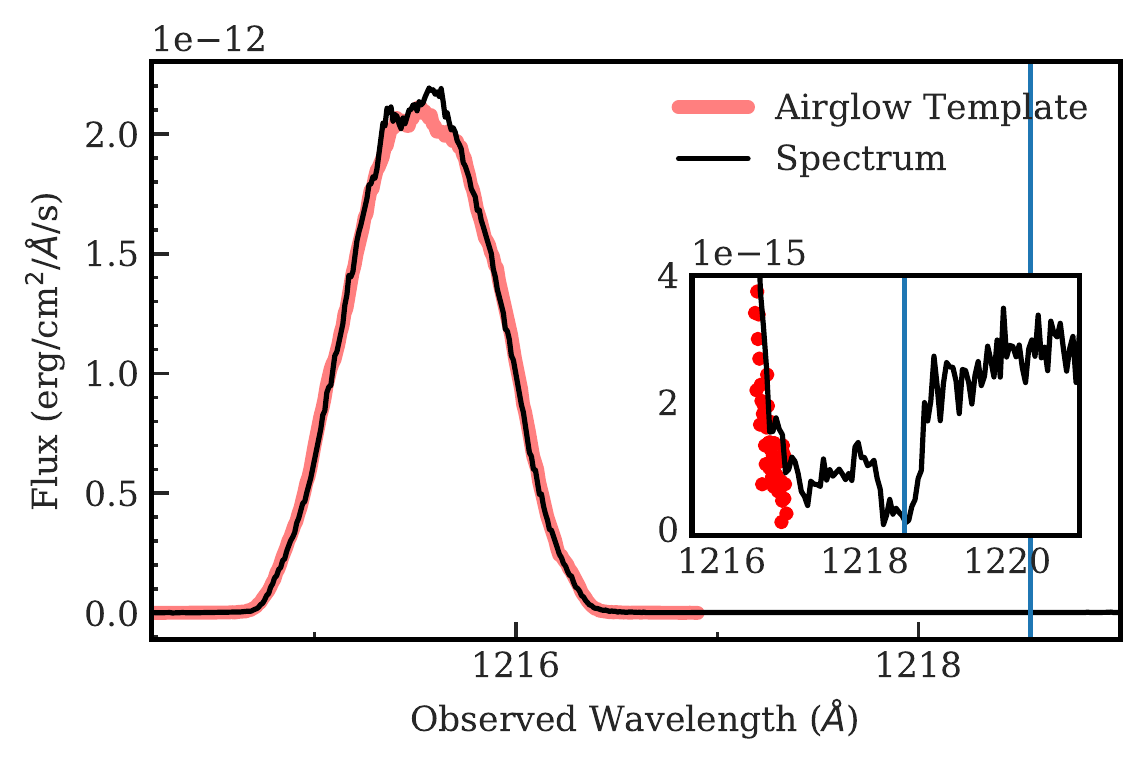}
\caption{Same as in Figure~\ref{fig:SBairglow}, but for SJ.}
\label{fig:SJairglow}
\end{center}
\end{figure*}

\subsection{Best VP fit to the Galactic {\lya} towards SJ}

\begin{figure*}
\begin{center}
\includegraphics[width=\linewidth]{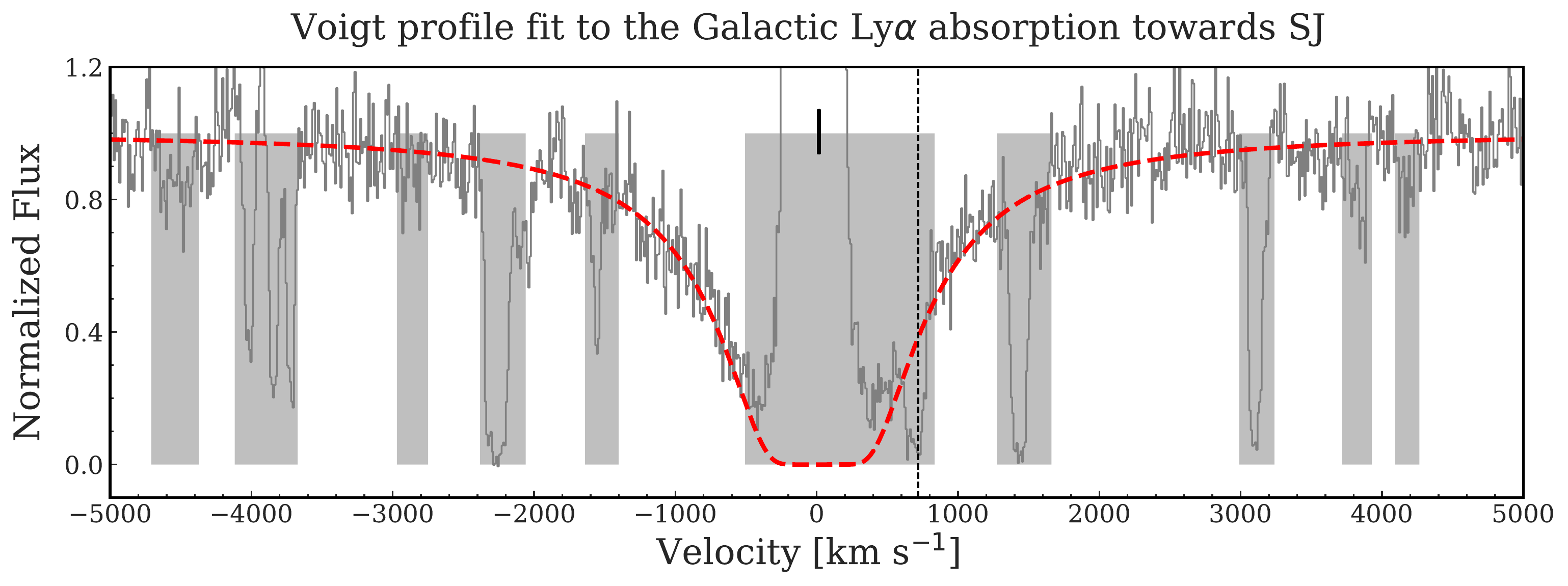}
\caption{Same as in Figure~\ref{fig:SBvpfitgal}, but for SJ.}
\label{fig:SJvpfitgal}
\end{center}
\end{figure*}

\subsection{Posterior distributions for the absorber properties towards SJ}

\label{appendix:SJparams}

\begin{figure*}
\begin{center}
\includegraphics[width=\linewidth]{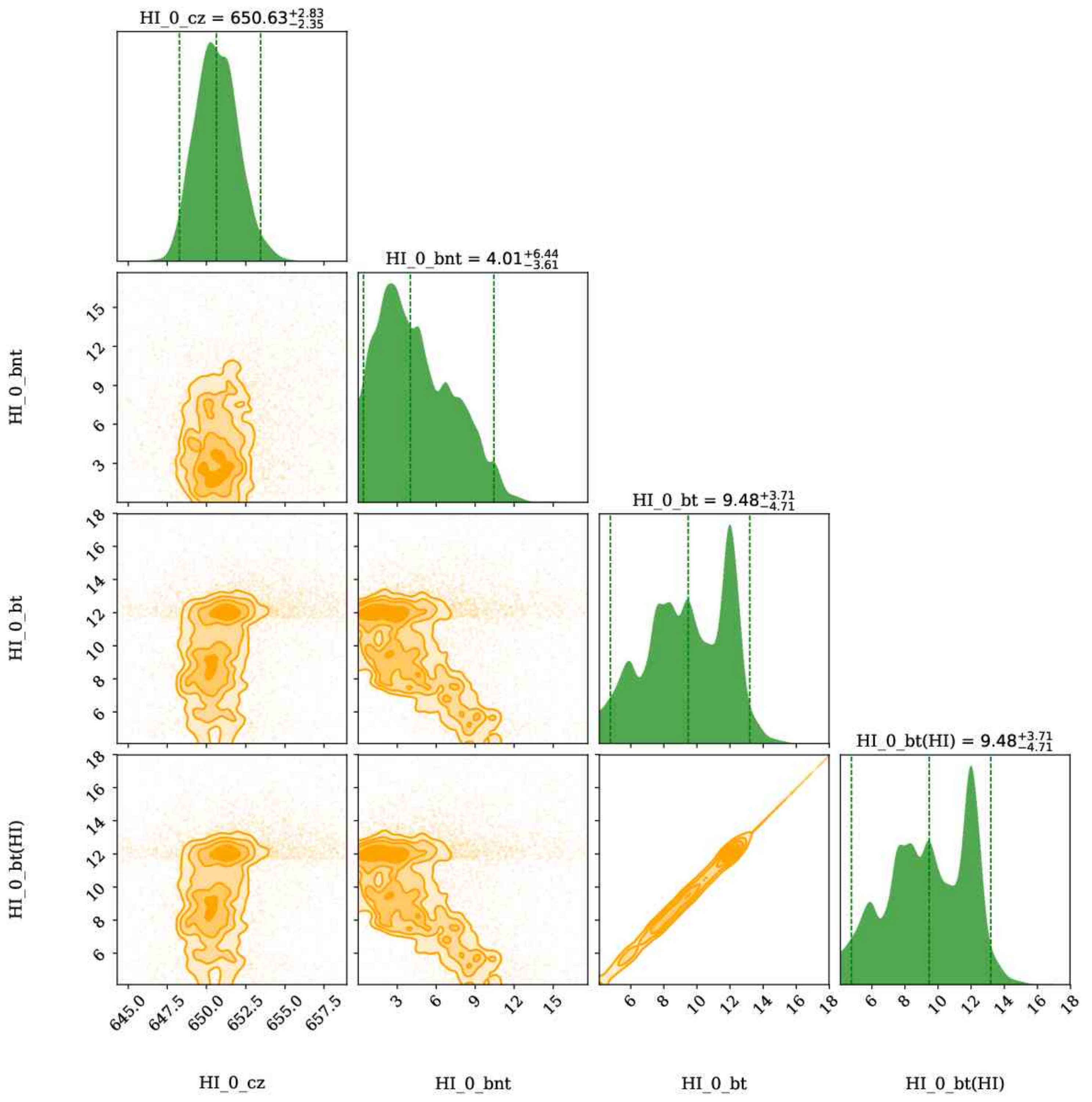}
\caption{The corner plot showing the marginalized posterior distributions for the absorption centroid ($z$), non-thermal Doppler broadening ($b_{nt}$), thermal Doppler broadening ($b_{t}$), total Doppler broadening ($b$), of the phase traced by the blueward cloud of the $z=0.0024$ absorber towards SJ. The over-plotted vertical lines in the posterior distribution span the 95\% credible interval. The contours indicate 0.5$\sigma$, 1$\sigma$, 1.5$\sigma$, and 2$\sigma$ levels. The model results are summarised in Table~\ref{tab:modelparams}, and the synthetic profiles based on these models are shown in Figure~\ref{fig:SJsysplot}}.
\label{fig:SJHI0b}
\end{center}
\end{figure*}

\begin{figure*}
\begin{center}
\includegraphics[width=\linewidth]{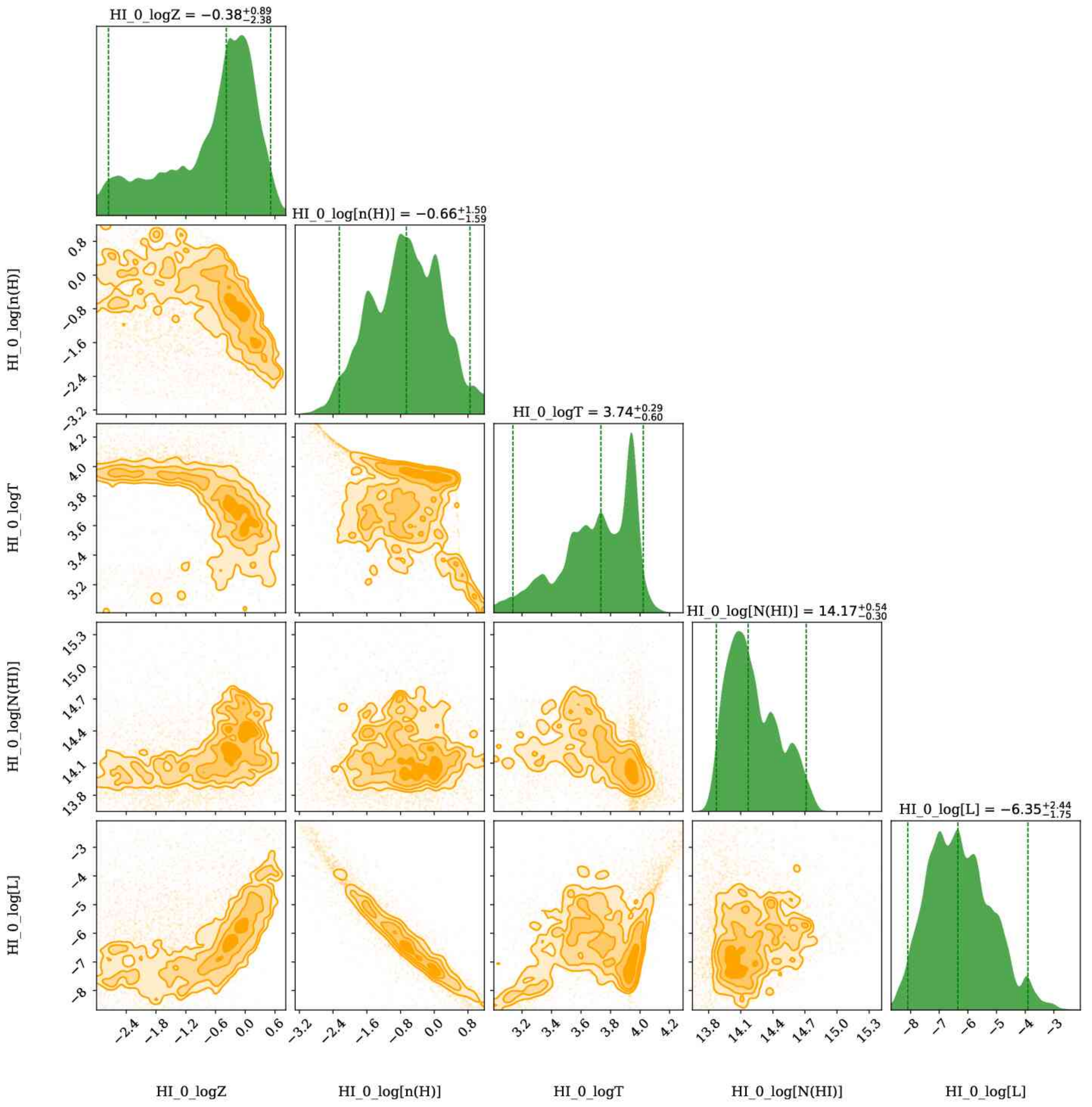}
\caption{The corner plot showing the marginalized posterior distributions for the metallicity ($\log Z$), hydrogen number density ($\log n_{H}$), temperature ($\log T$), neutral hydrogen column density ($\log N(\hi)$), and the line of sight thickness ($\log L$), of the phase traced by the blueward cloud of the $z=0.0024$ absorber towards SJ. The over-plotted vertical lines in the posterior distribution span the 95\% credible interval. The contours indicate 0.5$\sigma$, 1$\sigma$, 1.5$\sigma$, and 2$\sigma$ levels. The model results are summarised in Table~\ref{tab:modelparams}, and the synthetic profiles based on these models are shown in Figure~\ref{fig:SJsysplot}.}
\label{fig:SJHI0}
\end{center}
\end{figure*}

\begin{figure*}
\begin{center}
\includegraphics[width=\linewidth]{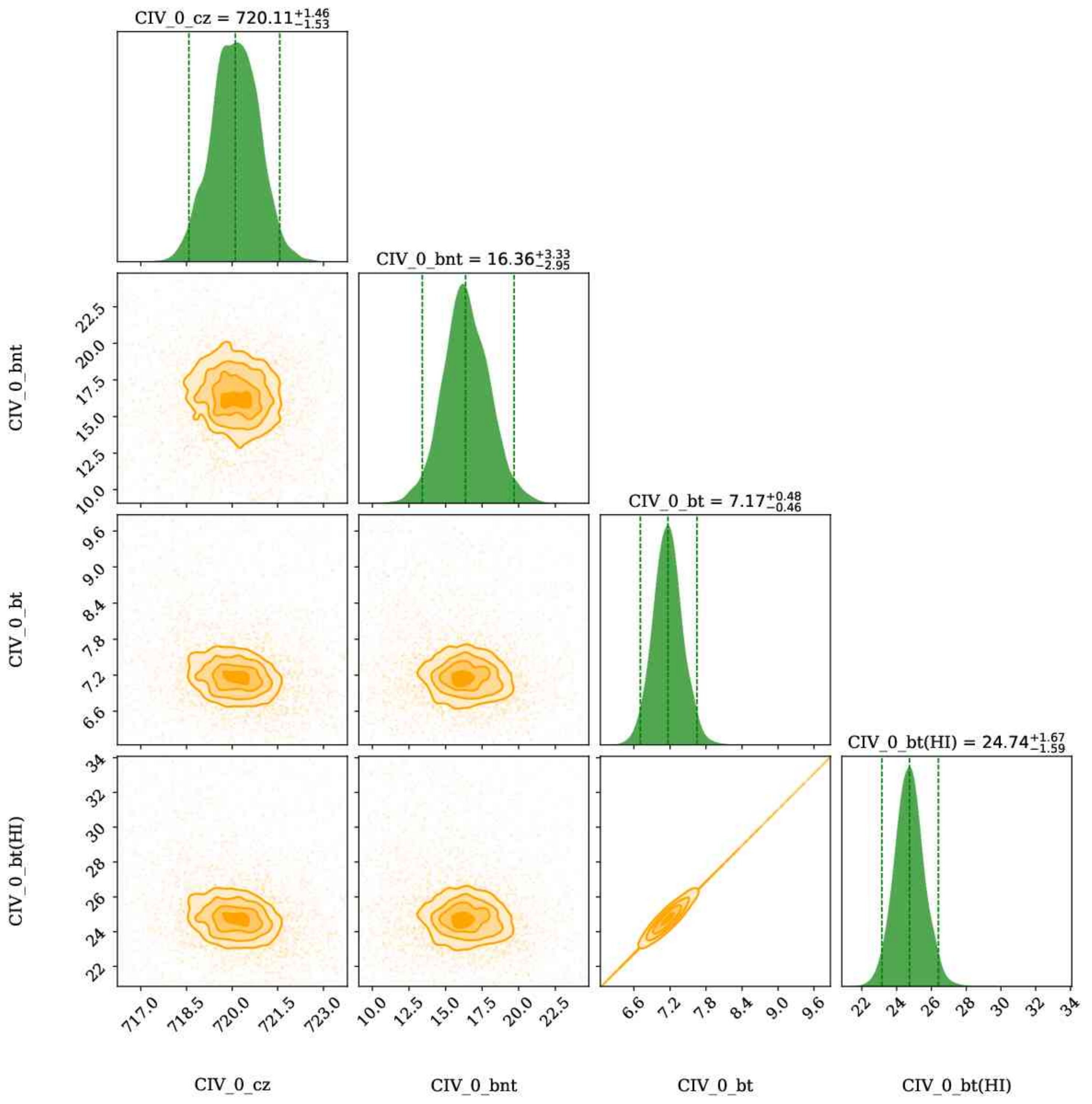}
\caption{The corner plot showing the marginalized posterior distributions for the absorption centroid ($z$), non-thermal Doppler broadening ($b_{nt}$), thermal Doppler broadening ($b_{t}$), total Doppler broadening ($b$), of the phase traced by the redward {\civ} cloud of the $z=0.0024$ absorber towards SJ. The over-plotted vertical lines in the posterior distribution span the 95\% credible interval. The contours indicate 0.5$\sigma$, 1$\sigma$, 1.5$\sigma$, and 2$\sigma$ levels. The model results are summarised in Table~\ref{tab:modelparams}, and the synthetic profiles based on these models are shown in Figure~\ref{fig:SJsysplot}.}
\label{fig:SJCIV0b}
\end{center}
\end{figure*}

\begin{figure*}
\begin{center}
\includegraphics[width=\linewidth]{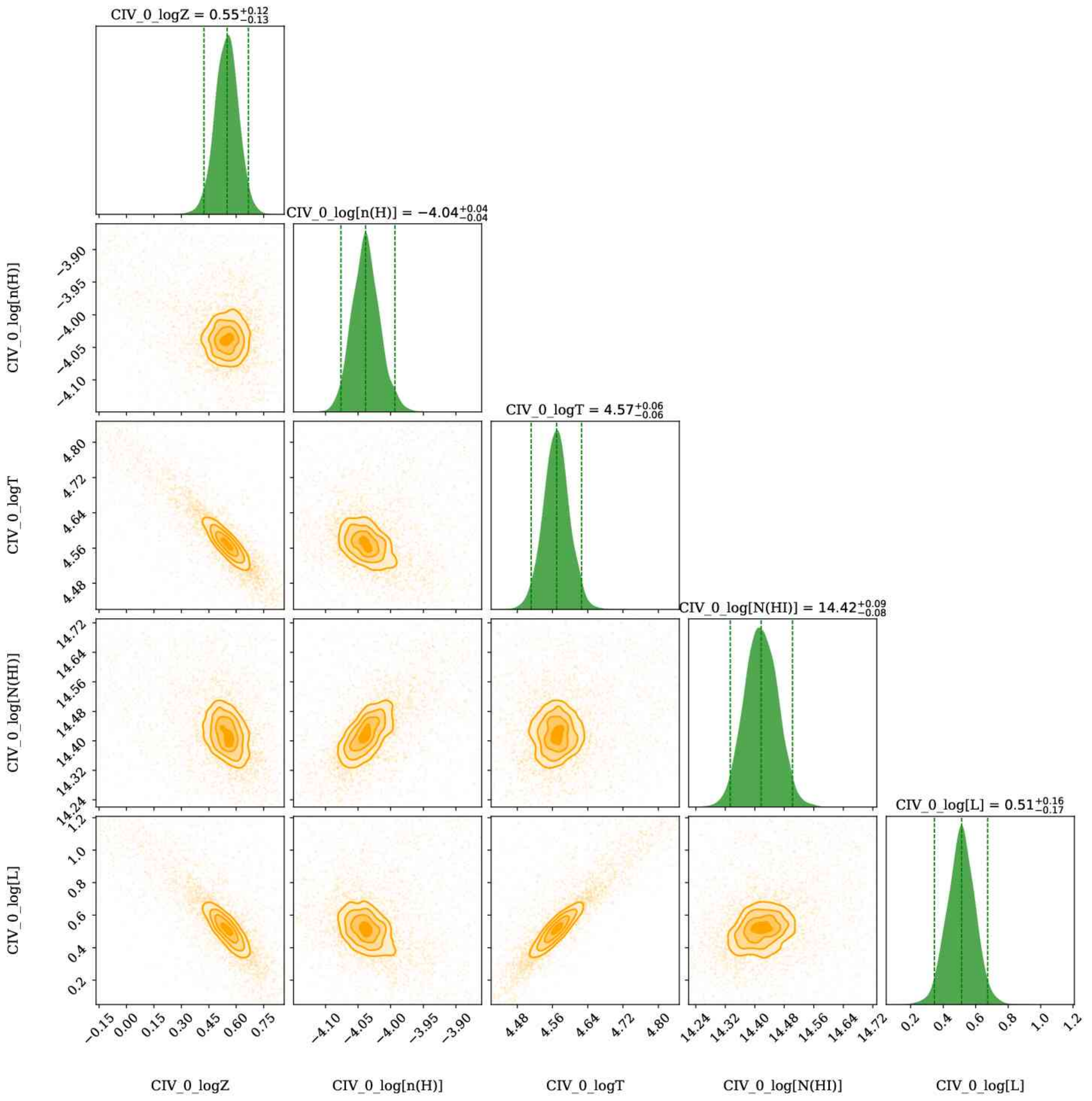}
\caption{The corner plot showing the marginalized posterior distributions for the metallicity ($\log Z$), hydrogen number density ($\log n_{H}$), temperature ($\log T$), neutral hydrogen column density ($\log N(\hi)$), and the line of sight thickness ($\log L$), of the phase traced by the redward {\civ} cloud of the $z=0.0024$ absorber towards SJ. The over-plotted vertical lines in the posterior distribution span the 95\% credible interval. The contours indicate 0.5$\sigma$, 1$\sigma$, 1.5$\sigma$, and 2$\sigma$ levels. The model results are summarised in Table~\ref{tab:modelparams}, and the synthetic profiles based on these models are shown in Figure~\ref{fig:SJsysplot}.}
\label{fig:SJCIV0}
\end{center}
\end{figure*}

\clearpage
\section{Plots for SK}
\label{appendix:SK}

\subsection{Airglow template fit towards SK}

\begin{figure*}
\begin{center}
\includegraphics[width=0.75\linewidth]{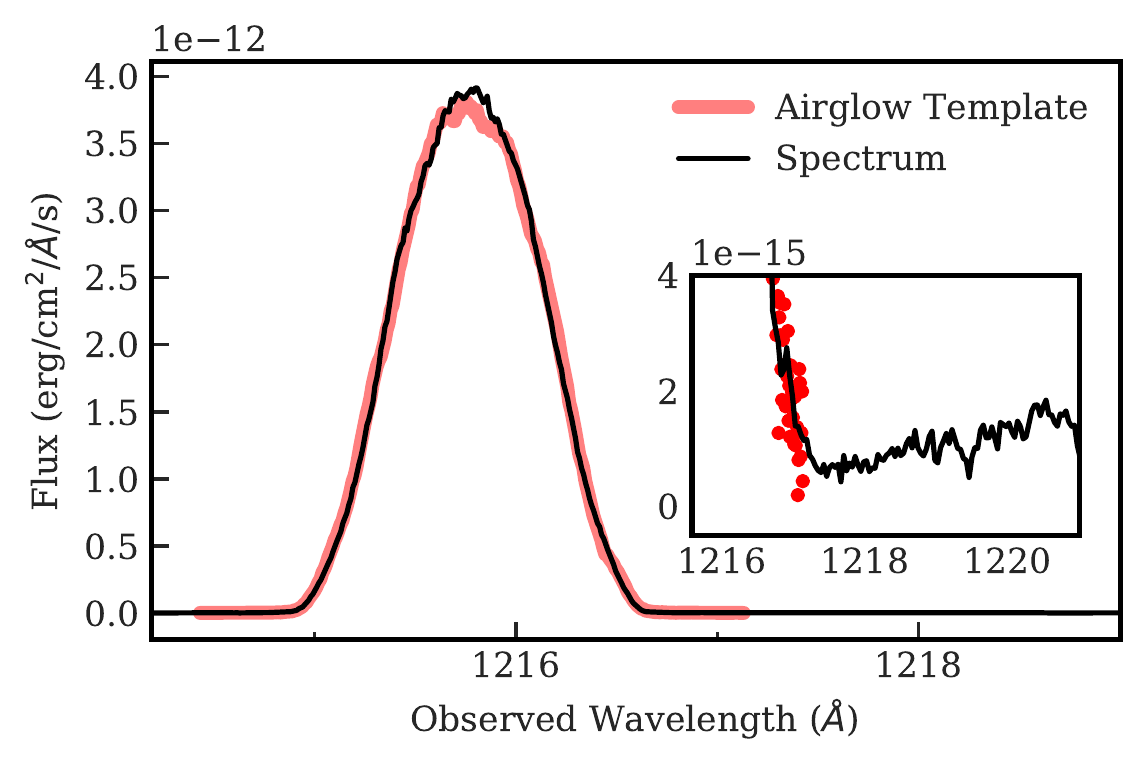}
\caption{Same as in Figure~\ref{fig:SBairglow}, but for SK.}
\label{fig:SKairglow}
\end{center}
\end{figure*}

\subsection{Best VP fit to the Galactic {\lya} towards SK}

\begin{figure*}
\begin{center}
\includegraphics[width=\linewidth]{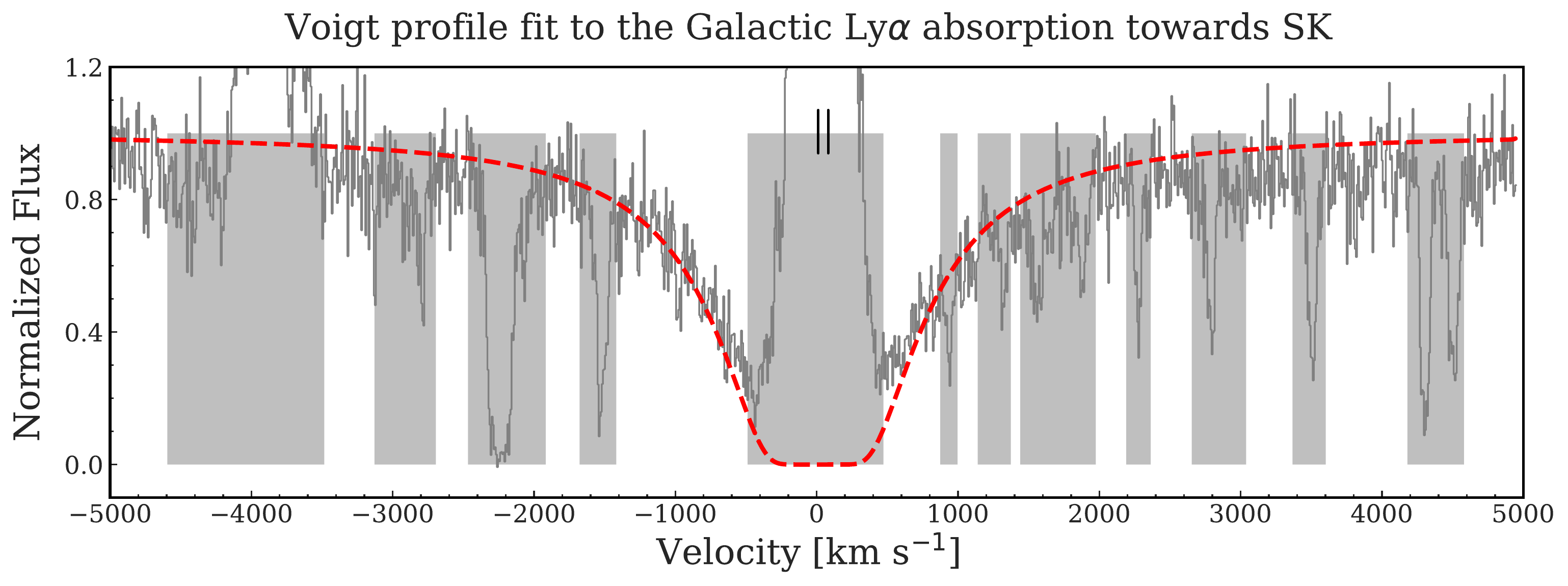}
\caption{Same as in Figure~\ref{fig:SBvpfitgal}, but for SK.}
\label{fig:SKvpfitgal}
\end{center}
\end{figure*}
\clearpage
\section{Plots for SL}
\label{appendix:SL}

\subsection{Airglow template fit towards SL}

\begin{figure*}
\begin{center}
\includegraphics[width=0.75\linewidth]{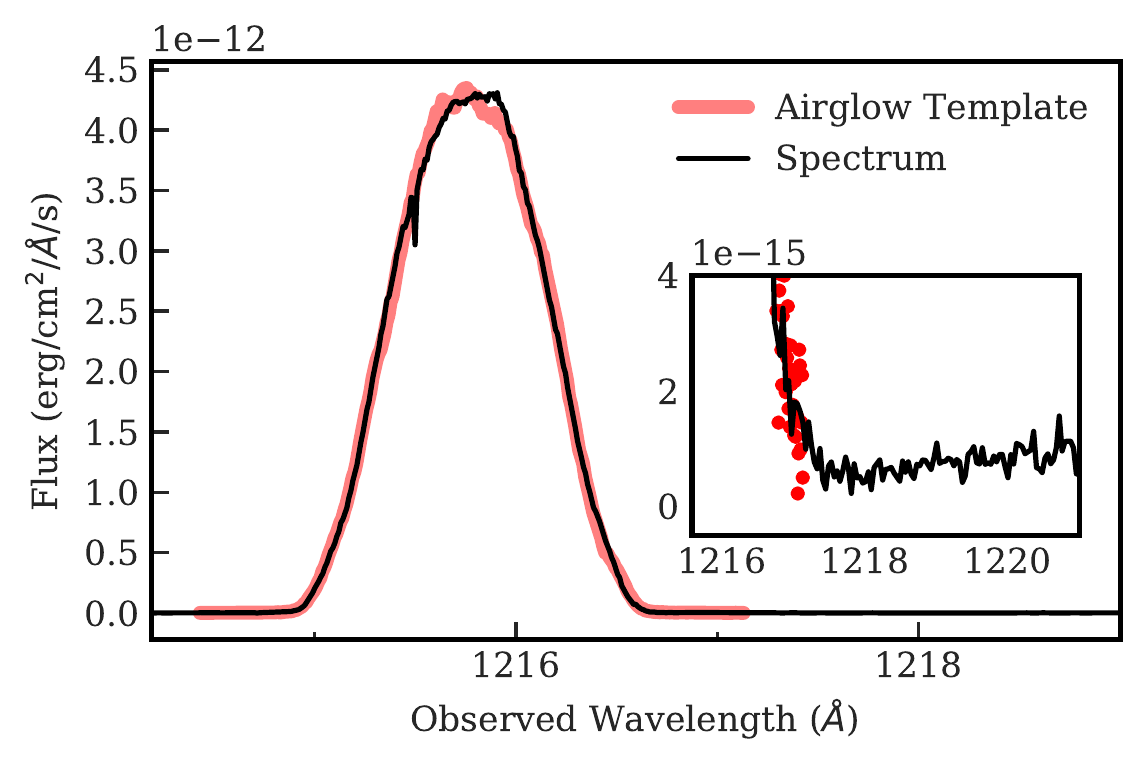}
\caption{Same as in Figure~\ref{fig:SBairglow}, but for SL.}
\label{fig:SLairglow}
\end{center}
\end{figure*}

\subsection{Best VP fit to the Galactic {\lya} towards SL}

\begin{figure*}
\begin{center}
\includegraphics[width=\linewidth]{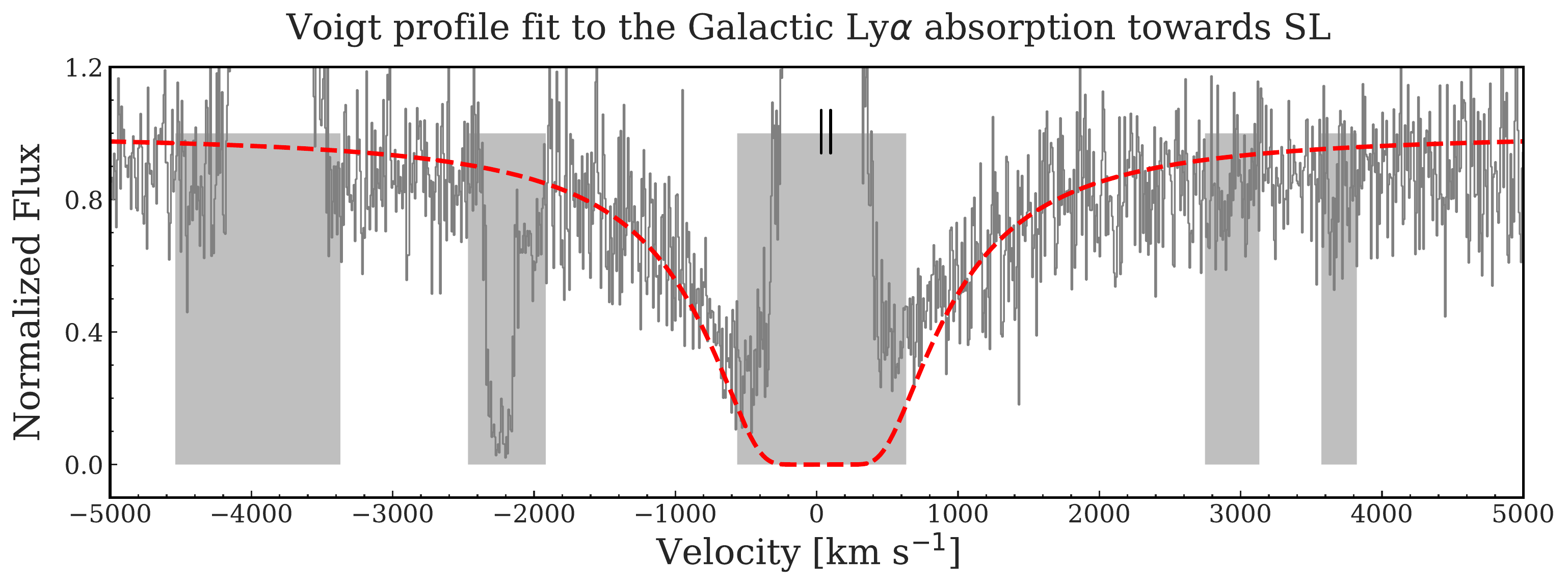}
\caption{Same as in Figure~\ref{fig:SBvpfitgal}, but for SL.}
\label{fig:SLvpfitgal}
\end{center}
\end{figure*}

\clearpage
\section{Column densities and total Doppler broadening parameter measurements}

In this section we present the column densities and total Doppler broadening parameter for various ions of interest across all model components, and along all the different sightlines.
\onecolumn 
\begin{table*}

\parbox{0.225\textwidth}{
\begin{tabular}{ccc}
\hline
 Ion & $b$  & $\log \frac{N}{\cmsq}$  \\ 
   & (\kms) &  \\
(1) & (2) & (3) \\\hline
SB & \textcolor{blue}{\hi_0} & ${722.1}_{-1.6}^{+1.9}$ \\
\hline
\AlII & ${4.6}_{-2.1}^{+11.5}$  & ${9.3}_{-2.5}^{+1.2}$  \\
\CI & ${5.4}_{-1.9}^{+10.8}$  & ${9.8}_{-2.1}^{+1.6}$  \\
\CII & ${5.4}_{-1.9}^{+10.8}$  & ${11.2}_{-2.5}^{+1.2}$  \\
\CIV & ${5.4}_{-1.9}^{+10.8}$  & ${6.4}_{-2.8}^{+2.8}$  \\
\FeII & ${4.2}_{-2.4}^{+11.8}$  & ${10.1}_{-2.2}^{+1.4}$  \\
\HI & ${14.4}_{-2.8}^{+4.9}$  & ${15.3}_{-0.8}^{+1.0}$  \\
\MgII & ${4.6}_{-2.1}^{+11.4}$  & ${10.2}_{-2.3}^{+1.3}$  \\
\NI & ${5.2}_{-1.9}^{+10.9}$  & ${9.8}_{-2.1}^{+1.6}$  \\
\NV & ${5.2}_{-1.9}^{+10.9}$  & ${4.2}_{-2.6}^{+2.7}$  \\
\OI & ${5.0}_{-1.9}^{+11.1}$  & ${10.9}_{-2.1}^{+1.5}$  \\
\OVI & ${5.0}_{-1.9}^{+11.1}$  & ${2.5}_{-1.1}^{+2.4}$  \\
\PII & ${4.5}_{-2.2}^{+11.5}$  & ${8.4}_{-2.5}^{+1.2}$  \\
\SI & ${4.5}_{-2.2}^{+11.6}$  & ${8.0}_{-2.1}^{+1.7}$  \\
\SII & ${4.5}_{-2.2}^{+11.6}$  & ${10.1}_{-2.4}^{+1.2}$  \\
\SiI & ${4.6}_{-2.1}^{+11.5}$  & ${7.8}_{-2.2}^{+1.5}$  \\
\SiII & ${4.6}_{-2.1}^{+11.5}$  & ${10.4}_{-2.4}^{+1.2}$  \\
\SiIII & ${4.6}_{-2.1}^{+11.5}$  & ${8.9}_{-3.3}^{+1.9}$  \\
\SiIV & ${4.6}_{-2.1}^{+11.5}$  & ${6.6}_{-3.4}^{+2.7}$  \\
\hline
\end{tabular}
}
\parbox{.225\textwidth}{
\begin{tabular}{ccc}
\hline
 Ion & $b$  & $\log \frac{N}{\cmsq}$  \\ 
   & (\kms) &  \\
(1) & (2) & (3) \\\hline
SC$^{a}$ & \textcolor{blue}{\nv_0} & ${710.0}_{-3.5}^{+2.9}$ \\
\hline

\AlII & ${50.7}_{-4.4}^{+4.5}$  & ${10.5}_{-0.1}^{+0.1}$  \\

\CI & ${51.0}_{-4.4}^{+4.4}$  & ${10.8}_{-0.3}^{+0.2}$  \\

\CII & ${51.0}_{-4.4}^{+4.4}$  & ${14.0}_{-0.2}^{+0.1}$  \\

\CIV & ${51.0}_{-4.4}^{+4.4}$  & ${14.7}_{-0.1}^{+0.1}$  \\

\FeII & ${50.6}_{-4.4}^{+4.5}$  & ${9.2}_{-0.1}^{+0.1}$  \\

\HI & ${55.9}_{-4.0}^{+4.0}$  & ${14.6}_{-0.1}^{+0.1}$  \\

\MgII & ${50.8}_{-4.4}^{+4.5}$  & ${10.2}_{-0.2}^{+0.2}$  \\

\NI & ${51.0}_{-4.4}^{+4.4}$  & ${10.0}_{-0.3}^{+0.2}$  \\

\NV & ${51.0}_{-4.4}^{+4.4}$  & ${13.8}_{-0.1}^{+0.1}$  \\

\OI & ${50.9}_{-4.4}^{+4.4}$  & ${9.9}_{-0.2}^{+0.2}$  \\

\OVI & ${50.9}_{-4.4}^{+4.4}$  & ${14.0}_{-0.1}^{+0.1}$  \\

\PII & ${50.7}_{-4.4}^{+4.5}$  & ${11.1}_{-0.1}^{+0.1}$  \\

\SI & ${50.7}_{-4.4}^{+4.5}$  & ${7.7}_{-0.2}^{+0.1}$  \\

\SII & ${50.7}_{-4.4}^{+4.5}$  & ${12.4}_{-0.2}^{+0.2}$  \\

\SiI & ${50.7}_{-4.4}^{+4.5}$  & ${8.1}_{-0.1}^{+0.1}$  \\

\SiII & ${50.7}_{-4.4}^{+4.5}$  & ${12.0}_{-0.1}^{+0.1}$  \\

\SiIII & ${50.7}_{-4.4}^{+4.5}$  & ${13.4}_{-0.1}^{+0.1}$  \\

\SiIV & ${50.7}_{-4.4}^{+4.5}$  & ${13.1}_{-0.1}^{+0.1}$  \\

\hline
\end{tabular}
}
\parbox{.225\textwidth}{
\begin{tabular}{ccc}
\hline
 Ion & $b$  & $\log \frac{N}{\cmsq}$  \\ 
   & (\kms) &  \\
(1) & (2) & (3) \\\hline
SC$^{a}$ & \textcolor{red}{\siiv_1} & ${716.4}_{-0.3}^{+0.2}$\\
\hline

\AlII & ${15.0}_{-0.6}^{+0.8}$  & ${13.6}_{-0.0}^{+0.0}$  \\

\CI & ${15.3}_{-0.6}^{+0.8}$  & ${13.0}_{-0.1}^{+0.1}$  \\

\CII & ${15.3}_{-0.6}^{+0.8}$  & ${15.5}_{-0.0}^{+0.0}$  \\

\CIV & ${15.3}_{-0.6}^{+0.8}$  & ${13.2}_{-0.1}^{+0.1}$  \\

\FeII & ${14.8}_{-0.6}^{+0.8}$  & ${13.9}_{-0.1}^{+0.1}$  \\

\HI & ${21.2}_{-0.6}^{+0.8}$  & ${18.0}_{-0.1}^{+0.1}$  \\

\MgII & ${15.0}_{-0.6}^{+0.8}$  & ${14.2}_{-0.1}^{+0.0}$  \\

\NI & ${15.2}_{-0.6}^{+0.8}$  & ${13.5}_{-0.1}^{+0.1}$  \\

\NV & ${15.2}_{-0.6}^{+0.8}$  & ${11.0}_{-0.1}^{+0.1}$  \\

\OI & ${15.2}_{-0.6}^{+0.8}$  & ${14.6}_{-0.1}^{+0.1}$  \\

\OVI & ${15.2}_{-0.6}^{+0.8}$  & ${8.3}_{-0.1}^{+0.6}$  \\

\PII & ${14.9}_{-0.6}^{+0.8}$  & ${12.7}_{-0.0}^{+0.0}$  \\

\SI & ${14.9}_{-0.6}^{+0.8}$  & ${11.1}_{-0.1}^{+0.1}$  \\

\SII & ${14.9}_{-0.6}^{+0.8}$  & ${14.2}_{-0.1}^{+0.0}$  \\

\SiI & ${14.9}_{-0.6}^{+0.8}$  & ${11.1}_{-0.1}^{+0.1}$  \\

\SiII & ${14.9}_{-0.6}^{+0.8}$  & ${14.5}_{-0.1}^{+0.1}$  \\

\SiIII & ${14.9}_{-0.6}^{+0.8}$  & ${14.7}_{-0.0}^{+0.0}$  \\

\SiIV & ${14.9}_{-0.6}^{+0.8}$  & ${13.4}_{-0.1}^{+0.1}$  \\

\hline
\end{tabular}
}
\parbox{.225\textwidth}{
\begin{tabular}{ccc}
\hline
 Ion & $b$  & $\log \frac{N}{\cmsq}$  \\ 
   & (\kms) &  \\
(1) & (2) & (3) \\\hline
SC$^{a}$ &\textcolor{green}{\cii_1} & ${878.6}_{-1.0}^{+0.8}$ \\
\hline

\AlII & ${4.5}_{-1.0}^{+1.7}$  & ${12.1}_{-0.2}^{+0.2}$  \\

\CI & ${5.4}_{-0.8}^{+1.5}$  & ${12.3}_{-0.4}^{+0.3}$  \\

\CII & ${5.4}_{-0.8}^{+1.5}$  & ${14.0}_{-0.2}^{+0.2}$  \\

\CIV & ${5.4}_{-0.8}^{+1.5}$  & ${9.5}_{-0.5}^{+0.5}$  \\

\FeII & ${4.1}_{-1.1}^{+1.8}$  & ${13.0}_{-0.3}^{+0.2}$  \\

\HI & ${14.1}_{-0.6}^{+1.0}$  & ${18.4}_{-0.0}^{+0.0}$  \\

\MgII & ${4.6}_{-1.0}^{+1.6}$  & ${13.0}_{-0.3}^{+0.2}$  \\

\NI & ${5.2}_{-0.9}^{+1.5}$  & ${12.9}_{-0.4}^{+0.3}$  \\

\NV & ${5.2}_{-0.9}^{+1.5}$  & ${5.2}_{-0.3}^{+1.1}$  \\

\OI & ${5.0}_{-0.9}^{+1.6}$  & ${13.9}_{-0.3}^{+0.3}$  \\

\OVI & ${5.0}_{-0.9}^{+1.6}$  & ${3.8}_{-0.1}^{+0.6}$  \\

\PII & ${4.4}_{-1.0}^{+1.7}$  & ${11.2}_{-0.2}^{+0.2}$  \\

\SI & ${4.4}_{-1.0}^{+1.7}$  & ${10.5}_{-0.4}^{+0.3}$  \\

\SII & ${4.4}_{-1.0}^{+1.7}$  & ${12.9}_{-0.2}^{+0.2}$  \\

\SiI & ${4.5}_{-1.0}^{+1.7}$  & ${10.3}_{-0.3}^{+0.3}$  \\

\SiII & ${4.5}_{-1.0}^{+1.7}$  & ${13.2}_{-0.2}^{+0.2}$  \\

\SiIII & ${4.5}_{-1.0}^{+1.7}$  & ${12.2}_{-0.3}^{+0.3}$  \\

\SiIV & ${4.5}_{-1.0}^{+1.7}$  & ${10.3}_{-0.4}^{+0.4}$  \\

\hline
\end{tabular}
}
\parbox{.225\textwidth}{
\begin{tabular}{ccc}
\hline
 Ion & $b$  & $\log \frac{N}{\cmsq}$  \\ 
   & (\kms) &  \\
(1) & (2) & (3) \\\hline
SC$^{a}$ & \textcolor{magenta}{\hi_0} & ${1035.8}_{-1.8}^{+1.8}$ \\
\hline

\AlII & ${29.1}_{-5.3}^{+2.9}$  & ${7.5}_{-1.3}^{+1.6}$  \\

\CI & ${29.3}_{-5.3}^{+2.9}$  & ${8.3}_{-1.1}^{+1.2}$  \\

\CII & ${29.3}_{-5.3}^{+2.9}$  & ${9.3}_{-1.4}^{+1.6}$  \\

\CIV & ${29.3}_{-5.3}^{+2.9}$  & ${4.8}_{-1.8}^{+3.6}$  \\

\FeII & ${29.1}_{-5.4}^{+2.9}$  & ${8.4}_{-1.2}^{+1.3}$  \\

\HI & ${31.7}_{-4.3}^{+2.5}$  & ${14.1}_{-0.1}^{+0.1}$  \\

\MgII & ${29.1}_{-5.3}^{+2.9}$  & ${8.5}_{-1.3}^{+1.4}$  \\

\NI & ${29.2}_{-5.3}^{+2.9}$  & ${8.3}_{-1.1}^{+1.2}$  \\

\NV & ${29.2}_{-5.3}^{+2.9}$  & ${3.8}_{-2.5}^{+3.0}$  \\

\OI & ${29.2}_{-5.3}^{+2.9}$  & ${9.3}_{-1.1}^{+1.2}$  \\

\OVI & ${29.2}_{-5.3}^{+2.9}$  & ${2.1}_{-1.9}^{+3.7}$  \\

\PII & ${29.1}_{-5.3}^{+2.9}$  & ${6.6}_{-1.3}^{+1.6}$  \\

\SI & ${29.1}_{-5.4}^{+2.9}$  & ${6.6}_{-1.2}^{+1.3}$  \\

\SII & ${29.1}_{-5.4}^{+2.9}$  & ${8.3}_{-1.3}^{+1.5}$  \\

\SiI & ${29.1}_{-5.3}^{+2.9}$  & ${6.2}_{-1.2}^{+1.3}$  \\

\SiII & ${29.1}_{-5.3}^{+2.9}$  & ${8.6}_{-1.3}^{+1.5}$  \\

\SiIII & ${29.1}_{-5.3}^{+2.9}$  & ${6.8}_{-2.7}^{+3.1}$  \\

\SiIV & ${29.1}_{-5.3}^{+2.9}$  & ${4.4}_{-2.5}^{+4.1}$  \\

\hline
\end{tabular}
}
\parbox{.225\textwidth}{
\begin{tabular}{ccc}
\hline
 Ion & $b$  & $\log \frac{N}{\cmsq}$  \\ 
   & (\kms) &  \\
(1) & (2) & (3) \\\hline
SC$^{b}$ & \textcolor{blue}{\nv_0} & ${706.8}_{-9.1}^{+12.3}$ \\
\hline

\AlII & ${44.3}_{-15.5}^{+15.6}$  & ${6.2}_{-1.1}^{+0.5}$  \\

\CI & ${46.9}_{-14.0}^{+15.0}$  & ${1.5}_{-2.8}^{+2.2}$  \\

\CII & ${46.9}_{-14.0}^{+15.0}$  & ${8.1}_{-1.0}^{+0.7}$  \\

\CIV & ${46.9}_{-14.0}^{+15.0}$  & ${13.2}_{-0.1}^{+0.2}$  \\

\FeII & ${43.1}_{-16.0}^{+15.9}$  & ${4.7}_{-1.6}^{+1.2}$  \\

\HI & ${85.3}_{-11.4}^{+12.8}$  & ${13.1}_{-0.6}^{+0.5}$  \\

\MgII & ${44.5}_{-15.3}^{+15.5}$  & ${7.6}_{-1.2}^{+0.7}$  \\

\NI & ${46.3}_{-14.3}^{+15.1}$  & ${3.3}_{-1.9}^{+1.7}$  \\

\NV & ${46.3}_{-14.3}^{+15.1}$  & ${13.8}_{-0.1}^{+0.1}$  \\

\OI & ${45.8}_{-14.6}^{+15.2}$  & ${4.9}_{-2.3}^{+1.7}$  \\

\OVI & ${45.8}_{-14.6}^{+15.2}$  & ${15.5}_{-0.7}^{+0.3}$  \\

\PII & ${44.0}_{-15.6}^{+15.6}$  & ${3.9}_{-0.7}^{+0.7}$  \\

\SI & ${43.9}_{-15.6}^{+15.6}$  & ${-2.5}_{-3.1}^{+2.5}$  \\

\SII & ${43.9}_{-15.6}^{+15.6}$  & ${5.5}_{-1.3}^{+1.2}$  \\

\SiI & ${44.2}_{-15.5}^{+15.6}$  & ${-0.2}_{-2.7}^{+2.2}$  \\

\SiII & ${44.2}_{-15.5}^{+15.6}$  & ${6.7}_{-0.7}^{+0.4}$  \\

\SiIII & ${44.2}_{-15.5}^{+15.6}$  & ${10.0}_{-0.5}^{+0.2}$  \\

\SiIV & ${44.2}_{-15.5}^{+15.6}$  & ${12.0}_{-0.3}^{+0.2}$  \\

\hline
\end{tabular}
}
\parbox{.225\textwidth}{
\begin{tabular}{ccc}
\hline
 Ion & $b$  & $\log \frac{N}{\cmsq}$  \\ 
   & (\kms) &  \\
(1) & (2) & (3) \\\hline
SC$^{b}$ & \textcolor{red}{\siiv_1} & ${716.2}_{-0.4}^{+0.3}$ \\
\hline

\AlII & ${16.7}_{-0.1}^{+0.1}$  & ${13.6}_{-0.0}^{+0.0}$  \\

\CI & ${17.2}_{-0.1}^{+0.1}$  & ${12.9}_{-0.0}^{+0.0}$  \\

\CII & ${17.2}_{-0.1}^{+0.1}$  & ${15.4}_{-0.0}^{+0.0}$  \\

\CIV & ${17.2}_{-0.1}^{+0.1}$  & ${13.6}_{-0.1}^{+0.1}$  \\

\FeII & ${16.6}_{-0.1}^{+0.1}$  & ${13.7}_{-0.1}^{+0.1}$  \\

\HI & ${23.8}_{-0.3}^{+0.3}$  & ${18.2}_{-0.1}^{+0.1}$  \\

\MgII & ${16.8}_{-0.1}^{+0.1}$  & ${14.1}_{-0.0}^{+0.1}$  \\

\NI & ${17.1}_{-0.1}^{+0.1}$  & ${13.4}_{-0.1}^{+0.1}$  \\

\NV & ${17.1}_{-0.1}^{+0.1}$  & ${11.6}_{-0.1}^{+0.1}$  \\

\OI & ${17.0}_{-0.1}^{+0.1}$  & ${14.5}_{-0.1}^{+0.1}$  \\

\OVI & ${17.0}_{-0.1}^{+0.1}$  & ${10.3}_{-0.2}^{+0.1}$  \\

\PII & ${16.7}_{-0.1}^{+0.1}$  & ${12.8}_{-0.0}^{+0.0}$  \\

\SI & ${16.7}_{-0.1}^{+0.1}$  & ${10.8}_{-0.1}^{+0.1}$  \\

\SII & ${16.7}_{-0.1}^{+0.1}$  & ${14.2}_{-0.0}^{+0.0}$  \\

\SiI & ${16.7}_{-0.1}^{+0.1}$  & ${10.9}_{-0.0}^{+0.1}$  \\

\SiII & ${16.7}_{-0.1}^{+0.1}$  & ${14.4}_{-0.1}^{+0.1}$  \\

\SiIII & ${16.7}_{-0.1}^{+0.1}$  & ${14.8}_{-0.0}^{+0.0}$  \\

\SiIV & ${16.7}_{-0.1}^{+0.1}$  & ${13.7}_{-0.0}^{+0.0}$  \\

\hline
\end{tabular}
}
\parbox{.225\textwidth}{
\begin{tabular}{ccc}
\hline
 Ion & $b$  & $\log \frac{N}{\cmsq}$  \\ 
   & (\kms) &  \\
(1) & (2) & (3) \\\hline
SC$^{b}$ & \textcolor{green}{\cii_1} & ${878.6}_{-1.0}^{+0.8}$ \\
\hline

\AlII & ${4.3}_{-0.8}^{+1.3}$  & ${12.1}_{-0.2}^{+0.2}$  \\

\CI & ${5.3}_{-0.7}^{+1.1}$  & ${12.2}_{-0.3}^{+0.2}$  \\

\CII & ${5.3}_{-0.7}^{+1.1}$  & ${14.1}_{-0.2}^{+0.2}$  \\

\CIV & ${5.3}_{-0.7}^{+1.1}$  & ${10.0}_{-0.5}^{+0.5}$  \\

\FeII & ${3.9}_{-1.0}^{+1.4}$  & ${13.0}_{-0.3}^{+0.2}$  \\

\HI & ${14.4}_{-0.5}^{+0.6}$  & ${18.3}_{-0.0}^{+0.0}$  \\

\MgII & ${4.4}_{-0.8}^{+1.3}$  & ${13.0}_{-0.3}^{+0.2}$  \\

\NI & ${5.0}_{-0.7}^{+1.1}$  & ${12.8}_{-0.3}^{+0.2}$  \\

\NV & ${5.0}_{-0.7}^{+1.1}$  & ${5.7}_{-0.6}^{+1.7}$  \\

\OI & ${4.8}_{-0.7}^{+1.2}$  & ${13.8}_{-0.3}^{+0.2}$  \\

\OVI & ${4.8}_{-0.7}^{+1.2}$  & ${4.1}_{-0.3}^{+0.4}$  \\

\PII & ${4.2}_{-0.9}^{+1.3}$  & ${11.2}_{-0.2}^{+0.2}$  \\

\SI & ${4.2}_{-0.9}^{+1.3}$  & ${10.4}_{-0.3}^{+0.3}$  \\

\SII & ${4.2}_{-0.9}^{+1.3}$  & ${12.9}_{-0.2}^{+0.2}$  \\

\SiI & ${4.3}_{-0.9}^{+1.3}$  & ${10.2}_{-0.3}^{+0.2}$  \\

\SiII & ${4.3}_{-0.9}^{+1.3}$  & ${13.2}_{-0.2}^{+0.2}$  \\

\SiIII & ${4.3}_{-0.9}^{+1.3}$  & ${12.5}_{-0.3}^{+0.3}$  \\

\SiIV & ${4.3}_{-0.9}^{+1.3}$  & ${10.7}_{-0.4}^{+0.4}$  \\

\hline
\end{tabular}
}

\caption{Basic properties of the different ions in different gas phases contributing to the absorption towards different sightlines summarized in Table~\ref{tab:modelparams}. The first line in each table denotes the sightline, the constraining ion cloud, and the heliocentric velocity of the corresponding gas phase. The basic properties of the different ions arising in the gas phase follow. Notes: (1) Ion name; (2) Total Doppler broadening parameter of the ion; (3) log column density of the ion.}
\label{tab:basicinfo}
\end{table*}

\begin{table*}
\parbox{.225\textwidth}{
\begin{tabular}{ccc}
\hline
 Ion & $b$  & $\log \frac{N}{\cmsq}$  \\ 
  & (\kms) &  \\
(1) & (2) & (3) \\\hline
SC$^{b}$ & \textcolor{magenta}{\hi_0} & ${1035.8}_{-1.8}^{+1.6}$ \\
\hline

\AlII & ${28.1}_{-3.9}^{+3.1}$  & ${8.8}_{-1.3}^{+1.1}$  \\

\CI & ${28.3}_{-3.9}^{+3.1}$  & ${8.9}_{-1.0}^{+0.9}$  \\

\CII & ${28.3}_{-3.9}^{+3.1}$  & ${10.7}_{-1.4}^{+1.2}$  \\

\CIV & ${28.3}_{-3.9}^{+3.1}$  & ${7.1}_{-3.0}^{+3.0}$  \\

\FeII & ${28.0}_{-4.0}^{+3.1}$  & ${9.3}_{-1.1}^{+1.2}$  \\

\HI & ${31.4}_{-3.4}^{+2.8}$  & ${14.1}_{-0.1}^{+0.1}$  \\

\MgII & ${28.1}_{-3.9}^{+3.1}$  & ${9.5}_{-1.2}^{+1.0}$  \\

\NI & ${28.2}_{-3.9}^{+3.1}$  & ${8.9}_{-1.0}^{+1.0}$  \\

\NV & ${28.2}_{-3.9}^{+3.1}$  & ${5.2}_{-3.3}^{+3.0}$  \\

\OI & ${28.2}_{-3.9}^{+3.1}$  & ${10.0}_{-1.0}^{+1.0}$  \\

\OVI & ${28.2}_{-3.9}^{+3.1}$  & ${2.8}_{-2.2}^{+4.4}$  \\

\PII & ${28.1}_{-4.0}^{+3.1}$  & ${7.9}_{-1.4}^{+1.2}$  \\

\SI & ${28.1}_{-4.0}^{+3.1}$  & ${7.1}_{-1.1}^{+1.0}$  \\

\SII & ${28.1}_{-4.0}^{+3.1}$  & ${9.5}_{-1.3}^{+1.1}$  \\

\SiI & ${28.1}_{-3.9}^{+3.1}$  & ${6.9}_{-1.0}^{+1.0}$  \\

\SiII & ${28.1}_{-3.9}^{+3.1}$  & ${9.8}_{-1.3}^{+1.1}$  \\

\SiIII & ${28.1}_{-3.9}^{+3.1}$  & ${9.3}_{-2.5}^{+1.9}$  \\

\SiIV & ${28.1}_{-3.9}^{+3.1}$  & ${7.5}_{-3.3}^{+2.5}$  \\

\hline
\end{tabular}
}
\parbox{.225\textwidth}{
\begin{tabular}{ccc}
\hline
 Ion & $b$  & $\log \frac{N}{\cmsq}$  \\ 
  & (\kms) &  \\
(1) & (2) & (3) \\\hline
SD & \textcolor{blue}{\siiv_0} & ${844.4}_{-1.7}^{+1.6}$ \\
\hline

\AlII & ${19.1}_{-2.7}^{+3.2}$  & ${13.1}_{-0.3}^{+0.2}$  \\

\CI & ${19.5}_{-2.6}^{+3.2}$  & ${12.3}_{-0.4}^{+0.3}$  \\

\CII & ${19.5}_{-2.6}^{+3.2}$  & ${14.9}_{-0.2}^{+0.2}$  \\

\CIV & ${19.5}_{-2.6}^{+3.2}$  & ${13.6}_{-0.2}^{+0.2}$  \\

\FeII & ${19.0}_{-2.7}^{+3.3}$  & ${12.7}_{-0.2}^{+0.3}$  \\

\HI & ${25.5}_{-2.2}^{+2.0}$  & ${16.9}_{-0.6}^{+0.5}$  \\

\MgII & ${19.2}_{-2.7}^{+3.2}$  & ${13.4}_{-0.2}^{+0.2}$  \\

\NI & ${19.4}_{-2.6}^{+3.2}$  & ${12.2}_{-0.4}^{+0.4}$  \\

\NV & ${19.4}_{-2.6}^{+3.2}$  & ${11.6}_{-0.3}^{+0.2}$  \\

\OI & ${19.3}_{-2.6}^{+3.2}$  & ${13.2}_{-0.3}^{+0.3}$  \\

\OVI & ${19.3}_{-2.6}^{+3.2}$  & ${10.5}_{-0.6}^{+0.4}$  \\

\PII & ${19.1}_{-2.7}^{+3.3}$  & ${12.3}_{-0.3}^{+0.2}$  \\

\SI & ${19.1}_{-2.7}^{+3.3}$  & ${10.2}_{-0.3}^{+0.3}$  \\

\SII & ${19.1}_{-2.7}^{+3.3}$  & ${13.6}_{-0.2}^{+0.2}$  \\

\SiI & ${19.1}_{-2.7}^{+3.2}$  & ${10.2}_{-0.3}^{+0.3}$  \\

\SiII & ${19.1}_{-2.7}^{+3.2}$  & ${13.7}_{-0.1}^{+0.2}$  \\

\SiIII & ${19.1}_{-2.7}^{+3.2}$  & ${14.5}_{-0.2}^{+0.2}$  \\

\SiIV & ${19.1}_{-2.7}^{+3.2}$  & ${13.5}_{-0.1}^{+0.1}$  \\

\hline
\end{tabular}
}
\parbox{.225\textwidth}{
\begin{tabular}{ccc}
\hline
 Ion & $b$  & $\log \frac{N}{\cmsq}$  \\ 
  & (\kms) &  \\
(1) & (2) & (3) \\\hline
SD & \textcolor{red}{\hi_0} & ${978.4}_{-8.0}^{+3.7}$ \\
\hline

\AlII & ${3.9}_{-2.2}^{+2.8}$  & ${10.2}_{-1.5}^{+1.0}$  \\

\CI & ${4.9}_{-3.0}^{+2.5}$  & ${10.8}_{-1.7}^{+1.8}$  \\

\CII & ${4.9}_{-3.0}^{+2.5}$  & ${12.0}_{-1.5}^{+1.0}$  \\

\CIV & ${4.9}_{-3.0}^{+2.5}$  & ${7.4}_{-2.1}^{+3.3}$  \\

\FeII & ${3.5}_{-2.2}^{+2.9}$  & ${10.9}_{-1.5}^{+1.2}$  \\

\HI & ${13.1}_{-8.0}^{+5.3}$  & ${16.8}_{-1.5}^{+0.9}$  \\

\MgII & ${4.0}_{-2.3}^{+2.8}$  & ${11.0}_{-1.5}^{+1.1}$  \\

\NI & ${4.7}_{-2.8}^{+2.5}$  & ${10.8}_{-1.7}^{+1.7}$  \\

\NV & ${4.7}_{-2.8}^{+2.5}$  & ${6.1}_{-2.4}^{+2.9}$  \\

\OI & ${4.5}_{-2.6}^{+2.6}$  & ${11.8}_{-1.7}^{+1.5}$  \\

\OVI & ${4.5}_{-2.6}^{+2.6}$  & ${4.4}_{-2.0}^{+3.9}$  \\

\PII & ${3.8}_{-2.2}^{+2.8}$  & ${9.2}_{-1.6}^{+1.0}$  \\

\SI & ${3.8}_{-2.2}^{+2.8}$  & ${9.1}_{-1.9}^{+2.0}$  \\

\SII & ${3.8}_{-2.2}^{+2.8}$  & ${10.9}_{-1.5}^{+1.0}$  \\

\SiI & ${3.8}_{-2.2}^{+2.8}$  & ${8.7}_{-1.6}^{+1.9}$  \\

\SiII & ${3.8}_{-2.2}^{+2.8}$  & ${11.2}_{-1.5}^{+1.0}$  \\

\SiIII & ${3.8}_{-2.2}^{+2.8}$  & ${9.3}_{-3.5}^{+2.7}$  \\

\SiIV & ${3.8}_{-2.2}^{+2.8}$  & ${6.8}_{-2.2}^{+3.9}$  \\

\hline
\end{tabular}
}
\parbox{.225\textwidth}{
\begin{tabular}{ccc}
\hline
 Ion & $b$  & $\log \frac{N}{\cmsq}$  \\ 
  & (\kms) &  \\
(1) & (2) & (3) \\\hline
SE$^{a}$ & \textcolor{blue}{\hi_0} & ${700.7}_{-1.0}^{+1.6}$ \\
\hline

\AlII & ${15.2}_{-3.4}^{+4.3}$  & ${10.0}_{-0.4}^{+0.4}$  \\

\CI & ${16.0}_{-3.1}^{+4.0}$  & ${10.0}_{-0.4}^{+0.2}$  \\

\CII & ${16.0}_{-3.1}^{+4.0}$  & ${12.5}_{-0.4}^{+0.3}$  \\

\CIV & ${16.0}_{-3.1}^{+4.0}$  & ${11.6}_{-0.7}^{+0.4}$  \\

\FeII & ${14.9}_{-3.6}^{+4.4}$  & ${9.0}_{-0.2}^{+0.7}$  \\

\HI & ${27.5}_{-2.6}^{+1.9}$  & ${14.1}_{-0.1}^{+0.1}$  \\

\MgII & ${15.3}_{-3.4}^{+4.3}$  & ${9.6}_{-0.5}^{+1.0}$  \\

\NI & ${15.8}_{-3.2}^{+4.1}$  & ${9.4}_{-0.3}^{+0.2}$  \\

\NV & ${15.8}_{-3.2}^{+4.1}$  & ${10.1}_{-0.8}^{+0.5}$  \\

\OI & ${15.6}_{-3.2}^{+4.1}$  & ${9.6}_{-0.2}^{+0.8}$  \\

\OVI & ${15.6}_{-3.2}^{+4.1}$  & ${9.8}_{-1.2}^{+0.5}$  \\

\PII & ${15.1}_{-3.5}^{+4.3}$  & ${9.6}_{-0.4}^{+0.2}$  \\

\SI & ${15.1}_{-3.5}^{+4.3}$  & ${7.0}_{-0.3}^{+0.4}$  \\

\SII & ${15.1}_{-3.5}^{+4.3}$  & ${11.0}_{-0.4}^{+0.2}$  \\

\SiI & ${15.2}_{-3.5}^{+4.3}$  & ${7.6}_{-0.3}^{+0.1}$  \\

\SiII & ${15.2}_{-3.5}^{+4.3}$  & ${10.8}_{-0.4}^{+0.2}$  \\

\SiIII & ${15.2}_{-3.5}^{+4.3}$  & ${12.1}_{-0.4}^{+0.2}$  \\

\SiIV & ${15.2}_{-3.5}^{+4.3}$  & ${11.2}_{-0.5}^{+0.3}$  \\

\hline
\end{tabular}
}
\parbox{.225\textwidth}{
\begin{tabular}{ccc}
\hline
 Ion & $b$  & $\log \frac{N}{\cmsq}$  \\ 
  & (\kms) &  \\
(1) & (2) & (3) \\\hline
SE$^{a}$ & \textcolor{red}{\civ_0} & ${704.6}_{-3.3}^{+2.8}$ \\
\hline

\AlII & ${27.4}_{-2.5}^{+2.5}$  & ${8.9}_{-0.3}^{+0.3}$  \\

\CI & ${27.8}_{-2.4}^{+2.5}$  & ${9.3}_{-0.3}^{+0.3}$  \\

\CII & ${27.8}_{-2.4}^{+2.5}$  & ${12.7}_{-0.2}^{+0.2}$  \\

\CIV & ${27.8}_{-2.4}^{+2.5}$  & ${14.0}_{-0.0}^{+0.0}$  \\

\FeII & ${27.2}_{-2.5}^{+2.6}$  & ${7.4}_{-0.3}^{+0.3}$  \\

\HI & ${36.2}_{-2.2}^{+2.5}$  & ${13.3}_{-0.2}^{+0.2}$  \\

\MgII & ${27.4}_{-2.5}^{+2.5}$  & ${8.5}_{-0.4}^{+0.4}$  \\

\NI & ${27.7}_{-2.5}^{+2.5}$  & ${8.5}_{-0.4}^{+0.3}$  \\

\NV & ${27.7}_{-2.5}^{+2.5}$  & ${13.4}_{-0.1}^{+0.1}$  \\

\OI & ${27.6}_{-2.5}^{+2.5}$  & ${8.4}_{-0.4}^{+0.3}$  \\

\OVI & ${27.6}_{-2.5}^{+2.5}$  & ${13.8}_{-0.1}^{+0.1}$  \\

\PII & ${27.3}_{-2.5}^{+2.5}$  & ${9.6}_{-0.3}^{+0.2}$  \\

\SI & ${27.3}_{-2.5}^{+2.5}$  & ${6.1}_{-0.4}^{+0.3}$  \\

\SII & ${27.3}_{-2.5}^{+2.5}$  & ${11.0}_{-0.3}^{+0.3}$  \\

\SiI & ${27.4}_{-2.5}^{+2.5}$  & ${6.3}_{-0.3}^{+0.3}$  \\

\SiII & ${27.4}_{-2.5}^{+2.5}$  & ${10.5}_{-0.2}^{+0.2}$  \\

\SiIII & ${27.4}_{-2.5}^{+2.5}$  & ${11.9}_{-0.2}^{+0.2}$  \\

\SiIV & ${27.4}_{-2.5}^{+2.5}$  & ${11.8}_{-0.2}^{+0.2}$  \\

\hline
\end{tabular}
}
\parbox{.225\textwidth}{
\begin{tabular}{ccc}
\hline
 Ion & $b$  & $\log \frac{N}{\cmsq}$  \\ 
  & (\kms) &  \\
(1) & (2) & (3) \\\hline
SE$^{b}$ & \textcolor{blue}{\hi_0} & ${701.7}_{-0.6}^{+0.6}$ \\
\hline

\AlII & ${8.4}_{-3.4}^{+3.4}$  & ${11.0}_{-0.2}^{+0.2}$  \\

\CI & ${9.1}_{-3.0}^{+3.3}$  & ${10.4}_{-0.3}^{+0.3}$  \\

\CII & ${9.1}_{-3.0}^{+3.3}$  & ${12.9}_{-0.2}^{+0.2}$  \\

\CIV & ${9.1}_{-3.0}^{+3.3}$  & ${10.9}_{-0.5}^{+0.4}$  \\

\FeII & ${8.2}_{-3.5}^{+3.5}$  & ${11.2}_{-0.5}^{+0.3}$  \\

\HI & ${17.1}_{-1.7}^{+2.3}$  & ${15.1}_{-0.4}^{+0.4}$  \\

\MgII & ${8.5}_{-3.3}^{+3.4}$  & ${11.5}_{-0.3}^{+0.2}$  \\

\NI & ${8.9}_{-3.1}^{+3.3}$  & ${10.6}_{-0.4}^{+0.3}$  \\

\NV & ${8.9}_{-3.1}^{+3.3}$  & ${8.6}_{-0.7}^{+0.7}$  \\

\OI & ${8.8}_{-3.1}^{+3.4}$  & ${11.7}_{-0.4}^{+0.3}$  \\

\OVI & ${8.8}_{-3.1}^{+3.4}$  & ${5.8}_{-1.2}^{+2.2}$  \\

\PII & ${8.4}_{-3.4}^{+3.5}$  & ${10.2}_{-0.1}^{+0.1}$  \\

\SI & ${8.4}_{-3.4}^{+3.5}$  & ${8.6}_{-0.4}^{+0.3}$  \\

\SII & ${8.4}_{-3.4}^{+3.5}$  & ${11.7}_{-0.2}^{+0.2}$  \\

\SiI & ${8.4}_{-3.4}^{+3.4}$  & ${8.5}_{-0.4}^{+0.3}$  \\

\SiII & ${8.4}_{-3.4}^{+3.4}$  & ${11.9}_{-0.3}^{+0.2}$  \\

\SiIII & ${8.4}_{-3.4}^{+3.4}$  & ${12.2}_{-0.2}^{+0.1}$  \\

\SiIV & ${8.4}_{-3.4}^{+3.4}$  & ${11.0}_{-0.4}^{+0.3}$  \\

\hline
\end{tabular}
}
\parbox{.225\textwidth}{
\begin{tabular}{ccc}
\hline
 Ion & $b$  & $\log \frac{N}{\cmsq}$  \\ 
  & (\kms) &  \\
(1) & (2) & (3) \\\hline
SE$^{b}$ & \textcolor{red}{\civ_0} & ${706.6}_{-2.8}^{+2.2}$ \\
\hline

\AlII & ${27.2}_{-2.4}^{+3.5}$  & ${7.6}_{-0.1}^{+0.0}$  \\

\CI & ${29.2}_{-2.3}^{+3.3}$  & ${7.3}_{-0.2}^{+0.2}$  \\

\CII & ${29.2}_{-2.3}^{+3.3}$  & ${11.1}_{-0.1}^{+0.1}$  \\

\CIV & ${29.2}_{-2.3}^{+3.3}$  & ${13.9}_{-0.0}^{+0.0}$  \\

\FeII & ${26.3}_{-2.5}^{+3.6}$  & ${7.9}_{-0.1}^{+0.1}$  \\

\HI & ${55.7}_{-1.3}^{+1.8}$  & ${12.6}_{-0.3}^{+0.2}$  \\

\MgII & ${27.3}_{-2.4}^{+3.5}$  & ${9.3}_{-0.2}^{+0.0}$  \\

\NI & ${28.7}_{-2.3}^{+3.3}$  & ${7.7}_{-0.2}^{+0.2}$  \\

\NV & ${28.7}_{-2.3}^{+3.3}$  & ${13.5}_{-0.1}^{+0.1}$  \\

\OI & ${28.3}_{-2.3}^{+3.4}$  & ${8.5}_{-0.2}^{+0.2}$  \\

\OVI & ${28.3}_{-2.3}^{+3.4}$  & ${12.0}_{-0.2}^{+0.5}$  \\

\PII & ${27.0}_{-2.5}^{+3.5}$  & ${7.2}_{-0.1}^{+0.1}$  \\

\SI & ${26.9}_{-2.5}^{+3.5}$  & ${4.2}_{-0.2}^{+0.2}$  \\

\SII & ${26.9}_{-2.5}^{+3.5}$  & ${9.7}_{-0.1}^{+0.1}$  \\

\SiI & ${27.1}_{-2.4}^{+3.5}$  & ${1.2}_{-1.5}^{+0.6}$  \\

\SiII & ${27.1}_{-2.4}^{+3.5}$  & ${8.7}_{-0.1}^{+0.1}$  \\

\SiIII & ${27.1}_{-2.4}^{+3.5}$  & ${11.3}_{-0.1}^{+0.1}$  \\

\SiIV & ${27.1}_{-2.4}^{+3.5}$  & ${12.2}_{-0.1}^{+0.0}$  \\

\hline
\end{tabular}
}
\parbox{.225\textwidth}{
\begin{tabular}{ccc}
\hline
 Ion & $b$  & $\log \frac{N}{\cmsq}$  \\ 
  & (\kms) &  \\
(1) & (2) & (3) \\\hline
SF & \textcolor{blue}{\siii_1} & ${909.9}_{-1.3}^{+1.3}$ \\
\hline

\AlII & ${12.0}_{-0.9}^{+0.8}$  & ${14.2}_{-0.2}^{+0.2}$  \\

\CI & ${12.1}_{-0.8}^{+0.8}$  & ${13.8}_{-0.1}^{+0.1}$  \\

\CII & ${12.1}_{-0.8}^{+0.8}$  & ${16.1}_{-0.1}^{+0.1}$  \\

\CIV & ${12.1}_{-0.8}^{+0.8}$  & ${13.4}_{-0.4}^{+0.4}$  \\

\FeII & ${12.0}_{-0.9}^{+0.8}$  & ${14.9}_{-0.2}^{+0.2}$  \\

\HI & ${13.9}_{-0.7}^{+0.9}$  & ${17.6}_{-0.1}^{+0.2}$  \\

\MgII & ${12.0}_{-0.9}^{+0.8}$  & ${15.1}_{-0.2}^{+0.1}$  \\

\NI & ${12.1}_{-0.9}^{+0.8}$  & ${14.2}_{-0.1}^{+0.1}$  \\

\NV & ${12.1}_{-0.9}^{+0.8}$  & ${11.7}_{-0.1}^{+0.0}$  \\

\OI & ${12.1}_{-0.9}^{+0.8}$  & ${15.2}_{-0.1}^{+0.2}$  \\

\OVI & ${12.1}_{-0.9}^{+0.8}$  & ${8.4}_{-0.2}^{+0.0}$  \\

\PII & ${12.0}_{-0.9}^{+0.8}$  & ${13.4}_{-0.2}^{+0.1}$  \\

\SI & ${12.0}_{-0.9}^{+0.8}$  & ${11.9}_{-0.1}^{+0.1}$  \\

\SII & ${12.0}_{-0.9}^{+0.8}$  & ${14.9}_{-0.1}^{+0.1}$  \\

\SiI & ${12.0}_{-0.9}^{+0.8}$  & ${12.2}_{-0.1}^{+0.1}$  \\

\SiII & ${12.0}_{-0.9}^{+0.8}$  & ${15.4}_{-0.1}^{+0.1}$  \\

\SiIII & ${12.0}_{-0.9}^{+0.8}$  & ${13.9}_{-0.3}^{+0.4}$  \\

\SiIV & ${12.0}_{-0.9}^{+0.8}$  & ${13.7}_{-0.2}^{+0.4}$  \\

\hline
\end{tabular}
}

\end{table*}

\begin{table*}
\parbox{.225\textwidth}{
\begin{tabular}{ccc}
\hline
 Ion & $b$  & $\log \frac{N}{\cmsq}$  \\ 
  & (\kms) &  \\
(1) & (2) & (3) \\\hline
SG & \textcolor{blue}{\civ_0} & ${817.2}_{-2.5}^{+1.4}$ \\
\hline

\AlII & ${18.1}_{-4.7}^{+4.9}$  & ${10.4}_{-0.2}^{+0.2}$  \\

\CI & ${18.4}_{-4.6}^{+4.9}$  & ${9.3}_{-0.2}^{+0.2}$  \\

\CII & ${18.4}_{-4.6}^{+4.9}$  & ${13.0}_{-0.1}^{+0.1}$  \\

\CIV & ${18.4}_{-4.6}^{+4.9}$  & ${14.0}_{-0.2}^{+0.2}$  \\

\FeII & ${18.0}_{-4.8}^{+5.0}$  & ${9.3}_{-0.4}^{+0.4}$  \\

\HI & ${24.0}_{-3.3}^{+4.2}$  & ${13.6}_{-0.1}^{+0.1}$  \\

\MgII & ${18.1}_{-4.7}^{+4.9}$  & ${10.7}_{-0.4}^{+0.4}$  \\

\NI & ${18.4}_{-4.6}^{+4.9}$  & ${8.8}_{-0.3}^{+0.3}$  \\

\NV & ${18.4}_{-4.6}^{+4.9}$  & ${12.9}_{-0.3}^{+0.3}$  \\

\OI & ${18.3}_{-4.6}^{+4.9}$  & ${9.6}_{-0.4}^{+0.4}$  \\

\OVI & ${18.3}_{-4.6}^{+4.9}$  & ${12.9}_{-0.4}^{+0.4}$  \\

\PII & ${18.1}_{-4.7}^{+5.0}$  & ${10.3}_{-0.2}^{+0.2}$  \\

\SI & ${18.1}_{-4.7}^{+5.0}$  & ${7.1}_{-0.3}^{+0.3}$  \\

\SII & ${18.1}_{-4.7}^{+5.0}$  & ${11.5}_{-0.2}^{+0.2}$  \\

\SiI & ${18.1}_{-4.7}^{+4.9}$  & ${7.0}_{-0.2}^{+0.2}$  \\

\SiII & ${18.1}_{-4.7}^{+4.9}$  & ${11.6}_{-0.2}^{+0.2}$  \\

\SiIII & ${18.1}_{-4.7}^{+4.9}$  & ${12.8}_{-0.2}^{+0.1}$  \\

\SiIV & ${18.1}_{-4.7}^{+4.9}$  & ${13.0}_{-0.2}^{+0.1}$  \\
\hline
\end{tabular}
}
\parbox{.225\textwidth}{
\begin{tabular}{ccc}
\hline
 Ion & $b$  & $\log \frac{N}{\cmsq}$  \\ 
  & (\kms) &  \\
(1) & (2) & (3) \\\hline
SG & \textcolor{red}{\siiv_0} & ${861.3}_{-3.8}^{+5.5}$ \\
\hline

\AlII & ${10.4}_{-2.1}^{+0.9}$  & ${11.0}_{-0.3}^{+0.4}$  \\

\CI & ${10.8}_{-2.0}^{+0.9}$  & ${10.0}_{-0.4}^{+0.4}$  \\

\CII & ${10.8}_{-2.0}^{+0.9}$  & ${13.3}_{-0.3}^{+0.2}$  \\

\CIV & ${10.8}_{-2.0}^{+0.9}$  & ${13.2}_{-0.3}^{+0.4}$  \\

\FeII & ${10.2}_{-2.1}^{+0.9}$  & ${10.7}_{-0.7}^{+0.5}$  \\

\HI & ${16.8}_{-2.1}^{+2.9}$  & ${14.1}_{-0.4}^{+0.6}$  \\

\MgII & ${10.4}_{-2.1}^{+0.9}$  & ${11.7}_{-0.5}^{+0.3}$  \\

\NI & ${10.7}_{-2.0}^{+0.9}$  & ${9.9}_{-0.5}^{+0.5}$  \\

\NV & ${10.7}_{-2.0}^{+0.9}$  & ${11.7}_{-0.5}^{+0.5}$  \\

\OI & ${10.6}_{-2.0}^{+0.9}$  & ${10.9}_{-0.6}^{+0.5}$  \\

\OVI & ${10.6}_{-2.0}^{+0.9}$  & ${11.3}_{-0.7}^{+0.6}$  \\

\PII & ${10.4}_{-2.1}^{+0.9}$  & ${10.7}_{-0.3}^{+0.3}$  \\

\SI & ${10.4}_{-2.1}^{+0.9}$  & ${8.1}_{-0.5}^{+0.4}$  \\

\SII & ${10.4}_{-2.1}^{+0.9}$  & ${12.0}_{-0.3}^{+0.3}$  \\

\SiI & ${10.4}_{-2.1}^{+0.9}$  & ${8.0}_{-0.5}^{+0.4}$  \\

\SiII & ${10.4}_{-2.1}^{+0.9}$  & ${12.3}_{-0.4}^{+0.3}$  \\

\SiIII & ${10.4}_{-2.1}^{+0.9}$  & ${13.1}_{-0.2}^{+0.2}$  \\

\SiIV & ${10.4}_{-2.1}^{+0.9}$  & ${12.9}_{-0.3}^{+0.2}$  \\
\hline
\end{tabular}
}
\parbox{.225\textwidth}{
\begin{tabular}{ccc}
\hline
 Ion & $b$  & $\log \frac{N}{\cmsq}$  \\ 
  & (\kms) &  \\
(1) & (2) & (3) \\\hline
SG & \textcolor{green}{\hi_1} & ${942.8}_{-4.5}^{+5.6}$ \\
\hline

\AlII & ${9.6}_{-6.7}^{+10.7}$  & ${9.0}_{-1.7}^{+1.8}$  \\

\CI & ${10.1}_{-5.9}^{+10.5}$  & ${9.6}_{-1.6}^{+1.5}$  \\

\CII & ${10.1}_{-5.9}^{+10.5}$  & ${10.8}_{-1.8}^{+1.9}$  \\

\CIV & ${10.1}_{-5.9}^{+10.5}$  & ${6.8}_{-2.8}^{+4.1}$  \\

\FeII & ${9.4}_{-7.3}^{+10.8}$  & ${9.7}_{-1.5}^{+1.5}$  \\

\HI & ${16.5}_{-3.8}^{+7.8}$  & ${15.4}_{-1.0}^{+1.0}$  \\

\MgII & ${9.7}_{-6.6}^{+10.7}$  & ${9.9}_{-1.6}^{+1.5}$  \\

\NI & ${10.0}_{-6.0}^{+10.5}$  & ${9.6}_{-1.6}^{+1.5}$  \\

\NV & ${10.0}_{-6.0}^{+10.5}$  & ${5.7}_{-3.4}^{+3.2}$  \\

\OI & ${9.9}_{-6.2}^{+10.6}$  & ${10.6}_{-1.6}^{+1.5}$  \\

\OVI & ${9.9}_{-6.2}^{+10.6}$  & ${4.2}_{-2.9}^{+4.0}$  \\

\PII & ${9.6}_{-6.8}^{+10.7}$  & ${8.1}_{-1.8}^{+1.9}$  \\

\SI & ${9.6}_{-6.8}^{+10.7}$  & ${7.8}_{-1.7}^{+1.5}$  \\

\SII & ${9.6}_{-6.8}^{+10.7}$  & ${9.7}_{-1.7}^{+1.7}$  \\

\SiI & ${9.6}_{-6.7}^{+10.7}$  & ${7.5}_{-1.6}^{+1.5}$  \\

\SiII & ${9.6}_{-6.7}^{+10.7}$  & ${10.0}_{-1.7}^{+1.7}$  \\

\SiIII & ${9.6}_{-6.7}^{+10.7}$  & ${8.8}_{-2.6}^{+3.1}$  \\

\SiIV & ${9.6}_{-6.7}^{+10.7}$  & ${6.6}_{-3.2}^{+4.3}$  \\

\hline
\end{tabular}
}
\parbox{.225\textwidth}{
\begin{tabular}{ccc}
\hline
 Ion & $b$  & $\log \frac{N}{\cmsq}$  \\ 
  & (\kms) &  \\
(1) & (2) & (3) \\\hline
SG & \textcolor{magenta}{\hi_2} & ${1056.7}_{-4.3}^{+4.2}$ \\
\hline

\AlII & ${11.3}_{-8.4}^{+11.1}$  & ${7.6}_{-1.5}^{+1.9}$  \\

\CI & ${11.7}_{-7.6}^{+10.9}$  & ${8.5}_{-1.5}^{+1.5}$  \\

\CII & ${11.7}_{-7.6}^{+10.9}$  & ${9.5}_{-1.7}^{+2.0}$  \\

\CIV & ${11.7}_{-7.6}^{+10.9}$  & ${5.5}_{-2.5}^{+4.1}$  \\

\FeII & ${11.2}_{-9.1}^{+11.2}$  & ${8.5}_{-1.4}^{+1.8}$  \\

\HI & ${16.9}_{-4.4}^{+9.1}$  & ${14.2}_{-0.4}^{+0.6}$  \\

\MgII & ${11.4}_{-8.4}^{+11.1}$  & ${8.6}_{-1.5}^{+1.8}$  \\

\NI & ${11.6}_{-7.8}^{+11.0}$  & ${8.5}_{-1.5}^{+1.6}$  \\

\NV & ${11.6}_{-7.8}^{+11.0}$  & ${4.4}_{-3.1}^{+3.6}$  \\

\OI & ${11.5}_{-7.9}^{+11.0}$  & ${9.4}_{-1.4}^{+1.6}$  \\

\OVI & ${11.5}_{-7.9}^{+11.0}$  & ${2.8}_{-2.6}^{+4.5}$  \\

\PII & ${11.3}_{-8.6}^{+11.1}$  & ${6.7}_{-1.5}^{+2.0}$  \\

\SI & ${11.3}_{-8.6}^{+11.1}$  & ${6.8}_{-1.6}^{+1.5}$  \\

\SII & ${11.3}_{-8.6}^{+11.1}$  & ${8.4}_{-1.5}^{+1.8}$  \\

\SiI & ${11.3}_{-8.5}^{+11.1}$  & ${6.3}_{-1.5}^{+1.6}$  \\

\SiII & ${11.3}_{-8.5}^{+11.1}$  & ${8.6}_{-1.5}^{+1.8}$  \\

\SiIII & ${11.3}_{-8.5}^{+11.1}$  & ${7.0}_{-2.7}^{+3.5}$  \\

\SiIV & ${11.3}_{-8.5}^{+11.1}$  & ${4.8}_{-3.0}^{+4.4}$  \\

\hline
\end{tabular}
}
\parbox{.225\textwidth}{
\begin{tabular}{ccc}
\hline
 Ion & $b$  & $\log \frac{N}{\cmsq}$  \\ 
  & (\kms) &  \\
(1) & (2) & (3) \\\hline
SG & \textcolor{brown}{\hi_3} & ${1297.2}_{-26.3}^{+27.1}$ \\
\hline

\AlII & ${6.2}_{-4.0}^{+4.8}$  & ${7.5}_{-3.0}^{+2.9}$  \\

\CI & ${7.0}_{-4.0}^{+6.2}$  & ${8.2}_{-2.9}^{+2.7}$  \\

\CII & ${7.0}_{-4.0}^{+6.2}$  & ${9.7}_{-3.4}^{+3.1}$  \\

\CIV & ${7.0}_{-4.0}^{+6.2}$  & ${6.9}_{-4.3}^{+6.3}$  \\

\FeII & ${5.8}_{-4.1}^{+4.9}$  & ${8.1}_{-2.8}^{+2.9}$  \\

\HI & ${14.8}_{-7.8}^{+25.5}$  & ${13.6}_{-2.4}^{+3.0}$  \\

\MgII & ${6.3}_{-4.0}^{+4.8}$  & ${8.3}_{-2.9}^{+3.0}$  \\

\NI & ${6.8}_{-4.0}^{+5.8}$  & ${8.1}_{-2.7}^{+2.8}$  \\

\NV & ${6.8}_{-4.0}^{+5.8}$  & ${6.4}_{-5.4}^{+6.3}$  \\

\OI & ${6.7}_{-3.9}^{+5.5}$  & ${9.0}_{-2.7}^{+3.0}$  \\

\OVI & ${6.7}_{-3.9}^{+5.5}$  & ${5.9}_{-6.6}^{+6.7}$  \\

\PII & ${6.1}_{-4.0}^{+4.8}$  & ${6.7}_{-3.1}^{+3.0}$  \\

\SI & ${6.1}_{-4.0}^{+4.8}$  & ${6.2}_{-3.1}^{+3.0}$  \\

\SII & ${6.1}_{-4.0}^{+4.8}$  & ${8.4}_{-3.1}^{+3.0}$  \\

\SiI & ${6.2}_{-4.0}^{+4.8}$  & ${6.0}_{-3.1}^{+2.8}$  \\

\SiII & ${6.2}_{-4.0}^{+4.8}$  & ${8.6}_{-3.0}^{+2.8}$  \\

\SiIII & ${6.2}_{-4.0}^{+4.8}$  & ${7.7}_{-5.0}^{+4.6}$  \\

\SiIV & ${6.2}_{-4.0}^{+4.8}$  & ${5.8}_{-5.2}^{+6.1}$  \\

\hline
\end{tabular}
}
\parbox{.225\textwidth}{
\begin{tabular}{ccc}
\hline
 Ion & $b$  & $\log \frac{N}{\cmsq}$  \\ 
  & (\kms) &  \\
(1) & (2) & (3) \\\hline
SH & \textcolor{blue}{\siiii_0} & ${766.8}_{-0.8}^{+0.3}$ \\
\hline

\AlII & ${17.6}_{-0.6}^{+0.2}$  & ${8.8}_{-0.1}^{+0.1}$  \\

\CI & ${19.2}_{-0.6}^{+0.2}$  & ${10.1}_{-0.2}^{+0.1}$  \\

\CII & ${19.2}_{-0.6}^{+0.2}$  & ${13.2}_{-0.1}^{+0.1}$  \\

\CIV & ${19.2}_{-0.6}^{+0.2}$  & ${13.5}_{-0.1}^{+0.1}$  \\

\FeII & ${16.9}_{-0.7}^{+0.2}$  & ${9.1}_{-0.1}^{+0.1}$  \\

\HI & ${38.8}_{-0.5}^{+0.4}$  & ${15.4}_{-0.1}^{+0.1}$  \\

\MgII & ${17.8}_{-0.6}^{+0.2}$  & ${9.0}_{-0.1}^{+0.1}$  \\

\NI & ${18.8}_{-0.6}^{+0.2}$  & ${9.7}_{-0.1}^{+0.1}$  \\

\NV & ${18.8}_{-0.6}^{+0.2}$  & ${11.6}_{-0.1}^{+0.1}$  \\

\OI & ${18.5}_{-0.6}^{+0.2}$  & ${9.8}_{-0.1}^{+0.1}$  \\

\OVI & ${18.5}_{-0.6}^{+0.2}$  & ${11.0}_{-0.1}^{+0.1}$  \\

\PII & ${17.5}_{-0.6}^{+0.2}$  & ${9.4}_{-0.1}^{+0.1}$  \\

\SI & ${17.4}_{-0.6}^{+0.2}$  & ${6.5}_{-0.2}^{+0.1}$  \\

\SII & ${17.4}_{-0.6}^{+0.2}$  & ${11.3}_{-0.1}^{+0.1}$  \\

\SiI & ${17.6}_{-0.6}^{+0.2}$  & ${7.6}_{-0.2}^{+0.1}$  \\

\SiII & ${17.6}_{-0.6}^{+0.2}$  & ${11.0}_{-0.1}^{+0.1}$  \\

\SiIII & ${17.6}_{-0.6}^{+0.2}$  & ${12.9}_{-0.1}^{+0.1}$  \\

\SiIV & ${17.6}_{-0.6}^{+0.2}$  & ${12.8}_{-0.0}^{+0.0}$  \\

\hline
\end{tabular}
}
\parbox{.225\textwidth}{
\begin{tabular}{ccc}
\hline
 Ion & $b$  & $\log \frac{N}{\cmsq}$  \\ 
  & (\kms) &  \\
(1) & (2) & (3) \\\hline
SH & \textcolor{red}{\siiii_1} & ${818.0}_{-0.9}^{+1.3}$ \\
\hline

\AlII & ${3.7}_{-0.5}^{+0.5}$  & ${12.2}_{-0.3}^{+0.4}$  \\

\CI & ${5.3}_{-0.6}^{+0.5}$  & ${11.1}_{-0.2}^{+0.2}$  \\

\CII & ${5.3}_{-0.6}^{+0.5}$  & ${14.8}_{-0.2}^{+0.2}$  \\

\CIV & ${5.3}_{-0.6}^{+0.5}$  & ${16.1}_{-0.3}^{+0.2}$  \\

\FeII & ${2.7}_{-0.4}^{+0.6}$  & ${10.7}_{-0.5}^{+0.6}$  \\

\HI & ${17.9}_{-1.8}^{+1.7}$  & ${15.4}_{-0.2}^{+0.2}$  \\

\MgII & ${3.9}_{-0.5}^{+0.5}$  & ${12.2}_{-0.5}^{+0.6}$  \\

\NI & ${5.0}_{-0.6}^{+0.5}$  & ${10.4}_{-0.2}^{+0.4}$  \\

\NV & ${5.0}_{-0.6}^{+0.5}$  & ${15.3}_{-0.4}^{+0.3}$  \\

\OI & ${4.7}_{-0.5}^{+0.5}$  & ${11.0}_{-0.4}^{+0.6}$  \\

\OVI & ${4.7}_{-0.5}^{+0.5}$  & ${15.3}_{-0.5}^{+0.4}$  \\

\PII & ${3.5}_{-0.5}^{+0.5}$  & ${12.0}_{-0.2}^{+0.3}$  \\

\SI & ${3.4}_{-0.5}^{+0.5}$  & ${8.6}_{-0.4}^{+0.5}$  \\

\SII & ${3.4}_{-0.5}^{+0.5}$  & ${13.2}_{-0.2}^{+0.3}$  \\

\SiI & ${3.6}_{-0.5}^{+0.5}$  & ${8.8}_{-0.2}^{+0.3}$  \\

\SiII & ${3.6}_{-0.5}^{+0.5}$  & ${13.3}_{-0.3}^{+0.3}$  \\

\SiIII & ${3.6}_{-0.5}^{+0.5}$  & ${14.7}_{-0.3}^{+0.3}$  \\

\SiIV & ${3.6}_{-0.5}^{+0.5}$  & ${14.8}_{-0.3}^{+0.3}$  \\

\hline
\end{tabular}
}
\parbox{.225\textwidth}{
\begin{tabular}{ccc}
\hline
 Ion & $b$  & $\log \frac{N}{\cmsq}$  \\ 
  & (\kms) &  \\
(1) & (2) & (3) \\\hline
SH & \textcolor{green}{\hi_2} & ${899.5}_{-2.9}^{+2.7}$ \\
\hline

\AlII & ${5.3}_{-2.7}^{+6.1}$  & ${7.7}_{-1.4}^{+2.0}$  \\

\CI & ${5.9}_{-2.2}^{+5.7}$  & ${8.7}_{-1.2}^{+1.3}$  \\

\CII & ${5.9}_{-2.2}^{+5.7}$  & ${9.5}_{-1.5}^{+2.1}$  \\

\CIV & ${5.9}_{-2.2}^{+5.7}$  & ${5.1}_{-1.7}^{+3.8}$  \\

\FeII & ${4.9}_{-3.1}^{+6.3}$  & ${8.6}_{-1.3}^{+1.7}$  \\

\HI & ${13.5}_{-1.6}^{+3.4}$  & ${14.5}_{-0.4}^{+0.4}$  \\

\MgII & ${5.3}_{-2.6}^{+6.1}$  & ${8.7}_{-1.3}^{+1.8}$  \\

\NI & ${5.8}_{-2.3}^{+5.8}$  & ${8.6}_{-1.2}^{+1.4}$  \\

\NV & ${5.8}_{-2.3}^{+5.8}$  & ${4.0}_{-2.6}^{+5.1}$  \\

\OI & ${5.6}_{-2.4}^{+5.9}$  & ${9.6}_{-1.2}^{+1.5}$  \\

\OVI & ${5.6}_{-2.4}^{+5.9}$  & ${3.0}_{-2.1}^{+4.9}$  \\

\PII & ${5.2}_{-2.8}^{+6.2}$  & ${6.7}_{-1.4}^{+2.0}$  \\

\SI & ${5.2}_{-2.8}^{+6.2}$  & ${7.0}_{-1.2}^{+1.3}$  \\

\SII & ${5.2}_{-2.8}^{+6.2}$  & ${8.4}_{-1.4}^{+2.0}$  \\

\SiI & ${5.2}_{-2.7}^{+6.1}$  & ${6.4}_{-1.2}^{+1.5}$  \\

\SiII & ${5.2}_{-2.7}^{+6.1}$  & ${8.7}_{-1.4}^{+2.0}$  \\

\SiIII & ${5.2}_{-2.7}^{+6.1}$  & ${6.8}_{-2.0}^{+3.2}$  \\

\SiIV & ${5.2}_{-2.7}^{+6.1}$  & ${4.3}_{-2.3}^{+4.1}$  \\

\hline
\end{tabular}
}
\end{table*}


\begin{table*}

\parbox{.225\textwidth}{
\begin{tabular}{ccc}
\hline
 Ion & $b$  & $\log \frac{N}{\cmsq}$  \\ 
  & (\kms) &  \\
(1) & (2) & (3) \\\hline
SH & \textcolor{magenta}{\hi_3} & ${976.8}_{-7.5}^{+5.1}$ \\
\hline

\AlII & ${40.7}_{-11.1}^{+14.0}$  & ${8.1}_{-0.9}^{+2.1}$  \\

\CI & ${41.4}_{-10.7}^{+13.6}$  & ${9.7}_{-1.7}^{+0.5}$  \\

\CII & ${41.4}_{-10.7}^{+13.6}$  & ${12.7}_{-2.3}^{+0.4}$  \\

\CIV & ${41.4}_{-10.7}^{+13.6}$  & ${13.3}_{-3.5}^{+0.3}$  \\

\FeII & ${40.3}_{-11.2}^{+14.1}$  & ${8.9}_{-0.9}^{+0.7}$  \\

\HI & ${53.9}_{-7.5}^{+9.7}$  & ${14.1}_{-0.0}^{+0.1}$  \\

\MgII & ${40.7}_{-11.0}^{+14.0}$  & ${8.3}_{-0.3}^{+1.9}$  \\

\NI & ${41.2}_{-10.8}^{+13.7}$  & ${9.5}_{-1.3}^{+0.4}$  \\

\NV & ${41.2}_{-10.8}^{+13.7}$  & ${11.8}_{-2.6}^{+0.7}$  \\

\OI & ${41.1}_{-10.9}^{+13.8}$  & ${9.8}_{-0.9}^{+0.4}$  \\

\OVI & ${41.1}_{-10.9}^{+13.8}$  & ${11.5}_{-2.9}^{+0.8}$  \\

\PII & ${40.6}_{-11.1}^{+14.0}$  & ${8.8}_{-1.8}^{+1.0}$  \\

\SI & ${40.6}_{-11.1}^{+14.0}$  & ${6.1}_{-1.4}^{+1.3}$  \\

\SII & ${40.6}_{-11.1}^{+14.0}$  & ${10.9}_{-1.8}^{+0.5}$  \\

\SiI & ${40.6}_{-11.1}^{+14.0}$  & ${6.8}_{-2.4}^{+0.8}$  \\

\SiII & ${40.6}_{-11.1}^{+14.0}$  & ${10.2}_{-1.5}^{+0.8}$  \\

\SiIII & ${40.6}_{-11.1}^{+14.0}$  & ${12.2}_{-3.6}^{+0.3}$  \\

\SiIV & ${40.6}_{-11.1}^{+14.0}$  & ${12.1}_{-5.1}^{+0.1}$  \\

\hline
\end{tabular}
}
\parbox{.225\textwidth}{
\begin{tabular}{ccc}
\hline
 Ion & $b$  & $\log \frac{N}{\cmsq}$  \\ 
  & (\kms) &  \\
(1) & (2) & (3) \\\hline
SJ & \textcolor{blue}{\hi_0} & ${650.6}_{-2.4}^{+2.8}$ \\
\hline

\AlII & ${4.5}_{-2.4}^{+6.1}$  & ${8.5}_{-2.6}^{+1.6}$  \\

\CI & ${5.0}_{-2.1}^{+5.6}$  & ${9.7}_{-2.5}^{+0.8}$  \\

\CII & ${5.0}_{-2.1}^{+5.6}$  & ${10.4}_{-2.7}^{+1.7}$  \\

\CIV & ${5.0}_{-2.1}^{+5.6}$  & ${6.0}_{-2.4}^{+2.7}$  \\

\FeII & ${4.2}_{-2.7}^{+6.3}$  & ${9.5}_{-2.6}^{+1.6}$  \\

\HI & ${10.9}_{-2.9}^{+3.4}$  & ${14.2}_{-0.3}^{+0.5}$  \\

\MgII & ${4.5}_{-2.4}^{+6.0}$  & ${9.6}_{-2.6}^{+1.5}$  \\

\NI & ${4.9}_{-2.2}^{+5.7}$  & ${9.7}_{-2.6}^{+0.9}$  \\

\NV & ${4.9}_{-2.2}^{+5.7}$  & ${6.7}_{-4.3}^{+1.2}$  \\

\OI & ${4.8}_{-2.2}^{+5.8}$  & ${10.6}_{-2.6}^{+1.0}$  \\

\OVI & ${4.8}_{-2.2}^{+5.8}$  & ${3.5}_{-2.2}^{+1.9}$  \\

\PII & ${4.4}_{-2.4}^{+6.2}$  & ${7.6}_{-2.6}^{+1.6}$  \\

\SI & ${4.4}_{-2.5}^{+6.2}$  & ${8.0}_{-2.5}^{+0.8}$  \\

\SII & ${4.4}_{-2.5}^{+6.2}$  & ${9.3}_{-2.6}^{+1.6}$  \\

\SiI & ${4.4}_{-2.4}^{+6.1}$  & ${7.6}_{-2.7}^{+1.1}$  \\

\SiII & ${4.4}_{-2.4}^{+6.1}$  & ${9.6}_{-2.6}^{+1.7}$  \\

\SiIII & ${4.4}_{-2.4}^{+6.1}$  & ${6.3}_{-3.7}^{+3.1}$  \\

\SiIV & ${4.4}_{-2.4}^{+6.1}$  & ${5.0}_{-3.4}^{+4.0}$  \\

\hline
\end{tabular}
}
\parbox{.225\textwidth}{
\begin{tabular}{ccc}
\hline
 Ion & $b$  & $\log \frac{N}{\cmsq}$  \\ 
  & (\kms) &  \\
(1) & (2) & (3) \\\hline
SJ & \textcolor{red}{\civ_0} & ${720.1}_{-1.5}^{+1.5}$ \\
\hline

\AlII & ${17.1}_{-2.8}^{+3.2}$  & ${10.3}_{-0.3}^{+0.3}$  \\

\CI & ${17.9}_{-2.7}^{+3.0}$  & ${10.7}_{-0.2}^{+0.1}$  \\

\CII & ${17.9}_{-2.7}^{+3.0}$  & ${13.7}_{-0.1}^{+0.1}$  \\

\CIV & ${17.9}_{-2.7}^{+3.0}$  & ${14.1}_{-0.1}^{+0.1}$  \\

\FeII & ${16.7}_{-2.9}^{+3.3}$  & ${9.1}_{-0.1}^{+0.1}$  \\

\HI & ${29.7}_{-2.1}^{+2.2}$  & ${14.4}_{-0.1}^{+0.1}$  \\

\MgII & ${17.1}_{-2.8}^{+3.2}$  & ${9.9}_{-0.4}^{+0.4}$  \\

\NI & ${17.7}_{-2.7}^{+3.0}$  & ${9.9}_{-0.2}^{+0.2}$  \\

\NV & ${17.7}_{-2.7}^{+3.0}$  & ${13.1}_{-0.1}^{+0.1}$  \\

\OI & ${17.5}_{-2.8}^{+3.1}$  & ${9.8}_{-0.1}^{+0.1}$  \\

\OVI & ${17.5}_{-2.8}^{+3.1}$  & ${13.2}_{-0.1}^{+0.1}$  \\

\PII & ${17.0}_{-2.8}^{+3.2}$  & ${10.8}_{-0.1}^{+0.1}$  \\

\SI & ${17.0}_{-2.8}^{+3.2}$  & ${7.6}_{-0.1}^{+0.1}$  \\

\SII & ${17.0}_{-2.8}^{+3.2}$  & ${12.1}_{-0.1}^{+0.1}$  \\

\SiI & ${17.0}_{-2.8}^{+3.2}$  & ${8.0}_{-0.1}^{+0.1}$  \\

\SiII & ${17.0}_{-2.8}^{+3.2}$  & ${11.8}_{-0.2}^{+0.2}$  \\

\SiIII & ${17.0}_{-2.8}^{+3.2}$  & ${13.1}_{-0.1}^{+0.1}$  \\

\SiIV & ${17.0}_{-2.8}^{+3.2}$  & ${12.7}_{-0.1}^{+0.1}$  \\

\hline
\end{tabular}
}

\end{table*}

\end{document}